%% file: gp_dla_detection.tex
\documentclass[usenatbib,a4paper,fleqn]{mnras}

\usepackage{newtxtext,newtxmath}
\usepackage[T1]{fontenc}
\usepackage{ae,aecompl}
\usepackage{amsmath}
\usepackage{textcomp}
\usepackage{url}
\usepackage{amssymb}
\usepackage{bm}
\usepackage{xspace}
\usepackage{nicefrac}
\usepackage{mathtools}
\usepackage{microtype}
\usepackage[caption=false]{subfig}
\usepackage{booktabs}

\robustify{\subref}

\usepackage{pgfplots}
\pgfplotsset{
  compat=newest,
  plot coordinates/math parser=false,
  tick label style={font=\footnotesize, /pgf/number format/fixed},
  label style={font=\small},
  legend style={font=\small, legend columns=-1},
  extra x tick style={
    tick align=inside,
    x tick label style={
      yshift=0.5ex,
      font=\footnotesize,
      anchor=south
    }
  },
  every axis/.append style={
    tick align=outside,
    clip mode=individual,
    scaled ticks=false,
    thick,
    tick style={semithick, black}
  }
}

\pgfkeys{/pgf/number format/.cd, set thousands separator={\,}}

\title{Detecting Damped Lyman-$\alpha$ Absorbers with Gaussian
  Processes}

\author[R.\ Garnett et al.]{Roman Garnett,$^1$\thanks{E-mail: garnett@wustl.edu (RG)}
  Shirley Ho,$^2$
  Simeon Bird,$^3$
  and Jeff Schneider$^4$
  \\
  $^1$Department of Computer Science and Engineering, Washington University in St.\ Louis, One Brookings Drive, St.\ Louis, MO 63130, USA\\
  $^2$Department of Physics, Carnegie Mellon University, 5000 Forbes Avenue, Pittsburgh, PA 15213, USA\\
  $^3$Department of Physics \& Astronomy, Johns Hopkins University, 3400 N.\ Charles Street, Baltimore, MD 21218, USA\\
  $^4$School of Computer Science, Carnegie Mellon University, 5000 Forbes Avenue, Pittsburgh, PA 15213, USA
}

\date{\today}
\pubyear{2017}

\usepgfplotslibrary{external}
\newlength\figurewidth
\newlength\figureheight

\newlength\sbsfigurewidth
\newlength\sbsfigureheight

\newlength\widefigurewidth
\newlength\widefigureheight

\newlength\squarefigurewidth
\newlength\squarefigureheight

\newcommand{\acro}[1]{\textsc{\MakeLowercase{#1}}}
\newcommand{\mc}[1]{\mathcal{#1}}
\newcommand{\data}{\mc{D}}
\newcommand{\model}{\mc{M}}
\newcommand{\st}[1]{_{\text{#1}}}
\newcommand{\sd}{\st{\acro{dla}}}
\newcommand{\snd}{_{\neg\text{\acro{dla}}}}
\newcommand{\zq}{z_{\text{\acro{qso}}}}
\newcommand{\zd}{z_{\text{\acro{dla}}}}
\newcommand{\lm}{\lambda}
\newcommand{\lr}{\lm_{\text{rest}}}
\newcommand{\lo}{\lm_{\text{obs}}}
\newcommand{\nhi}{N_{\text{\acro{hi}}}}
\newcommand{\lni}{\log_{10} \nhi}
\newcommand{\given}{\mid}
\newcommand{\intd}[1]{\,\mathrm{d}{#1}}
\newcommand{\R}{\mathbb{R}}
\newcommand{\Mnd}{\model\snd}
\newcommand{\Md}{\model\sd}
\newcommand{\Mdk}[1]{\model\st{\acro{dla(}{\ensuremath{#1}}\acro{)}}}
\newcommand{\inv}{^{-1}}
\newcommand{\trans}{^\top}
\newcommand{\mat}[1]{\bm{\mathrm{#1}}}
\renewcommand{\vec}[1]{\bm{#1}}
\newcommand{\NaN}{NaN\xspace}
\newcommand{\Npix}{N_{\text{pixels}}}
\newcommand{\Nspec}{N_{\text{spec}}}
\newcommand{\Ndla}{N_{\text{\acro{dla}}}}
\newcommand{\noise}{v}
\newcommand{\Noise}{V}

\DeclareMathOperator{\diag}{diag}
\DeclareMathOperator{\cov}{cov}

\DeclareMathOperator*{\argmax}{arg\,max}

\begin{document}
\label{firstpage}
\pagerange{1--16}
\maketitle

\begin{abstract}
  We develop an automated technique for detecting damped
  Lyman-$\alpha$ absorbers (\acro{dla}s) along spectroscopic lines of
  sight to quasi-stellar objects (\acro{qso}s or quasars).  The
  detection of \acro{dla}s in large-scale spectroscopic surveys such
  as \acro{sdss-iii} sheds light on galaxy formation at high redshift,
  showing the nucleation of galaxies from diffuse gas.  We use nearly
  50\,000 \acro{qso} spectra to learn a novel tailored Gaussian
  process model for quasar emission spectra, which we apply to the
  \acro{dla} detection problem via Bayesian model selection.  We
  propose models for identifying an arbitrary number of \acro{dla}s
  along a given line of sight.  We demonstrate our method's
  effectiveness using a large-scale validation experiment, with
  excellent performance.  We also provide a catalog of our results
  applied to 162\,858 spectra from \acro{sdss--iii} data release 12.
\end{abstract}

\section{Introduction}
The damped Lyman-$\alpha$ (Ly\,$\alpha$) systems (\acro{dla}s)
\citep{wolfe86,wolfe05}) define the class of absorption-line systems
discovered in the restframe \acro{uv} spectra of distant quasars, with
H\,\acro{i} column densities $\nhi > 2 \times 10^{20}$\,cm$^{-2}$, as
measured from the analysis of damping wings in the Ly\,$\alpha$
profile.  Recent spectroscopic quasar surveys such as the Sloan
Digital Sky Survey (\acro{sdss}) \citep{york00} have produced a vast
sample of quasar spectra showing Ly\,$\alpha$ absorption at $z>2$.
\acro{sdss--iii} has measured nearly 300\,000 quasar spectra over its
brief history.  Even larger surveys, such as the Dark Energy
Spectroscopic Instrument
\acro{desi}\footnote{\url{http://desi.lbl.gov}}) survey, soon plan to
observe 1--2 million quasars. Finding \acro{dla}s in these surveys has
historically involved a combination of automated template fitting and
visual inspection, but visual inspection is clearly infeasible with
the size of upcoming datasets.  Furthermore, \acro{sdss} data trades
off low signal-to-noise ratios for statistical power, making detection
of even distinctive signals such as \acro{dla}s substantially harder,
and making noise-induced systematic error hard to control.

There have been several previous \acro{dla} searches in \acro{sdss}.
These include a visual-inspection survey
\citep{slosar_et_al_jcap_2011}, visually guided Voigt-profile fitting
\citep{prochaska2005, prochaska2009} and two automated approaches: a
template-matching approach \citep{noterdaeme_et_al_aa_2012}, and an
unpublished machine-learning approach using Fisher discriminant
analysis \citep{carithers_unpublished_2012}.  Although these methods
have had some success in detecting large \acro{dla} catalogs, their
reliance entirely on templates made them subject to hard-to-quantify
systematic biases. In particular, these methods lack an explicit
global model of quasar emission beyond simple continuum estimation,
and the lack of such a model may give rise to unexpected false
positives.

We present a new, completely automated method based on a rigorous
Bayesian model-selection framework. We model the quasar spectra,
including the continuum and non-\acro{dla} absorption, using Gaussian
process \citep{gpml} models with a bespoke covariance
function. Earlier catalogs are used as prior information to train the
covariance.  We provide a catalog of our results on 162\,858
\acro{qso}s with $z \geq 2.15$ from data release 12 of
\acro{sdss--iii}, demonstrating that our method scales to very large
datasets, making it ideally suited for future surveys. Furthermore, as
our method relies on a well-defined probabilistic framework, it allows
us to estimate the probability that each system is indeed a
\acro{dla}, rather than a noise fluctuation, degrading gracefully for
low signal-to-noise observations. This property allows us to obtain
substantially more-reliable measurements of the statistics of the
\acro{dla} population in situations with reliable uncertainties even
where systematic uncertainty dominates \citep{bird_analysis}. We are
also able to extend our catalog to high redshift even with low-quality
data.

Our method is applicable not just to \acro{dla}s, but also to other
classes of absorption systems, such as Lyman limit systems and metal
absorbers, which we intend to examine in future work.  We focus on
\acro{dla}s here both because of the large body of prior work which
enables us to thoroughly verify our catalogs, and the intrinsic
importance of these systems.

\acro{dla}s are a direct probe of neutral gas at densities close to
those required to form stars \citep{Cen:2012}.  The exact nature of
the systems hosting \acro{dla}s was initially debated, with kinematic
data combined with simple semi-analytic models appearing to indicate
objects similar in size to present day star-forming galaxies
\citep{Prochaska:1997, Jedamzik:1998, Maller:2001}, whereas early
simulations produced clumps closer in size to dwarf galaxies
\citep{Haehnelt:1998, Okoshi:2005}.  Recent numerical simulations are
able to reproduce most observations with neutral hydrogen clouds
stretching almost to the virial radius of objects larger than dwarfs,
but smaller than present day star-forming galaxies
\citep{Pontzen:2008, Rahmati:2013, Bird:2014a}.  Associated galactic
stellar components have been detected in a few, particularly neutral
hydrogen and metal-rich systems at low redshift
\citep{lebrun97,rao03,chen05}. However, unbiased surveys have placed
strong upper limits on the star-formation rates of the median
\acro{dla} \citep{Fumagalli:2015}, indicating that \acro{dla}s are
associated with low star-formation rate objects.

\acro{dla}s represent our only probe of small- to moderate-sized
galaxies at high redshift, and are known to have dominated the
neutral-gas content of the Universe from redshift $z = 5$ (when the
Universe was 1.2\,Gyr old) to today \citep{Gardner:1997, wolfe05}. The
neutral gas in these systems ultimately accretes onto galactic halos
and fuels star formation. Thus their abundance as a function of
redshift provides strong constraints on models of galaxy formation
\citep{Bird:2014}.  Our work, including publicly available software,
will not only provide observers with a new automated tool for
detecting these objects, but also provide theorists with a reliable
catalog on which to base theoretical models.

\section{Notation}

We will briefly establish some notation.  Consider a \acro{qso} with
redshift $\zq$; we will always assume that $\zq$ is known, allowing us
to work in the quasar restframe.  We will notate a \acro{qso}'s true
emission spectrum by a function $f\colon \R \to \R$, where $f(\lm)$
represents the flux corresponding to rest wavelength $\lm$. Without
subscript, $\lm$ will always refer to quasar rest wavelengths, $\lr$,
rather than observed wavelengths, $\lo$.  Note that the spectral
emission function $f$ is never directly observed, both due to
measurement error and due to absorption by intervening matter along
the line of sight.  We will denote the observed flux by a
corresponding function $y(\lm)$, which will again be a function of the
rest wavelengths.

Spectrographic observations of a \acro{qso} are made at a discrete set
of wavelengths $\vec{\lm}$, for which we observe a corresponding
vector of flux measurements $\vec{y}$, where we have defined $y_i =
y(\lm_i)$.  For a given \acro{qso}, we will represent the set of
observation locations and values $(\vec{\lm}, \vec{y})$ by $\data$.

We will often encounter data with missing values due to
observation-dependent pixel masking.  When required, we will
represent these in the text with a special value called \NaN (for
`not a number').  Calculations on data containing \NaN\/s will
always ignore these values.

\section{Bayesian Model Selection}

Our approach to \acro{dla} detection will depend on \emph{Bayesian
  model selection,} which will allow us to directly compute the
probability that a given quasar sightline contains a \acro{dla}.  We
will develop two probabilistic models for a given set of
spectroscopic observations $\data$: one for sightlines with
intervening \acro{dla}s, and one for those without.  Then, given the
available data, we will compute the posterior probability that the
former model is correct.  We will give a high-level overview of Bayesian
model selection below, then proceed to describe our models for
\acro{dla} detection below.

Let $\model$ be a probabilistic model, and let $\theta$ represent a
vector of parameters for this model (if any).  Given a set of observed
data $\data$ and a set of candidate models $\{ \model_i \}$ containing
$\model$, we wish to compute the probability of $\model$ being the
correct model to explain $\data$.  The key quantity of interest to
model selection is the so-called \emph{model evidence}:
\begin{equation}
  p(\data \given \model) =
  \int p(\data \given \model, \theta) p(\theta \given \model) \intd{\theta},
\end{equation}
which represents the probability of having generating the observed
data with the model, after having integrated out any uncertainty in
the parameter vector $\theta$.  Given the model evidence, we can apply
Bayes' rule to compute the posterior probability of the model given the data:
\begin{equation}
  \Pr(\model \given \data)
  = \frac{p(\data \given \model)\Pr(\model)}
         {p(\data)}
  = \frac{p(\data \given \model)\Pr(\model)}
         {\sum_i p(\data \given \model_i) \Pr(\model_i)},
\end{equation}
where $\Pr(\model)$ represents the prior probability of the model.
Notice that computing the posterior probability of $\model$ requires
computing the normalizing constant in the denominator.

We will develop two models for spectroscopic observations of
\acro{qso}s, $\Mnd$, for lines of sight that do not contain
intervening \acro{dla}s, and $\Md$, for those that do.  Both of these
models will rely heavily on Gaussian processes, which we will
introduce below.

\section{Gaussian Processes}

The main object of interest we wish to perform inference about is a
given \acro{qso}'s emission function $f(\lm)$.  This is in general a
complicated function with no simple parametric form available, so we
will instead use nonparametric inference techniques to reason about
it.  \emph{Gaussian processes} (\acro{gp}s) provide a powerful
nonparametric framework for modeling unknown functions, which we will
adopt.  See \cite{gpml} for an extensive introduction to \acro{gp}s.

\subsection{Definition and prior distribution}

Let $\mc{X}$ be an arbitrary input space, for example the real line
$\R$, and let $f\colon \mc{X} \to \R$ be a real-valued function on
$\mc{X}$ we wish to model.  We will continue to use $\lm$ to indicate
inputs to the function $f$.  A Gaussian process is an extension of the
multivariate Gaussian distribution $\mc{N}(\mu, \Sigma)$ to infinite
domains.  Like the multivariate Gaussian distribution, a \acro{gp} is
fully specified by its first two central moments: a mean function
$\mu(\lm)$ and a positive semidefinite covariance function $K(\lm,
\lm')$:\footnote{A function $K\colon \mc{X}^2 \to \R$ is
  \emph{positive semidefinite} if, for every finite subset $\Lambda =
  \{\lm_i\}_{i = 1}^n \subset \mc{X}$, the $n \times n$ \emph{Gram matrix}
  $\mat{A}$, defined by $A_{ij} = K(\lm_i, \lm_j)$, satisfies
  $\vec{c}\trans \mat{A} \vec{c} \geq 0$ for all $\vec{c} \in \R^n$.}
\begin{align*}
  \mu(\lm) &= \mspace{14.5mu}\mathbb{E}\bigl[f(\lm) \given \lm \bigr]; \\
  K(\lm, \lm') &= \cov\bigl[f(\lm), f(\lm') \given \lm, \lm' \bigr].
\end{align*}
The former describes the pointwise expected value of the function and
the latter describes the correlation around the mean.  Given $\mu$ and
$K$, we may endow the function space $f$ with a Gaussian process prior
probability distribution:
\begin{equation}
  p(f) = \mc{GP}(f; \mu, K).
  \label{prior}
\end{equation}
The defining characteristic of a Gaussian process is that given a
finite set of inputs $\vec{\lm}$, the corresponding vector of function
values $\vec{f} = f(\vec{\lm})$ is multivariate Gaussian distributed:
\begin{equation}
  \label{gp_prior}
  p(\vec{f}) =
  \mc{N}\bigl(\vec{f}; \mu(\vec{\lm}), K(\vec{\lm}, \vec{\lm}) \bigr),
\end{equation}
where the mean vector and covariance matrix are derived simply by
evaluating the mean and covariance functions at the inputs
$\vec{\lm}$, and the multivariate Gaussian probability distribution
function is given by
\begin{equation}
  \mc{N}(\vec{f}; \vec{\mu}, \mat{K})
  =
  \frac{1}{\sqrt{(2\pi)^d \det \mat{K}}}
  \exp\biggl(
  -\frac{1}{2} (\vec{f} - \vec{\mu})\trans \mat{K}\inv (\vec{f} - \vec{\mu})
  \biggr),
\end{equation}
where $d$ is the dimension of $\vec{f}$.

\subsection{Observation model}
\label{sec:observation_model}

Consider a set of noisy observations $\data = (\vec{\lm}, \vec{y})$
made at input locations $\vec{\lm}$.  Our Gaussian process prior on
$f$ implies a multivariate Gaussian distribution for the corresponding
(unknown, so-called \emph{latent}) function values $\vec{f} =
f(\vec{\lm})$, but does not specify the relationship between these
values and our observations $\vec{y}$.  Instead we must further model
the mechanism generating our observations, which we will encode by a
distribution
\begin{equation}
  p(\vec{y} \given \vec{\lm}, \vec{f}).
\end{equation}
In general this can be any arbitrary probabilistic model, but here we
will assume additive Gaussian noise.

Given a single input location $\lm$, we assume that the corresponding
observed value $y$ is realized by corrupting the true value of the
latent function $f(\lm)$ by zero-mean additive Gaussian noise with
known variance $\sigma(\lm)^2$:
\begin{equation}
  \label{individual_observation_model}
  p\bigl(y \given \lm, f(\lm), \sigma(\lm)\bigr)
  = \mc{N}\bigl(y; f(\lm), \sigma(\lm)^2\bigr).
\end{equation}
We assume the noise process is independent for every $\lm$, but note
that we do not make a homoskedasticity assumption; rather, we allow
the noise variance to depend on $\lm$.  This capability to handle
heteroskedastic noise is critical for the analysis of spectroscopic
measurements, where the noise associated with flux measurements can
vary widely as a function of wavelength.

Returning to our entire set of observations $\data = (\vec{\lm},
\vec{y})$, we assume that the noise variance associated with each of
these measurements is known and given by a corresponding vector
$\vec{\noise}$, with $\noise_i = \sigma(\lm_i)^2$.  Given our model
for individual observations \eqref{individual_observation_model} and
the noise independence assumption, the entire observation model is
given by
\begin{equation}
  \label{observation_model}
  p(\vec{y} \given \vec{\lm}, \vec{f}, \vec{\noise})
  =
  \mc{N}(\vec{y}; \vec{f}, \mat{\Noise}),
\end{equation}
where $\mat{\Noise} = \diag \vec{\noise}$, and we use the $\diag$
notation applied to a vector to refer to a square diagonal matrix with
leading diagonal equal to the specified vector.

\subsubsection{Prior of noisy observations}

Given a set of observations locations $\vec{\lm}$ and a corresponding
vector of noise variances $\vec{\noise}$, we may use the above to
compute the prior distribution for a corresponding vector of
observations $\vec{y}$ by marginalizing the latent function values
$\vec{f}$:
\begin{align}
  p(\vec{y} \given \vec{\lm}, \vec{\noise})
  &=
  \int p(\vec{y} \given \vec{\lm}, \vec{f}, \vec{\noise})
  p(\vec{f} \given \vec{\lm}) \intd{\vec{f}}
  \nonumber
  \\
  &=
  \int \mc{N}(\vec{y}; \vec{f}, \mat{\Noise})
  \, \mc{N}\bigl(\vec{f}; \mu(\vec{\lm}), K(\vec{\lm}, \vec{\lm})\bigr) \intd{\vec{f}}
  \nonumber
  \\
  &=
  \mc{N}\bigl(\vec{y}; \mu(\vec{\lm}), K(\vec{\lm}, \vec{\lm}) + \mat{\Noise}\bigr),
  \label{noisy_prior}
\end{align}
where we have used the fact that Gaussians are closed under
convolution to compute the integral in closed form.

In typical applications of \acro{gp} inference, the prior mean
function $\mu$ and prior covariance function $K$ would be selected
from numerous several off-the-shelf solutions available for this
purpose; however, none of these would be directly appropriate for
modeling \acro{qso} emission spectra, due to their somewhat complex
nature.  Typical parametric covariance functions, for example, tend to
be translation invariant and encode strictly decreasing covariance as
a function of the distance between inputs.\footnote{The Wiener
  process, modeling the sample paths of Brownian motion, is a Gaussian
  process with such a covariance function.}  \acro{qso} emission
spectra, however, are neither stationary, nor should we expect the
covariance to be diagonal dominant.  For example, strong off-diagonal
correlations must exist between potentially distant emission lines,
such as members of the Lyman series.  Rather, below we will construct
a custom \acro{gp} prior distribution for modeling these spectra in
the next section.

\section{Learning a GP Prior for QSO Spectra}
\label{sec:null_model}

We wish to construct a Gaussian process prior for \acro{qso} spectra,
specifically, those that do not contain an intervening \acro{dla}
along the line of sight.  This will form the basis for our null model
$\Mnd$.  We will later extend this to form our \acro{dla} model $\Md$.

As described in the previous section, a Gaussian process is defined
entirely by its first two moments: a mean function $\mu(\lm)$ and a
covariance function $K(\lm, \lm')$.  Therefore, our goal in this
section will be to derive reasonable prior choices for these functions.  Due
to the complex structure of \acro{qso} emission spectra, our approach
will be to make as few assumptions as possible.  Instead, we adopt a
data-driven approach and learn an appropriate model given over 48\,000
examples contained in a previously compiled catalog of quasar spectra
recorded by the \acro{boss} spectrograph \citep{smee13}.

\subsection{Data}

Together, \acro{sdss--i}, \acro{--ii} \citep{abazajian09}, and
\acro{-iii} \citep{eisenstein11} used a drift-scanning mosaic
\acro{ccd} camera \citep{gunn98} to image over one-third of the sky
(14\,555 square degrees) in five photometric bandpasses
\citep{fukugita96, smith02, doi10} to a limiting magnitude of $r <
22.5$ using the dedicated $2.5$\,m Sloan Telescope \citep{gunn06}
located at Apache Point Observatory in New Mexico.

The Baryon Oscillation Spectroscopic Survey (\acro{boss}), a part of
the \acro{sdss--iii} survey \citep{eisenstein11} has obtained spectra
of 1.5 million galaxies approximately volume limited out to $z\sim
0.6$ \citep{reid16}, and an additional 150\,000 spectra of
high-redshift quasars and ancillary sources.  \acro{boss} has measured
the characteristic scale imprinted by baryon acoustic oscillations
(\acro{bao}s) in the early Universe from the spatial distribution of
galaxies at $z\sim 0.5$ and the H\,\acro{i} absorption lines in the
intergalactic medium at $z\sim 2.3$
\citep{anderson13,anderson14,auborg15}.  The quasar target selection
is described in \citep{ross12,bovy11}.  Here we use data included in
data releases 9 (\acro{dr9}) \citep{ahn12} and 12 (\acro{dr12})
\citep{ahn14} of \acro{sdss--iii}; in particular, we primarily use the
associated quasar catalogs from various data
releases\footnote{\url{http://www.sdss.org/dr12/algorithms/boss-dr12-quasar-catalog/}}
\citep{paris12,paris14}.

\subsubsection{Description of data}

We used the \acro{qso} spectra from the \acro{boss dr9} Lyman-$\alpha$
forest sample \citep{lee_et_al_aj_2013} to train our \acro{gp} model.
This sample comprises 54\,468 \acro{qso} spectra with $\zq > 2.15$
from the \acro{dr9} release appropriate for Lyman-$\alpha$ forest
analysis.  An analogous model built from the entire \acro{dr12} sample
will be published along with manuscript for general-purpose use, along
with the source code (in \acro{matlab}) we used to train our model
and conduct our investigation.

The Lyman-$\alpha$ forest sample was augmented with a previously
compiled `concordance' \acro{dla} catalog
\citep{carithers_unpublished_2012}, combining the results of three
previous \acro{dla} searches.  These include a visual-inspection
survey \citep{slosar_et_al_jcap_2011} and two previous automated
approaches: a template-matching approach
\citep{noterdaeme_et_al_aa_2012}, and an unpublished machine-learning
approach using Fisher discriminant analysis
\citep{carithers_unpublished_2012}.  Any line of sight flagged in at
least two of these catalogs as containing a \acro{dla} is included in
the concordance catalog.  Both previous automated \acro{dla} searches
also produced estimates of the absorber redshift $\zd$ and column
density $\lni$. The concordance catalog also includes these estimates
for flagged sightlines; when a sightline is included in both automated
catalogs, the arithmetic mean of the associated estimates was
recorded.  A total of 5\,854 lines of sight are flagged as containing
an intervening \acro{dla} in the catalog (10.7\%).

\begin{figure*}
  \centering
  \subfloat[]{
    \hspace*{0.2em}
    \input{figures/raw_data.tex}
    \label{raw_data}
  }
  \\
  \subfloat[]{
    \input{figures/normalized_data.tex}
    \label{normalized_data}
  }
  \caption{An illustration of the data preprocessing procedure for
    object \acro{sdss} 020712.80+052753.4, (plate, \acro{mjd}, fiber)
    = (4401, 55510, 338); $\zq = 3.741$.  This \acro{qso} is included
    in the \acro{dla} concordance catalog with $(\zd, \lni) = (3.283,
    20.39)$, corresponding to central absorption wavelength $\lo =
    5\,206$\,\AA\/ or $\lr = 1\,098$\,\AA\/ in the \acro{qso}
    restframe.  The wavelengths are shifted to the \acro{qso}
    restframe and pixels outside $\lr \in [911.75\,\text{\AA},
      1215.75\,\text{\AA}]$ are discarded.  Finally, the flux and
    noise estimates are normalized by dividing by the median flux
    observed in the range $[1310\,\text{\AA}, 1325\,\text{\AA}]$.  The
    final result is shown in \subref{normalized_data}.  }
  \label{preprocessing}
\end{figure*}

\subsection{Modeling decisions}

To avoid effects due to redshift, we will build our emission model for
wavelengths in the rest frame of the \acro{qso}.  Furthermore, to
account for arbitrary scaling of flux measurements, we will build a
\acro{gp} prior for normalized flux. Specifically, given the observed
flux of a \acro{qso}, we normalize all flux measurements by dividing
by the median flux observed between 1310\,\AA\/ and 1325\,\AA\/ in the
rest frame of the \acro{qso}, a region redwards of the Ly\,$\alpha$
forest and void of major emission features.

Because this study is concerned with identifying \acro{dla}s, we will
only model the flux bluewards of the Ly\,$\alpha$ emission in the rest
frame of a given \acro{qso}.\footnote{One could consider an extension
  of our approach where metal absorption lines corresponding to
  wavelengths redwards of Ly\,$\alpha$ were considered, requiring
  modeling spectra over a larger range of wavelengths; however, we
  will not do so here.}  Specifically, we model emissions in the range
spanning from the Lyman limit to the Lyman-$\alpha$ line in the
\acro{qso} restframe.\footnote{We stop at the Lyman limit to avoid
  being confused by the Lyman break associated with Lyman limit
  systems.}  Our approach will be to learn a mean vector and
covariance matrix on a dense grid of wavelengths in this range, which
we will then interpolate as required by a particular set of observed
wavelengths.\footnote{Such interpolation introduces minor correlation
  between pixels; however, this effect is unlikely to be large.}  The
chosen grid was the set of wavelengths
\begin{equation}
  \lm \in [911.75\,\text{\AA}, 1215.75\,\text{\AA}],
\end{equation}
with a linearly equal spacing of $\Delta \lm =
0.25\,\text{\AA}$.\footnote{This represents about 3--4 times the
  maximum pixel separation of the \acro{boss} spectrograph; the
  minimum separation in a single \acro{boss} spectrum's measured
  wavelengths is approximately $(10^{\log_{10} 3600 + 0.0001} -
  3600)\,\text{\AA} \approx 0.83\,\text{\AA}$.  Note, however, that we
  have tens of thousands of observations corresponding to each of the
  wavelengths in our chosen grid.}  This resulted in a vector of input
locations $\vec{\lm}$ with $\lvert \vec{\lm} \rvert = \Npix = 1\,217$
pixels.

Given a \acro{gp} prior for \acro{qso} emission spectra, $p(f) =
\mc{GP}(f; \mu, K)$, the prior distribution for emissions on the
chosen grid $\vec{\lm}$, $\vec{f} = f(\vec{\lm})$ is a multivariate
Gaussian:
\begin{equation}
  \label{f_prior}
  p(\vec{f} \given \vec{\lm}, \zq) = \mc{N}(\vec{f}; \vec{\mu}, \mat{K}),
\end{equation}
where $\vec{\mu} = \mu(\vec{\lm})$ and $\mat{K} = K(\vec{\lm},
\vec{\lm})$.  Note that we must condition on the \acro{qso} redshift
$\zq$ because it is required for shifting into the quasar restframe.

As mentioned previously, however, we can never observe $f$ directly,
both due to measurement error and due to absorption by intervening
matter along the line of sight.  The former can be handled easily for
our spectra by using the pipeline error estimates in the role of the
noise vector $\vec{\noise}$ (see Section \ref{sec:observation_model}).
However, the latter is more problematic, especially in our chosen
region, which includes the Lyman-$\alpha$ forest.  In principle, if we
knew the exact nature of the intervening matter, we could model this
absorption explicitly; however, this is unrealistic.  We will instead
model the effect of small absorption phenomena (absorption by objects
with column density below the \acro{dla} limit, $\lni < 20.3$) by an
additional additive wavelength- and redshift-dependent Gaussian noise
term, which we will learn.  Therefore the characteristic `dips' of the
Lyman-$\alpha$ forest will be modeled as noisy deviations from the
true underlying smooth continuum.  Later we will explicitly model
larger absorption phenomena (\acro{dla}s with $\lni \geq 20.3$) to
build our \acro{dla} model $\Md$.

The mathematical consequence of this modeling decision is as follows.
Consider the arbitrary \acro{gp} model in \eqref{f_prior}.  We wish to
model the associated spectroscopic observation values on the chosen
grid, $\vec{y} = y(\vec{\lm})$.  Suppose that the measurement noise
vector $\vec{\noise} = \sigma(\vec{\lm})^2$ has been specified.
During our exposition on \acro{gp}s, we described the additive
Gaussian noise observation model \eqref{observation_model}.  The model
we adapt here will involve a shared non-\acro{dla} absorption
`noise' vector $\vec{\omega}$, defined in the quasar restframe,
modeling absorption deviations from the \acro{qso} continuum.

Due to the evolution of the Lyman-$\alpha$ forest flux with redshift,
we additionally incorporate a simple power-law redshift dependence
into this absorption noise model. Namely, the absorption noise
standard deviation we incorporate at an observed wavelength $\lo$ is
defined to be
\begin{align}
  \omega'(\lo, \lr) &=
  \omega(\lr) s\bigl(z(\lo)\bigr)^2;
  \\
  s(z) &=
  1 - \exp\bigl(-\tau_0 (1 + z)^\beta\bigr) + c_0, \label{sofz}
\end{align}
where $\omega(\lr)$ is the shared absorption noise corresponding to
the wavelength in the quasar restframe, $c_0$, $\tau_0$, and $\beta$
are constants, and $z(\lo)$ is the redshift of Lyman-$\alpha$ at the
observed wavelength. Hence our model depends on the redshift of the
quasar as well as the redshift of Lyman-$\alpha$ along the line of
sight.

The resulting observation model is
\begin{equation}
  p(\vec{y} \given \vec{f}, \vec{\noise}, \vec{\omega}, \zq, \Mnd)
  =
  \mc{N}(\vec{y}; \vec{f}, \mat{\Omega} + \mat{\Noise}),
\end{equation}
where $\mat{\Omega} = \diag \vec{\omega}'$, and $\vec{\omega}'$
incorporates the redshift dependence as defined above. Therefore,
given our chosen grid $\vec{\lm}$, the prior distribution of
associated spectroscopic observations $\vec{y}$ is
\begin{equation}
  \label{spectra_observation_prior}
  p(\vec{y} \given \vec{\noise}, \vec{\omega}, \zq, \Mnd)
  =
  \mc{N}(\vec{y}; \vec{\mu}, \mat{K} + \mat{\Omega} + \mat{\Noise}),
\end{equation}
derived analogously to \eqref{noisy_prior}.  Our goal now is to learn
appropriate values for $\vec{\mu}$, $\vec{K}$, $\vec{\omega}$, $c_0$,
$\tau_0$, and $\beta$, which will fully specify our null model $\Mnd$.

\subsection{Learning appropriate parameters}

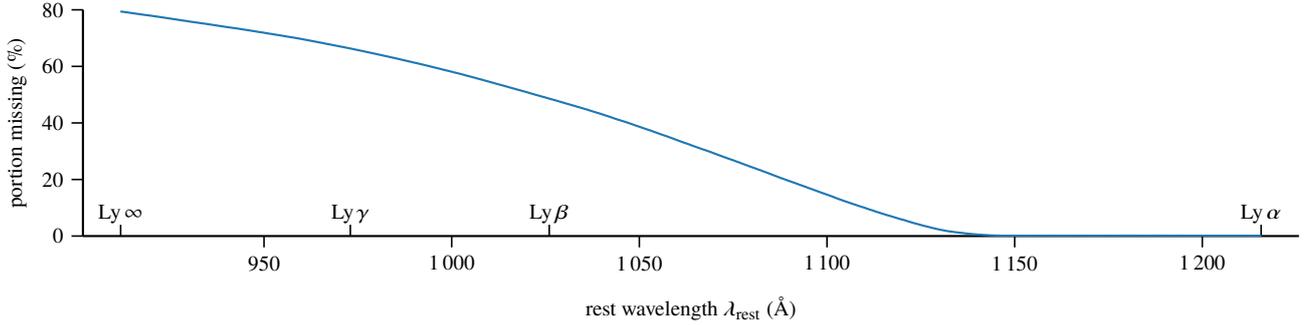
\begin{figure*}
  \centering
  \input{figures/missing_pixels.tex}
  \caption{The portion of missing pixels as a function of wavelength
    for the 48\,614 \acro{qso}s in the \acro{boss dr9} Lyman-$\alpha$
    forest sample used for learning our \acro{gp} model.}
  \label{missing_pixels}
\end{figure*}

To build our null model, we took the $\Nspec = 48\,614$ spectra from
the \acro{boss dr9} Lyman-$\alpha$ forest sample that are putatively
absent of intervening \acro{dla}s.  We prepared each of these spectra
for processing in an identical manner as follows.
\begin{itemize}
\item The augmented spectrum file was loaded and the $($wavelength,
  observed flux, pipeline noise variance$) = (\lambda, y, v)$
  measurements in the chosen modeled region were extracted.
\item The wavelengths were shifted to the rest frame of the
  \acro{qso}.
\item Flux measurements with serious pixel mask bit flags
  (\texttt{FULLREJECT}, \texttt{NOSKY}, \texttt{BRIGHTSKY},
  \texttt{NODATA}) set by the \acro{sdss} pipeline were masked
  (replaced by \NaN).
\item The flux normalizer was determined by examining the region
  corresponding to $[1310, 1325]\,\AA$ in the restframe of the quasar;
  the median nonmasked value in this range was used for normalization.
\item The flux and noise variance were normalized with the value
  computed in the last step.
\end{itemize}

Finally, we linearly interpolated the resulting flux and noise
variance measurements of each spectrum onto the chosen wavelength grid
$\vec{\lm}$. Note that this interpolation preserved \NaN\/s; we did
not `interpolate through' masked pixels.  We also did not
extrapolate beyond the range of wavelengths present in each spectrum.
The preprocessing procedure is illustrated in Figure
\ref{preprocessing} on a spectrum we will use as a running example.

We collect the resulting interpolated vectors into $(\Nspec \times
\Npix)$ matrices $\mat{Y}$ and $\mat{\Noise}$, containing the normalized
flux and noise variance vectors, respectively.  For \acro{qso} $i$, we
will write $\vec{y}_i$ and $\vec{\noise}_i$ to represent the
corresponding observed flux and noise variance vectors, and will
define $\mat{\Noise}_i = \diag \vec{\noise}_i$.

Due to masked pixels and varying redshifts of each \acro{qso}, the
$\mat{Y}$ and $\mat{\Noise}$ matrices contain numerous missing values,
especially on the blue end.  Figure \ref{missing_pixels} shows the
portion of available data as a function of wavelength.

\subsubsection{Learning the mean}

\begin{figure*}
  \centering
  \input{figures/learned_mean.tex}
  \caption{The learned mean vector $\vec{\mu}$ derived by taking the
    median across the stacked spectra.  The vector has been smoothed
    with a 4-pixel ($1$\,\AA) boxcar function for clarity on the
    blue end.}
  \label{learned_mean}
\end{figure*}
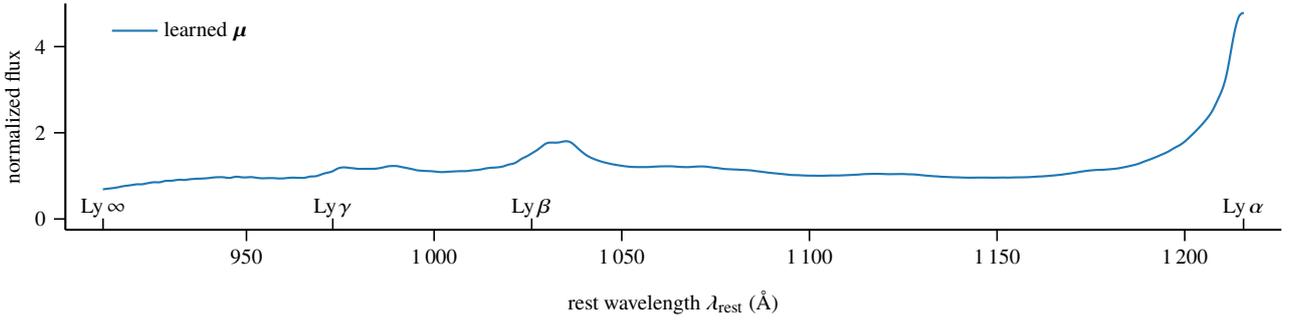

Identifying an appropriate mean vector $\vec{\mu}$ is straightforward
with so many example spectra.  We simply found the mean recorded value
for each rest wavelength in our grid across the available
measurements:
\begin{equation}
  \mu_j = \frac{1}{N_{\neg\text{NaN}}} \sum_{y_{ij} \neq \text{NaN}} y_{ij}.
\end{equation}
Note that the sample mean is the maximum-likelihood estimator for
$\vec{\mu}$.  The learned mean vector $\vec{\mu}$ is plotted in Figure
\ref{learned_mean}.  Several emission features are obvious.

\subsubsection{Learning the flux covariance and additional absorption noise}

We will use standard unconstrained optimization techniques to learn
the covariance matrix $\mat{K}$ and absorption `noise' vector
$\vec{\omega}$.  Without further structural assumptions on $\mat{K}$,
however, we would be forced to learn $\Npix^2 \approx 1.5 \times 10^6$
entries.  Instead we will use a low-rank decomposition to limit the
number of free variables in our model:
\begin{equation}
  \mat{K} = \mat{M}\mat{M}\trans,
\end{equation}
where $\mat{M}$ is an $(\Npix \times k)$ matrix with $k \ll \Npix$.
This decomposition guarantees that $\mat{K}$ will be positive
semidefinite (and thus a valid covariance matrix) for any $\mat{M}$.
Note that this decomposition is similar to that encountered in
principal component analysis (\acro{pca}); however, note that we do
not constrain the columns of $\mat{M}$ (the `eigenspectra') to be
orthogonal.  Here we took $k = 20$.

We assume that each of our measured flux vectors is an independent
realization drawn from a common observation prior
\eqref{spectra_observation_prior}:
\begin{equation}
  p(\mat{Y} \given \vec{\lm}, \mat{\Noise}, \mat{M}, \vec{\omega}, \vec{z}_{\text{\acro{qso}}}, \Mnd)
  =
  \prod_{i = 1}^{\Nspec}
  \mc{N}(\vec{y}_i; \vec{\mu}, \mat{K} + \mat{\Omega} + \mat{\Noise}_i),
\end{equation}
where $\vec{z}_{\text{\acro{qso}}}$ is a vector concatenating the
redshifts of the quasars, and all \NaN values are ignored in the
computation. That is, in the $i$th entry of the product, we only use
the entries of $\vec{\mu}$, $\vec{v}_i$, and $\vec{w}$', and only the
rows of $\mat{M}$, corresponding to the non-masked values in
$\vec{y}_i$.

We define the log likelihood of the data, $\mc{L}$, as a function of
the covariance parameters $\mat{M}$ and $\vec{\omega}$. To simplify
the notation, we first define the following quantities:
\begin{align}
  \mat{\Sigma}_i &= \mat{K} + \mat{\Omega} + \mat{\Noise}_i; \\
  \vec{\alpha}_i &= \mat{\Sigma}_i\inv(\vec{y}_i - \vec{\mu}).
\end{align}
Now the log likelihood is
\begin{align}
  \mc{L}(\mat{M}, \vec{\omega})
  &=
  \log p(\mat{Y} \given \vec{\lm}, \mat{\Noise}, \mat{M}, \vec{\omega}, \vec{z}_{\text{\acro{qso}}}, \Mnd)
  \nonumber
  \\
  &=
  \sum_{i = 1}^{\Nspec}
  \log \mc{N}(\vec{y}_i; \vec{\mu}, \mat{\Sigma}_i)
  \nonumber
  \\
  &=
  \sum_{i = 1}^{\Nspec}
  -\frac{1}{2}
  \bigl(
  \vec{\alpha}_i\trans
  (\vec{y}_i - \vec{\mu})
  +
  \log \det \mat{\Sigma}_i
  +
  N_i \log 2\pi
  \bigr),
\end{align}
where $N_i$ is the number of non-\NaN pixels in $\vec{y}_i$.  We will
maximize $\mc{L}(\mat{M}, \vec{\omega})$ with respect to the
covariance parameters to derive our model, giving the emission model
most likely to have generated our data.  To enable unconstrained
optimization, we parameterize the $\vec{\omega}$ parameter by its
natural logarithm, guaranteeing every entry of $\vec{\omega}$ is
positive after exponentiation.  In the context of its role in our
model, this is equivalent to reasoning about the optical depth $\tau$
rather than the absorption $\exp(-\tau)$.

An important feature of our particular choice of model is that we can
compute the matrix inverse and the log determinant of $(\mat{K} +
\mat{\Omega} + \mat{\Noise})$ quickly.  Namely, this matrix has the form
$\mat{M}\mat{M}\trans + \mat{D}$, where $\mat{D}$ is diagonal.  We may
apply the Woodbury identity to derive
\begin{equation}
  \label{efficient_inv}
  (\mat{M}\mat{M}\trans + \mat{D})\inv
  =
  \mat{D}\inv
  -
  \mat{D}\inv\mat{M}
  (\mat{I} + \mat{M}\trans \mat{D}\inv \mat{M})\inv
  \mat{M}\trans\mat{D}\inv,
\end{equation}
where $\mat{I}$ is the identity matrix. Note the nominally $\Npix
\times \Npix$ inverse can be computed via a much less expensive $k
\times k$ inverse.  Similarly, we may use the Sylvester determinant
theorem to derive
\begin{equation}
  \label{efficient_det}
  \log \det (\mat{M}\mat{M}\trans + \mat{D})
  =
  \log \det \mat{D}
  +
  \log \det (\mat{I} + \mat{M}\trans \mat{D}\inv \mat{M}),
\end{equation}
again reducing the problem to a determinant on a $k \times k$ matrix.

To maximize our joint log likelihood, we applied the \acro{l-bfgs}
algorithm, a quasi-Newton algorithm for unconstrained
optimization. The required partial derivatives are:
\begin{align}
  \frac{\partial{\mc{L}_i}}{\partial \mat{M}}
  &=
  (\vec{\alpha}_i \vec{\alpha}_i\trans - \mat{\Sigma}_i\inv) \mat{M};
  \\
  \frac{\partial{\mc{L}_i}}{\partial \log \vec{\omega}}
  &=
  \vec{\omega}'
  \circ
  (\vec{\alpha}_i^2 - \diag \mat{\Sigma}_i\inv),
\end{align}
where $\circ$ is the Hadamard (elementwise) product, and we define
$\diag$ applied to a square matrix to return its leading diagonal as a
vector. The partial derivatives with respect to $\log c_0$, $\log
\tau_0$, and $\log \beta$ all have the same form:
\begin{align}
  \frac{\partial{\mc{L}_i}}{\partial \log x}
  &=
  \vec{\alpha}_i\trans
  \biggl(
  \diag \frac{\partial \vec{\omega}'}{\partial x}
  \biggr)
  \vec{\alpha}_i
  +
  \biggl(\frac{\partial \vec{\omega}'}{\partial x}\biggr)\trans
  \diag \mat{\Sigma}\inv; \\
  \frac{\partial{\vec{\omega}'}}{\partial \log c_0}
  &=
  c_0\, \vec{\omega} \circ s(\vec{z});
  \\
  \frac{\partial{\vec{\omega}'}}{\partial \log \tau_0}
  &=
  \tau_0\,
  \vec{\omega}
  \circ
  s(\vec{z})
  \circ
  (1 + \vec{z})^\beta
  \circ
  \exp\bigl(-\tau_0 (1 + \vec{z})^\beta \bigr);
  \\
  \frac{\partial{\vec{\omega}'}}{\partial \log \beta}
  &=
  \beta
  \log \vec{z}
  \circ
  \frac{\partial{\vec{\omega}'}}{\partial \log \tau_0},
\end{align}
where $\vec{z}$ is a vector of the Lyman-$\alpha$ redshifts
corresponding to the observations, and the redshift contribution $s$
is defined in \eqref{sofz}.

We learned the decomposed covariance matrix $\mat{M}$, $\vec{\omega}$,
$c_0$, $\tau_0$, and $\beta$ via \acro{l-bfgs} on the selected
training spectra.  For this model learning phase only, we masked all
pixels with noise variance larger than unity after normalization (that
is, pixels with signal to noise ratios below approximately 1).  Note
that these pixels were only masked here and at no other point in this
study. The initial value for $\mat{M}$ was taken to be the top-20
principal components of $\mat{Y}$, estimated entrywise using available
data.  Masking low-\acro{snr} pixels was required here because
\acro{pca}, in its most basic form, does not account for noise in
measured values, and our heteroskedastic noise is especially
troublesome.  The initial value of each entry in $\vec{\omega}$ was
taken to be the sample variance of the corresponding column of
$\mat{Y}$, ignoring \NaN\/s.

\begin{figure*}
  \centering
  \subfloat[]{
    \hspace*{0.6em}
    \input{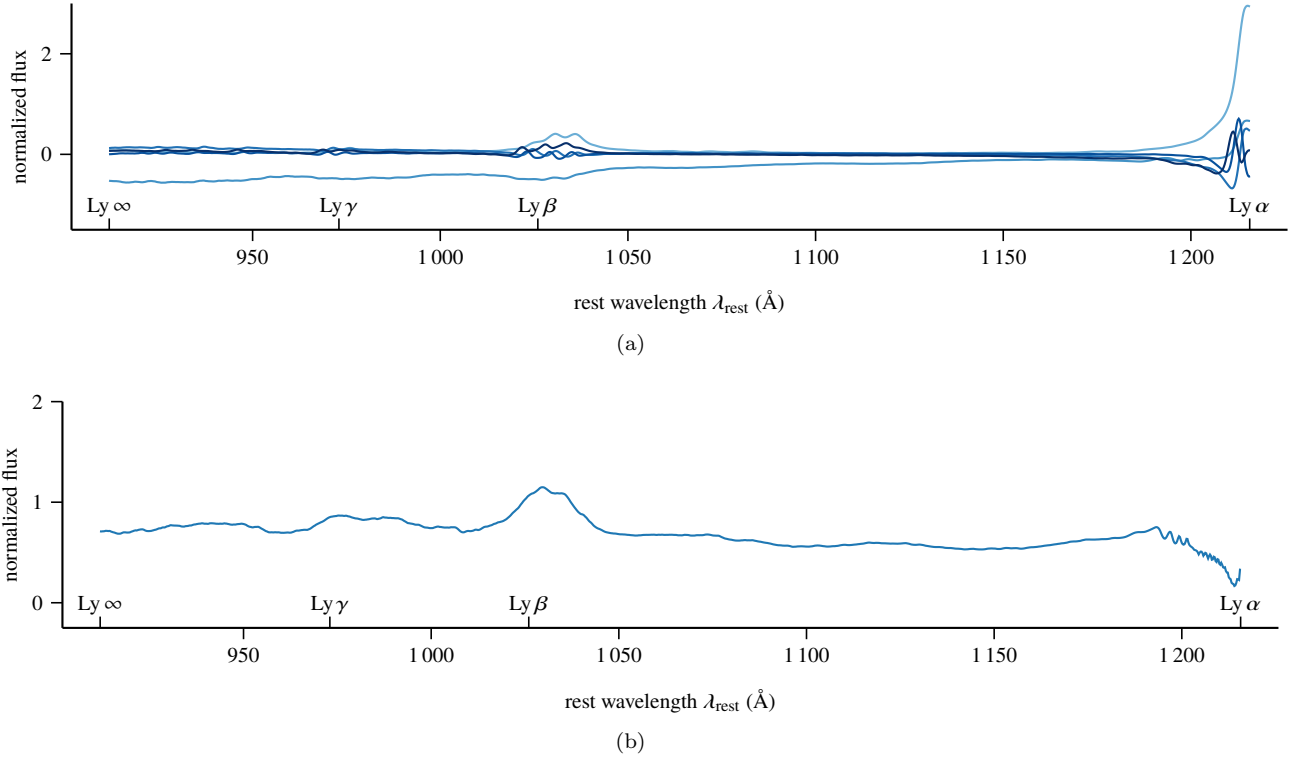}
    \label{learned_top_A}
  }
  \\
  \subfloat[]{
    \hspace*{-0.2em}
    \input{figures/learned_sigma.tex}
    \label{learned_sigma}
  }
  \caption{\subref{learned_top_A}: The first five columns of the
    learned \mat{M} and \subref{learned_sigma}: the learned
    absorption noise vector $\vec{\omega}$, both learned from the
    48\,614 \acro{qso}s in the \acro{boss dr9} Lyman-$\alpha$ forest
    sample. Both have been smoothed with a 4-pixel ($1$\,\AA)
    boxcar function for clarity on the blue end.}
  \label{learned_K_parameters}
\end{figure*}

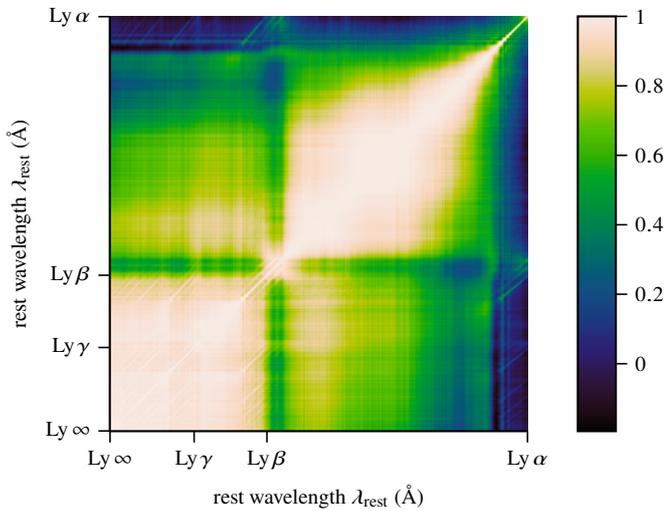
\begin{figure}
  \centering
  \input{figures/normalized_K.tex}
  \caption{The observation covariance matrix $\mat{K}$ corresponding
    to the learned parameters shown in Figure
    \ref{learned_K_parameters}.  The entries have been normalized to
    give unit diagonal; the entries are therefore correlations rather
    than raw covariances.}
  \label{learned_K}
\end{figure}

The first five columns of the learned $\mat{M}$ and the learned
absorption noise vector $\mat{\omega}$ are shown in Figure
\ref{learned_K_parameters}.  The corresponding covariance matrix
$\mat{M}\mat{M}\trans$ is shown in Figure \ref{learned_K}.  Features
corresponding to the Lyman series are clearly visible, including
strong off-diagonal correlations between pairs of emission lines.  At
least seven members of the Lyman series can be identified in the
covariance entries corresponding to Lyman-$\alpha$ emission.  This
complex (and physically correct) structure was automatically learned
from the data. The parameters for the redshift-dependent component of
the absorption noise vector were
\begin{equation}
  c_0 = 0.3371; \qquad \tau_0 = 0.01178; \qquad \beta = 1.797.
\end{equation}

We have now fully specified our \acro{gp} prior for \acro{qso}
emission spectra in the range $\lm \in [911.75\,\text{\AA},
  1215.75\,\text{\AA}]$.  Figure \ref{model_sample} demonstrates our
model by showing an example sample from the prior distribution on
\acro{qso} continuua $\vec{f}$, as well as a corresponding sample from
the prior distribution on observations $\vec{y}$ incorporating our
absorption `noise' vector $\vec{\omega}$.  The samples closely
resemble actual observations.

Note that to apply our model to observations corresponding to a set of
input wavelengths differing from the grid we used to learn the model,
we simply interpolate (linearly) the learned $\vec{\mu}$, $\mat{K}$,
and $\vec{\omega}$ onto the desired wavelengths. We may also
account for redshift trivially should we wish to work with observed
rather than rest wavelengths.

\begin{figure*}
  \centering
  \input{figures/model_sample.tex}
  \caption{An example sample from our learned \acro{qso} emission
    spectrum model $\mc{GP}(\vec{f}; \vec{\mu}, \mat{K})$ (in red), and the
    corresponding sample after incorporating our additional absorption
    correction into the model, a draw from $p(\vec{y} \given
    \vec{\lm}, \vec{\noise}, \Mnd) = \mc{GP}(\vec{y}; \vec{\mu}, \mat{K} +
    \mat{\Omega} + \mat{\Noise})$ (in blue).  Constant observation noise
    with variance $\noise = 0.1^2$ was simulated for the $\vec{y}$
    sample.}
  \label{model_sample}
\end{figure*}

\subsection{Model evidence}

We note that our null model $\Mnd$ has no parameters.  Consider a set
of observations of a \acro{qso} $\data = (\vec{\lm}, \vec{y})$ with
known observation noise variance vector $\vec{\noise}$.  The model
evidence for $\Mnd$ given by observations can be computed directly:
\begin{equation}
  p(\data \given \Mnd, \vec{\noise}, \zq)
  \propto
  p(\vec{y} \given \vec{\lm}, \vec{\noise}, \zq, \Mnd).
\end{equation}
The constant of proportionality is $p(\vec{\lm} \given \Mnd)$, a
quantity that we do not model here.  Rather, we will assume that
$p(\vec{\lm} \given \model)$ is constant across models, causing it to
to cancel during the calculation of the model posterior.  Therefore
for the purposes of model comparison, we need only compute
\begin{equation}
  \label{ndla_evidence}
  p(\vec{y} \given \vec{\lm}, \vec{\noise}, \zq, \Mnd)
  =
  \mc{N}(\vec{y}; \vec{\mu}, \mat{K} + \mat{\Omega} + \mat{\Noise}),
\end{equation}
where the $\vec{\mu}$, $\mat{K}$, and $\vec{\omega}$ learned above
have been interpolated onto $\vec{\lm}$.

\section{A GP Model for QSO Spectral Sightlines with Intervening DLAs}

In the previous section, we learned an appropriate \acro{gp} model for
\acro{qso} spectra without intervening \acro{dla}s, forming our null
model $\Mnd$.  Here we will extend that model to create a model for
sightlines containing intervening \acro{dla}s.  We will first fully
describe the model for spectra containing exactly one intervening
\acro{dla}, then extend this model to the case of two-or-more
\acro{dla}s along a line of sight.  We will call our model for lines
of sight containing exactly $k$ intervening \acro{dla}s $\Mdk{k}$;
here we describe $\Mdk{1}$.  Taking the conjunction of these models
$\{\Mdk{i}\}_{i=1}^\infty$ gives our complete \acro{dla} model $\Md$.

Consider a quasar with redshift $\zq$, and suppose that there is an
intervening \acro{dla} along the line of sight with redshift $\zd$ and
column density $\nhi$. The effect of this on our observations is to
multiply the emitted flux $f(\lambda)$ by an appropriate absorption
function:
\begin{equation}
  y(\lm) = f(\lm)\exp\bigl(-\tau(\lm; \zd, \nhi)\bigr) + \varepsilon,
\end{equation}
where $\varepsilon$ is additive Gaussian noise due to measurement
error and $\tau$ is the absorption cross section, which has a
contribution corresponding to each transition we wish to model.  Here
we model absorption for several members of the Lyman series:
\begin{equation}
  \tau(\lm; \zd, \nhi) =
  \nhi
  \frac{\pi e^2 f \lm'}{m_e c}
  \phi(v, b, \gamma),
\end{equation}
where $e$ is the elementary charge, $\lm'$ is the transition
wavelength ($\lm' = 1215.6701\,\text{\AA}$ for Lyman-$\alpha$), and
$f$ is the oscillator strength of the transition ($f = 0.4164$ for
Lyman-$\alpha$).  The line profile function $\phi$ is a Voigt profile,
where $v$ is the relative velocity:
\begin{equation}
  v = c \biggl(\frac{\lambda}{\lambda' (1 + \zd)} - 1\biggr),
\end{equation}
$b / \sqrt{2}$ is the standard deviation of the Gaussian (Maxwellian)
broadening contribution:
\begin{equation}
  b = \sqrt{2\frac{kT}{m_p}},
\end{equation}
and $\gamma$ is the width of the Lorenztian broadening contribution:
\begin{equation}
  \gamma = \frac{\Gamma \lm'}{4\pi},
\end{equation}
where $\Gamma$ is a damping constant ($\Gamma = 6.265 \times
10^8\,\text{s}^{-1}$ for Lyman-$\alpha$).  The gas temperature $T$ is
fixed to $10^4$\,K. This imparts a thermal broadening of
13\,$\text{km\,s}^{-1}$, which is negligible compared to broadening of
the \acro{DLA} profile from Lorentzian damping wings. We neglect
broadening due to any turbulence of the gas within the \acro{DLA},
which could potentially contribute at lower column densities.  We
considered line profiles corresponding to Lyman-$\alpha$, -$\beta$,
and -$\gamma$ absorption, which we may compute for a given set of
wavelengths given the known transition parameters, the temperature
$T$, and $\zd$ and $\nhi$.

Gaussian processes provide a simple mechanism to model the
multiplicative effect introduced by the absorption function
$\exp(-\tau)$. Let a function $f$ have a Gaussian process prior
distribution $p(f) = \mc{GP}(f; \mu, K)$, and let $a(\lm)$ be a known
function.  Then the distribution of the product $g(\lm) =
a(\lm)f(\lm)$ is also a Gaussian process (\acro{gp}s are closed under
affine transformations):
\begin{equation}
  p(g) = \mc{GP}(f; \mu', K'),
\end{equation}
where
\begin{equation}
  \mu'(\lm) = a(\lm)\mu(\lm); \qquad
  K'(\lm, \lm') = a(\lm)K(\lm, \lm')a(\lm').
\end{equation}

\begin{figure*}
  \centering
  \subfloat[]{
    \hspace*{0.6em}
    \input{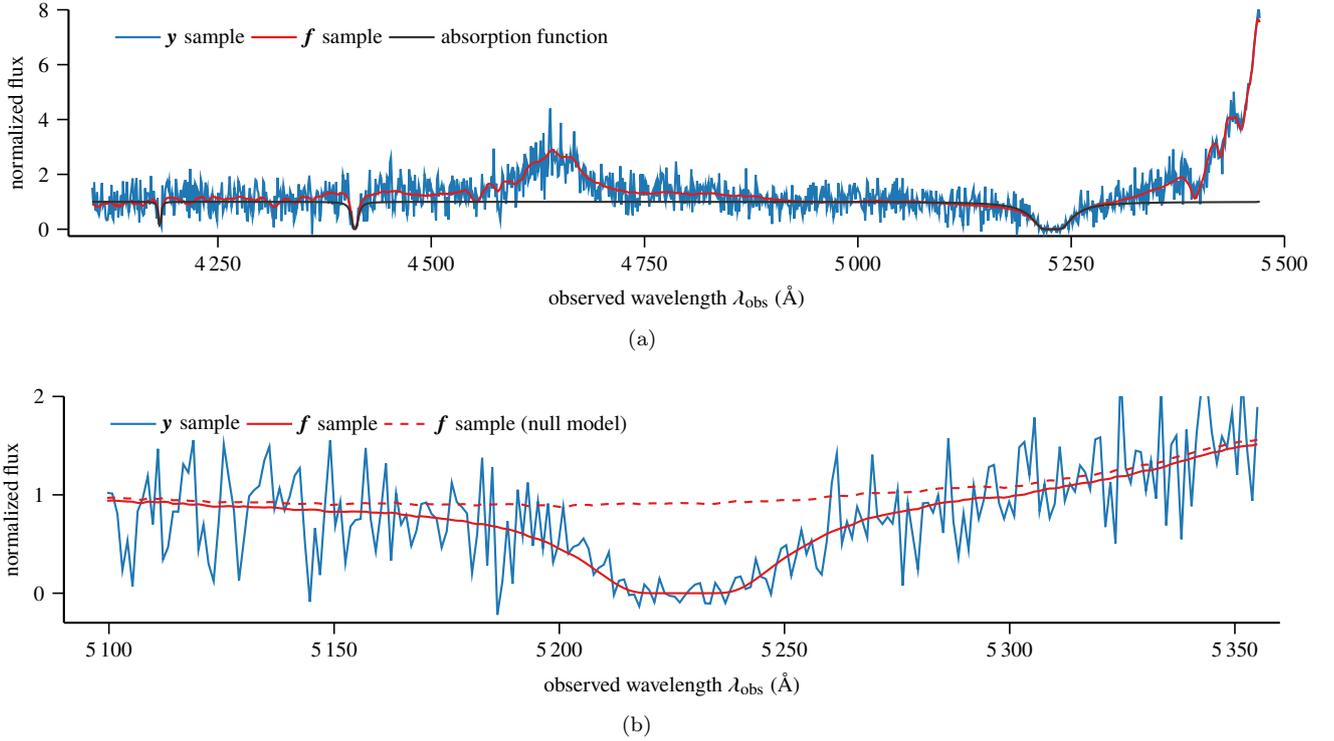}
    \label{dla_model_sample}
  }
  \\
  \subfloat[]{
    \hspace*{-0.6em}
    \input{figures/dla_model_sample_closeup.tex}
    \label{dla_model_sample_closeup}
  }
  \caption{An example sample from our model for \acro{qso} emission
    spectra with one \acro{dla} along the line of sight.  Here we
    simulate a \acro{qso} with $\zq = 2.5$ with a \acro{dla} at $\zd =
    2.2$ and $\lni = 20.8$.  This sample corresponds to that in Figure
    \ref{model_sample}, but is instead drawn from the \acro{dla} model
    with the appropriate absorption profile (plotted in grey).  In
    \subref{dla_model_sample}, we show the entire simulated
    observations, and in \subref{dla_model_sample_closeup} we show
    detail in the region of the Lyman-$\alpha$ absorption central
    wavelength, with the continuum sample from Figure
    \ref{model_sample} for comparison.  Note that the full sample also
    reflects corresponding Lyman-$\beta$ and Lyman-$\gamma$
    absorption.}
  \label{dla_model_samples}
\end{figure*}

Therefore, given the parameters $(\zd, \nhi)$ of a putative
\acro{dla}, we compute the appropriate absorption function
$\exp\bigl(-\tau(\lm; \zd, \nhi)\bigr)$ and modify the null \acro{gp}
model from the previous section as above. Specifically, consider
observations of a \acro{qso} sightline at rest wavelengths
$\vec{\lm}$.  Our model for the corresponding emitted flux $\vec{f}$
remains as in \eqref{f_prior}.  Given the observation noise variance
vector $\vec{\noise}$, the prior for the observation vector $\vec{y}$
without intervening \acro{dla}s is
\begin{equation}
  p(\vec{y} \given \vec{\lm}, \vec{\noise}, \zq, \Mnd)
  =
  \mc{N}(\vec{y}; \vec{\mu}, \mat{K} + \mat{\Omega} + \mat{\Noise}).
\end{equation}
Suppose now that we wish to model the observed flux with a \acro{dla}
at known redshift $\zd$ and column density $\nhi$.  First we compute
the theoretical absorption function with these parameters at
$\vec{\lm}$; call this vector $\vec{a}$:
\begin{equation}
  \vec{a} = \exp\bigl(-\tau(\vec{\lm}; \zd, \nhi)\bigr).
\end{equation}
Now, applying the result above, the prior for $\vec{y}$ with the
specified \acro{dla} is
\begin{multline}
  \label{absorption_posterior}
  p(\vec{y} \given \vec{\lm}, \vec{\noise}, \zq, \zd, \nhi, \Mdk{1})
  \\
  =
  \mc{N}\bigl(\vec{y}; \vec{a} \circ \vec{\mu}, \mat{A}(\mat{K} + \mat{\Omega})\mat{A} + \mat{\Noise}\bigr),
\end{multline}
where $\vec{a} = \diag \mat{A}$.

Figure \ref{dla_model_samples} displays a draw from our \acro{dla}
prior corresponding to the null model sample in Figure
\ref{model_sample}.

An important feature of this model is that it is not in any way
specific to \acro{dla}s, nor to data from the \acro{BOSS} instrument.
Our \acro{gp} model for quasar emission spectra could be modified in
an identical manner to model observed flux associated with any desired
absorption feature.

\subsection{Model evidence}

Unlike our null model, which was parameter free, our \acro{dla} model
$\Mdk{1}$ contains two parameters describing a putative \acro{dla}:
the redshift $\zd$ and column density $\nhi$.  We will denote the
model parameter vector by $\theta = (\zd, \nhi)$.  To compute the
model evidence, we must compute the following integral:
\begin{multline}
  \label{dla_evidence}
  p(\data \given \Mdk{1}, \vec{\noise}, \zq)
  \propto
  p(\vec{y} \given \vec{\lm}, \vec{\noise}, \zq, \Mdk{1})
  =
  \\
  \int
  p(\vec{y} \given \vec{\lm}, \vec{\noise}, \zq, \theta, \Mdk{1})
  p(\theta \given \zq, \Mdk{1})
  \intd\theta,
\end{multline}
where we have marginalized the parameters given a prior distribution
$p(\theta \given \zq, \Mdk{1})$.  Before we describe the approximation of
this integral, we will first describe the prior distribution used in
our experiments.

\subsection{Parameter prior}

First, we make the assumption that absorber redshift and column density
are conditionally independent given $\zq$ and that the column density is
independent of the \acro{qso} redshift:
\begin{multline}
  p(\theta \given \zq, \Mdk{1}) = \\ p(\zd \given \zq, \Mdk{1})p(\nhi \given \Mdk{1}).
\end{multline}
For the distribution $p(\zd \given \zq, \Mdk{1})$, we define the
following range of allowable $\zd$:
\begin{align}
  z_{\text{min}} &=
  \max \begin{cases}
    \frac{\lm_{\text{Ly}\infty}}{\lm_{\text{Ly}\alpha}}(1 + \zq) - 1 + 3\,000\,\text{km\,s}^{-1} / c
    \\
    \frac{\min \lo}{\lm_{\text{Ly}\alpha}} - 1
  \end{cases}
  \\
  z_{\text{max}} &= \zq - 3\,000\,\text{km\,s}^{-1} / c;
\end{align}
that is, we insist the absorber center be within the range of observed
wavelengths (after restricting to $\lr \in [911.75\,\text{\AA},
  1216.75\,\text{\AA}]$).  We also apply a conservative cutoff of
$3\,000\,\text{km\,s}^{-1}$ in the immediate vicinity of the
\acro{qso} to avoid proximity ionization effects, and in the immediate
vicinity of the Lyman limit in the quasar restframe (if visible) to
avoid problems caused by possible incorrect determination of $\zq$.

Given these, we simply take a uniform prior distribution on this
range:
\begin{equation}
  p(\zd \given \zq, \Mdk{1}) = \mc{U}[z_{\text{min}}, z_{\text{max}}].
\end{equation}

The column density prior $p(\nhi \given \Md)$ is slightly more
complicated. We first make a nonparametric estimate of the density
given the examples contained in the \acro{dla} catalog provided with
the \acro{boss dr9} Lyman-$\alpha$ forest sample.  Due to the large
dynamic range of column densities, we instead choose a prior on its
base-10 logarithm, $\lni$.

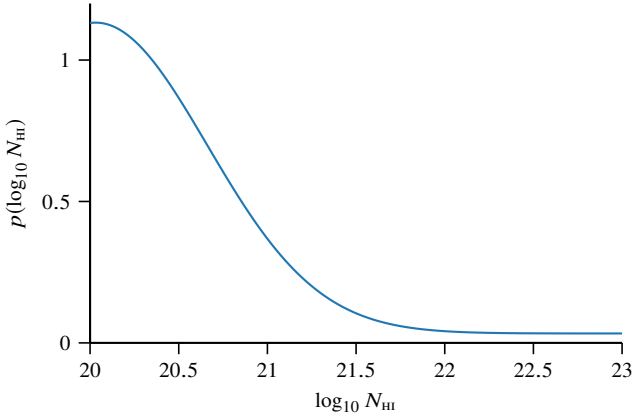
\begin{figure}
  \centering
  \input{figures/lni_kde.tex}
  \caption{The probability density function of the log column density
    prior used in the experiments, $p(\lni \given \Mdk{1})$.}
  \label{lni_kde}
\end{figure}

We use the reported $\lni$ values for the $\Ndla = 5\,854$ \acro{dla}s
contained in the \acro{dr9} sample to make a kernel density estimate
of the density $p(\lni \given \Mdk{1})$.  Kernel density estimation
entails centering small so-called `kernel' functions on each
observation and summing them to form our estimate.  Here we selected
the univariate Gaussian probability density function for our kernels,
with bandwidth selected via a plug-in estimator. The final estimate
is:
\begin{equation}
  \label{plni_estimate}
  p_{\text{\acro{kde}}}(\lni \given \Mdk{1})
  =
  \frac{1}{\Ndla}
  \sum_{i = 1}^{\Ndla}
  \mc{N}(\lni; \ell_i, \sigma^2),
\end{equation}
where $\ell_i$ is the base-10 logarithm of the $i$th observed column
density. To account for some possible systematic bias in estimating
this distribution, such as preferred numbers during visual inspection
or underestimation of the probability of high-density systems due to
low sample size, we make two adjustments.  First, we simplify the form
of the distribution by fitting a parametric prior to the nonparametric
kernel density estimate of the form
\begin{multline}
  p_{\text{\acro{kde}}}(\lni = N \given \Mdk{1})
  \approx
  \\
  q(\lni = N)
  \propto
  \exp(aN^2 + bN + c);
\end{multline}
the values we learned, via least-squares fitting to the log
probability over the range $\lni \in [20, 22]$, were
\begin{equation}
  a = -1.2695; \qquad
  b = 50.863; \qquad
  c = -509.33.
\end{equation}
Finally, to account for some possible observation bias in the
concordance catalog, we take a mixture of this this approximate column
density prior with a simple log-uniform prior over a wide dynamic
range:
\begin{equation}
  \label{lnhi_prior}
  p(\lni \given \Mdk{1}) =
  \alpha
  q(\lni = N)
  + (1 - \alpha)
  \mc{U}[20, 23].
\end{equation}
Here the mixture coefficient $\alpha = 0.9$ favors the
data-driven prior.  The upper limit of $\lni = 23$ is more than
sufficient to model all thus-far observed \acro{dla}s.  The final
prior $ p(\lni \given \Mdk{1})$ is shown in Figure
\ref{lni_kde}), showing the expected bias towards
smaller column densities.

\subsection{Approximating the model evidence}

Given our choice of parameter prior, the integral in
\eqref{dla_evidence} is unfortunately intractable, so we will result
to numerical integration.  In particular, we will use quasi-Monte
Carlo (\acro{qmc}) integration \citep{qmc}.  In \acro{qmc}, we select
$N$ parameter samples $\lbrace \theta_i \rbrace$, evaluate the model
likelihood given each of these samples, and approximate the integral
in \eqref{dla_evidence} by the sample mean:
\begin{equation}
  p(\vec{y} \given \vec{\lm}, \vec{\noise}, \zq, \Mdk{1})
  \approx
  \frac{1}{N}
  \sum_{i = 1}^N
  p(\vec{y} \given \vec{\lm}, \vec{\noise}, \zq, \theta_i, \Mdk{1}).
\end{equation}
This is the same estimator encountered in standard Monte Carlo
integration, which selects the samples by sampling independently from
the parameter prior $p(\theta \given \zq, \Mdk{1})$.  Quasi-Monte
Carlo differs from normal Monte Carlo integration in that the samples
$\lbrace \theta_i \rbrace$ are taken from a so-called
\emph{low-discrepancy sequence}, which guarantees the chosen samples
are evenly distributed, leading to faster convergence.  Here we used
$N = 10\,000$ samples generated from a scrambled Halton sequence
\citep{halton} to define our parameter samples.  Note that the Halton
sequence gives values approximately uniformly distributed on the unit
square $[0, 1]^2$, which (after a trivial transformation) agrees in
density with our redshift prior $p(\zd \given \zq, \Mdk{1})$, but not
our column density prior $p(\lni \given \Mdk{1})$.
To correct for this, we used inverse transform sampling to endow the
generated samples with the appropriate distribution.  For the inverse
transformation, we used the approximated inverse cumulative
distribution function corresponding to our prior in \eqref{lnhi_prior}.

Note that we can use the same technique to approximate other
quantities of interest.  For example, if we wish to restrict our
search to only \acro{dla}s with a certain minimum column density (for
example, $\lni > 22$), we can simply discard all parameter samples
out of range, giving an unbiased estimate of the desired integral:
\begin{multline}
  \int_{z_{\text{min}}}^{z_{\text{max}}}
  \int_{22}^\infty
  p(\vec{y} \given \vec{\lm}, \vec{\noise}, \zq, \theta, \Mdk{1})
  \\
  p(\theta \given \zq, \Mdk{1})
  \intd\zd \intd\lni.
\end{multline}
Note such estimators will, however, have higher variance due to the
discarded parameter samples.

\subsection{Multiple DLAs}

While the catalog we produce considers only one \acro{dla} per sightline,
our model for \acro{qso} sightlines containing \acro{dla}s can readily
model sightlines containing two or more intervening \acro{dla}s.
Again, given the parameters $(\zd, \nhi)$ of each absorber along the
line of sight, we may compute the corresponding absorption function
$a$ and compute the observation posterior as in
\eqref{absorption_posterior}.

Let $\Mdk{k}$ represent a model explaining exactly $k$ \acro{dla}s
along the line of sight; we described $\Mdk{1}$ in the preceding
sections.  The model evidence integral \eqref{dla_evidence} for
$\Mdk{k}$ remains the same; however $\theta$ will have dimension $2k$.
Furthermore, we must consider the joint parameter prior $p(\theta
\given \Mdk{k})$.

We propose a (nearly) independent prior between each set of \acro{dla}
parameters; the dependence will be discussed later.  Rather than
generating a $2k$-dimensional low-discrepancy sequence these
parameters, we propose a stepwise approach.  Given a spectrum, we
first use the $\Mdk{1}$ parameter samples $\{ \theta_i \}$ described
above to approximate the model evidence \eqref{dla_evidence}.  We can
then approximate the posterior distribution of the single-\acro{dla}
parameters by normalization:
\begin{equation}
  p(\theta \given \zq, \data, \Mdk{1})
  \propto
  p(\vec{y} \given \vec{\lm}, \vec{\noise}, \zq, \theta_i, \Mdk{1}).
\end{equation}
We may decompose the $\Mdk{2}$ parameters as $\theta = [\theta_1,
  \theta_2]\trans$, where each $\theta_i$ component describes a single
\acro{dla}.  We propose the following prior for the $\Mdk{2}$ model:
\begin{multline}
  p(\theta_1, \theta_2 \given \zq, \data, \Mdk{2})
  =
  \\
  p(\theta_1 \given \zq, \data, \Mdk{1})
  p(\theta_2 \given \zq, \Mdk{1}).
\end{multline}
That is, we use the posterior probabilities from the analysis of the
$\Mdk{1}$ model as the prior for the parameters of \emph{one} of the
\acro{dla}s when considering the $\Mdk{2}$ model.  The prior for the
parameters of the other \acro{dla} remains the noninformative prior as
described above.  For models $\Mdk{k}$ with $k > 2$, we apply a
similar approach, where we combine a noninformative prior for
$\theta_k$ with an informed prior for $\{ \theta_i \}_{i = 1}^{k -1}$:
\begin{multline}
  p\bigl(\{\theta\} \given \zq, \data, \Mdk{k}\bigr)
  =
  \\
  p\bigl(\{\theta_i\}_{i = 1}^{k - 1} \given \zq, \data, \Mdk{k - 1}\bigr)
  p(\theta_k \given \zq, \Mdk{1}).
\end{multline}

We do suggest injecting a small amount of dependence between the
\acro{dla} parameters; specifically, any samples where any pair of
$\zd$ values correspond to a small relative velocity should be
discarded to avoid samples describing two discrete \acro{dla}s in the
same region of space.

In practice, the above scheme can be realized by first processing the
spectrum with model $\Mdk{1}$; we then approximate the $\theta_1$
posterior by renormalizing.  To process the spectrum with model
$\Mdk{2}$, we loop through the generated samples, each providing
$\theta_2$.  For each sample, we sample a corresponding $\theta_1$
sample from the approximate posterior.  If the $\zd$ values are too
close, we discard the sample; otherwise, we have a valid $\theta$
sample with which to approximate the model evidence for $\Mdk{2}$.
For $\Mdk{k}$ we proceed in a similar way, using some minor
bookkeeping to approximate the $\{\theta_i\}_{i=1}^{k-1}$ posterior.

Note that the catalog we produce considers only $\Mdk{1}$ to maintain
statistical reliability with the low-\acro{SNR} spectra from
\acro{SDSS}; however, the techniques we introduce are not tied to any
particular source of data.

\section{Model Prior}

Given a set of spectroscopic observations $\data$, our ultimate goal
is to compute the probability the \acro{qso} sightline contains a
\acro{dla}: $p(\Md \given \data)$.  As described above, the Bayesian
model selection approach requires two components: the data-independent
prior probability that sightline contains a \acro{dla}, $\Pr(\Md)$,
and the ability to compute the ratio of model evidences $p(\data
\given \Mnd)$ and $p(\data \given \Md)$.  The \acro{gp} model built
above allows us to compute the latter; in this section we focus on the
former.

Only approximately 10\% of the \acro{qso} sightlines in the \acro{dr9}
release contain \acro{dla}s.  A simple approach to prior specification
would be to use a fixed value of $\Pr(\Md) \approx \nicefrac{1}{10}$.
However, it is less likely to observe a \acro{dla} in low-redshift
\acro{qso}s due to the wavelength coverage of the \acro{sdss} and
\acro{boss} spectrographs being limited to $\lo = 3\,800$\,\AA\/ and
$\lo = 3\,650$\,\AA, respectively, on the blue end.  Therefore, here
we will use a slightly more-sophisticated approach and derive a
redshift-dependent prior $\Pr(\Md \given \zq)$.

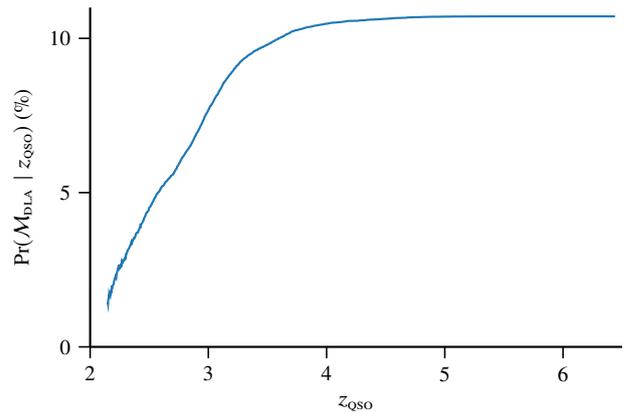
\begin{figure}
  \centering
  \input{figures/dla_prior.tex}
  \caption{The redshift-dependent model prior $\Pr(\Md \given \zq)$
    computed from the \acro{boss dr9} Lyman-$\alpha$ forest sample
    with parameter $z' = 30\,000\,\text{km\,s}^{-1}/c$.}
  \label{dla_prior}
\end{figure}

Our prior is simple and data driven.  Consider a \acro{qso} with
redshift $\zq$.  Let $N$ be the number of \acro{qso}s in the training
sample with redshift less than $\zq + z'$, where $z'$ is a small
constant.  Here we took $z' = 30\,000\,\text{km\,s}^{-1}/c$. Let $M$
be the number of the sightlines of these containing \acro{dla}s within
the range of quasar rest wavelengths we search here.  We define
\begin{equation}
  \label{model_prior}
  \Pr(\Md \given \zq)
  =
  \frac{M}{N}.
\end{equation}
The constant $z'$ serves to ensure that \acro{qso}s with very small
redshift have sufficient data for estimating the prior. The resulting
prior $\Pr(\Md \given \zq)$ calculated from the \acro{dr9} sample is
plotted in Figure \ref{dla_prior}.

If we wish to break down our \acro{dla} prior $\Pr(\Md \given \zq)$
into its component parts, for example to find $\Pr(\Mdk{1} \given
\zq)$, we assume that \acro{dla} occurrence is independent.  If
$\frac{M}{N}$ of sightlines contain at least one \acro{dla}, then
$\frac{M^2}{N^2}$ contain at least two \acro{dla}s, etc., giving:
\begin{equation}
  \Pr(\Mdk{k} \given \zq)
  \approx
  \biggl(\frac{M}{N}\biggr)^k - \biggl(\frac{M}{N}\biggr)^{k + 1}.
\end{equation}

\section{Example}

We have now developed all of the mathematical machinery required to
compute the posterior odds that a given quasar sightline contains an
intervening \acro{dla}, given a set of noisy spectroscopic
observations $\data$.  Briefly, we summarize the steps below, using
the example from Figure \ref{preprocessing}.  We limit this example to
searching for a single \acro{dla}, using only $\Mdk{1}$.

\begin{figure*}
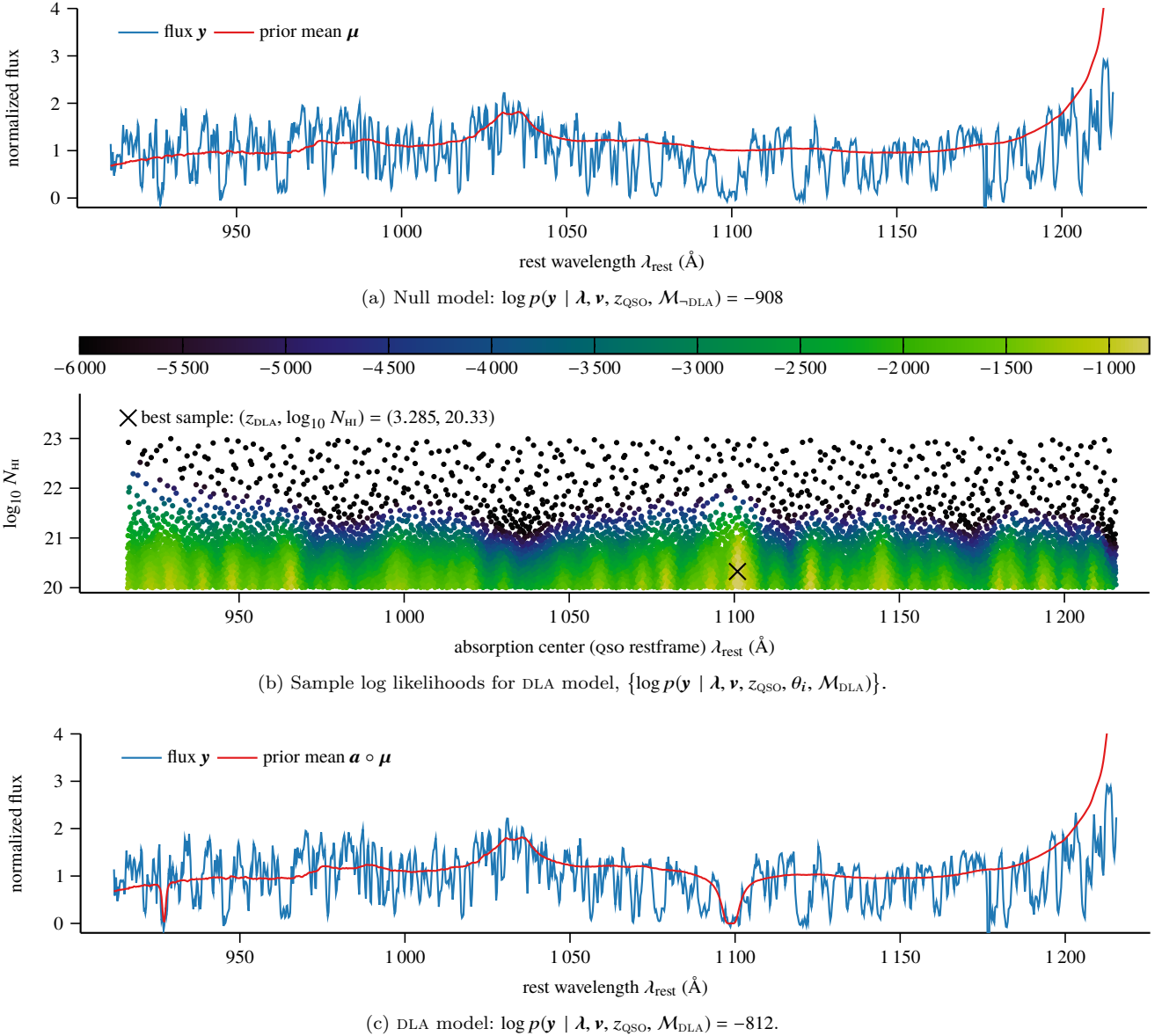

  \centering \subfloat[Null model: $\log p(\vec{y} \given \vec{\lm},
    \vec{\noise}, \zq, \Mnd) = -908$]{
    \includegraphics{tikz/example_null.pdf}
    \label{example_null}
  }
  \\
  \subfloat[Sample log likelihoods for \acro{dla} model, $\bigl\lbrace \log
      p(\vec{y} \given \vec{\lm}, \vec{\noise}, \zq, \theta_i, \Md) \bigr\rbrace.$]{
    \includegraphics{tikz/example_samples.pdf}
    \label{example_samples}
  }
  \\
  \subfloat[\acro{dla} model: $\log p(\vec{y} \given \vec{\lm}, \vec{\noise}, \zq, \Md) = -812$.]{
    \input{figures/example_dla.tex}
    \label{example_dla}
  }
  \caption{An illustration of the proposed \acro{dla}-finding
    procedure for the quasar sightline in Figure \ref{preprocessing}.
    \subref{example_null} shows the normalized flux with the prior
    \acro{gp} mean for our learned null model
    $\Mnd$. \subref{example_samples} shows the log likelihoods for
    each of the parameter samples used to approximate the marginal
    likelihood of our \acro{dla} model $\Md$. \subref{example_dla}
    shows the normalized flux with the prior \acro{gp} mean associated
    with the best \acro{dla} sample, $(\zd, \lni) = (3.285, 20.33)$.
    Notice the Lyman-$\beta$ absorption feature corresponding to this
    sample.}
  \label{example_processing}
\end{figure*}

Consider a quasar with known redshift $\zq$, and suppose we have made
spectroscopic observations of the object $\data = (\vec{\lm},
\vec{y})$, with known observation noise variance vector $\vec{\noise}$.
First, we compute the prior probability of the \acro{dla} model $\Md$,
$\Pr(\Md \given \zq)$ \eqref{model_prior}.  This allows us to compute
the prior odds in favor of the \acro{dla} model:
\begin{equation}
  \label{dla_prior_odds}
  \frac{\Pr(\Md \given \zq)}{\Pr(\Mnd \given \zq)}
  =
  \frac{\Pr(\Md \given \zq)}{1 - \Pr(\Md \given \zq)}.
\end{equation}
For our example, $\Pr(\Md \given \zq) = 10.3\%$, giving prior odds of
0.114 (9-to-1 against the \acro{dla} model).  Next, we compute the
Bayes factor in favor of the \acro{dla} model:
\begin{equation}
  \label{dla_bayes_factor}
  \frac{p(\vec{y} \given \vec{\lm}, \vec{\noise}, \zq, \Mdk{1})}
       {p(\vec{y} \given \vec{\lm}, \vec{\noise}, \zq, \Mnd)}.
\end{equation}
See \eqref{ndla_evidence} for how to compute the model likelihood for
the null model and \eqref{dla_evidence} for our approximation to the
\acro{dla} model likelihood.  For our \acro{dla} example, the Bayes
factor overwhelmingly supports the \acro{dla} model, with a value of
$\exp(96) \approx 5 \times 10^{41}$.  The computation of the Bayes
factor is illustrated in Figure \ref{example_processing}, which shows
the prior \acro{gp} mean for the null model (Figure
\ref{example_null}), the log likelihoods for the \acro{dla} model
parameter samples (Figure \ref{example_samples}), and the prior
\acro{gp} mean for the best \acro{dla} model parameter sample (Figure
\ref{example_dla}).

Finally, the posterior odds in favor of the sightline containing an
intervening \acro{dla} is the product of \eqref{dla_prior_odds} and
\eqref{dla_bayes_factor}.  In practice, due to the typically large
dynamic range of these quantities, it is numerically more convenient
to compute the log odds.  The log odds in favor of $\Md$ for the
example from Figure \ref{preprocessing} are 94\,nats,\footnote{Nats
  are the logarithmic unit analogous to bits or dex corresponding to
  the base of the natural logarithm.} and the probability of the
sightline containing a \acro{dla} is effectively unity. The \acro{dla}
parameter sample with the highest likelihood was $(\zd, \lni) =
(3.285, 20.33)$, closely matching the values reported in the
\acro{dla} concordance catalog $(\zd, \lni) = (3.283, 20.39)$.

We may also compute the evidence for higher-order models to derive a
posterior distribution over the number of \acro{DLA}s. In this case,
the log model evidence for models $\Mdk{2}$, $\Mdk{3}$, $\Mdk{4}$, and
$\Mdk{5}$, respectively, are $-840$, $-977$, $-1141$, and $-1385$;
incorporating the model prior \eqref{model_prior} and normalizing, the
single-\acro{DLA} model dominates.

\section{Catalog}

To verify the validity of our proposed method, we computed the
posterior probability of $\Md$ for 162\,858 quasar sightlines in the
\acro{dr12q} release of \acro{sdss---iii}.  Our catalog and data
products will be made available publicly at
\url{http://tiny.cc/dla_catalog_gp_dr12q}, and the code to reproduce
the entire catalog from raw \acro{SDSS} spectra will be posted under a
permissive license at
\url{https://github.com/rmgarnett/gp_dla_detection}.

The full \acro{dr12q} catalog contains 297\,301 quasars, to which we
applied the following cuts:
\begin{itemize}
\item
  We eliminate low-redshift ($\zq < 2.15$) quasars. A total of 113\,030
  quasars in \acro{dr12q} satisfy this removal condition.
\item
  We eliminate broad absorption line (\acro{bal}) quasars, determined
  by the \acro{bal} visual inspection survey results in the
  \texttt{BAL\_VI} field of the catalog.  A total of 29\,580 quasars
  in \acro{dr12q} satisfy this removal condition.
\item
  We eliminate quasars that we cannot normalize due to no non-masked
  pixels in the range $\lr \in [1310, 1325]\,\text{\AA}$.  A total of
  125 quasars in \acro{dr12q} satisfy this removal condition.
\item
  We eliminate quasars that have fewer than 200 non-masked pixels in
  the range $\lr \in [911.75, 1216.75]\,\text{\AA}$.  A total of 35
  quasars in \acro{dr12q} satisfy this removal condition.
\end{itemize}

For each of the remaining spectra, we computed the posterior
probability of the $\Mnd$ and $\Mdk{1}$ models, given the
observations, as described in the previous sections.  We produce a
full catalog of our results, comprising two tables, the first rows of
which are shown in Tables \ref{los_table} and \ref{dla_table}.  The
full catalog will be available electronically alongside this
manuscript.

When computing the likelihoods for the \acro{dla} model, we convolved
the computed Voigt profile corresponding to each parameter sample with
a Gaussian broadening profile with $\text{\acro{fwhm}} = c / 2000 =
150\,\text{km\,s}^{-1}$, corresponding to the \acro{BOSS} instrument's
spectral resolution of $R \approx 2\,000$.

For each spectrum analyzed, the results catalog includes:
\begin{itemize}
\item
  the range of redshifts searched for \acro{dla}s, $[z_\text{min},
    z_\text{max}]$,
\item
  the log model prior, $\log \Pr(\model \given \zq)$, for each model
  considered,
\item
  the log model evidence, $\log p(\vec{y} \given \vec{\lm}, \vec{\noise},
  \zq, \model)$, for each model considered,
\item
  the model posterior, $\Pr(\model \given \data, \zq)$, and
\item
  the \acro{map} estimates of the $\Mdk{1}$ model's parameters.
\end{itemize}

\begin{table*}
  \centering
  \caption{The 297\,301 objects in the \acro{sdss--iii dr12q} catalog,
    and the results of our cuts.}
  \begin{tabular}{llllllllll}
  \toprule
  thing id &
  \acro{sdss} name &
  plate &
  \acro{mjd} &
  fiber \acro{id} &
  right ascension &
  declination &
  $\zq$ &
  \acro{snr} &
  cut flags \\
  \midrule
  268514930 &
  000000.45+174625.4 &
  6173 &
  56238 &
  0528 &
  0.0018983 &
  +17.7737391 &
  2.3091 &
  0.7795 &
  0000 \\
  \multicolumn{10}{l}{(297\,300 rows removed)}\\
  \bottomrule
  \end{tabular}
  \label{los_table}
\end{table*}

\begin{table*}
  \raggedright
  \caption{The 162\,858 objects in the \acro{sdss--iii dr12q} catalog
    processed by our proposed \acro{gp} \acro{dla} detection method,
    and a summary of derived quantities of interest.  Note: the first
    nine columns match Table \ref{los_table} for the included
    objects (those with all cut flags equal to zero).}
  \begin{tabular}{lllllll}
    \toprule
    &
    \multicolumn{2}{c}{search range} &
    \multicolumn{2}{c}{model prior} &
    \multicolumn{2}{c}{model evidence} \\
    \cmidrule(r){2-3}
    \cmidrule{4-5}
    \cmidrule(l){6-7}
    thing id &
    $z_{\text{min}}$ &
    $z_{\text{max}}$ &
    $\log \Pr(\Mnd \given \zq)$ &
    $\log \Pr(\Md \given \zq)$ &
    $\log p(\vec{y} \given \vec{\lm}, \vec{\noise}, \zq, \Mnd)$ &
    $\log p(\vec{y} \given \vec{\lm}, \vec{\noise}, \zq, \Mdk{1})$ \\
    \midrule
    268514930 &
    1.9654 &
    2.2989 &
    -0.03081 &
    -3.49537 &
    -1.04359e+03 &
    -1.04256e+03 \\
    \multicolumn{7}{l}{(162\,857 rows removed)}\\
    \bottomrule
  \end{tabular}
  \\\vspace*{1ex}(cont.)
  \begin{tabular}{llll}
    \toprule
    \multicolumn{2}{c}{model posterior} &
    \multicolumn{2}{c}{$\argmax_{\theta} p(\vec{y} \given \vec{\lm}, \vec{\noise}, \zq, \Mdk{1})$} \\
    \cmidrule(r){1-2}
    \cmidrule(l){3-4}
    $\Pr(\Mnd \given \data, \zq)$ &
    $\Pr(\Md \given \data, \zq)$ &
    $\zd$ &
    $\lni$ \\
    \midrule
    9.19661e-001 &
    8.03389e-002 &
    2.2160 &
    20.0077 \\
    \multicolumn{2}{l}{(162\,857 rows removed)}\\
    \bottomrule
  \end{tabular}
  \label{dla_table}
\end{table*}

\subsection{Running time}

The running time of our approach allows it to easily scale to
extremely large surveys and/or larger sample sizes.  Our
implementation is able to compute the model posterior over $\Mnd$ and
$\Mdk{1}$ in 0.5--2 seconds per spectrum on a standard Apple iMac
desktop machine.  For each spectrum we must compute 10\,001 log
likelihoods of the form \eqref{ndla_evidence} (one for
\eqref{ndla_evidence} and 10\,000 for the $\Mdk{1}$ model
\eqref{dla_evidence}); however, the low-rank structure of our
covariance allows us to compute each rapidly using the identities in
\eqref{efficient_inv} and \eqref{efficient_det}.

\subsection{Analysis of results}

To evaluate our results, we examined the ranking induced on the
sightlines by the log posterior odds in favor of the \acro{dla} model
$\Md$.  If our method is performing correctly, true \acro{dla}s should
be at the top of this list, above the non-\acro{dla}-containing
sightlines.  To visualize the quality of our ranking, we created a
receiver--operating characteristic (\acro{roc}) plot, which, for every
possible threshold on the log posterior odds, plots the false positive
rate (portion of non-\acro{dla}s with larger posterior odds) against
the true positive rate (portion of \acro{dla}s with larger posterior
odds).

\begin{figure}
  \centering
  \input{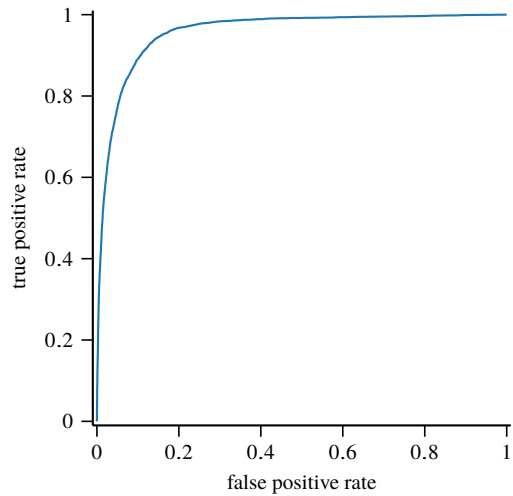}
  \caption{The \acro{roc} plot for the ranking of the 54\,360
    \acro{qso} sightlines contained in the \acro{boss dr9}
    Lyman-$\alpha$ forest sample (that were not filtered by our cuts),
    induced by the log posterior odds of containing a
    \acro{dla}. Ground-truth labelings were derived from the
    corresponding \acro{dla} concordance catalog.}
  \label{rocs}
\end{figure}

Notice that creating an \acro{roc} plot requires knowledge of the
ground-truth labels for each of our objects, which of course we do not
have. Instead, we use the \acro{dla} concordance catalog distributed
with the \acro{boss dr9} Lyman-$\alpha$ forest catalog as surrogate
ground truth, and restrict our analysis to lines of sight that both
appear in that catalog and were not removed by our cuts.  A total of
54\,360 objects comprise this intersection (99.9\% of the catalog).
The resulting \acro{roc} plot is displayed in Figure \ref{rocs}.  The
top 1\%, 2\%, 5\%, 10\%, and 20\% of our ranked list, respectively,
recover 42.7\%, 57.5\%, 77.0\%, 89.1\%, and 96.8\% of the \acro{dla}s
listed in the concordance catalog.  Thus even presorting the list by
the posterior probability of $\Md$ can dramatically speed up visual
inspection.

A useful summary of the \acro{roc} plot is the area under the curve
(\acro{auc}) statistic.  The \acro{auc} has a natural interpretation:
if we select a positive example and a negative example uniformly at
random from those available, the \acro{auc} is the probability that
the positive example would be ranked higher than the negative example.
For the \acro{dr9} \acro{dla} concordance catalog surrogate, our
\acro{auc} was 95.8\%.  Clearly our approach is effective at
identifying \acro{dla}s.

\begin{figure*}
  \centering
  \input{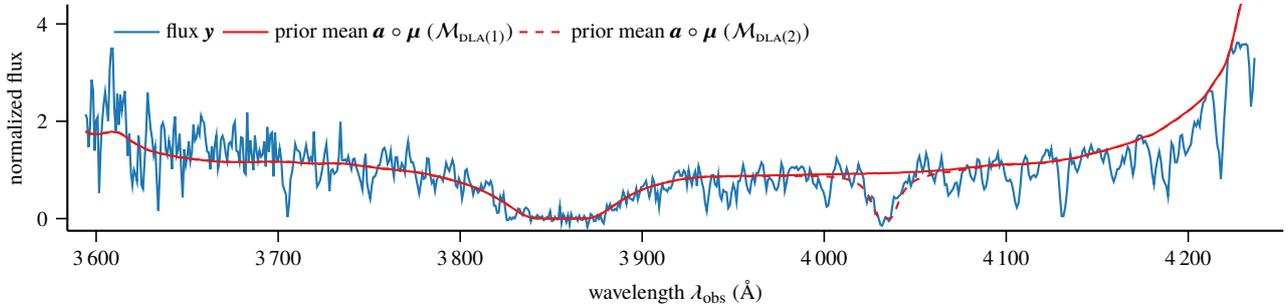}
  \caption{The spectrum appearing in the \acro{boss dr9}
    Lyman-$\alpha$ forest sample, not contained in the corresponding
    \acro{dla} concordance catalog, with the highest posterior
    probability of containing an \acro{dla} according to our model.
    The object is \acro{sdss} 170023.94+205331.7, (plate, \acro{mjd},
    fiber) = (4175, 55680, 764), $\zq = 2.4852$. We overwhelmingly
    believe there to be two \acro{dla}s along the line of sight;
    $\Pr(\Mdk{2} \given \data, \zq) = 1 - 2.1 \times 10^{-22}$.  The
    prior means corresponding to the highest-likelihood parameter
    sample for $\Mdk{1}$ and $\Mdk{2}$ are plotted, corresponding to
    $(\zd, \lni) = (2.1717, 21.414)$ and $(\zd, \lni) = \{(2.1715,
    21.519), (2.3179, 20.075)\}$.}
  \label{false_positive}
\end{figure*}

An important caveat to all of the results above is that none of the
surrogates is likely to represent the true ground truth, and many
`false positive' sightlines could in fact contain as-yet
undiscovered \acro{dla}s.  Figure \ref{false_positive} gives an
example of such a `false positive,' showing the spectrum not
contained in the \acro{dla} concordance catalog that we rank the
highest according to our model posterior ranking.

In fact, this spectrum appears to contain two \acro{dla}s along the
line of sight.  As a demonstration of our ability to detect multiple
\acro{dla}s, we reprocessed this spectrum using the two-\acro{dla}
model $\Mdk{2}$. The data overwhelmingly support $\Mdk{2}$ over either
$\Mdk{1}$ or $\Mnd$; $\Pr(\Mdk{2} \given \data, \zq) = 1 - 2.1 \times
10^{-22}$.  Despite this line of sight not appearing in the
\acro{dr9q} \acro{dla} concordance catalog, we do note that it was
flagged during the \acro{dr12q} visual inspection.

We have visually inspected several of these `confident false
positives;' of the top-30 such examples, 29 appear to contain large
absorption features at the location indicated by the maximum
likelihood parameter sample.  The other is a very low \acro{SNR}
spectrum that appears not to have been normalized satisfactorily.

The observation corresponding to our most-egregious false negative,
that is, the spectrum flagged in the concordance catalog that we
assign the greatest confidence to being \acro{DLA}-free, is
\acro{SDSS} 081807.84+520935.1. There is a \acro{DLA} along this line
of sight, but outside the range of redshifts we search.

\begin{figure*}
  \centering
  \subfloat[]{
    \input{figures/hat_z_dlas_kde.tex}
    \label{hat_z_dlas_kde}
  }
  \subfloat[]{
    \input{figures/hat_log_nhis_kde.tex}
    \label{hat_log_nhis_kde}
  }
  \caption{Kernel density estimate of the difference between the
    \acro{map} estimates of the \acro{dla} parameters $(\zd, \lni)$
    for \acro{dla}s listed in the \acro{boss dr9} Lyman-$\alpha$
    forest sample, against the catalog-reported values.}
\end{figure*}
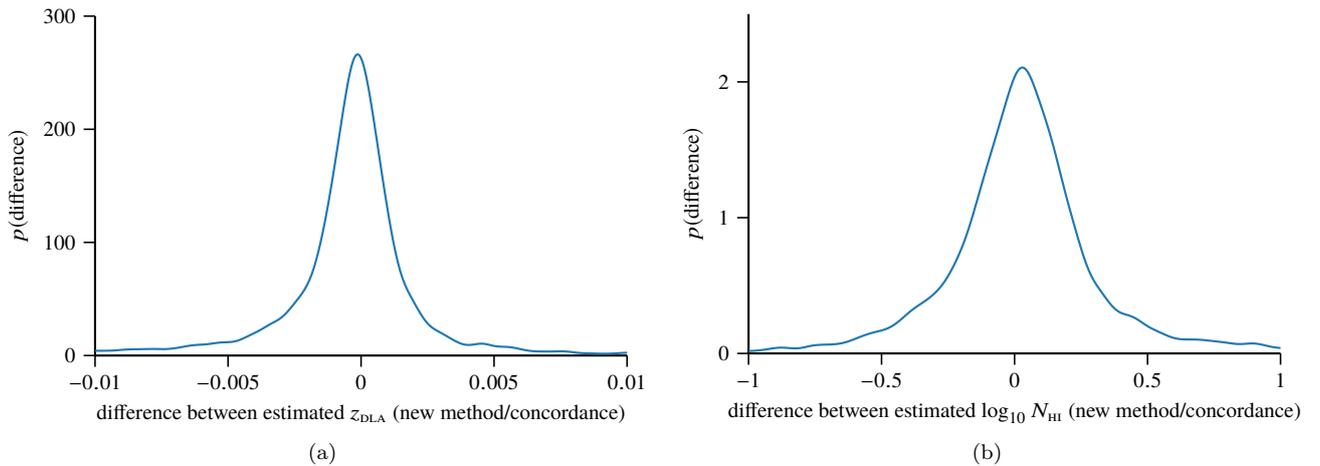

\subsection{DLA parameter estimation analysis}

The main goal of our \acro{dla}-detection method is to rank \acro{qso}
sightlines by their probability of containing \acro{dla}s.  The
computation of the evidence of our \acro{dla} model $\Md$ requires
averaging over many samples of the \acro{dla} parameters $(\zd,
\lni)$.  We may use these samples to further derive point estimates of
these parameters for presumed \acro{dla}s, if desired.  The simplest
approach is to report the sample with the highest likelihood:
\begin{equation}
  \argmax_{\theta_i}
  p(\vec{y} \given \vec{\lm}, \vec{\noise}, \zq, \theta_i, \Md);
\end{equation}
this represents the \emph{maximum a posteriori} (\acro{map}) estimate
of the parameters. We analyze the behavior of the \acro{map} estimate
by comparing it with the reported values in the \acro{dr9} concordance
\acro{dla} catalog.

The \acro{map} estimates of the absorber redshift $\zd$ are remarkably
close to the catalog figures.  The median difference between the two
is $-2.7 \times 10^{-4}$ ($-80.6\,\text{km\,s}^{-1}$) and the
interquartile range is $2.5 \times 10^{-3}$
($742\,\text{km\,s}^{-1}$). Figure \ref{hat_z_dlas_kde} displays a
kernel density estimate of the distribution of the difference between
the \acro{map} $\zd$ estimates and the values reported in the
concordance catalog.

Examining the larger `errors' in our estimation of $\zd$, we make an
interesting observation that several $\zd$ values reported in the
concordance catalog correspond exactly to the central wavelength of
Lyman-$\beta$ absorption for our redshift estimates.  There does not
seem to be an obvious pattern in the reverse direction, indicating
that our method is less susceptible that previous techniques to
mistaking Lyman-$\beta$ absorption for Lyman-$\alpha$ absorption.
Unlike previous approaches, which involve Voigt profile fitting to
Lyman-$\alpha$ absorption only, we model the entire spectrum jointly,
as well as the entire absorption profile corresponding to a given set
of object parameters.  Samples incorrectly setting $\zd$ corresponding
to a Lyman-$\beta$ absorption feature should explain the observed
spectrum worse than a sample setting $\zd$ corresponding to a
Lyman-$\alpha$ absorption feature, which in our setup can better
explain both the larger Lyman-$\alpha$ absorption as well as the
corresponding Lyman-$\beta$ feature.

The \acro{map} estimates of the log column density $\lni$ show more
variation with the catalog figures.  The median difference between the
two is quite small, only $0.030$\,dex.  The interquartile range,
however, is nontrivial at approximately $0.27$\,dex. Figure
\ref{hat_log_nhis_kde} displays a kernel density estimate of the
distribution of the difference between the \acro{map} $\lni$ values
versus the values reported in the concordance catalog.

In practice, for suspected \acro{dla}s, we suggest standard procedures
for Voigt-profile fitting, if an accurate estimate of the parameters
is desired.  Our \acro{dla}-detection procedure is primarily concerned
with the evidence contained in the entire set of parameter samples,
and the \acro{map} estimate carries no special significance.  In
particular, several parameter ranges might have large likelihood,
corresponding to several potential absorption features.  The
\acro{map} estimate alone cannot convey such information.

\section*{Acknowledgements}
RG was supported by the National Science Foundation under Award Number
IIA-1355406.  SH and JS was supported by the US Department of Energy
under Award Number DOE-DESC0011114. SB was supported by a McWilliams
Fellowship from Carnegie Mellon University and by NASA through
Einstein Postdoctoral Fellowship Award Number PF5-160133.

\bibliographystyle{mnras}
\bibliography{astro}
\bsp
\label{lastpage}

\end{document}

%% file: figures/missing_pixels.tex
%
\tikzsetnextfilename{missing_pixels}
\definecolor{mycolor1}{rgb}{0.12157,0.47059,0.70588}%
\begin{tikzpicture}

\begin{axis}[%
width=\widefigurewidth,
height=\widefigureheight,
at={(0\widefigurewidth,0\widefigureheight)},
scale only axis,
xmin=901.75,
xmax=1225.75,
xtick={ 900,  950, 1000, 1050, 1100, 1150, 1200, 1250},
xlabel={rest wavelength $\lambda_{\text{rest}}$ (\AA)},
ymin=0,
ymax=80,
ylabel={portion missing (\%)},
axis background/.style={fill=white},
axis x line*=bottom,
axis y line*=left,
extra x ticks={911.8,973,1026,1215.67}, extra x tick labels={{Ly\,$\infty$},{Ly\,$\gamma$},{Ly\,$\beta$},{Ly\,$\alpha$}},
]
\addplot [color=mycolor1,solid,forget plot]
  table[row sep=crcr]{%
911.75	79.4250205700672\\
912	79.3703717348856\\
912.25	79.3243193456877\\
912.5	79.2758108290658\\
912.75	79.2383548858515\\
913	79.1867762099498\\
913.25	79.137039629616\\
913.5	79.0891451448501\\
913.75	79.0498471060679\\
914	79.0056368124378\\
914.25	78.9516020091122\\
914.5	78.8969531739307\\
914.75	78.8545849758685\\
915	78.8054624273907\\
915.25	78.7618661656167\\
915.5	78.709059426003\\
915.75	78.6611649412371\\
916	78.6163406157511\\
916.25	78.5653759717054\\
916.5	78.5199376143634\\
916.75	78.4732711933095\\
917	78.41678026256\\
917.25	78.3707278733621\\
917.5	78.3326578982918\\
917.75	78.2878335728058\\
918	78.2423952154638\\
918.25	78.1865183165703\\
918.5	78.1496764052119\\
918.75	78.1079222390058\\
919	78.0772206462071\\
919.25	78.0280980977293\\
919.5	77.9869579633791\\
919.75	77.9384494467573\\
920	77.8917830257034\\
920.25	77.8469587002174\\
920.5	77.7978361517395\\
920.75	77.7450294121259\\
921	77.6946787999361\\
921.25	77.6467843151703\\
921.5	77.6019599896843\\
921.75	77.5534514730624\\
922	77.5037148927286\\
922.25	77.4656449176583\\
922.5	77.4085399550529\\
922.75	77.3526630561594\\
923	77.3109088899532\\
923.25	77.2611723096194\\
923.5	77.203453315158\\
923.75	77.1598570533839\\
924	77.1064362819143\\
924.25	77.0511734148768\\
924.5	77.0167876309423\\
924.75	76.9535423497771\\
925	76.8964373871716\\
925.25	76.8509990298297\\
925.5	76.8037185769198\\
925.75	76.7509118373061\\
926	76.6987191295484\\
926.25	76.6465264217908\\
926.5	76.6047722555846\\
926.75	76.5495093885471\\
927	76.4991587763573\\
927.25	76.4531063871594\\
927.5	76.3990715838338\\
927.75	76.349949035356\\
928	76.2885458497587\\
928.25	76.2498618428324\\
928.5	76.1903007528031\\
928.75	76.1417922361812\\
929	76.1031082292549\\
929.25	76.0533716489211\\
929.5	75.9938105588918\\
929.75	75.9471441378379\\
930	75.8974075575041\\
930.25	75.8378464674747\\
930.5	75.7813555367252\\
930.75	75.740829434231\\
931	75.696619140601\\
931.25	75.6481106239792\\
931.5	75.5959179162215\\
931.75	75.5566198774392\\
932	75.5111815200973\\
932.25	75.4639010671874\\
932.5	75.4092522320058\\
932.75	75.360129683528\\
933	75.3079369757703\\
933.25	75.2667968414201\\
933.5	75.2225865477901\\
933.75	75.1777622223041\\
934	75.1323238649621\\
934.25	75.0960959854597\\
934.5	75.0383769909983\\
934.75	74.9923246018003\\
935	74.9346056073389\\
935.25	74.8928514411328\\
935.5	74.8400447015191\\
935.75	74.7841678026256\\
936	74.730747031156\\
936.25	74.684694641958\\
936.5	74.640484348328\\
936.75	74.5876776087143\\
937	74.5318007098208\\
937.25	74.4888184799027\\
937.5	74.4347836765771\\
937.75	74.3776787139717\\
938	74.3273281017819\\
938.25	74.2726792666003\\
938.5	74.2253988136905\\
938.75	74.1738201377888\\
939	74.1191713026072\\
939.25	74.0737329452652\\
939.5	74.0166279826597\\
939.75	73.9736457527417\\
940	73.9282073953997\\
940.25	73.876014687642\\
940.5	73.8262781073082\\
940.75	73.7783836225423\\
941	73.7323312333444\\
941.25	73.6850507804345\\
941.5	73.6396124230925\\
941.75	73.5929460020386\\
942	73.5438234535608\\
942.25	73.5045254147785\\
942.5	73.4572449618686\\
942.75	73.403824190399\\
943	73.3553156737772\\
943.25	73.3123334438591\\
943.5	73.2650529909492\\
943.75	73.2147023787594\\
944	73.1526851613062\\
944.25	73.1047906765403\\
944.5	73.0495278095028\\
944.75	72.9936509106092\\
945	72.9475985214113\\
945.25	72.8947917817976\\
945.5	72.8389148829041\\
945.75	72.7879502388584\\
946	72.7412838178045\\
946.25	72.6866349826229\\
946.5	72.6319861474413\\
946.75	72.5785653759717\\
947	72.5325129867738\\
947.25	72.4797062471601\\
947.5	72.4207591889867\\
947.75	72.3599700352454\\
948	72.3065492637758\\
948.25	72.2543565560181\\
948.5	72.1966375615567\\
948.75	72.1413746945192\\
949	72.0922521460413\\
949.25	72.0523400754031\\
949.5	72.0124280047649\\
949.75	71.9620773925751\\
950	71.8902356654263\\
950.25	71.8276044161171\\
950.5	71.7809379950632\\
950.75	71.7133944909062\\
951	71.6691841972762\\
951.25	71.6151493939506\\
951.5	71.5537462083533\\
951.75	71.5033955961635\\
952	71.4567291751096\\
952.25	71.4069925947758\\
952.5	71.3455894091785\\
952.75	71.2823441280134\\
953	71.2313794839676\\
953.25	71.1834849992018\\
953.5	71.1288361640202\\
953.75	71.0821697429663\\
954	71.0189244618011\\
954.25	70.9667317540434\\
954.5	70.9151530781417\\
954.75	70.8703287526557\\
955	70.8095395989144\\
955.25	70.7475223814612\\
955.5	70.6941016099915\\
955.75	70.63945277481\\
956	70.5768215255007\\
956.25	70.5154183399035\\
956.5	70.4699799825615\\
956.75	70.4079627651083\\
957	70.3551560254946\\
957.25	70.3066475088728\\
957.5	70.256296896683\\
957.75	70.2022620933574\\
958	70.1519114811676\\
958.25	70.0923503911383\\
958.5	70.0254209188373\\
958.75	69.96401773324\\
959	69.9173513121861\\
959.25	69.8645445725724\\
959.5	69.8019133232632\\
959.75	69.739282073954\\
960	69.6821771113485\\
960.25	69.6201598938953\\
960.5	69.5550725171622\\
960.75	69.4893711085731\\
961	69.4273538911199\\
961.25	69.3634945780987\\
961.5	69.2984072013656\\
961.75	69.2370040157683\\
962	69.1792850213069\\
962.25	69.1227940905574\\
962.5	69.041127853713\\
962.75	68.9852509548195\\
963	68.9220056736544\\
963.25	68.8513920102175\\
963.5	68.7930589839001\\
963.75	68.7371820850066\\
964	68.6641122941458\\
964.25	68.6008670129806\\
964.5	68.5382357636714\\
964.75	68.4682361320905\\
965	68.4049908509253\\
965.25	68.3552542705916\\
965.5	68.2932370531383\\
965.75	68.2361320905328\\
966	68.1477115032728\\
966.25	68.0770978398359\\
966.5	68.0113964312469\\
966.75	67.9475371182257\\
967	67.8849058689165\\
967.25	67.8320991293028\\
967.5	67.7651696570018\\
967.75	67.6976261528448\\
968	67.6386790946714\\
968.25	67.5692934949465\\
968.5	67.4986798315097\\
968.75	67.4317503592086\\
969	67.3764874921711\\
969.25	67.312014147294\\
969.5	67.243856611281\\
969.75	67.176313107124\\
970	67.1112257303909\\
970.25	67.0485944810817\\
970.5	66.9804369450687\\
970.75	66.9165776320475\\
971	66.8539463827383\\
971.25	66.7919291652851\\
971.5	66.7280698522639\\
971.75	66.6507018384114\\
972	66.5880705891022\\
972.25	66.5285094990728\\
972.5	66.4542116445001\\
972.75	66.3915803951909\\
973	66.3246509228899\\
973.25	66.2601775780127\\
973.5	66.186493755296\\
973.75	66.119564282995\\
974	66.0440383647104\\
974.25	65.9721966375616\\
974.5	65.9071092608285\\
974.75	65.8383376929595\\
975	65.7720222525145\\
975.25	65.6989524616537\\
975.5	65.6221984796571\\
975.75	65.5534269117882\\
976	65.4957079173268\\
976.25	65.4330766680175\\
976.5	65.3575507497329\\
976.75	65.2795687040244\\
977	65.2064989131636\\
977.25	65.1334291223029\\
977.5	65.0542190128824\\
977.75	64.9811492220216\\
978	64.9105355585848\\
978.25	64.8472902774196\\
978.5	64.7754485502708\\
978.75	64.704220854978\\
979	64.635449287109\\
979.25	64.5531690184087\\
979.5	64.4899237372435\\
979.75	64.4162399145268\\
980	64.3358017413943\\
980.25	64.2633459823896\\
980.5	64.1896621596728\\
980.75	64.1233467192278\\
981	64.0539611195029\\
981.25	63.983961487922\\
981.5	63.9078215377814\\
981.75	63.827383364649\\
982	63.7524714782203\\
982.25	63.691682324479\\
982.5	63.6112441513466\\
982.75	63.5492269338933\\
983	63.4718589200408\\
983.25	63.3908067150524\\
983.5	63.31036854192\\
983.75	63.2379127829152\\
984	63.1771236291739\\
984.25	63.1034398064572\\
984.5	63.0211595377568\\
984.75	62.9554581291677\\
985	62.8793181790271\\
985.25	62.8056343563104\\
985.5	62.7331785973056\\
985.75	62.6490562330374\\
986	62.5643198369131\\
986.25	62.4930921416203\\
986.5	62.4169521914797\\
986.75	62.3322157953555\\
987	62.2456373036633\\
987.25	62.1719534809466\\
987.5	62.094585467094\\
987.75	62.0251998673691\\
988	61.9515160446524\\
988.25	61.8778322219357\\
988.5	61.802920335507\\
988.75	61.7298505446463\\
989	61.646342212234\\
989.25	61.5616058161097\\
989.5	61.4934482800968\\
989.75	61.4111680113964\\
990	61.3295017745521\\
990.25	61.2478355377077\\
990.5	61.1888884795343\\
990.75	61.1084503064019\\
991	61.0421348659568\\
991.25	60.9616966928244\\
991.5	60.8720480418524\\
991.75	60.7940659961439\\
992	60.7166979822913\\
992.25	60.6374878728708\\
992.5	60.5564356678824\\
992.75	60.4741553991821\\
993	60.3986294808975\\
993.25	60.320033403333\\
993.5	60.2352970072087\\
993.75	60.1536307703644\\
994	60.07196453352\\
994.25	59.9945965196674\\
994.5	59.9160004421029\\
994.75	59.8318780778347\\
995	59.7545100639821\\
995.25	59.6654754448661\\
995.5	59.5739846983262\\
995.75	59.4990728118975\\
996	59.4278451166047\\
996.25	59.3418806567685\\
996.5	59.262056515492\\
996.75	59.1883726927753\\
997	59.1030222647951\\
997.25	59.0078473271193\\
997.5	58.9145144850115\\
997.75	58.8310061525992\\
998	58.7505679794668\\
998.25	58.6652175514866\\
998.5	58.5976740473296\\
998.75	58.5190779697651\\
999	58.4361836692088\\
999.25	58.362499846492\\
999.5	58.2722371636641\\
999.75	58.1973252772354\\
1000	58.1248695182306\\
1000.25	58.0235542619951\\
1000.5	57.9406599614388\\
1000.75	57.8559235653146\\
1001	57.7601345957828\\
1001.25	57.6790823907944\\
1001.5	57.6121529184934\\
1001.75	57.5268024905132\\
1002	57.4439081899569\\
1002.25	57.3640840486804\\
1002.5	57.2676810472927\\
1002.75	57.1823306193125\\
1003	57.0994363187562\\
1003.25	57.0110157314962\\
1003.5	56.9170688575323\\
1003.75	56.8163676331528\\
1004	56.7144383450613\\
1004.25	56.6358422674968\\
1004.5	56.5627724766361\\
1004.75	56.4835623672156\\
1005	56.3791769517003\\
1005.25	56.2895283007282\\
1005.5	56.2078620638839\\
1005.75	56.1206695403358\\
1006	56.0408453990593\\
1006.25	55.9499686843753\\
1006.5	55.8474253644279\\
1006.75	55.7540925223201\\
1007	55.66567193506\\
1007.25	55.5741811885201\\
1007.5	55.4919009198197\\
1007.75	55.4010242051358\\
1008	55.3187439364354\\
1008.25	55.216814648344\\
1008.5	55.1204116469562\\
1008.75	55.0246226774245\\
1009	54.9398862813003\\
1009.25	54.8385710250648\\
1009.5	54.7501504378047\\
1009.75	54.6605017868327\\
1010	54.5745373269965\\
1010.25	54.4818185167447\\
1010.5	54.3927838976286\\
1010.75	54.2920826732491\\
1011	54.2005919267092\\
1011.25	54.1109432757371\\
1011.5	54.0114701150696\\
1011.75	53.9076987314102\\
1012	53.821120239718\\
1012.25	53.7271733657542\\
1012.5	53.6369106829262\\
1012.75	53.5570865416498\\
1013	53.4606835402621\\
1013.25	53.3759471441378\\
1013.5	53.2875265568778\\
1013.75	53.1911235554901\\
1014	53.0947205541023\\
1014.25	53.0020017438505\\
1014.5	52.9080548698866\\
1014.75	52.831914919746\\
1015	52.721389185671\\
1015.25	52.6329685984109\\
1015.5	52.5445480111508\\
1015.75	52.4561274238908\\
1016	52.3591103906471\\
1016.25	52.2639354529713\\
1016.5	52.1681464834396\\
1016.75	52.0870942784512\\
1017	51.9974456274792\\
1017.25	51.8998145623795\\
1017.5	51.797885274288\\
1017.75	51.7070085596041\\
1018	51.6259563546157\\
1018.25	51.5332375443638\\
1018.5	51.4417467978239\\
1018.75	51.348413955716\\
1019	51.2464846676246\\
1019.25	51.1599061759324\\
1019.5	51.0628891426887\\
1019.75	50.9750825872846\\
1020	50.8835918407447\\
1020.25	50.787188839357\\
1020.5	50.6699087548662\\
1020.75	50.5802601038942\\
1021	50.472804529099\\
1021.25	50.3764015277113\\
1021.5	50.2885949723072\\
1021.75	50.1964901939113\\
1022	50.1000871925235\\
1022.25	49.9938596814403\\
1022.5	49.9048250623242\\
1022.75	49.8219307617679\\
1023	49.7267558240922\\
1023.25	49.6260545997126\\
1023.5	49.5308796620369\\
1023.75	49.4485993933365\\
1024	49.3411438185413\\
1024.25	49.2490390401454\\
1024.5	49.1722850581488\\
1024.75	49.0795662478969\\
1025	48.9776369598055\\
1025.25	48.8855321814096\\
1025.5	48.7891291800219\\
1025.75	48.6939542423461\\
1026	48.606147686942\\
1026.25	48.4968500165789\\
1026.5	48.3955347603434\\
1026.75	48.3193948102028\\
1027	48.2291321273748\\
1027.25	48.1370273489789\\
1027.5	48.0522909528546\\
1027.75	47.9565019833229\\
1028	47.8515025359516\\
1028.25	47.7544855027079\\
1028.5	47.6629947561679\\
1028.75	47.5536970858048\\
1029	47.4628203711209\\
1029.25	47.3529086689017\\
1029.5	47.2638740497857\\
1029.75	47.1686991121099\\
1030	47.0594014417468\\
1030.25	46.9629984403591\\
1030.5	46.8776480123789\\
1030.75	46.7781748517113\\
1031	46.6774736273318\\
1031.25	46.583526753368\\
1031.5	46.4865097201243\\
1031.75	46.4011592921441\\
1032	46.3041422589004\\
1032.25	46.2040550663768\\
1032.5	46.1088801287011\\
1032.75	46.0130911591693\\
1033	45.9154600940697\\
1033.25	45.8080045192745\\
1033.5	45.7085313586069\\
1033.75	45.615812548355\\
1034	45.5298480885188\\
1034.25	45.4230065455796\\
1034.5	45.3137088752164\\
1034.75	45.2222181286765\\
1035	45.1374817325523\\
1035.25	45.031254221469\\
1035.5	44.9262547740977\\
1035.75	44.8378341868376\\
1036	44.7420452173059\\
1036.25	44.649326407054\\
1036.5	44.5388006729789\\
1036.75	44.4202925247762\\
1037	44.3232754915325\\
1037.25	44.2195041078731\\
1037.5	44.1065222463741\\
1037.75	44.0027508627148\\
1038	43.8959093197755\\
1038.25	43.79459406354\\
1038.5	43.6945068710165\\
1038.75	43.593191614781\\
1039	43.4924903904015\\
1039.25	43.4003856120056\\
1039.5	43.2910879416424\\
1039.75	43.1971410676786\\
1040	43.0878433973155\\
1040.25	42.9822299180882\\
1040.5	42.8612656424615\\
1040.75	42.7654766729298\\
1041	42.6451264291591\\
1041.25	42.5370568225079\\
1041.5	42.4381976936964\\
1041.75	42.3282859914772\\
1042	42.2103918751305\\
1042.25	42.1053924277592\\
1042.5	41.9893404069803\\
1042.75	41.8665340357858\\
1043	41.749867983151\\
1043.25	41.6405703127878\\
1043.5	41.5202200690172\\
1043.75	41.4023259526704\\
1044	41.2881160274595\\
1044.25	41.1886428667919\\
1044.5	41.0836434194206\\
1044.75	40.9976789595844\\
1045	40.8705743653981\\
1045.25	40.7809257144261\\
1045.5	40.6753122351988\\
1045.75	40.580751329379\\
1046	40.4702255953039\\
1046.25	40.3670682435005\\
1046.5	40.2497881590097\\
1046.75	40.1429466160704\\
1047	40.0342629775633\\
1047.25	39.9188249886404\\
1047.5	39.8113694138452\\
1047.75	39.6971594886343\\
1048	39.5841776271353\\
1048.25	39.4632133515087\\
1048.5	39.3404069803141\\
1048.75	39.2157585135517\\
1049	39.0917240786452\\
1049.25	38.9818123764261\\
1049.5	38.8854093750384\\
1049.75	38.7644450994118\\
1050	38.6447088874971\\
1050.25	38.5390954082698\\
1050.5	38.421815323779\\
1050.75	38.3057633030002\\
1051	38.1909393459333\\
1051.25	38.0699750703066\\
1051.5	37.9600633680875\\
1051.75	37.8440113473087\\
1052	37.7279593265299\\
1052.25	37.6223458473026\\
1052.5	37.5142762406514\\
1052.75	37.4117329207039\\
1053	37.3042773459087\\
1053.25	37.1722604968746\\
1053.5	37.0519102531039\\
1053.75	36.9327880730452\\
1054	36.8198062115463\\
1054.25	36.6853332350882\\
1054.5	36.5803337877169\\
1054.75	36.4581414483783\\
1055	36.3420894275995\\
1055.25	36.2235812793968\\
1055.5	36.0872662073708\\
1055.75	35.9748983777278\\
1056	35.8619165162289\\
1056.25	35.7550749732896\\
1056.5	35.6408650480787\\
1056.75	35.520514804308\\
1057	35.4136732613688\\
1057.25	35.3062176865736\\
1057.5	35.1926217932186\\
1057.75	35.0747276768719\\
1058	34.9420967959818\\
1058.25	34.8119220425156\\
1058.5	34.7020103402965\\
1058.75	34.57797590539\\
1059	34.4490292156357\\
1059.25	34.3354333222808\\
1059.5	34.1997322821108\\
1059.75	34.0763118790603\\
1060	33.9504353485859\\
1060.25	33.836225423375\\
1060.5	33.7066647017647\\
1060.75	33.5857004261381\\
1061	33.4635080867995\\
1061.25	33.3388596200371\\
1061.5	33.2148251851306\\
1061.75	33.0981591324958\\
1062	32.9839492072849\\
1062.25	32.8715813776419\\
1062.5	32.7671959621265\\
1062.75	32.6493018457798\\
1063	32.5289516020091\\
1063.25	32.418425867934\\
1063.5	32.3036019108671\\
1063.75	32.1568482972897\\
1064	32.0285156393914\\
1064.25	31.9124636186125\\
1064.5	31.7908853111299\\
1064.75	31.6815876407668\\
1065	31.5532549828685\\
1065.25	31.4261503886822\\
1065.5	31.3285193235825\\
1065.75	31.205098920532\\
1066	31.092731090889\\
1066.25	30.9760650382542\\
1066.5	30.8538726989156\\
1066.75	30.7230839135934\\
1067	30.6088739883825\\
1067.25	30.4897518083238\\
1067.5	30.4148399218951\\
1067.75	30.3012440285402\\
1068	30.1796657210576\\
1068.25	30.0685259551266\\
1068.5	29.9518599024917\\
1068.75	29.8321236905771\\
1069	29.7093173193825\\
1069.25	29.5736162792126\\
1069.5	29.4581782902897\\
1069.75	29.3237053138317\\
1070	29.1929165285095\\
1070.25	29.0750224121627\\
1070.5	28.957742327672\\
1070.75	28.8257254786378\\
1071	28.7016910437313\\
1071.25	28.5715162902651\\
1071.5	28.4505520146385\\
1071.75	28.3455525672672\\
1072	28.2190620049368\\
1072.25	28.080290805487\\
1072.5	27.967922975844\\
1072.75	27.8500288594972\\
1073	27.7450294121259\\
1073.25	27.6160827223716\\
1073.5	27.4883640963293\\
1073.75	27.3747682029744\\
1074	27.2777511697307\\
1074.25	27.1500325436884\\
1074.5	27.0376647140454\\
1074.75	26.9252968844024\\
1075	26.79757825836\\
1075.25	26.6882805879969\\
1075.5	26.5624040575225\\
1075.75	26.4340713996242\\
1076	26.2977563275983\\
1076.25	26.1878446253792\\
1076.5	26.0705645408884\\
1076.75	25.9459160741259\\
1077	25.8096010021\\
1077.25	25.6855665671935\\
1077.5	25.5572339092952\\
1077.75	25.4436380159403\\
1078	25.3208316447457\\
1078.25	25.1820604452959\\
1078.5	25.0733768067887\\
1078.75	24.9628510727137\\
1079	24.8492551793587\\
1079.25	24.7110980117649\\
1079.5	24.5710987486031\\
1079.75	24.4525906004003\\
1080	24.3463630893171\\
1080.25	24.2309251003942\\
1080.5	24.1062766336318\\
1080.75	23.9920667084208\\
1081	23.8686463053703\\
1081.25	23.7470679978877\\
1081.5	23.6150511488536\\
1081.75	23.4922447776591\\
1082	23.3854032347198\\
1082.25	23.2810178192045\\
1082.5	23.1582114480099\\
1082.75	23.0077736432966\\
1083	22.8954058136536\\
1083.25	22.7689152513232\\
1083.5	22.6485650075526\\
1083.75	22.5251446045021\\
1084	22.4041803288755\\
1084.25	22.2869002443847\\
1084.5	22.1690061280379\\
1084.75	22.0296208967321\\
1085	21.8951479202741\\
1085.25	21.7784818676393\\
1085.5	21.6397106681895\\
1085.75	21.5242726792666\\
1086	21.3855014798168\\
1086.25	21.2602389811983\\
1086.5	21.1312922914441\\
1086.75	21.0140122069533\\
1087	20.8924338994707\\
1087.25	20.7548907637328\\
1087.5	20.6265581058345\\
1087.75	20.50375173464\\
1088	20.3821734271574\\
1088.25	20.2759459160741\\
1088.5	20.1574377678714\\
1088.75	20.0413857470926\\
1089	19.919193407754\\
1089.25	19.7945449409915\\
1089.5	19.6705105060851\\
1089.75	19.5526163897383\\
1090	19.4334942096796\\
1090.25	19.3235825074605\\
1090.5	19.2026182318339\\
1090.75	19.0712154146557\\
1091	18.974198381412\\
1091.25	18.8563042650653\\
1091.5	18.7328838620148\\
1091.75	18.614989745668\\
1092	18.4835869284898\\
1092.25	18.3601665254393\\
1092.5	18.2355180586769\\
1092.75	18.1182379741861\\
1093	18.0040280489752\\
1093.25	17.8904321556202\\
1093.5	17.7700819118496\\
1093.75	17.6411352220953\\
1094	17.5392059340039\\
1094.25	17.4108732761056\\
1094.5	17.2856107774871\\
1094.75	17.1621903744366\\
1095	17.0442962580899\\
1095.25	16.9202618231834\\
1095.5	16.8017536749807\\
1095.75	16.6832455267779\\
1096	16.5665794741431\\
1096.25	16.4376327843889\\
1096.5	16.3166685087622\\
1096.75	16.2012305198394\\
1097	16.0759680212209\\
1097.25	15.9513195544585\\
1097.5	15.8389517248155\\
1097.75	15.7247417996046\\
1098	15.5921109187145\\
1098.25	15.468690515664\\
1098.5	15.3520244630291\\
1098.75	15.2230777732749\\
1099	15.1008854339363\\
1099.25	15.0007982414128\\
1099.5	14.8736936472264\\
1099.75	14.7551854990237\\
1100	14.6495720197964\\
1100.25	14.5316779034496\\
1100.5	14.4070294366872\\
1100.75	14.2805388743568\\
1101	14.163872821722\\
1101.25	14.050276928367\\
1101.5	13.9354529713002\\
1101.75	13.7972958037063\\
1102	13.6529983175527\\
1102.25	13.5277358189343\\
1102.5	13.3957189699002\\
1102.75	13.265544216434\\
1103	13.1451939726633\\
1103.25	13.0205455059008\\
1103.5	12.9137039629616\\
1103.75	12.7958098466148\\
1104	12.6815999214039\\
1104.25	12.5397585626742\\
1104.5	12.4304608923111\\
1104.75	12.3254614449398\\
1105	12.208181360449\\
1105.25	12.1050240086456\\
1105.5	11.9865158604428\\
1105.75	11.8796743175036\\
1106	11.7525697233172\\
1106.25	11.645114148522\\
1106.5	11.5253779366074\\
1106.75	11.4148522025323\\
1107	11.3153790418647\\
1107.25	11.1980989573739\\
1107.5	11.0955556374265\\
1107.75	10.9758194255118\\
1108	10.8640656277248\\
1108.25	10.7345049061145\\
1108.5	10.6190669171917\\
1108.75	10.5288042343637\\
1109	10.4385415515357\\
1109.25	10.3224895307569\\
1109.5	10.2162620196736\\
1109.75	10.1210870819978\\
1110	10.0111753797787\\
1110.25	9.89266723157597\\
1110.5	9.77170295594935\\
1110.75	9.6605631900183\\
1111	9.54328310552751\\
1111.25	9.43030124402854\\
1111.5	9.3210035736654\\
1111.75	9.22337250856574\\
1112	9.11775902933844\\
1112.25	9.01460167753503\\
1112.5	8.90223384789203\\
1112.75	8.77758538112957\\
1113	8.66091932849476\\
1113.25	8.54916553070773\\
1113.5	8.45214849746405\\
1113.75	8.35328936865245\\
1114	8.24092153900944\\
1114.25	8.13100983679033\\
1114.5	8.02171216642719\\
1114.75	7.91855481462378\\
1115	7.82522197251593\\
1115.25	7.72267865256849\\
1115.5	7.61215291849341\\
1115.75	7.49855702513846\\
1116	7.39724176890297\\
1116.25	7.31434746834666\\
1116.5	7.22408478551867\\
1116.75	7.12829581598693\\
1117	7.01961217747977\\
1117.25	6.92750739908386\\
1117.5	6.8188237605767\\
1117.75	6.73163123702858\\
1118	6.63093001264906\\
1118.25	6.50935170516646\\
1118.5	6.43014159574599\\
1118.75	6.31408957496715\\
1119	6.22198479657125\\
1119.25	6.11452922177603\\
1119.5	6.01751218853234\\
1119.75	5.91128467744907\\
1120	5.84005698215623\\
1120.25	5.76330300015965\\
1120.5	5.67488241289958\\
1120.75	5.57540925223201\\
1121	5.47532205970846\\
1121.25	5.36049810264157\\
1121.5	5.25304252784634\\
1121.75	5.1394466344914\\
1122	5.0381313782559\\
1122.25	4.94050031315625\\
1122.5	4.84778150290437\\
1122.75	4.75137850151666\\
1123	4.64883518156922\\
1123.25	4.54629186162178\\
1123.5	4.45173095580199\\
1123.75	4.36945068710165\\
1124	4.28901251396922\\
1124.25	4.19322354443749\\
1124.5	4.10111876604158\\
1124.75	4.01454027434943\\
1125	3.9162951773938\\
1125.25	3.82357636714193\\
1125.5	3.72778739761019\\
1125.75	3.65471760674944\\
1126	3.5589286372177\\
1126.25	3.48585884635695\\
1126.5	3.40173648208869\\
1126.75	3.31638605410849\\
1127	3.23717594468801\\
1127.25	3.16901840867504\\
1127.5	3.08366798069484\\
1127.75	3.00200174385047\\
1128	2.89884439204706\\
1128.25	2.81533605963477\\
1128.5	2.74349433248597\\
1128.75	2.66674035048938\\
1129	2.59244249591669\\
1129.25	2.50340787680065\\
1129.5	2.4303380859399\\
1129.75	2.36770683663068\\
1130	2.2977072050498\\
1130.25	2.21788306377335\\
1130.5	2.16323422859178\\
1130.75	2.09937491557062\\
1131	2.03060334770168\\
1131.25	1.97227032138427\\
1131.5	1.90042859423547\\
1131.75	1.83595524935834\\
1132	1.77455206376107\\
1132.25	1.71192081445185\\
1132.5	1.67876309422933\\
1132.75	1.63271070503138\\
1133	1.59034250696926\\
1133.25	1.54920237261909\\
1133.5	1.49086934630169\\
1133.75	1.44236082967984\\
1134	1.40490488646551\\
1134.25	1.35209814685186\\
1134.5	1.31709833106142\\
1134.75	1.2882388338307\\
1135	1.24402854020067\\
1135.25	1.21087081997814\\
1135.5	1.169116653772\\
1135.75	1.13288877426961\\
1136	1.0948187991993\\
1136.25	1.04631028257746\\
1136.5	1.00639821193924\\
1136.75	0.960345822741284\\
1137	0.935784548502376\\
1137.25	0.89771457343207\\
1137.5	0.849206056810227\\
1137.75	0.82464478257132\\
1138	0.787802871212958\\
1138.25	0.760171437694188\\
1138.5	0.723943558191799\\
1138.75	0.693241965393165\\
1139	0.665610531874394\\
1139.25	0.634294907219787\\
1139.5	0.614031855972688\\
1139.75	0.593154772869616\\
1140	0.55508479779931\\
1140.25	0.524383205000675\\
1140.5	0.502892090041631\\
1140.75	0.478944847658696\\
1141	0.469734369819106\\
1141.25	0.453769541563816\\
1141.5	0.429822299180881\\
1141.75	0.400348770094193\\
1142	0.377629591423203\\
1142.25	0.365348954303749\\
1142.5	0.343243807488733\\
1142.75	0.32666494737747\\
1143	0.308243991698289\\
1143.25	0.281226590035491\\
1143.5	0.264033698068256\\
1143.75	0.240086455685321\\
1144	0.235788232693512\\
1144.25	0.231490009701703\\
1144.5	0.215525181446413\\
1144.75	0.191577939063479\\
1145	0.181139397511943\\
1145.25	0.183595524935834\\
1145.5	0.174385047096243\\
1145.75	0.167016664824571\\
1146	0.160876346264844\\
1146.25	0.138157167593855\\
1146.5	0.140613295017746\\
1146.75	0.131402817178155\\
1147	0.127104594186346\\
1147.25	0.121578307482592\\
1147.5	0.116052020778838\\
1147.75	0.118508148202729\\
1148	0.122806371194538\\
1148.25	0.12342040305051\\
1148.5	0.128946689754264\\
1148.75	0.135087008313991\\
1149	0.135087008313991\\
1149.25	0.133244912746073\\
1149.5	0.132016849034128\\
1149.75	0.144297486153582\\
1150	0.150437804713309\\
1150.25	0.147367645433445\\
1150.5	0.141227326873718\\
1150.75	0.133244912746073\\
1151	0.129560721610237\\
1151.25	0.135087008313991\\
1151.5	0.138157167593855\\
1151.75	0.132016849034128\\
1152	0.126490562330374\\
1152.25	0.128946689754264\\
1152.5	0.135087008313991\\
1152.75	0.124648466762456\\
1153	0.12096427562662\\
1153.25	0.125262498618428\\
1153.5	0.124034434906483\\
1153.75	0.132016849034128\\
1154	0.132630880890101\\
1154.25	0.1393852313058\\
1154.5	0.131402817178155\\
1154.75	0.128946689754264\\
1155	0.125876530474401\\
1155.25	0.137543135737882\\
1155.5	0.136315072025937\\
1155.75	0.137543135737882\\
1156	0.133858944602046\\
1156.25	0.135701040169964\\
1156.5	0.127718626042319\\
1156.75	0.128332657898292\\
1157	0.126490562330374\\
1157.25	0.132630880890101\\
1157.5	0.135701040169964\\
1157.75	0.139999263161773\\
1158	0.145525549865527\\
1158.25	0.145525549865527\\
1158.5	0.149823772857336\\
1158.75	0.143683454297609\\
1159	0.138771199449827\\
1159.25	0.14859570914539\\
1159.5	0.1461395817215\\
1159.75	0.145525549865527\\
1160	0.141227326873718\\
1160.25	0.1393852313058\\
1160.5	0.135701040169964\\
1160.75	0.140613295017746\\
1161	0.141841358729691\\
1161.25	0.136315072025937\\
1161.5	0.133244912746073\\
1161.75	0.127104594186346\\
1162	0.125876530474401\\
1162.25	0.136929103881909\\
1162.5	0.141227326873718\\
1162.75	0.137543135737882\\
1163	0.130788785322183\\
1163.25	0.131402817178155\\
1163.5	0.132630880890101\\
1163.75	0.134472976458019\\
1164	0.135701040169964\\
1164.25	0.130788785322183\\
1164.5	0.1393852313058\\
1164.75	0.145525549865527\\
1165	0.144297486153582\\
1165.25	0.143683454297609\\
1165.5	0.137543135737882\\
1165.75	0.143683454297609\\
1166	0.137543135737882\\
1166.25	0.144911518009554\\
1166.5	0.136929103881909\\
1166.75	0.128332657898292\\
1167	0.136315072025937\\
1167.25	0.136929103881909\\
1167.5	0.136929103881909\\
1167.75	0.138157167593855\\
1168	0.136315072025937\\
1168.25	0.140613295017746\\
1168.5	0.141841358729691\\
1168.75	0.143683454297609\\
1169	0.145525549865527\\
1169.25	0.143069422441636\\
1169.5	0.136929103881909\\
1169.75	0.130788785322183\\
1170	0.128332657898292\\
1170.25	0.118508148202729\\
1170.5	0.116052020778838\\
1170.75	0.122806371194538\\
1171	0.121578307482592\\
1171.25	0.133858944602046\\
1171.5	0.132630880890101\\
1171.75	0.125876530474401\\
1172	0.125262498618428\\
1172.25	0.12096427562662\\
1172.5	0.114823957066893\\
1172.75	0.119736211914674\\
1173	0.126490562330374\\
1173.25	0.130788785322183\\
1173.5	0.145525549865527\\
1173.75	0.149209741001363\\
1174	0.147981677289418\\
1174.25	0.143069422441636\\
1174.5	0.143069422441636\\
1174.75	0.14859570914539\\
1175	0.140613295017746\\
1175.25	0.140613295017746\\
1175.5	0.136929103881909\\
1175.75	0.126490562330374\\
1176	0.12342040305051\\
1176.25	0.117894116346756\\
1176.5	0.130788785322183\\
1176.75	0.132016849034128\\
1177	0.136929103881909\\
1177.25	0.138771199449827\\
1177.5	0.131402817178155\\
1177.75	0.134472976458019\\
1178	0.144297486153582\\
1178.25	0.142455390585664\\
1178.5	0.143069422441636\\
1178.75	0.1461395817215\\
1179	0.147981677289418\\
1179.25	0.152893932137199\\
1179.5	0.150437804713309\\
1179.75	0.149823772857336\\
1180	0.143069422441636\\
1180.25	0.147981677289418\\
1180.5	0.145525549865527\\
1180.75	0.140613295017746\\
1181	0.141841358729691\\
1181.25	0.14859570914539\\
1181.5	0.14859570914539\\
1181.75	0.145525549865527\\
1182	0.149823772857336\\
1182.25	0.144297486153582\\
1182.5	0.143683454297609\\
1182.75	0.1393852313058\\
1183	0.135701040169964\\
1183.25	0.1393852313058\\
1183.5	0.141841358729691\\
1183.75	0.1461395817215\\
1184	0.14859570914539\\
1184.25	0.1461395817215\\
1184.5	0.145525549865527\\
1184.75	0.139999263161773\\
1185	0.137543135737882\\
1185.25	0.132016849034128\\
1185.5	0.124648466762456\\
1185.75	0.134472976458019\\
1186	0.131402817178155\\
1186.25	0.127718626042319\\
1186.5	0.133858944602046\\
1186.75	0.144297486153582\\
1187	0.138771199449827\\
1187.25	0.135701040169964\\
1187.5	0.144911518009554\\
1187.75	0.149823772857336\\
1188	0.152893932137199\\
1188.25	0.152893932137199\\
1188.5	0.149209741001363\\
1188.75	0.141227326873718\\
1189	0.141227326873718\\
1189.25	0.142455390585664\\
1189.5	0.149823772857336\\
1189.75	0.151665868425254\\
1190	0.144911518009554\\
1190.25	0.1461395817215\\
1190.5	0.143683454297609\\
1190.75	0.134472976458019\\
1191	0.133244912746073\\
1191.25	0.136929103881909\\
1191.5	0.131402817178155\\
1191.75	0.126490562330374\\
1192	0.133244912746073\\
1192.25	0.143069422441636\\
1192.5	0.144297486153582\\
1192.75	0.142455390585664\\
1193	0.142455390585664\\
1193.25	0.136315072025937\\
1193.5	0.1393852313058\\
1193.75	0.141841358729691\\
1194	0.143683454297609\\
1194.25	0.151665868425254\\
1194.5	0.152279900281227\\
1194.75	0.143069422441636\\
1195	0.130788785322183\\
1195.25	0.115437988922865\\
1195.5	0.127104594186346\\
1195.75	0.137543135737882\\
1196	0.135701040169964\\
1196.25	0.13017475346621\\
1196.5	0.124034434906483\\
1196.75	0.119122180058701\\
1197	0.120350243770647\\
1197.25	0.122806371194538\\
1197.5	0.115437988922865\\
1197.75	0.117894116346756\\
1198	0.127104594186346\\
1198.25	0.12342040305051\\
1198.5	0.132630880890101\\
1198.75	0.134472976458019\\
1199	0.132630880890101\\
1199.25	0.12342040305051\\
1199.5	0.130788785322183\\
1199.75	0.130788785322183\\
1200	0.127104594186346\\
1200.25	0.125876530474401\\
1200.5	0.120350243770647\\
1200.75	0.135087008313991\\
1201	0.141841358729691\\
1201.25	0.141227326873718\\
1201.5	0.135701040169964\\
1201.75	0.125262498618428\\
1202	0.13017475346621\\
1202.25	0.136315072025937\\
1202.5	0.135087008313991\\
1202.75	0.132016849034128\\
1203	0.135701040169964\\
1203.25	0.128332657898292\\
1203.5	0.120350243770647\\
1203.75	0.127104594186346\\
1204	0.127718626042319\\
1204.25	0.125876530474401\\
1204.5	0.128332657898292\\
1204.75	0.124034434906483\\
1205	0.127718626042319\\
1205.25	0.124648466762456\\
1205.5	0.115437988922865\\
1205.75	0.116666052634811\\
1206	0.12096427562662\\
1206.25	0.116052020778838\\
1206.5	0.115437988922865\\
1206.75	0.11420992521092\\
1207	0.119736211914674\\
1207.25	0.112981861498975\\
1207.5	0.110525734075084\\
1207.75	0.108683638507166\\
1208	0.117280084490783\\
1208.25	0.109297670363138\\
1208.5	0.109297670363138\\
1208.75	0.112981861498975\\
1209	0.116052020778838\\
1209.25	0.117280084490783\\
1209.5	0.112981861498975\\
1209.75	0.109911702219111\\
1210	0.106841542939248\\
1210.25	0.104385415515357\\
1210.5	0.096403001387712\\
1210.75	0.0976310650996574\\
1211	0.0976310650996574\\
1211.25	0.0988591288116027\\
1211.5	0.0994731606675754\\
1211.75	0.108683638507166\\
1212	0.109297670363138\\
1212.25	0.108683638507166\\
1212.5	0.109297670363138\\
1212.75	0.112367829643002\\
1213	0.106841542939248\\
1213.25	0.106227511083275\\
1213.5	0.115437988922865\\
1213.75	0.121578307482592\\
1214	0.117280084490783\\
1214.25	0.112981861498975\\
1214.5	0.10499944737133\\
1214.75	0.104385415515357\\
1215	0.102543319947439\\
1215.25	0.101315256235494\\
1215.5	0.109297670363138\\
1215.75	0.106841542939248\\
};
\end{axis}
\end{tikzpicture}%

%% file: figures/learned_mean.tex
%
\tikzsetnextfilename{learned_mean}
\definecolor{mycolor1}{rgb}{0.12157,0.47059,0.70588}%
\begin{tikzpicture}

\begin{axis}[%
width=\widefigurewidth,
height=\widefigureheight,
at={(0\widefigurewidth,0\widefigureheight)},
scale only axis,
xmin=901.75,
xmax=1225.75,
xtick={ 900,  950, 1000, 1050, 1100, 1150, 1200, 1250},
xlabel={rest wavelength $\lr$ (\AA)},
ymin=-0.25,
ymax=5,
ylabel={normalized flux},
axis background/.style={fill=white},
axis x line*=bottom,
axis y line*=left,
legend style={at={(0.03,0.97)},anchor=north west,legend cell align=left,align=left,fill=none,draw=none},
extra x ticks={911.8,973,1026,1215.67}, extra x tick labels={{Ly\,$\infty$},{Ly\,$\gamma$},{Ly\,$\beta$},{Ly\,$\alpha$}},
]
\addplot [color=mycolor1,solid]
  table[row sep=crcr]{%
911.75	0.686333184771422\\
912	0.690350357630175\\
912.25	0.694722650304614\\
912.5	0.697237674412643\\
912.75	0.700135842179731\\
913	0.701759591312927\\
913.25	0.704646656694654\\
913.5	0.70751731431018\\
913.75	0.709009157653358\\
914	0.711878094660329\\
914.25	0.714639278543013\\
914.5	0.717660678066854\\
914.75	0.720280093797\\
915	0.722713226160096\\
915.25	0.727552739580756\\
915.5	0.731342642368675\\
915.75	0.735014962413007\\
916	0.739466380535175\\
916.25	0.743281159734849\\
916.5	0.749601162137443\\
916.75	0.75390574297844\\
917	0.758049230209562\\
917.25	0.761697151263536\\
917.5	0.765124440582382\\
917.75	0.769184667016332\\
918	0.771178540301608\\
918.25	0.773377803778564\\
918.5	0.775218860260104\\
918.75	0.779044746240084\\
919	0.781767519067078\\
919.25	0.784427735542112\\
919.5	0.787672653781131\\
919.75	0.791513020000778\\
920	0.796226684099307\\
920.25	0.798862849512744\\
920.5	0.801433981035219\\
920.75	0.804061041730039\\
921	0.80595996534393\\
921.25	0.805400330822295\\
921.5	0.803612964384887\\
921.75	0.804383808946655\\
922	0.805879733271386\\
922.25	0.806720338247782\\
922.5	0.808954752363581\\
922.75	0.812741895418241\\
923	0.818264659983643\\
923.25	0.822292812451412\\
923.5	0.825923315288105\\
923.75	0.831780536632556\\
924	0.836443485025839\\
924.25	0.839764541651973\\
924.5	0.843329404301759\\
924.75	0.846875393055935\\
925	0.850167117590955\\
925.25	0.851328273654046\\
925.5	0.852466817878533\\
925.75	0.853377962056764\\
926	0.85221127731815\\
926.25	0.850430708216889\\
926.5	0.850177513049708\\
926.75	0.850988600796939\\
927	0.853620315405801\\
927.25	0.857702084936826\\
927.5	0.863342263620669\\
927.75	0.870096877059233\\
928	0.876010840668718\\
928.25	0.880014304957089\\
928.5	0.883701098622712\\
928.75	0.88575079850438\\
929	0.886412368868749\\
929.25	0.885630278045801\\
929.5	0.88533811980731\\
929.75	0.886826847002469\\
930	0.887166717199629\\
930.25	0.889006618287107\\
930.5	0.891873332562931\\
930.75	0.895840270714634\\
931	0.899635650387003\\
931.25	0.902151945118789\\
931.5	0.905082538196556\\
931.75	0.90753020856552\\
932	0.909812933346208\\
932.25	0.910731298754567\\
932.5	0.910314255207582\\
932.75	0.909126967172953\\
933	0.906922178592979\\
933.25	0.906099420270711\\
933.5	0.905213313332074\\
933.75	0.906613262459408\\
934	0.908698213671991\\
934.25	0.91308145820648\\
934.5	0.916969534762428\\
934.75	0.920002781783409\\
935	0.922825419972733\\
935.25	0.924747538567821\\
935.5	0.927023472230886\\
935.75	0.928238447319235\\
936	0.929979988879514\\
936.25	0.931262131142355\\
936.5	0.933226754983498\\
936.75	0.934162820501479\\
937	0.934220456211171\\
937.25	0.934676526726415\\
937.5	0.935780094095799\\
937.75	0.936711868916224\\
938	0.936807821496183\\
938.25	0.937785741587744\\
938.5	0.940166220599716\\
938.75	0.940659612171205\\
939	0.941122405128682\\
939.25	0.942940383971859\\
939.5	0.945850406361623\\
939.75	0.948033597215888\\
940	0.949375130933113\\
940.25	0.952586543632172\\
940.5	0.955742315074659\\
940.75	0.959027094916113\\
941	0.961350716332529\\
941.25	0.963446607146063\\
941.5	0.964951452485496\\
941.75	0.966994618735509\\
942	0.969123943541823\\
942.25	0.968568170379265\\
942.5	0.968914368668064\\
942.75	0.969840200989124\\
943	0.970860261415412\\
943.25	0.971353295721588\\
943.5	0.96982349520892\\
943.75	0.968938204985006\\
944	0.967616277530752\\
944.25	0.964477300414766\\
944.5	0.960165556278523\\
944.75	0.955972374162354\\
945	0.953973693887951\\
945.25	0.952471280141981\\
945.5	0.951716006155735\\
945.75	0.953840819175783\\
946	0.959288629579459\\
946.25	0.964419360231926\\
946.5	0.969530829511415\\
946.75	0.973911470964433\\
947	0.977731543016248\\
947.25	0.980339323264438\\
947.5	0.979906012845702\\
947.75	0.978454961518698\\
948	0.975784855516899\\
948.25	0.97337389088364\\
948.5	0.970543233895136\\
948.75	0.966942885569187\\
949	0.965585705544925\\
949.25	0.965220413161521\\
949.5	0.964455818947996\\
949.75	0.964345101553157\\
950	0.965098032692089\\
950.25	0.967826370869273\\
950.5	0.969387104763536\\
950.75	0.970431843576258\\
951	0.971287890413865\\
951.25	0.972134161857383\\
951.5	0.971281427663467\\
951.75	0.968516134430164\\
952	0.964999709435764\\
952.25	0.961693552850804\\
952.5	0.959165173335586\\
952.75	0.955292361712699\\
953	0.952388596383342\\
953.25	0.95027522819884\\
953.5	0.948793176510845\\
953.75	0.946300148034027\\
954	0.943895452408772\\
954.25	0.94411835980718\\
954.5	0.944592238656582\\
954.75	0.9443973448235\\
955	0.945303197202546\\
955.25	0.947002957136459\\
955.5	0.94810767460791\\
955.75	0.948743742140162\\
956	0.948602820803674\\
956.25	0.949324140136142\\
956.5	0.948908347559198\\
956.75	0.948761993708661\\
957	0.949130155942803\\
957.25	0.948095178179491\\
957.5	0.945815398956427\\
957.75	0.94491209004843\\
958	0.943669533548048\\
958.25	0.941226444493857\\
958.5	0.940225885134787\\
958.75	0.939760296402814\\
959	0.940675888723978\\
959.25	0.939232647181314\\
959.5	0.939618879559311\\
959.75	0.941605398681514\\
960	0.943238688377933\\
960.25	0.945347812465534\\
960.5	0.947610544245541\\
960.75	0.950255651313457\\
961	0.95181907871141\\
961.25	0.954351877527351\\
961.5	0.955945060774939\\
961.75	0.956807548698857\\
962	0.957734897314056\\
962.25	0.958708635087469\\
962.5	0.959062313324791\\
962.75	0.958650649132113\\
963	0.957414403182235\\
963.25	0.956488435939209\\
963.5	0.956070360625006\\
963.75	0.955484158428031\\
964	0.954764555531095\\
964.25	0.953541404802158\\
964.5	0.953590745942627\\
964.75	0.953818097253483\\
965	0.953485298229995\\
965.25	0.953417464919544\\
965.5	0.956104736839597\\
965.75	0.960328762764302\\
966	0.964995357129759\\
966.25	0.969511433256334\\
966.5	0.975066891275983\\
966.75	0.981337611300767\\
967	0.984745722737793\\
967.25	0.985865983998319\\
967.5	0.986219144776531\\
967.75	0.986786278895498\\
968	0.987774050721088\\
968.25	0.989114368902685\\
968.5	0.992729144599766\\
968.75	0.996915571433899\\
969	1.00276573627879\\
969.25	1.01069991563118\\
969.5	1.01827078330552\\
969.75	1.02698494187183\\
970	1.03449030677803\\
970.25	1.04257805717468\\
970.5	1.05005860549468\\
970.75	1.05627493265592\\
971	1.06259408291746\\
971.25	1.06741045208661\\
971.5	1.07169483870485\\
971.75	1.07657517811863\\
972	1.08185735305562\\
972.25	1.0867739638242\\
972.5	1.09230744290135\\
972.75	1.09941066499411\\
973	1.10946823176154\\
973.25	1.12084813653868\\
973.5	1.13254897280314\\
973.75	1.1451744523702\\
974	1.15717331830566\\
974.25	1.16762858467704\\
974.5	1.17647492201566\\
974.75	1.18332827886107\\
975	1.18823138841018\\
975.25	1.19218902419611\\
975.5	1.19492619922899\\
975.75	1.19709065063804\\
976	1.19764658020824\\
976.25	1.1968630747412\\
976.5	1.19558209409647\\
976.75	1.19336764173464\\
977	1.19056345518162\\
977.25	1.18720177875106\\
977.5	1.18419817358669\\
977.75	1.18229014256425\\
978	1.18079591446411\\
978.25	1.17856348960582\\
978.5	1.17630058642735\\
978.75	1.17383300415562\\
979	1.17122467028199\\
979.25	1.16741483791931\\
979.5	1.16454658230246\\
979.75	1.16296464539683\\
980	1.16242455044863\\
980.25	1.16207625541187\\
980.5	1.16143777665072\\
980.75	1.16243881590126\\
981	1.16365347327057\\
981.25	1.1636640984298\\
981.5	1.16282899082702\\
981.75	1.16326653468506\\
982	1.16386894016339\\
982.25	1.16404437423345\\
982.5	1.16322934672849\\
982.75	1.16352779381152\\
983	1.16390682571777\\
983.25	1.16369097485718\\
983.5	1.16384629038932\\
983.75	1.16443671464464\\
984	1.16506194314455\\
984.25	1.16559019940598\\
984.5	1.16747526847103\\
984.75	1.1702077279309\\
985	1.17289588948639\\
985.25	1.17585630162416\\
985.5	1.17987999179938\\
985.75	1.18460043149722\\
986	1.18901350051762\\
986.25	1.1927697016194\\
986.5	1.19807039301802\\
986.75	1.20356118319471\\
987	1.20889293582817\\
987.25	1.21353779672322\\
987.5	1.21827674358179\\
987.75	1.2224177751591\\
988	1.224804878649\\
988.25	1.22671688089594\\
988.5	1.22785542992432\\
988.75	1.22951670965707\\
989	1.22996872065118\\
989.25	1.2296073242493\\
989.5	1.22974368195518\\
989.75	1.22882424453597\\
990	1.227486624882\\
990.25	1.22504277461672\\
990.5	1.22110107379821\\
990.75	1.21701850807336\\
991	1.21186669306658\\
991.25	1.2073010563778\\
991.5	1.20290349595968\\
991.75	1.19754182151982\\
992	1.19384367458911\\
992.25	1.19003933784368\\
992.5	1.18625833434384\\
992.75	1.18196764406166\\
993	1.17710610675052\\
993.25	1.17320659700547\\
993.5	1.16827729769917\\
993.75	1.16431486874907\\
994	1.16048063709407\\
994.25	1.15501313837532\\
994.5	1.14910705561644\\
994.75	1.14367652453349\\
995	1.13863990803356\\
995.25	1.1333720741087\\
995.5	1.12876873341461\\
995.75	1.12660891242616\\
996	1.12499715160882\\
996.25	1.12220385725701\\
996.5	1.12033143212588\\
996.75	1.11938366883474\\
997	1.11796895723116\\
997.25	1.11592555755339\\
997.5	1.114787242378\\
997.75	1.11433722557775\\
998	1.11392788610081\\
998.25	1.11225872373325\\
998.5	1.11128965863977\\
998.75	1.11057143117234\\
999	1.10971658823477\\
999.25	1.10833328679753\\
999.5	1.10627135277039\\
999.75	1.1047312727321\\
1000	1.10286647272155\\
1000.25	1.10079353468954\\
1000.5	1.09758837802192\\
1000.75	1.09490985224324\\
1001	1.09286724681838\\
1001.25	1.09157839055471\\
1001.5	1.09027622292817\\
1001.75	1.08871769055608\\
1002	1.08811073682244\\
1002.25	1.08828868539069\\
1002.5	1.08933990724202\\
1002.75	1.09002754554162\\
1003	1.09080505063761\\
1003.25	1.09242233459106\\
1003.5	1.09475739664549\\
1003.75	1.09655684724449\\
1004	1.097190302871\\
1004.25	1.09811969326659\\
1004.5	1.09960099217974\\
1004.75	1.10086731383226\\
1005	1.10163619153336\\
1005.25	1.10241893602327\\
1005.5	1.1036259885837\\
1005.75	1.1059398301494\\
1006	1.10691779298106\\
1006.25	1.10730193894712\\
1006.5	1.10869532941451\\
1006.75	1.10947282348709\\
1007	1.11017528890583\\
1007.25	1.10955235918559\\
1007.5	1.11003532144408\\
1007.75	1.1104194311752\\
1008	1.10959712222619\\
1008.25	1.11016586492576\\
1008.5	1.1112279331306\\
1008.75	1.11287793586289\\
1009	1.1141475719373\\
1009.25	1.1160033333148\\
1009.5	1.11944233291635\\
1009.75	1.12285469771732\\
1010	1.1263050593296\\
1010.25	1.12859630176687\\
1010.5	1.13055429142425\\
1010.75	1.13261061701624\\
1011	1.13432821349508\\
1011.25	1.13609908108768\\
1011.5	1.13669031451087\\
1011.75	1.13855915874457\\
1012	1.14199411690981\\
1012.25	1.14577092432731\\
1012.5	1.14942747912556\\
1012.75	1.15343286874331\\
1013	1.15900766123677\\
1013.25	1.1638558810431\\
1013.5	1.16800198046552\\
1013.75	1.17233728877543\\
1014	1.17624986253935\\
1014.25	1.17935832252731\\
1014.5	1.18198257359474\\
1014.75	1.18421003331273\\
1015	1.18634078143677\\
1015.25	1.18784147632651\\
1015.5	1.18883408425802\\
1015.75	1.19045583502491\\
1016	1.19154119484959\\
1016.25	1.1929716104753\\
1016.5	1.19468575550783\\
1016.75	1.19695945852723\\
1017	1.19906201395503\\
1017.25	1.20127407654146\\
1017.5	1.20429567141093\\
1017.75	1.20758514995488\\
1018	1.21170322635915\\
1018.25	1.21633214628102\\
1018.5	1.22220129843975\\
1018.75	1.22950529039608\\
1019	1.23671083968066\\
1019.25	1.24502938172502\\
1019.5	1.25247962169692\\
1019.75	1.25902841011889\\
1020	1.26495194786627\\
1020.25	1.26925034650415\\
1020.5	1.27363463340818\\
1020.75	1.27776457361618\\
1021	1.28221292356393\\
1021.25	1.2886224469036\\
1021.5	1.29724892328364\\
1021.75	1.30796738969279\\
1022	1.32031101280281\\
1022.25	1.33350970790759\\
1022.5	1.3487771244088\\
1022.75	1.36483977354226\\
1023	1.37956976093424\\
1023.25	1.39269585842522\\
1023.5	1.40488515060727\\
1023.75	1.41600157213063\\
1024	1.42655540151101\\
1024.25	1.43577420626351\\
1024.5	1.44562832802595\\
1024.75	1.45680624724932\\
1025	1.46842634409065\\
1025.25	1.48161682172285\\
1025.5	1.49502761027104\\
1025.75	1.50854904173475\\
1026	1.52160843835108\\
1026.25	1.53437034686317\\
1026.5	1.54789430530945\\
1026.75	1.56135720055447\\
1027	1.57403935266338\\
1027.25	1.58784425044359\\
1027.5	1.6026061901269\\
1027.75	1.618511148909\\
1028	1.63542972918323\\
1028.25	1.65194533780767\\
1028.5	1.66919688165144\\
1028.75	1.68622177325734\\
1029	1.70352582275047\\
1029.25	1.72023592330675\\
1029.5	1.73473499194133\\
1029.75	1.74769172398999\\
1030	1.75769322013375\\
1030.25	1.76526430257793\\
1030.5	1.77009353883756\\
1030.75	1.77207543430504\\
1031	1.77230661258837\\
1031.25	1.77218921696793\\
1031.5	1.77256731747242\\
1031.75	1.77211150960522\\
1032	1.77141005193006\\
1032.25	1.77072259449829\\
1032.5	1.77103015975627\\
1032.75	1.77221262721689\\
1033	1.77386717469331\\
1033.25	1.77643541038701\\
1033.5	1.77993878866061\\
1033.75	1.78459616142141\\
1034	1.7901061333568\\
1034.25	1.79481957944608\\
1034.5	1.79813014047932\\
1034.75	1.80112030360469\\
1035	1.80345297911058\\
1035.25	1.80455823304952\\
1035.5	1.8024959777836\\
1035.75	1.79835643313949\\
1036	1.79239220743404\\
1036.25	1.7840755046156\\
1036.5	1.77320752059603\\
1036.75	1.75873414651276\\
1037	1.74249903439247\\
1037.25	1.72408183571482\\
1037.5	1.70531546137732\\
1037.75	1.68538543826151\\
1038	1.66464554121794\\
1038.25	1.64445252670275\\
1038.5	1.62456298340197\\
1038.75	1.60565374876882\\
1039	1.58646876658401\\
1039.25	1.5681321847784\\
1039.5	1.55012435152899\\
1039.75	1.53235790611312\\
1040	1.51553240337774\\
1040.25	1.49928284289533\\
1040.5	1.48432075794643\\
1040.75	1.47006713363084\\
1041	1.45737604348438\\
1041.25	1.4457443730897\\
1041.5	1.43414302581033\\
1041.75	1.42347586155626\\
1042	1.41334671946279\\
1042.25	1.40385007345448\\
1042.5	1.39495928692423\\
1042.75	1.38622485247362\\
1043	1.37807324665535\\
1043.25	1.37029488360205\\
1043.5	1.36202591458871\\
1043.75	1.35474408296459\\
1044	1.34760816731706\\
1044.25	1.34101939053315\\
1044.5	1.33388536644734\\
1044.75	1.32683809838021\\
1045	1.32116613281352\\
1045.25	1.31532210054234\\
1045.5	1.30921875936066\\
1045.75	1.30305180023224\\
1046	1.2987191221952\\
1046.25	1.29418250643813\\
1046.5	1.28915042468934\\
1046.75	1.28391754514118\\
1047	1.27940693958256\\
1047.25	1.27488031217438\\
1047.5	1.27007046124105\\
1047.75	1.26618518361095\\
1048	1.26272110055323\\
1048.25	1.25903858952638\\
1048.5	1.25467276524603\\
1048.75	1.25118892778592\\
1049	1.24784823036187\\
1049.25	1.24413403039795\\
1049.5	1.24049519942847\\
1049.75	1.23731852422058\\
1050	1.2343940857191\\
1050.25	1.23116312279932\\
1050.5	1.22772184069289\\
1050.75	1.22496107562638\\
1051	1.222131267411\\
1051.25	1.21952694064134\\
1051.5	1.21722175723248\\
1051.75	1.2151503112096\\
1052	1.21305829241456\\
1052.25	1.21110843938689\\
1052.5	1.20954976572359\\
1052.75	1.20836087216621\\
1053	1.20711303254668\\
1053.25	1.20612115387816\\
1053.5	1.20576649722368\\
1053.75	1.2052797974669\\
1054	1.20505762486086\\
1054.25	1.20433390813896\\
1054.5	1.20472365221894\\
1054.75	1.20477189860317\\
1055	1.20427926055245\\
1055.25	1.20425420625549\\
1055.5	1.20445189507492\\
1055.75	1.20485653640507\\
1056	1.20419849493124\\
1056.25	1.20435109012857\\
1056.5	1.20503761756898\\
1056.75	1.20531503307533\\
1057	1.2055929719111\\
1057.25	1.20591425546559\\
1057.5	1.2073214207762\\
1057.75	1.20908231823245\\
1058	1.21027623537569\\
1058.25	1.21165656789849\\
1058.5	1.2126991787083\\
1058.75	1.21467767841819\\
1059	1.21605276097523\\
1059.25	1.21634394417845\\
1059.5	1.21746795199784\\
1059.75	1.21849414283613\\
1060	1.21974708581194\\
1060.25	1.22001018952735\\
1060.5	1.22036152612651\\
1060.75	1.22146575657375\\
1061	1.22182074560725\\
1061.25	1.22205558098505\\
1061.5	1.22200192572022\\
1061.75	1.22222370340841\\
1062	1.22263257637568\\
1062.25	1.22250948175775\\
1062.5	1.22250514100286\\
1062.75	1.22268413023498\\
1063	1.22299225626476\\
1063.25	1.22261470438301\\
1063.5	1.220955509987\\
1063.75	1.21978629136686\\
1064	1.21872696492799\\
1064.25	1.216926313485\\
1064.5	1.21441900259151\\
1064.75	1.21274995042912\\
1065	1.21192879900774\\
1065.25	1.21032968656151\\
1065.5	1.20855400793334\\
1065.75	1.20751223523837\\
1066	1.20673454609901\\
1066.25	1.205928667354\\
1066.5	1.20497286293558\\
1066.75	1.2038199776206\\
1067	1.20325970726484\\
1067.25	1.20285278899939\\
1067.5	1.20335343935693\\
1067.75	1.20377446621225\\
1068	1.20433204163358\\
1068.25	1.20581663507628\\
1068.5	1.20797798101856\\
1068.75	1.20935548088378\\
1069	1.21084707588999\\
1069.25	1.21245397317186\\
1069.5	1.21379304640354\\
1069.75	1.21548439391904\\
1070	1.21598519139079\\
1070.25	1.21691883185113\\
1070.5	1.21818488257946\\
1070.75	1.2186756099442\\
1071	1.21886522989599\\
1071.25	1.21850149819915\\
1071.5	1.21861325046946\\
1071.75	1.2184930244268\\
1072	1.21713346055114\\
1072.25	1.215469013849\\
1072.5	1.21424228671777\\
1072.75	1.21229487029947\\
1073	1.20940311547896\\
1073.25	1.20654048411624\\
1073.5	1.20307411576646\\
1073.75	1.20000655191848\\
1074	1.19684931122757\\
1074.25	1.193968879293\\
1074.5	1.19162880706664\\
1074.75	1.1893547671598\\
1075	1.18722474513377\\
1075.25	1.18515065899117\\
1075.5	1.18254279186811\\
1075.75	1.17961613729421\\
1076	1.17654359274741\\
1076.25	1.17337370522434\\
1076.5	1.17016152484307\\
1076.75	1.16677138732087\\
1077	1.16393973380771\\
1077.25	1.16175974348491\\
1077.5	1.16003108475863\\
1077.75	1.15860648428634\\
1078	1.15697038685055\\
1078.25	1.15596338394307\\
1078.5	1.15542924479098\\
1078.75	1.15406219051288\\
1079	1.15260526317727\\
1079.25	1.15075300639702\\
1079.5	1.15031257805965\\
1079.75	1.14955104008872\\
1080	1.14816853928162\\
1080.25	1.14767313082613\\
1080.5	1.14689299879793\\
1080.75	1.14605354421456\\
1081	1.14460406522965\\
1081.25	1.14303915971004\\
1081.5	1.14174443908207\\
1081.75	1.14050735849113\\
1082	1.139588379808\\
1082.25	1.1388105975556\\
1082.5	1.13776054063853\\
1082.75	1.13678490350425\\
1083	1.13605929498151\\
1083.25	1.13480426602313\\
1083.5	1.13323975338129\\
1083.75	1.13167228420368\\
1084	1.12997693243501\\
1084.25	1.12862308779839\\
1084.5	1.12572408805043\\
1084.75	1.12260060383373\\
1085	1.12014065346659\\
1085.25	1.11747888224174\\
1085.5	1.1146652270323\\
1085.75	1.11148126469867\\
1086	1.10880314222155\\
1086.25	1.10657321158186\\
1086.5	1.10408940924233\\
1086.75	1.10118442055312\\
1087	1.09842446355104\\
1087.25	1.09565265769921\\
1087.5	1.093293546549\\
1087.75	1.09059025736721\\
1088	1.08713647555445\\
1088.25	1.08414286368115\\
1088.5	1.0817164276289\\
1088.75	1.07857978364389\\
1089	1.07507411726939\\
1089.25	1.07243108101125\\
1089.5	1.07041949759384\\
1089.75	1.06839186125252\\
1090	1.06604109235657\\
1090.25	1.06410767203059\\
1090.5	1.06264153265063\\
1090.75	1.06051168356333\\
1091	1.05811196483271\\
1091.25	1.0556568476984\\
1091.5	1.05319568614023\\
1091.75	1.05119054305037\\
1092	1.04914761819235\\
1092.25	1.04733088065827\\
1092.5	1.04517252166066\\
1092.75	1.04297784095843\\
1093	1.04086769177949\\
1093.25	1.03869013169452\\
1093.5	1.03618698727877\\
1093.75	1.03352998183912\\
1094	1.03137436779925\\
1094.25	1.02929300171069\\
1094.5	1.02676757248057\\
1094.75	1.02420393964505\\
1095	1.02243513277921\\
1095.25	1.02070585976032\\
1095.5	1.01951051034657\\
1095.75	1.01877317613606\\
1096	1.01833973720584\\
1096.25	1.0183714247914\\
1096.5	1.01782360618325\\
1096.75	1.01732722104097\\
1097	1.01660087929813\\
1097.25	1.01508117192503\\
1097.5	1.01406591789913\\
1097.75	1.01243526932825\\
1098	1.01157302146456\\
1098.25	1.01062670501682\\
1098.5	1.00900129554154\\
1098.75	1.00845484226093\\
1099	1.00755634114658\\
1099.25	1.0070246661122\\
1099.5	1.00628836685144\\
1099.75	1.0055580604286\\
1100	1.00572201109334\\
1100.25	1.00561212824313\\
1100.5	1.00548762577234\\
1100.75	1.0052677442122\\
1101	1.00490725552498\\
1101.25	1.00497209851886\\
1101.5	1.00462580437284\\
1101.75	1.00391861013814\\
1102	1.00329997791937\\
1102.25	1.00312548905973\\
1102.5	1.00280226188212\\
1102.75	1.0023277813971\\
1103	1.00226617431639\\
1103.25	1.0025019275711\\
1103.5	1.00311612050058\\
1103.75	1.00321924222094\\
1104	1.00356675082924\\
1104.25	1.00440278279332\\
1104.5	1.00489846273779\\
1104.75	1.00576450650604\\
1105	1.00645835044881\\
1105.25	1.0072576516728\\
1105.5	1.00820367928982\\
1105.75	1.00864494250174\\
1106	1.009061971372\\
1106.25	1.00858552100959\\
1106.5	1.00866712700954\\
1106.75	1.00869129317135\\
1107	1.00814082009013\\
1107.25	1.0082197362302\\
1107.5	1.00861907873287\\
1107.75	1.00931257231904\\
1108	1.00948819532961\\
1108.25	1.01027760783704\\
1108.5	1.0114785082506\\
1108.75	1.01240230074164\\
1109	1.01294678559472\\
1109.25	1.01402725004291\\
1109.5	1.01549664133129\\
1109.75	1.01625854576783\\
1110	1.01706643743553\\
1110.25	1.01793946452051\\
1110.5	1.0189708741144\\
1110.75	1.02020899720371\\
1111	1.02107706788383\\
1111.25	1.02183690535311\\
1111.5	1.02345494468048\\
1111.75	1.02473475410033\\
1112	1.02573423850134\\
1112.25	1.02673519232136\\
1112.5	1.02755400941973\\
1112.75	1.02909150377743\\
1113	1.03028651802248\\
1113.25	1.03141701573143\\
1113.5	1.03360729677962\\
1113.75	1.03518934211697\\
1114	1.03683904870479\\
1114.25	1.03903494318097\\
1114.5	1.04068211420094\\
1114.75	1.04191427518966\\
1115	1.04234489022892\\
1115.25	1.04300762186924\\
1115.5	1.04388421971348\\
1115.75	1.0440434234267\\
1116	1.04440764523571\\
1116.25	1.04509463130608\\
1116.5	1.04577412975889\\
1116.75	1.04644889954479\\
1117	1.0468403234243\\
1117.25	1.04700591605961\\
1117.5	1.04727422955792\\
1117.75	1.04765434189933\\
1118	1.04765787507341\\
1118.25	1.04761604056671\\
1118.5	1.04754991385465\\
1118.75	1.04693786825083\\
1119	1.04606141418109\\
1119.25	1.04523991598347\\
1119.5	1.04472326095788\\
1119.75	1.04378474247074\\
1120	1.04277177831238\\
1120.25	1.04216980739081\\
1120.5	1.04141688296528\\
1120.75	1.04068293616487\\
1121	1.04019900938756\\
1121.25	1.03998431034325\\
1121.5	1.03953991954338\\
1121.75	1.03989677548184\\
1122	1.04060609443644\\
1122.25	1.04082333569196\\
1122.5	1.04081860552719\\
1122.75	1.0412073131639\\
1123	1.04173500385708\\
1123.25	1.04195488011902\\
1123.5	1.04209821686319\\
1123.75	1.0425051252978\\
1124	1.0429187855143\\
1124.25	1.04312433672959\\
1124.5	1.04362633484688\\
1124.75	1.0434390636678\\
1125	1.04271125347972\\
1125.25	1.04221862073433\\
1125.5	1.04118599301376\\
1125.75	1.04001498030456\\
1126	1.0387162244055\\
1126.25	1.03752985308582\\
1126.5	1.03704959877749\\
1126.75	1.03635215176426\\
1127	1.03592619279454\\
1127.25	1.03529511246601\\
1127.5	1.03429177199833\\
1127.75	1.03372941453753\\
1128	1.03245024996836\\
1128.25	1.03062350334888\\
1128.5	1.02885875031272\\
1128.75	1.02708571068443\\
1129	1.02535560689069\\
1129.25	1.02279095014215\\
1129.5	1.02003436707799\\
1129.75	1.01777956921098\\
1130	1.01555597708211\\
1130.25	1.01331705033722\\
1130.5	1.01073365289478\\
1130.75	1.00847181943591\\
1131	1.00685071841905\\
1131.25	1.00526532284024\\
1131.5	1.00368608324309\\
1131.75	1.0016435144627\\
1132	1.00016739885526\\
1132.25	0.998473742145533\\
1132.5	0.996380151942082\\
1132.75	0.994634061055138\\
1133	0.99315192734777\\
1133.25	0.991977850049932\\
1133.5	0.990567566577757\\
1133.75	0.989502940156611\\
1134	0.988478086440567\\
1134.25	0.986993475993569\\
1134.5	0.984909711703945\\
1134.75	0.982675674686485\\
1135	0.981172232309819\\
1135.25	0.979814149475713\\
1135.5	0.978423931498857\\
1135.75	0.977169645317935\\
1136	0.97651194651358\\
1136.25	0.976238905887391\\
1136.5	0.975505025250621\\
1136.75	0.974656297371529\\
1137	0.973760651967327\\
1137.25	0.972664629399659\\
1137.5	0.971352872361237\\
1137.75	0.970308896744583\\
1138	0.969224513301864\\
1138.25	0.968237143652411\\
1138.5	0.967478562721216\\
1138.75	0.967207533909189\\
1139	0.966937958827597\\
1139.25	0.966189808660287\\
1139.5	0.965653468489863\\
1139.75	0.964945934830765\\
1140	0.9644215163389\\
1140.25	0.963417534447791\\
1140.5	0.962122044557776\\
1140.75	0.961467787236058\\
1141	0.960781858808828\\
1141.25	0.960198535771679\\
1141.5	0.959520690140092\\
1141.75	0.959080241446615\\
1142	0.958830085786489\\
1142.25	0.958655622113334\\
1142.5	0.958462374908514\\
1142.75	0.958064298873875\\
1143	0.958101635681371\\
1143.25	0.958074201979147\\
1143.5	0.958338872449904\\
1143.75	0.958579133445008\\
1144	0.958851863506297\\
1144.25	0.959621305588366\\
1144.5	0.959722381963366\\
1144.75	0.959931053697991\\
1145	0.960376221846643\\
1145.25	0.960601342328713\\
1145.5	0.960472643735441\\
1145.75	0.960122064296968\\
1146	0.96030491273826\\
1146.25	0.960454139114929\\
1146.5	0.9600591114708\\
1146.75	0.95985634129897\\
1147	0.960206332506134\\
1147.25	0.960482740618103\\
1147.5	0.960020589521087\\
1147.75	0.959528760619253\\
1148	0.959463179580288\\
1148.25	0.959077161035772\\
1148.5	0.958340407661105\\
1148.75	0.957902162976333\\
1149	0.957924525749406\\
1149.25	0.95838152987867\\
1149.5	0.958836052902095\\
1149.75	0.959259112704004\\
1150	0.9596914578456\\
1150.25	0.959901988157629\\
1150.5	0.960409347988324\\
1150.75	0.96075343186425\\
1151	0.961107732350328\\
1151.25	0.961383683366847\\
1151.5	0.961463806015989\\
1151.75	0.961621601333718\\
1152	0.961458397531025\\
1152.25	0.961282012984537\\
1152.5	0.960619042412705\\
1152.75	0.959924843810741\\
1153	0.959833596873759\\
1153.25	0.960338667673765\\
1153.5	0.961015455480014\\
1153.75	0.961261705600523\\
1154	0.962065650123126\\
1154.25	0.963352796117517\\
1154.5	0.964656497983729\\
1154.75	0.964978664431026\\
1155	0.964934507861475\\
1155.25	0.965533605480256\\
1155.5	0.965660845092278\\
1155.75	0.965715217502337\\
1156	0.965665888725066\\
1156.25	0.965919890439398\\
1156.5	0.966536440875801\\
1156.75	0.966729908380674\\
1157	0.96743647614969\\
1157.25	0.967985021876537\\
1157.5	0.968531614931306\\
1157.75	0.968974110120347\\
1158	0.969491000436806\\
1158.25	0.970353119675534\\
1158.5	0.9709931301214\\
1158.75	0.971954907332487\\
1159	0.973181048485897\\
1159.25	0.974603074442337\\
1159.5	0.97622454609254\\
1159.75	0.977683152272801\\
1160	0.979464114137828\\
1160.25	0.981060882912127\\
1160.5	0.982227538187295\\
1160.75	0.98358223249536\\
1161	0.984781554767267\\
1161.25	0.985929270018414\\
1161.5	0.986500770010967\\
1161.75	0.987309937924838\\
1162	0.988579881542975\\
1162.25	0.990046786000272\\
1162.5	0.991620902750423\\
1162.75	0.992888203815796\\
1163	0.994781336270847\\
1163.25	0.996714129902868\\
1163.5	0.998491845720837\\
1163.75	1.00010230413919\\
1164	1.00152875504746\\
1164.25	1.00372519827519\\
1164.5	1.00577461424104\\
1164.75	1.00722685201025\\
1165	1.00883337494364\\
1165.25	1.01096438099367\\
1165.5	1.0136009199144\\
1165.75	1.01587684101912\\
1166	1.01772471579032\\
1166.25	1.02020655339819\\
1166.5	1.02249212844915\\
1166.75	1.02460356000191\\
1167	1.02684337003701\\
1167.25	1.02924487516888\\
1167.5	1.03235312722878\\
1167.75	1.03568884680961\\
1168	1.03911355299499\\
1168.25	1.04207368858784\\
1168.5	1.0448687329376\\
1168.75	1.04796126539458\\
1169	1.05108407235063\\
1169.25	1.05389029530606\\
1169.5	1.05698259276781\\
1169.75	1.06054580078257\\
1170	1.06412532923791\\
1170.25	1.06741270825609\\
1170.5	1.07063216787411\\
1170.75	1.07411146684559\\
1171	1.07779585346193\\
1171.25	1.08158133734366\\
1171.5	1.08511261931265\\
1171.75	1.08886693789801\\
1172	1.0927864421439\\
1172.25	1.0965093098748\\
1172.5	1.09962128918182\\
1172.75	1.10254173106234\\
1173	1.1058865857326\\
1173.25	1.10912866134379\\
1173.5	1.11199078856616\\
1173.75	1.11465065528124\\
1174	1.11756949663638\\
1174.25	1.12043973438533\\
1174.5	1.12273367316202\\
1174.75	1.12472365710697\\
1175	1.12690226946815\\
1175.25	1.12922478342551\\
1175.5	1.13109252469532\\
1175.75	1.13263315158999\\
1176	1.13435510596436\\
1176.25	1.13632104237909\\
1176.5	1.13777418198008\\
1176.75	1.13852789651896\\
1177	1.13934218183679\\
1177.25	1.13968586765407\\
1177.5	1.1401902556105\\
1177.75	1.1405428607772\\
1178	1.14081617449129\\
1178.25	1.14183333755201\\
1178.5	1.14308152032794\\
1178.75	1.14490908635329\\
1179	1.14656468840196\\
1179.25	1.14828184899836\\
1179.5	1.15026306586658\\
1179.75	1.15198897721088\\
1180	1.15376721823955\\
1180.25	1.15578153253391\\
1180.5	1.15770686372506\\
1180.75	1.15932676849456\\
1181	1.16116571453279\\
1181.25	1.16343163109717\\
1181.5	1.1657291001362\\
1181.75	1.16827472862299\\
1182	1.17096120330028\\
1182.25	1.17443423252025\\
1182.5	1.17824316090433\\
1182.75	1.1823173034366\\
1183	1.18663983440076\\
1183.25	1.19057138972881\\
1183.5	1.19473055887063\\
1183.75	1.19913306124243\\
1184	1.20372933901177\\
1184.25	1.20795760742684\\
1184.5	1.21240088195055\\
1184.75	1.21734390059091\\
1185	1.22231147511385\\
1185.25	1.22728476499251\\
1185.5	1.23212780308532\\
1185.75	1.23705749206435\\
1186	1.2418953910552\\
1186.25	1.24712336212868\\
1186.5	1.2523817935044\\
1186.75	1.25766202499754\\
1187	1.26369623647627\\
1187.25	1.27010920491636\\
1187.5	1.27692518299396\\
1187.75	1.28433353785139\\
1188	1.29242831438139\\
1188.25	1.3009390913187\\
1188.5	1.30923861737204\\
1188.75	1.31776977972078\\
1189	1.32647339073696\\
1189.25	1.33462769334034\\
1189.5	1.34265174105987\\
1189.75	1.35056262491389\\
1190	1.35856749024362\\
1190.25	1.36625675777852\\
1190.5	1.37391142155485\\
1190.75	1.38189304067187\\
1191	1.39017758181584\\
1191.25	1.39844266716231\\
1191.5	1.40665749241313\\
1191.75	1.41555396658841\\
1192	1.42465349384529\\
1192.25	1.43369974600999\\
1192.5	1.44265884383745\\
1192.75	1.45180264886669\\
1193	1.46130794824114\\
1193.25	1.47077577443274\\
1193.5	1.48013823207237\\
1193.75	1.48968780664582\\
1194	1.49936481752876\\
1194.25	1.50900536858633\\
1194.5	1.51830419866913\\
1194.75	1.52791746903724\\
1195	1.53783573490977\\
1195.25	1.54813982429166\\
1195.5	1.5591573316748\\
1195.75	1.57085848351817\\
1196	1.58316466682816\\
1196.25	1.59612009251934\\
1196.5	1.60951811940804\\
1196.75	1.62288934342996\\
1197	1.6358333215868\\
1197.25	1.64828522291169\\
1197.5	1.66037467956026\\
1197.75	1.67187422978842\\
1198	1.68298605844436\\
1198.25	1.69430296087965\\
1198.5	1.70591635932708\\
1198.75	1.71854105561619\\
1199	1.73217500655166\\
1199.25	1.74690109633992\\
1199.5	1.76287996230025\\
1199.75	1.7794220074364\\
1200	1.79710402895983\\
1200.25	1.81561927870112\\
1200.5	1.8346992772824\\
1200.75	1.85432699256807\\
1201	1.87447475591105\\
1201.25	1.89550736413164\\
1201.5	1.91637902605781\\
1201.75	1.93700219239364\\
1202	1.95798943285313\\
1202.25	1.97906844195989\\
1202.5	2.00046032745915\\
1202.75	2.02201481796517\\
1203	2.04433357194941\\
1203.25	2.06731092662937\\
1203.5	2.09022676064248\\
1203.75	2.11332295532781\\
1204	2.13674046336729\\
1204.25	2.1601006780361\\
1204.5	2.1837007993055\\
1204.75	2.20766269886018\\
1205	2.23246964675983\\
1205.25	2.25809337059749\\
1205.5	2.28376160577042\\
1205.75	2.31025470871056\\
1206	2.33732803849523\\
1206.25	2.36521448156122\\
1206.5	2.39458359008665\\
1206.75	2.42525759597711\\
1207	2.45854035267621\\
1207.25	2.4947242557784\\
1207.5	2.53379441716434\\
1207.75	2.57525928639759\\
1208	2.61889568370526\\
1208.25	2.66425119878276\\
1208.5	2.71033701016333\\
1208.75	2.75671812755266\\
1209	2.80351671982304\\
1209.25	2.85129170205102\\
1209.5	2.90017571075761\\
1209.75	2.951769600894\\
1210	3.00729951513321\\
1210.25	3.06803011769749\\
1210.5	3.13619631141348\\
1210.75	3.21297134969202\\
1211	3.29979814279086\\
1211.25	3.39631021778156\\
1211.5	3.50256719341587\\
1211.75	3.61836406626427\\
1212	3.74033331201207\\
1212.25	3.86513398328346\\
1212.5	3.9896599664258\\
1212.75	4.11150488803851\\
1213	4.22735226750163\\
1213.25	4.33421171542538\\
1213.5	4.43115492426473\\
1213.75	4.51752274528393\\
1214	4.59317215804895\\
1214.25	4.65720524528313\\
1214.5	4.70985351803421\\
1214.75	4.73713259100585\\
1215	4.7578321550294\\
1215.25	4.7722506191691\\
1215.5	4.77877631346236\\
1215.75	4.77693442045853\\
};
\addlegendentry{learned $\vec{\mu}$};

\end{axis}
\end{tikzpicture}%

%% file: figures/learned_sigma.tex
%
\tikzsetnextfilename{learned_sigma}
\definecolor{mycolor1}{rgb}{0.12157,0.47059,0.70588}%
\begin{tikzpicture}

\begin{axis}[%
width=\widefigurewidth,
height=\widefigureheight,
at={(0\widefigurewidth,0\widefigureheight)},
scale only axis,
xmin=901.75,
xmax=1225.75,
xtick={ 900,  950, 1000, 1050, 1100, 1150, 1200, 1250},
xlabel={rest wavelength $\lambda_{\text{rest}}$ (\AA)},
ymin=-0.25,
ymax=2,
ylabel={normalized flux},
axis background/.style={fill=white},
axis x line*=bottom,
axis y line*=left,
extra x ticks={911.8,973,1026,1215.67}, extra x tick labels={{Ly\,$\infty$},{Ly\,$\gamma$},{Ly\,$\beta$},{Ly\,$\alpha$}},
]
\addplot [color=mycolor1,solid,forget plot]
  table[row sep=crcr]{%
911.75	0.709920505016394\\
912	0.709009074336427\\
912.25	0.709175221736592\\
912.5	0.710585804819908\\
912.75	0.710642854800424\\
913	0.712306809184819\\
913.25	0.71520575703594\\
913.5	0.714952975040743\\
913.75	0.713915214157083\\
914	0.712986987454303\\
914.25	0.714275796806872\\
914.5	0.712014404887479\\
914.75	0.706960252454593\\
915	0.704269405883824\\
915.25	0.700901422110395\\
915.5	0.698992648060202\\
915.75	0.69591081462026\\
916	0.692115200088934\\
916.25	0.689334919727653\\
916.5	0.686787120451155\\
916.75	0.686659127050854\\
917	0.686716454212881\\
917.25	0.687881028932269\\
917.5	0.692001548269729\\
917.75	0.696201339985418\\
918	0.699756454757421\\
918.25	0.700176799150404\\
918.5	0.700067911012011\\
918.75	0.698841379805043\\
919	0.697930974118626\\
919.25	0.698492475689576\\
919.5	0.698639373904942\\
919.75	0.700506410206543\\
920	0.705122517915115\\
920.25	0.708459954218181\\
920.5	0.711292714436919\\
920.75	0.715440798920901\\
921	0.718050619752781\\
921.25	0.720112165909831\\
921.5	0.718625622589952\\
921.75	0.721248055732149\\
922	0.724858086303827\\
922.25	0.726457680508308\\
922.5	0.728735994505374\\
922.75	0.729453885955228\\
923	0.730334216214724\\
923.25	0.728547986383757\\
923.5	0.72499717746286\\
923.75	0.719619917922969\\
924	0.716036905001205\\
924.25	0.713363811343496\\
924.5	0.712926223470803\\
924.75	0.712542232258388\\
925	0.713190902336774\\
925.25	0.713994947766317\\
925.5	0.715311092258385\\
925.75	0.718561617099907\\
926	0.720645639430604\\
926.25	0.724116744581181\\
926.5	0.725678359502351\\
926.75	0.728286844603267\\
927	0.732240883492495\\
927.25	0.736285312953575\\
927.5	0.739686031885578\\
927.75	0.740003005042491\\
928	0.741385465696774\\
928.25	0.746164885345184\\
928.5	0.749135637261988\\
928.75	0.748998672156963\\
929	0.748971189650758\\
929.25	0.752933139933554\\
929.5	0.758506660360619\\
929.75	0.760616614325357\\
930	0.762591678379005\\
930.25	0.766110546365104\\
930.5	0.770925401694474\\
930.75	0.770820472197027\\
931	0.769165077023903\\
931.25	0.768757696492986\\
931.5	0.767328036506552\\
931.75	0.76801573285827\\
932	0.766796525249317\\
932.25	0.767451836960298\\
932.5	0.766138796254436\\
932.75	0.765725393664634\\
933	0.764878223019552\\
933.25	0.761333068219582\\
933.5	0.758664166425783\\
933.75	0.757787304102153\\
934	0.757335101716552\\
934.25	0.75785583609502\\
934.5	0.761181598154835\\
934.75	0.762224287229141\\
935	0.762874477222506\\
935.25	0.764203949540867\\
935.5	0.766997429402005\\
935.75	0.768679981722221\\
936	0.767949849662298\\
936.25	0.770919122739087\\
936.5	0.77449804043642\\
936.75	0.777331178319067\\
937	0.77872313806812\\
937.25	0.779302478598529\\
937.5	0.782196291640573\\
937.75	0.783583295932099\\
938	0.785518938695347\\
938.25	0.786498245442828\\
938.5	0.786811793234412\\
938.75	0.789355897428655\\
939	0.790146534299171\\
939.25	0.788918579904882\\
939.5	0.786478893733675\\
939.75	0.785821384110148\\
940	0.786445199514089\\
940.25	0.785422239848176\\
940.5	0.784534480586778\\
940.75	0.786213828072194\\
941	0.789879625563288\\
941.25	0.790395603533048\\
941.5	0.79060993157776\\
941.75	0.790017628405872\\
942	0.789215077020542\\
942.25	0.788387089663457\\
942.5	0.787030453561817\\
942.75	0.78573170484846\\
943	0.785766850117393\\
943.25	0.78501906420231\\
943.5	0.784823022580407\\
943.75	0.786889231361114\\
944	0.785335015445872\\
944.25	0.78578711080874\\
944.5	0.785950147336962\\
944.75	0.788504981933933\\
945	0.789514819440424\\
945.25	0.787213362343703\\
945.5	0.787123485389913\\
945.75	0.784409578617461\\
946	0.782652798362066\\
946.25	0.780347715569834\\
946.5	0.777916229822946\\
946.75	0.776714829320183\\
947	0.775840292832603\\
947.25	0.77754558780233\\
947.5	0.778171545178317\\
947.75	0.777372807048196\\
948	0.778868570177977\\
948.25	0.781053967132428\\
948.5	0.782130723739824\\
948.75	0.782771697193056\\
949	0.783576175572866\\
949.25	0.785305112063051\\
949.5	0.786354698863936\\
949.75	0.784648583755715\\
950	0.783032811807017\\
950.25	0.780412432537814\\
950.5	0.777112248348155\\
950.75	0.772923718752146\\
951	0.765933032715769\\
951.25	0.76148456081145\\
951.5	0.757133947097764\\
951.75	0.754709069788537\\
952	0.753166347875259\\
952.25	0.750227256980369\\
952.5	0.749628234509629\\
952.75	0.751783433242562\\
953	0.753863495740864\\
953.25	0.754168841730802\\
953.5	0.75420643686347\\
953.75	0.755212408045536\\
954	0.754550797157662\\
954.25	0.749644428392815\\
954.5	0.744355782244591\\
954.75	0.738019875864399\\
955	0.733521389791125\\
955.25	0.728400879120655\\
955.5	0.722627408372566\\
955.75	0.716816358429983\\
956	0.711768991747531\\
956.25	0.709044693661958\\
956.5	0.704398322966852\\
956.75	0.700663733616245\\
957	0.70009336109268\\
957.25	0.699947797983704\\
957.5	0.70079488776425\\
957.75	0.699898034888868\\
958	0.699394200677719\\
958.25	0.700684568601026\\
958.5	0.700960778916\\
958.75	0.700123910976061\\
959	0.699797461738103\\
959.25	0.700541779549886\\
959.5	0.699648343842656\\
959.75	0.697742437505827\\
960	0.695494452642348\\
960.25	0.696208338169781\\
960.5	0.69554406961552\\
960.75	0.695652204226033\\
961	0.696418876005024\\
961.25	0.697125168986574\\
961.5	0.698061151962742\\
961.75	0.698764567764669\\
962	0.698011178922357\\
962.25	0.696769649811744\\
962.5	0.697104511436732\\
962.75	0.698423180775865\\
963	0.701872387097709\\
963.25	0.704523501928933\\
963.5	0.708628898949962\\
963.75	0.712888283393667\\
964	0.716149681516584\\
964.25	0.71773503790587\\
964.5	0.717422482022516\\
964.75	0.717474204806353\\
965	0.719599900211016\\
965.25	0.721143127756733\\
965.5	0.722777670246076\\
965.75	0.723415392165111\\
966	0.725060058843564\\
966.25	0.72560806240642\\
966.5	0.725366161510856\\
966.75	0.724920172326922\\
967	0.724782746449885\\
967.25	0.728273667359656\\
967.5	0.732064481224267\\
967.75	0.736295878872762\\
968	0.742148439290501\\
968.25	0.749028547488549\\
968.5	0.7558736747268\\
968.75	0.762546488053866\\
969	0.767100334208758\\
969.25	0.774432688740647\\
969.5	0.779632978882703\\
969.75	0.784097165125402\\
970	0.789115616796939\\
970.25	0.792842464581117\\
970.5	0.79850455753914\\
970.75	0.801991607561238\\
971	0.80601475224318\\
971.25	0.811002107537091\\
971.5	0.816667214957982\\
971.75	0.82345933359858\\
972	0.829957312848484\\
972.25	0.836657820318157\\
972.5	0.842480588101577\\
972.75	0.847543632859764\\
973	0.851392884102016\\
973.25	0.854639353780954\\
973.5	0.856141397297156\\
973.75	0.857540256670157\\
974	0.859556844053729\\
974.25	0.860463882993703\\
974.5	0.861941704924814\\
974.75	0.862591296985211\\
975	0.865535141638609\\
975.25	0.867251685717286\\
975.5	0.866864118555208\\
975.75	0.866200355218924\\
976	0.865912101251459\\
976.25	0.866707325582987\\
976.5	0.865085337782238\\
976.75	0.86353362996003\\
977	0.864046708491796\\
977.25	0.865685855352875\\
977.5	0.864585929817364\\
977.75	0.8608978081198\\
978	0.85873488564449\\
978.25	0.857347837957957\\
978.5	0.854214394970121\\
978.75	0.849856139662746\\
979	0.847186881280112\\
979.25	0.844675330818887\\
979.5	0.842138504611394\\
979.75	0.839381879144501\\
980	0.838159363421319\\
980.25	0.837722572429548\\
980.5	0.835793215198245\\
980.75	0.833468944975431\\
981	0.830448129148742\\
981.25	0.829159810782965\\
981.5	0.825947904036833\\
981.75	0.824278124157465\\
982	0.823619487075589\\
982.25	0.824775760225595\\
982.5	0.82611421732429\\
982.75	0.826616283425312\\
983	0.828689827691085\\
983.25	0.829738368794232\\
983.5	0.83162047888121\\
983.75	0.833717302296093\\
984	0.835903030074255\\
984.25	0.837681714605759\\
984.5	0.838021786657544\\
984.75	0.838442831186775\\
985	0.838890010742308\\
985.25	0.836328690394244\\
985.5	0.832695354010974\\
985.75	0.832177158156116\\
986	0.835002304871632\\
986.25	0.837243260032107\\
986.5	0.839661193221881\\
986.75	0.843782237241902\\
987	0.849319763475148\\
987.25	0.850970864518706\\
987.5	0.848508965783754\\
987.75	0.846938084233779\\
988	0.846933496525791\\
988.25	0.845872671918775\\
988.5	0.844211936776598\\
988.75	0.84372772300143\\
989	0.844381742812292\\
989.25	0.844438533296921\\
989.5	0.842271727911848\\
989.75	0.842187762668261\\
990	0.84197312150427\\
990.25	0.840728142901086\\
990.5	0.840125238927765\\
990.75	0.839772854727443\\
991	0.840295749396906\\
991.25	0.83957789847955\\
991.5	0.836739916656489\\
991.75	0.833155973376672\\
992	0.830065118087672\\
992.25	0.827242529209551\\
992.5	0.821069246365418\\
992.75	0.814634097060518\\
993	0.809843298040468\\
993.25	0.804469672242984\\
993.5	0.798306857487447\\
993.75	0.792542984967235\\
994	0.791153706439542\\
994.25	0.789989100263779\\
994.5	0.787714447019836\\
994.75	0.786511635992544\\
995	0.784802492199518\\
995.25	0.78259718549396\\
995.5	0.779508055899924\\
995.75	0.776660687076034\\
996	0.77615996743604\\
996.25	0.775873580631836\\
996.5	0.776196902201467\\
996.75	0.775042673520847\\
997	0.770486868436013\\
997.25	0.766236835212769\\
997.5	0.761685407192188\\
997.75	0.757011641093117\\
998	0.752620998544567\\
998.25	0.749301967307739\\
998.5	0.748207785694978\\
998.75	0.747517524584706\\
999	0.746913707439094\\
999.25	0.745064277742432\\
999.5	0.743133860724794\\
999.75	0.74198579484208\\
1000	0.742386201836503\\
1000.25	0.742145557104644\\
1000.5	0.743355032532376\\
1000.75	0.747401691962425\\
1001	0.750171857189794\\
1001.25	0.753224748909351\\
1001.5	0.756048729696749\\
1001.75	0.757501655395843\\
1002	0.757142822167185\\
1002.25	0.75386985669953\\
1002.5	0.753537645995443\\
1002.75	0.752356114812729\\
1003	0.75048396491386\\
1003.25	0.749316738058954\\
1003.5	0.74784984598255\\
1003.75	0.747818797021121\\
1004	0.747654997349091\\
1004.25	0.748397895316642\\
1004.5	0.74738723898678\\
1004.75	0.747262286525699\\
1005	0.748338929420361\\
1005.25	0.751066951445724\\
1005.5	0.750797653694436\\
1005.75	0.750529963654032\\
1006	0.75073701310445\\
1006.25	0.750436259177013\\
1006.5	0.748687907943817\\
1006.75	0.743542699329271\\
1007	0.737008518285511\\
1007.25	0.728832262760218\\
1007.5	0.719800837970456\\
1007.75	0.711558249988269\\
1008	0.705280956521988\\
1008.25	0.701459934414722\\
1008.5	0.700110783304857\\
1008.75	0.700902797517907\\
1009	0.704408813542882\\
1009.25	0.708669586586159\\
1009.5	0.712038932645627\\
1009.75	0.713991361170986\\
1010	0.713862447518555\\
1010.25	0.712407235477171\\
1010.5	0.712046313602434\\
1010.75	0.713197034478481\\
1011	0.71535961715374\\
1011.25	0.717821716356182\\
1011.5	0.722493533897262\\
1011.75	0.726318768382137\\
1012	0.726602569560171\\
1012.25	0.725057533542263\\
1012.5	0.722982303695317\\
1012.75	0.724619127172212\\
1013	0.7283083087286\\
1013.25	0.733430409011991\\
1013.5	0.739572657482049\\
1013.75	0.745028971895381\\
1014	0.751014071364268\\
1014.25	0.75446096510823\\
1014.5	0.756586807130256\\
1014.75	0.75714317988029\\
1015	0.758429496179302\\
1015.25	0.761310220271091\\
1015.5	0.764993889047769\\
1015.75	0.768316048656167\\
1016	0.77045204632511\\
1016.25	0.77420195212845\\
1016.5	0.779384086040447\\
1016.75	0.784164630325328\\
1017	0.786194897428509\\
1017.25	0.788856448580778\\
1017.5	0.793739781150131\\
1017.75	0.798652596497034\\
1018	0.800820603044253\\
1018.25	0.802353358326644\\
1018.5	0.805147635074394\\
1018.75	0.807568614067297\\
1019	0.810349557583793\\
1019.25	0.8141794953684\\
1019.5	0.819533647360422\\
1019.75	0.82498297032463\\
1020	0.830358500926123\\
1020.25	0.838230928295617\\
1020.5	0.846437765591171\\
1020.75	0.854155058467153\\
1021	0.86278378712274\\
1021.25	0.872264834167353\\
1021.5	0.883451900836879\\
1021.75	0.893438124034919\\
1022	0.901855492733493\\
1022.25	0.910006373335443\\
1022.5	0.918862483318\\
1022.75	0.927862103905285\\
1023	0.935293550219516\\
1023.25	0.944513715182752\\
1023.5	0.956576137709533\\
1023.75	0.970430350598609\\
1024	0.982545763302699\\
1024.25	0.99366603034981\\
1024.5	1.00706121192303\\
1024.75	1.01769708498843\\
1025	1.0275838234442\\
1025.25	1.03637056736932\\
1025.5	1.04759667400022\\
1025.75	1.05856560419227\\
1026	1.06548071432615\\
1026.25	1.07290136697777\\
1026.5	1.07880179408197\\
1026.75	1.08449472446963\\
1027	1.08729065596307\\
1027.25	1.09033084896147\\
1027.5	1.09599229903281\\
1027.75	1.10160973526864\\
1028	1.10870188623435\\
1028.25	1.11637113474042\\
1028.5	1.1239590058836\\
1028.75	1.13307996068132\\
1029	1.14124019554301\\
1029.25	1.14641434294981\\
1029.5	1.14861051131693\\
1029.75	1.14866409145647\\
1030	1.1472400322487\\
1030.25	1.14311464978065\\
1030.5	1.13666716838241\\
1030.75	1.13013533169458\\
1031	1.12280501195297\\
1031.25	1.11502773998765\\
1031.5	1.10855538455151\\
1031.75	1.10062668194402\\
1032	1.09452666707134\\
1032.25	1.09142822562961\\
1032.5	1.08969923662494\\
1032.75	1.08860726048762\\
1033	1.08656952910615\\
1033.25	1.08774322034116\\
1033.5	1.08949480950253\\
1033.75	1.08873545161446\\
1034	1.08760131104432\\
1034.25	1.08669228651739\\
1034.5	1.0879962084584\\
1034.75	1.08858155297606\\
1035	1.08688584579566\\
1035.25	1.08461333086404\\
1035.5	1.08119469823976\\
1035.75	1.07474667641271\\
1036	1.06491079068796\\
1036.25	1.05298986867576\\
1036.5	1.04055433499532\\
1036.75	1.02709079678588\\
1037	1.01165047377038\\
1037.25	0.997372155889466\\
1037.5	0.984422844951245\\
1037.75	0.971536370296485\\
1038	0.957911418817741\\
1038.25	0.945442292008656\\
1038.5	0.933665813525715\\
1038.75	0.921671692475864\\
1039	0.910534274774083\\
1039.25	0.900044561997287\\
1039.5	0.89293688941118\\
1039.75	0.884871440795082\\
1040	0.880890866003637\\
1040.25	0.879387071943675\\
1040.5	0.875066304904318\\
1040.75	0.868580831510476\\
1041	0.857706385311402\\
1041.25	0.849839215383743\\
1041.5	0.839555785406926\\
1041.75	0.827356178097244\\
1042	0.815721897894415\\
1042.25	0.806149020105144\\
1042.5	0.799336176484524\\
1042.75	0.790264779442401\\
1043	0.779830011274007\\
1043.25	0.768913496796764\\
1043.5	0.760720733714283\\
1043.75	0.753991741369119\\
1044	0.747357722521641\\
1044.25	0.742132852977647\\
1044.5	0.73764107069009\\
1044.75	0.734532879962593\\
1045	0.73075826949228\\
1045.25	0.726035185447899\\
1045.5	0.721575344888323\\
1045.75	0.715916433300147\\
1046	0.710997408585632\\
1046.25	0.706470641698594\\
1046.5	0.703163350061538\\
1046.75	0.700667856146782\\
1047	0.698688725667042\\
1047.25	0.696667366509285\\
1047.5	0.694800204686728\\
1047.75	0.692825995730066\\
1048	0.691319120582685\\
1048.25	0.690771691416858\\
1048.5	0.688877449354721\\
1048.75	0.688588775335414\\
1049	0.68814291720863\\
1049.25	0.687126477950843\\
1049.5	0.685505069709366\\
1049.75	0.683605005624387\\
1050	0.682350262068791\\
1050.25	0.68096051713408\\
1050.5	0.680298145036182\\
1050.75	0.680419802789606\\
1051	0.67994583174486\\
1051.25	0.679059256766666\\
1051.5	0.678927542796612\\
1051.75	0.678012191342197\\
1052	0.676867583992702\\
1052.25	0.675484965066139\\
1052.5	0.67501988863459\\
1052.75	0.674671142300063\\
1053	0.673123938862141\\
1053.25	0.672048207104011\\
1053.5	0.671525435546708\\
1053.75	0.67076458657938\\
1054	0.669885932513738\\
1054.25	0.66863923677448\\
1054.5	0.668261991999151\\
1054.75	0.668595308785679\\
1055	0.668805725518391\\
1055.25	0.670379814996921\\
1055.5	0.670689999584447\\
1055.75	0.67081103952254\\
1056	0.671814999156345\\
1056.25	0.673191282943938\\
1056.5	0.674308740569707\\
1056.75	0.674554808471721\\
1057	0.674780289114548\\
1057.25	0.675164339691458\\
1057.5	0.67533931821529\\
1057.75	0.674926653535732\\
1058	0.675473406999863\\
1058.25	0.675514500844306\\
1058.5	0.676061814274817\\
1058.75	0.676661802453386\\
1059	0.676376722042081\\
1059.25	0.676885025145798\\
1059.5	0.676571523786103\\
1059.75	0.675367186283919\\
1060	0.674414167996315\\
1060.25	0.674884234402702\\
1060.5	0.676277270397938\\
1060.75	0.676486398493647\\
1061	0.675538138826874\\
1061.25	0.676125057514411\\
1061.5	0.677003708737486\\
1061.75	0.676148725041637\\
1062	0.675015357187101\\
1062.25	0.674771603945534\\
1062.5	0.674974214028035\\
1062.75	0.675357614210606\\
1063	0.675194940012545\\
1063.25	0.675524784914338\\
1063.5	0.67631648606827\\
1063.75	0.676021434749251\\
1064	0.676340858813058\\
1064.25	0.676065550245447\\
1064.5	0.675230123381586\\
1064.75	0.674671724923182\\
1065	0.674471436845021\\
1065.25	0.674592790425447\\
1065.5	0.672861632682828\\
1065.75	0.671165465782225\\
1066	0.670725915952847\\
1066.25	0.671141243912348\\
1066.5	0.670822413295207\\
1066.75	0.67027802255588\\
1067	0.671451791380449\\
1067.25	0.673400480435364\\
1067.5	0.675335852058433\\
1067.75	0.675064066057721\\
1068	0.674540425693938\\
1068.25	0.674235170382626\\
1068.5	0.673393017826581\\
1068.75	0.671646095032174\\
1069	0.669084685060163\\
1069.25	0.668113362989094\\
1069.5	0.66789310552294\\
1069.75	0.667946544908131\\
1070	0.668064005841912\\
1070.25	0.668525798478072\\
1070.5	0.670418831021881\\
1070.75	0.671612854614059\\
1071	0.67182843048498\\
1071.25	0.671682065820842\\
1071.5	0.671420655243983\\
1071.75	0.672027320206345\\
1072	0.671669940013467\\
1072.25	0.671830531051603\\
1072.5	0.673040881920051\\
1072.75	0.674665298443429\\
1073	0.677182486375277\\
1073.25	0.678196879703486\\
1073.5	0.678797150850908\\
1073.75	0.678636501188535\\
1074	0.677520941746222\\
1074.25	0.67617348648679\\
1074.5	0.674164070495396\\
1074.75	0.671728973125913\\
1075	0.669608874884456\\
1075.25	0.668293305656686\\
1075.5	0.667937048363874\\
1075.75	0.667418058364656\\
1076	0.667060682902859\\
1076.25	0.667133674198467\\
1076.5	0.666995924175971\\
1076.75	0.665016604879527\\
1077	0.660290681246779\\
1077.25	0.655396274102734\\
1077.5	0.651159539454485\\
1077.75	0.647360919658794\\
1078	0.642901916188684\\
1078.25	0.639246490804737\\
1078.5	0.637071292704182\\
1078.75	0.635563839307631\\
1079	0.633132486668004\\
1079.25	0.630732400110081\\
1079.5	0.62887726505029\\
1079.75	0.627801426161816\\
1080	0.626615106872382\\
1080.25	0.624381849834052\\
1080.5	0.622449348277763\\
1080.75	0.62078231405128\\
1081	0.619229830588885\\
1081.25	0.618356188455197\\
1081.5	0.618234555008676\\
1081.75	0.617966074488923\\
1082	0.618028754400824\\
1082.25	0.618269366947535\\
1082.5	0.619787762681606\\
1082.75	0.620549813766122\\
1083	0.620649049785835\\
1083.25	0.622133361180522\\
1083.5	0.622813473369835\\
1083.75	0.622853119790504\\
1084	0.621424978925732\\
1084.25	0.620729170963717\\
1084.5	0.621139815144835\\
1084.75	0.620907525762505\\
1085	0.620755628767361\\
1085.25	0.621061315349974\\
1085.5	0.622072182272837\\
1085.75	0.621147996073343\\
1086	0.61898484861394\\
1086.25	0.616555954764077\\
1086.5	0.615033161980426\\
1086.75	0.613095174191491\\
1087	0.610738592571759\\
1087.25	0.6090928238828\\
1087.5	0.607640247905185\\
1087.75	0.606754836437686\\
1088	0.604519640313052\\
1088.25	0.602519344107148\\
1088.5	0.600434663828954\\
1088.75	0.598643774307921\\
1089	0.597250610453415\\
1089.25	0.595620224505324\\
1089.5	0.594965600991168\\
1089.75	0.593012121989358\\
1090	0.590991673868303\\
1090.25	0.589201362070075\\
1090.5	0.587195436942966\\
1090.75	0.584570675842633\\
1091	0.581904713901581\\
1091.25	0.579714314180776\\
1091.5	0.578095656637891\\
1091.75	0.576309029052936\\
1092	0.574555637258448\\
1092.25	0.573153358278577\\
1092.5	0.571806641538695\\
1092.75	0.571841206017142\\
1093	0.571128009505337\\
1093.25	0.570573853287332\\
1093.5	0.56907686624066\\
1093.75	0.56821266426079\\
1094	0.568001248156382\\
1094.25	0.567482671618379\\
1094.5	0.567954325726498\\
1094.75	0.567966053998997\\
1095	0.56774446998464\\
1095.25	0.566311234661725\\
1095.5	0.564555420515523\\
1095.75	0.562010005294635\\
1096	0.558918768616423\\
1096.25	0.556693703834097\\
1096.5	0.556391051759954\\
1096.75	0.557166587638018\\
1097	0.557444850261831\\
1097.25	0.55825357290949\\
1097.5	0.559618032080174\\
1097.75	0.560437662696227\\
1098	0.560614852023869\\
1098.25	0.561103096982756\\
1098.5	0.561475269475938\\
1098.75	0.561436517114394\\
1099	0.560929141393328\\
1099.25	0.560339288620095\\
1099.5	0.559793157267228\\
1099.75	0.558930026671702\\
1100	0.558585763282503\\
1100.25	0.558434988222054\\
1100.5	0.558197684599562\\
1100.75	0.558379926642205\\
1101	0.559704782183987\\
1101.25	0.560743593473774\\
1101.5	0.561437351383119\\
1101.75	0.563428361978381\\
1102	0.565324299418148\\
1102.25	0.567641991333653\\
1102.5	0.568360351635095\\
1102.75	0.569084672728469\\
1103	0.569811373712538\\
1103.25	0.56889247211582\\
1103.5	0.568547509433385\\
1103.75	0.56710076450209\\
1104	0.565819720212969\\
1104.25	0.564139276754015\\
1104.5	0.562793680603066\\
1104.75	0.562066703007484\\
1105	0.560613393238572\\
1105.25	0.559341545817125\\
1105.5	0.558627750191003\\
1105.75	0.558686772863922\\
1106	0.559090744256577\\
1106.25	0.559496578901694\\
1106.5	0.560442571676666\\
1106.75	0.561641093613975\\
1107	0.56228416772517\\
1107.25	0.562674719329666\\
1107.5	0.562597134976113\\
1107.75	0.563783498530337\\
1108	0.564797885415314\\
1108.25	0.565547523153951\\
1108.5	0.567115654847744\\
1108.75	0.569108349821854\\
1109	0.571309288840137\\
1109.25	0.572939783185212\\
1109.5	0.57441819306109\\
1109.75	0.575546960488033\\
1110	0.576147548517319\\
1110.25	0.576162902091015\\
1110.5	0.575593686383791\\
1110.75	0.574161342796814\\
1111	0.573454580446307\\
1111.25	0.573153280396311\\
1111.5	0.573193868303567\\
1111.75	0.573931920928978\\
1112	0.574691134973812\\
1112.25	0.575740716994268\\
1112.5	0.576943830888681\\
1112.75	0.578039875411259\\
1113	0.578807138713267\\
1113.25	0.579606212193921\\
1113.5	0.580931541799732\\
1113.75	0.582498064154146\\
1114	0.583075741782949\\
1114.25	0.584483080715529\\
1114.5	0.585843232196803\\
1114.75	0.586566640011536\\
1115	0.587571400186174\\
1115.25	0.588731977520069\\
1115.5	0.591020201402684\\
1115.75	0.592958587902203\\
1116	0.59485229736873\\
1116.25	0.596805658605448\\
1116.5	0.598141372891412\\
1116.75	0.598333185060566\\
1117	0.597462838401135\\
1117.25	0.596145552987011\\
1117.5	0.595263690435161\\
1117.75	0.594642412200734\\
1118	0.594156952130052\\
1118.25	0.594955415262532\\
1118.5	0.595757231875189\\
1118.75	0.595634955233839\\
1119	0.595493626129483\\
1119.25	0.595085830919628\\
1119.5	0.594304213775027\\
1119.75	0.592778436929677\\
1120	0.591589569818346\\
1120.25	0.59153071059619\\
1120.5	0.591281927284925\\
1120.75	0.591149216135402\\
1121	0.591018432533655\\
1121.25	0.590388291314241\\
1121.5	0.589841618719466\\
1121.75	0.589413827737089\\
1122	0.588939246262953\\
1122.25	0.588231175996083\\
1122.5	0.587977796239003\\
1122.75	0.588545266534785\\
1123	0.588688062170203\\
1123.25	0.588429002485277\\
1123.5	0.58808917060247\\
1123.75	0.587898986099457\\
1124	0.587636466740063\\
1124.25	0.58710952260501\\
1124.5	0.587793170884233\\
1124.75	0.588244023391086\\
1125	0.588599417919159\\
1125.25	0.589426404326053\\
1125.5	0.590201649548855\\
1125.75	0.591402709329067\\
1126	0.591871533405216\\
1126.25	0.592694207303486\\
1126.5	0.592661917097961\\
1126.75	0.590907368597081\\
1127	0.589144420255452\\
1127.25	0.58732708368182\\
1127.5	0.584773240834711\\
1127.75	0.581636384575835\\
1128	0.579141961384232\\
1128.25	0.577982408907164\\
1128.5	0.576751079966525\\
1128.75	0.576181632063334\\
1129	0.575861292762249\\
1129.25	0.576076868949848\\
1129.5	0.57701212029464\\
1129.75	0.577775052386714\\
1130	0.578684668260828\\
1130.25	0.578309010576167\\
1130.5	0.576950486919573\\
1130.75	0.575570385738158\\
1131	0.573994995660068\\
1131.25	0.572055111667052\\
1131.5	0.569487560972522\\
1131.75	0.567074429412881\\
1132	0.565749373842486\\
1132.25	0.564393143114184\\
1132.5	0.562775918637167\\
1132.75	0.561250310363102\\
1133	0.559936417589144\\
1133.25	0.558285193093165\\
1133.5	0.557137584213344\\
1133.75	0.555783808768887\\
1134	0.554732048153285\\
1134.25	0.554282312070753\\
1134.5	0.553811843005936\\
1134.75	0.554058978131195\\
1135	0.553991417663358\\
1135.25	0.554151139861342\\
1135.5	0.553795960091428\\
1135.75	0.552852036696701\\
1136	0.552190804583238\\
1136.25	0.551243178140606\\
1136.5	0.549576269660189\\
1136.75	0.547388590817526\\
1137	0.545987039070074\\
1137.25	0.545407629984877\\
1137.5	0.544954760124537\\
1137.75	0.544456001360425\\
1138	0.544159507142415\\
1138.25	0.544496682580065\\
1138.5	0.544058454266742\\
1138.75	0.542971111605346\\
1139	0.541961805726434\\
1139.25	0.540697914315758\\
1139.5	0.539341441642496\\
1139.75	0.53825337522636\\
1140	0.537824723158021\\
1140.25	0.537632041272803\\
1140.5	0.537245699299024\\
1140.75	0.536409970048032\\
1141	0.53581026461789\\
1141.25	0.53490510277884\\
1141.5	0.53350837421398\\
1141.75	0.532110359050425\\
1142	0.530862941523199\\
1142.25	0.53086953948846\\
1142.5	0.531175071012825\\
1142.75	0.531746752803573\\
1143	0.532092552080837\\
1143.25	0.533531695665097\\
1143.5	0.534781745553796\\
1143.75	0.535381479201294\\
1144	0.53569204101357\\
1144.25	0.535991630194902\\
1144.5	0.536355567920003\\
1144.75	0.534754695717397\\
1145	0.533446584457119\\
1145.25	0.532442389013234\\
1145.5	0.531649854626021\\
1145.75	0.530894152381858\\
1146	0.530420153472105\\
1146.25	0.530501407589386\\
1146.5	0.530237274604781\\
1146.75	0.530176962871268\\
1147	0.530572438824001\\
1147.25	0.531254144922728\\
1147.5	0.532193510056618\\
1147.75	0.533648099905215\\
1148	0.535829141195886\\
1148.25	0.537573404431807\\
1148.5	0.538640055217077\\
1148.75	0.539141229829912\\
1149	0.539636677033301\\
1149.25	0.540175283376446\\
1149.5	0.539815628605494\\
1149.75	0.53925417698994\\
1150	0.538978009523873\\
1150.25	0.538683279425305\\
1150.5	0.537327168134063\\
1150.75	0.535450403783892\\
1151	0.534713211696465\\
1151.25	0.534654581602709\\
1151.5	0.53465480319\\
1151.75	0.535343943258733\\
1152	0.537041640457696\\
1152.25	0.539412818318011\\
1152.5	0.541463798537223\\
1152.75	0.543095854574656\\
1153	0.544524750398668\\
1153.25	0.544889950896395\\
1153.5	0.544643857517568\\
1153.75	0.544123465389757\\
1154	0.543496044873694\\
1154.25	0.542190993819219\\
1154.5	0.541281559678325\\
1154.75	0.541442433247718\\
1155	0.542086934000247\\
1155.25	0.542727150267313\\
1155.5	0.542722137521719\\
1155.75	0.543263548018724\\
1156	0.542946270115625\\
1156.25	0.541997608704768\\
1156.5	0.540477282508431\\
1156.75	0.538774226213781\\
1157	0.537738956448044\\
1157.25	0.537452748545037\\
1157.5	0.538180180501945\\
1157.75	0.539028629746087\\
1158	0.540503695294538\\
1158.25	0.542375608486611\\
1158.5	0.544308132878154\\
1158.75	0.546158904216706\\
1159	0.547643377802626\\
1159.25	0.549197656086658\\
1159.5	0.550616403956474\\
1159.75	0.551373316759166\\
1160	0.55214765445309\\
1160.25	0.552148303659556\\
1160.5	0.552497150279764\\
1160.75	0.553457349314867\\
1161	0.554145899588846\\
1161.25	0.555063968996962\\
1161.5	0.556125324754529\\
1161.75	0.557730715933341\\
1162	0.559133130945842\\
1162.25	0.559766765921239\\
1162.5	0.561133897271469\\
1162.75	0.562718397573385\\
1163	0.564178676652525\\
1163.25	0.56598279543059\\
1163.5	0.567345020379906\\
1163.75	0.568530675631793\\
1164	0.569559063257956\\
1164.25	0.570862099449745\\
1164.5	0.571619699105017\\
1164.75	0.572355153392183\\
1165	0.573488477474345\\
1165.25	0.575219223983935\\
1165.5	0.576748763084002\\
1165.75	0.578294644618117\\
1166	0.579869168088982\\
1166.25	0.581053306386392\\
1166.5	0.582257804475815\\
1166.75	0.583543754551905\\
1167	0.585363310953052\\
1167.25	0.586342490761543\\
1167.5	0.587551296355381\\
1167.75	0.589439586974329\\
1168	0.590916263526947\\
1168.25	0.59170098409612\\
1168.5	0.59237632633989\\
1168.75	0.593845379455214\\
1169	0.59550351084877\\
1169.25	0.596796816900163\\
1169.5	0.598893600459689\\
1169.75	0.601696897182089\\
1170	0.604483457580566\\
1170.25	0.60655840626352\\
1170.5	0.607642566797729\\
1170.75	0.608475834079254\\
1171	0.609058398263986\\
1171.25	0.609750618139041\\
1171.5	0.609674840861928\\
1171.75	0.60985045373587\\
1172	0.610661813827017\\
1172.25	0.61173831390718\\
1172.5	0.61314930226329\\
1172.75	0.613953997663099\\
1173	0.61583836085143\\
1173.25	0.618783790991444\\
1173.5	0.621579448385481\\
1173.75	0.623467417936433\\
1174	0.625288160259976\\
1174.25	0.627463847896779\\
1174.5	0.628612340819815\\
1174.75	0.628686311564764\\
1175	0.628531803492853\\
1175.25	0.628808704769877\\
1175.5	0.628751449621827\\
1175.75	0.62824370629365\\
1176	0.628274965430961\\
1176.25	0.628137684426732\\
1176.5	0.628143814061738\\
1176.75	0.628184749310334\\
1177	0.628470186007752\\
1177.25	0.62915551365029\\
1177.5	0.629401774630532\\
1177.75	0.6296957358129\\
1178	0.630282616809511\\
1178.25	0.631118719351031\\
1178.5	0.631769311416764\\
1178.75	0.63242601263782\\
1179	0.63228079277414\\
1179.25	0.632477892400615\\
1179.5	0.632650598345755\\
1179.75	0.632659243903856\\
1180	0.633253532066344\\
1180.25	0.633825715901079\\
1180.5	0.635251525735931\\
1180.75	0.636613097751964\\
1181	0.638835742009687\\
1181.25	0.641278040548448\\
1181.5	0.642517259041604\\
1181.75	0.64344747066156\\
1182	0.644227604621178\\
1182.25	0.644626530115952\\
1182.5	0.644119389372028\\
1182.75	0.643829001442605\\
1183	0.644029953976565\\
1183.25	0.644239176001841\\
1183.5	0.644900136697616\\
1183.75	0.645921593437304\\
1184	0.64716149033313\\
1184.25	0.648463371847165\\
1184.5	0.650298857396357\\
1184.75	0.652358822249846\\
1185	0.655131699687476\\
1185.25	0.658528096038152\\
1185.5	0.662531875285075\\
1185.75	0.666970745333837\\
1186	0.671304962748288\\
1186.25	0.675970624013499\\
1186.5	0.680492443478862\\
1186.75	0.685075240514521\\
1187	0.689368474644203\\
1187.25	0.693283601848355\\
1187.5	0.696596463666433\\
1187.75	0.698936768301752\\
1188	0.700575646359279\\
1188.25	0.701843133969207\\
1188.5	0.702570483311436\\
1188.75	0.702381083748914\\
1189	0.702103449435683\\
1189.25	0.702394579227086\\
1189.5	0.703089970096552\\
1189.75	0.703244728846433\\
1190	0.70419198840936\\
1190.25	0.706233907627606\\
1190.5	0.709205436323877\\
1190.75	0.712748786682848\\
1191	0.716084759556201\\
1191.25	0.720190712856379\\
1191.5	0.724243533127167\\
1191.75	0.728671728035663\\
1192	0.732844721625142\\
1192.25	0.737420221524943\\
1192.5	0.741914621787592\\
1192.75	0.746515892851923\\
1193	0.750650041407569\\
1193.25	0.75199802907777\\
1193.5	0.749829615353201\\
1193.75	0.743063340814173\\
1194	0.73144123590907\\
1194.25	0.713066795296475\\
1194.5	0.690368644826288\\
1194.75	0.668335692746424\\
1195	0.651071960161814\\
1195.25	0.640332040929681\\
1195.5	0.637373117264727\\
1195.75	0.643950324993799\\
1196	0.658210810115805\\
1196.25	0.676564652320058\\
1196.5	0.69389915363791\\
1196.75	0.704609408748177\\
1197	0.703557439330981\\
1197.25	0.684722918181446\\
1197.5	0.650505234471498\\
1197.75	0.618303189612592\\
1198	0.598205531568736\\
1198.25	0.590455055202138\\
1198.5	0.594221305998808\\
1198.75	0.612009825507508\\
1199	0.641438631448927\\
1199.25	0.661873332115677\\
1199.5	0.653708759049839\\
1199.75	0.620981011582471\\
1200	0.595018287627698\\
1200.25	0.581697903945954\\
1200.5	0.577751569327047\\
1200.75	0.584244160737876\\
1201	0.605915065436405\\
1201.25	0.635594429486078\\
1201.5	0.634156225960484\\
1201.75	0.599381991353218\\
1202	0.576301711942517\\
1202.25	0.564768358476597\\
1202.5	0.557365040924012\\
1202.75	0.550516670115623\\
1203	0.541511552496687\\
1203.25	0.528005418189713\\
1203.5	0.524009338833273\\
1203.75	0.52101850701575\\
1204	0.514901516402897\\
1204.25	0.501711290442917\\
1204.5	0.481137822501241\\
1204.75	0.497623598455406\\
1205	0.508501322421829\\
1205.25	0.507971798510674\\
1205.5	0.494647140080176\\
1205.75	0.467662674423853\\
1206	0.481826786792536\\
1206.25	0.496490837037019\\
1206.5	0.491723426376867\\
1206.75	0.464395279548\\
1207	0.439815983773572\\
1207.25	0.463306151861733\\
1207.5	0.479975024208238\\
1207.75	0.466427192592094\\
1208	0.427911672080936\\
1208.25	0.431163424792072\\
1208.5	0.457891303963371\\
1208.75	0.453134081482196\\
1209	0.414039148003401\\
1209.25	0.404989037754675\\
1209.5	0.433415389675726\\
1209.75	0.428739324753248\\
1210	0.387202302240602\\
1210.25	0.377921761389629\\
1210.5	0.397271567351885\\
1210.75	0.378442691193028\\
1211	0.352324714289069\\
1211.25	0.347863001283878\\
1211.5	0.352329819553239\\
1211.75	0.315320498387105\\
1212	0.301025766433739\\
1212.25	0.299335100260754\\
1212.5	0.255162660521651\\
1212.75	0.241244264426296\\
1213	0.234275689584525\\
1213.25	0.199200312562132\\
1213.5	0.190191851783027\\
1213.75	0.186233653371321\\
1214	0.166597510105861\\
1214.25	0.172531745344781\\
1214.5	0.195539566082313\\
1214.75	0.234953130499046\\
1215	0.232649267575745\\
1215.25	0.226692166427258\\
1215.5	0.328368431007124\\
1215.75	0.326419261913051\\
};
\end{axis}
\end{tikzpicture}%

%% file: figures/normalized_K.tex
%
\tikzsetnextfilename{normalized_K}
\begin{tikzpicture}

\begin{axis}[%
width=\squarefigurewidth,
height=\squarefigureheight,
at={(0\squarefigurewidth,0\squarefigureheight)},
scale only axis,
point meta min=-0.194238921916826,
point meta max=1,
axis on top,
xmin=911.625,
xmax=1215.875,
xtick={911.8,973,1026,1215.67},
xticklabels={{Ly\,$\infty$},{Ly\,$\gamma$},{Ly\,$\beta$},{Ly\,$\alpha$}},
xlabel={rest wavelength $\lr$ (\AA)},
ymin=911.625,
ymax=1215.875,
ytick={911.8,973,1026,1215.67},
yticklabels={{Ly\,$\infty$},{Ly\,$\gamma$},{Ly\,$\beta$},{Ly\,$\alpha$}},
ylabel={rest wavelength $\lr$ (\AA)},
axis background/.style={fill=white},
axis x line*=bottom,
axis y line*=left,
colorbar style={ytick pos=right},
colormap={mymap}{[1pt] rgb(0pt)=(0.0143,0.0143,0.0143); rgb(1pt)=(0.0244084,0.0168792,0.0197143); rgb(2pt)=(0.0343475,0.0194402,0.02529); rgb(3pt)=(0.0440851,0.0219827,0.031023); rgb(4pt)=(0.0535893,0.0245066,0.0369095); rgb(5pt)=(0.062828,0.0270114,0.0429454); rgb(6pt)=(0.071769,0.0294971,0.0491269); rgb(7pt)=(0.0803802,0.0319633,0.05545); rgb(8pt)=(0.0886297,0.0344097,0.0619107); rgb(9pt)=(0.0964852,0.0368363,0.0685051); rgb(10pt)=(0.103915,0.0392426,0.0752293); rgb(11pt)=(0.110886,0.0416284,0.0820793); rgb(12pt)=(0.117368,0.0439936,0.0890511); rgb(13pt)=(0.123327,0.0463378,0.0961409); rgb(14pt)=(0.128732,0.0486608,0.103345); rgb(15pt)=(0.13355,0.0509623,0.110658); rgb(16pt)=(0.13775,0.0532421,0.118078); rgb(17pt)=(0.1413,0.0555,0.1256); rgb(18pt)=(0.144494,0.0577222,0.133297); rgb(19pt)=(0.147635,0.0598991,0.141234); rgb(20pt)=(0.150707,0.0620365,0.149395); rgb(21pt)=(0.15369,0.0641403,0.157763); rgb(22pt)=(0.156566,0.0662164,0.166324); rgb(23pt)=(0.159318,0.0682706,0.175061); rgb(24pt)=(0.161926,0.0703088,0.183957); rgb(25pt)=(0.164372,0.0723369,0.192998); rgb(26pt)=(0.166638,0.0743606,0.202166); rgb(27pt)=(0.168707,0.076386,0.211447); rgb(28pt)=(0.170558,0.0784188,0.220823); rgb(29pt)=(0.172176,0.0804649,0.230279); rgb(30pt)=(0.17354,0.0825301,0.2398); rgb(31pt)=(0.174632,0.0846204,0.249368); rgb(32pt)=(0.175436,0.0867415,0.258968); rgb(33pt)=(0.175931,0.0888995,0.268584); rgb(34pt)=(0.1761,0.0911,0.2782); rgb(35pt)=(0.17608,0.0932893,0.288236); rgb(36pt)=(0.17602,0.0954203,0.299026); rgb(37pt)=(0.175922,0.0975092,0.310428); rgb(38pt)=(0.175786,0.0995723,0.322297); rgb(39pt)=(0.175613,0.101626,0.334492); rgb(40pt)=(0.175404,0.103685,0.346867); rgb(41pt)=(0.17516,0.105768,0.359281); rgb(42pt)=(0.174882,0.107889,0.37159); rgb(43pt)=(0.174571,0.110066,0.383651); rgb(44pt)=(0.174227,0.112313,0.39532); rgb(45pt)=(0.173853,0.114648,0.406454); rgb(46pt)=(0.173448,0.117087,0.41691); rgb(47pt)=(0.173013,0.119645,0.426544); rgb(48pt)=(0.172551,0.122339,0.435213); rgb(49pt)=(0.17206,0.125186,0.442775); rgb(50pt)=(0.171543,0.128201,0.449085); rgb(51pt)=(0.171,0.1314,0.454); rgb(52pt)=(0.170031,0.134957,0.458091); rgb(53pt)=(0.16829,0.139002,0.46202); rgb(54pt)=(0.165863,0.143483,0.465783); rgb(55pt)=(0.162835,0.148348,0.469374); rgb(56pt)=(0.15929,0.153547,0.47279); rgb(57pt)=(0.155314,0.159029,0.476026); rgb(58pt)=(0.150992,0.164743,0.479077); rgb(59pt)=(0.146408,0.170638,0.481938); rgb(60pt)=(0.141647,0.176662,0.484605); rgb(61pt)=(0.136795,0.182765,0.487073); rgb(62pt)=(0.131936,0.188896,0.489338); rgb(63pt)=(0.127156,0.195003,0.491395); rgb(64pt)=(0.122538,0.201036,0.49324); rgb(65pt)=(0.118169,0.206943,0.494867); rgb(66pt)=(0.114133,0.212673,0.496273); rgb(67pt)=(0.110515,0.218176,0.497452); rgb(68pt)=(0.1074,0.2234,0.4984); rgb(69pt)=(0.104662,0.228422,0.49922); rgb(70pt)=(0.102104,0.233358,0.500015); rgb(71pt)=(0.099699,0.238221,0.500783); rgb(72pt)=(0.0974245,0.24302,0.50152); rgb(73pt)=(0.0952557,0.247765,0.502223); rgb(74pt)=(0.0931681,0.252468,0.502888); rgb(75pt)=(0.0911374,0.257138,0.503513); rgb(76pt)=(0.0891393,0.261787,0.504093); rgb(77pt)=(0.0871493,0.266426,0.504626); rgb(78pt)=(0.0851432,0.271063,0.505109); rgb(79pt)=(0.0830965,0.275711,0.505538); rgb(80pt)=(0.0809849,0.280379,0.50591); rgb(81pt)=(0.0787839,0.285079,0.506222); rgb(82pt)=(0.0764694,0.28982,0.50647); rgb(83pt)=(0.0740168,0.294614,0.506651); rgb(84pt)=(0.0714018,0.29947,0.506762); rgb(85pt)=(0.0686,0.3044,0.5068); rgb(86pt)=(0.0653572,0.309404,0.506262); rgb(87pt)=(0.0615042,0.314469,0.504714); rgb(88pt)=(0.0571445,0.319589,0.502254); rgb(89pt)=(0.0523815,0.324755,0.498978); rgb(90pt)=(0.0473184,0.32996,0.494986); rgb(91pt)=(0.0420588,0.335197,0.490374); rgb(92pt)=(0.036706,0.340457,0.48524); rgb(93pt)=(0.0313634,0.345734,0.479682); rgb(94pt)=(0.0261344,0.351019,0.473798); rgb(95pt)=(0.0211225,0.356305,0.467685); rgb(96pt)=(0.0164309,0.361585,0.461442); rgb(97pt)=(0.0121631,0.366851,0.455165); rgb(98pt)=(0.00842253,0.372095,0.448952); rgb(99pt)=(0.00531253,0.377309,0.442902); rgb(100pt)=(0.00293651,0.382486,0.437111); rgb(101pt)=(0.00139787,0.387619,0.431678); rgb(102pt)=(0.0008,0.3927,0.4267); rgb(103pt)=(0.000709623,0.397742,0.42201); rgb(104pt)=(0.000624546,0.402766,0.417367); rgb(105pt)=(0.000544794,0.407771,0.412761); rgb(106pt)=(0.000470387,0.412759,0.408183); rgb(107pt)=(0.00040135,0.417729,0.403621); rgb(108pt)=(0.000337705,0.422684,0.399065); rgb(109pt)=(0.000279475,0.427622,0.394505); rgb(110pt)=(0.000226682,0.432546,0.38993); rgb(111pt)=(0.00017935,0.437456,0.385331); rgb(112pt)=(0.000137501,0.442352,0.380696); rgb(113pt)=(0.000101158,0.447235,0.376016); rgb(114pt)=(7.03435e-05,0.452106,0.37128); rgb(115pt)=(4.50806e-05,0.456965,0.366478); rgb(116pt)=(2.5392e-05,0.461813,0.361599); rgb(117pt)=(1.13005e-05,0.466652,0.356634); rgb(118pt)=(2.82893e-06,0.47148,0.351571); rgb(119pt)=(0,0.4763,0.3464); rgb(120pt)=(0,0.481102,0.341072); rgb(121pt)=(0,0.485879,0.335558); rgb(122pt)=(0,0.490634,0.329884); rgb(123pt)=(0,0.49537,0.324072); rgb(124pt)=(0,0.50009,0.318147); rgb(125pt)=(0,0.504796,0.312132); rgb(126pt)=(0,0.509492,0.306052); rgb(127pt)=(0,0.514179,0.29993); rgb(128pt)=(0,0.518862,0.293791); rgb(129pt)=(0,0.523544,0.287657); rgb(130pt)=(0,0.528226,0.281554); rgb(131pt)=(0,0.532912,0.275504); rgb(132pt)=(0,0.537604,0.269532); rgb(133pt)=(0,0.542307,0.263662); rgb(134pt)=(0,0.547021,0.257917); rgb(135pt)=(0,0.551752,0.252322); rgb(136pt)=(0,0.5565,0.2469); rgb(137pt)=(0,0.561307,0.241711); rgb(138pt)=(0,0.5662,0.236772); rgb(139pt)=(0,0.571161,0.232039); rgb(140pt)=(0,0.576172,0.22747); rgb(141pt)=(0,0.581218,0.223024); rgb(142pt)=(0,0.58628,0.218657); rgb(143pt)=(0,0.591341,0.214328); rgb(144pt)=(0,0.596384,0.209994); rgb(145pt)=(0,0.601392,0.205613); rgb(146pt)=(0,0.606347,0.201143); rgb(147pt)=(0,0.611231,0.196542); rgb(148pt)=(0,0.616029,0.191766); rgb(149pt)=(0,0.620721,0.186774); rgb(150pt)=(0,0.625292,0.181524); rgb(151pt)=(0,0.629724,0.175973); rgb(152pt)=(0,0.633999,0.170079); rgb(153pt)=(0,0.6381,0.1638); rgb(154pt)=(0.00153448,0.642042,0.156421); rgb(155pt)=(0.00594177,0.645859,0.147433); rgb(156pt)=(0.0129276,0.649563,0.137101); rgb(157pt)=(0.0221978,0.653167,0.125691); rgb(158pt)=(0.033458,0.656683,0.113468); rgb(159pt)=(0.046414,0.660125,0.100697); rgb(160pt)=(0.0607716,0.663503,0.0876444); rgb(161pt)=(0.0762365,0.666832,0.0745752); rgb(162pt)=(0.0925145,0.670123,0.0617548); rgb(163pt)=(0.109311,0.673389,0.0494487); rgb(164pt)=(0.126333,0.676643,0.0379222); rgb(165pt)=(0.143285,0.679896,0.0274408); rgb(166pt)=(0.159872,0.683162,0.0182699); rgb(167pt)=(0.175802,0.686453,0.010675); rgb(168pt)=(0.19078,0.689781,0.00492137); rgb(169pt)=(0.20451,0.693159,0.00127458); rgb(170pt)=(0.2167,0.6966,0); rgb(171pt)=(0.227739,0.700166,0); rgb(172pt)=(0.238267,0.703889,0); rgb(173pt)=(0.248364,0.707737,0); rgb(174pt)=(0.258112,0.711676,0); rgb(175pt)=(0.267591,0.715675,0); rgb(176pt)=(0.276883,0.7197,0); rgb(177pt)=(0.286067,0.723718,0); rgb(178pt)=(0.295225,0.727696,0); rgb(179pt)=(0.304438,0.731602,0); rgb(180pt)=(0.313786,0.735402,0); rgb(181pt)=(0.32335,0.739065,0); rgb(182pt)=(0.333211,0.742556,0); rgb(183pt)=(0.34345,0.745844,0); rgb(184pt)=(0.354148,0.748895,0); rgb(185pt)=(0.365385,0.751677,0); rgb(186pt)=(0.377242,0.754156,0); rgb(187pt)=(0.3898,0.7563,0); rgb(188pt)=(0.403574,0.758195,0); rgb(189pt)=(0.418885,0.759957,0); rgb(190pt)=(0.435525,0.761599,0); rgb(191pt)=(0.453286,0.763136,0); rgb(192pt)=(0.471958,0.764579,0); rgb(193pt)=(0.491333,0.765943,0); rgb(194pt)=(0.511203,0.767241,0); rgb(195pt)=(0.531358,0.768485,0); rgb(196pt)=(0.551591,0.76969,0); rgb(197pt)=(0.571692,0.770869,0); rgb(198pt)=(0.591453,0.772035,0); rgb(199pt)=(0.610666,0.773201,0); rgb(200pt)=(0.629122,0.77438,0); rgb(201pt)=(0.646612,0.775587,0); rgb(202pt)=(0.662927,0.776833,0); rgb(203pt)=(0.677859,0.778133,0); rgb(204pt)=(0.6912,0.7795,0); rgb(205pt)=(0.70344,0.780892,0.00343064); rgb(206pt)=(0.715224,0.782261,0.0132589); rgb(207pt)=(0.72658,0.783614,0.0287895); rgb(208pt)=(0.737533,0.784956,0.0493267); rgb(209pt)=(0.748107,0.786293,0.0741753); rgb(210pt)=(0.758328,0.787632,0.10264); rgb(211pt)=(0.768223,0.788977,0.134025); rgb(212pt)=(0.777816,0.790335,0.167635); rgb(213pt)=(0.787133,0.791712,0.202774); rgb(214pt)=(0.7962,0.793114,0.238748); rgb(215pt)=(0.805042,0.794545,0.27486); rgb(216pt)=(0.813684,0.796013,0.310415); rgb(217pt)=(0.822153,0.797523,0.344718); rgb(218pt)=(0.830473,0.799081,0.377074); rgb(219pt)=(0.838671,0.800692,0.406786); rgb(220pt)=(0.846771,0.802363,0.43316); rgb(221pt)=(0.8548,0.8041,0.4555); rgb(222pt)=(0.863049,0.805883,0.475346); rgb(223pt)=(0.871713,0.807693,0.494702); rgb(224pt)=(0.880675,0.809539,0.513574); rgb(225pt)=(0.889816,0.811426,0.531964); rgb(226pt)=(0.899019,0.813361,0.549875); rgb(227pt)=(0.908164,0.815352,0.567311); rgb(228pt)=(0.917134,0.817406,0.584275); rgb(229pt)=(0.925811,0.819529,0.600771); rgb(230pt)=(0.934077,0.821728,0.616801); rgb(231pt)=(0.941812,0.824011,0.63237); rgb(232pt)=(0.9489,0.826383,0.64748); rgb(233pt)=(0.955221,0.828853,0.662135); rgb(234pt)=(0.960658,0.831427,0.676338); rgb(235pt)=(0.965093,0.834111,0.690093); rgb(236pt)=(0.968407,0.836914,0.703402); rgb(237pt)=(0.970482,0.839841,0.71627); rgb(238pt)=(0.9712,0.8429,0.7287); rgb(239pt)=(0.9712,0.846163,0.740814); rgb(240pt)=(0.971197,0.84969,0.752722); rgb(241pt)=(0.971189,0.853471,0.764409); rgb(242pt)=(0.971174,0.857495,0.775859); rgb(243pt)=(0.971149,0.861753,0.787056); rgb(244pt)=(0.971112,0.866234,0.797986); rgb(245pt)=(0.97106,0.870928,0.808633); rgb(246pt)=(0.970992,0.875824,0.81898); rgb(247pt)=(0.970903,0.880913,0.829014); rgb(248pt)=(0.970793,0.886183,0.838717); rgb(249pt)=(0.970658,0.891625,0.848075); rgb(250pt)=(0.970497,0.897228,0.857072); rgb(251pt)=(0.970306,0.902981,0.865692); rgb(252pt)=(0.970083,0.908876,0.873921); rgb(253pt)=(0.969826,0.914901,0.881742); rgb(254pt)=(0.969533,0.921046,0.88914); rgb(255pt)=(0.9692,0.9273,0.8961)},
colorbar
]
\addplot [forget plot] graphics [xmin=911.625,xmax=1215.875,ymin=911.625,ymax=1215.875] {figures/normalized_K-1.png};
\end{axis}
\end{tikzpicture}%

%% file: figures/dla_model_sample_closeup.tex
%
\tikzsetnextfilename{dla_model_sample_closeup}
\definecolor{mycolor1}{rgb}{0.12157,0.47059,0.70588}%
\definecolor{mycolor2}{rgb}{0.89020,0.10196,0.10980}%
\begin{tikzpicture}

\begin{axis}[%
width=\widefigurewidth,
height=\widefigureheight,
at={(0\widefigurewidth,0\widefigureheight)},
scale only axis,
xmin=5090,
xmax=5360,
xtick={5100, 5150, 5200, 5250, 5300, 5350},
xlabel={observed wavelength $\lo$ (\AA)},
ymin=-0.3,
ymax=2,
ylabel={normalized flux},
axis background/.style={fill=white},
axis x line*=bottom,
axis y line*=left,
legend style={at={(0.03,0.97)},anchor=north west,legend cell align=left,align=left,fill=none,draw=none}
]
\addplot [color=mycolor1,solid]
  table[row sep=crcr]{%
5099.625	1.02159917384353\\
5100.75	1.01280088114062\\
5101.875	0.809319073230802\\
5103	0.234550565882252\\
5104.125	0.534640017520752\\
5105.25	0.0685500849557911\\
5106.375	0.832787344217755\\
5107.5	0.964300461919037\\
5108.625	1.18899368609248\\
5109.75	0.699983015283192\\
5110.875	1.46780011883363\\
5112	0.343118127572183\\
5113.125	0.466750260849617\\
5114.25	0.830906212055051\\
5115.375	0.827975170433875\\
5116.5	1.31131740155939\\
5117.625	1.22799681675344\\
5118.75	1.55659720761561\\
5119.875	0.441352010923205\\
5121	0.697591186533521\\
5122.125	0.29774216058131\\
5123.25	0.123684957124003\\
5124.375	0.730692742452933\\
5125.5	1.51365953581749\\
5126.625	1.16735653109307\\
5127.75	0.842523707942403\\
5128.875	0.129398061863967\\
5130	0.554125410229294\\
5131.125	0.986528064920423\\
5132.25	1.0488563113434\\
5133.375	1.09084635056353\\
5134.5	1.34884044369178\\
5135.625	1.49016115263846\\
5136.75	0.856952479218683\\
5137.875	1.06454002049934\\
5139	0.842388014324972\\
5140.125	0.981475866643515\\
5141.25	1.19577273569093\\
5142.375	1.2716460377713\\
5143.5	0.527112190179448\\
5144.625	-0.0871024770111798\\
5145.75	0.661409596920413\\
5146.875	0.185131830719414\\
5148	0.910540356811641\\
5149.125	1.55457948041958\\
5150.25	0.684362572386437\\
5151.375	0.929638637434595\\
5152.5	0.307570522525381\\
5153.625	0.67805531281774\\
5154.75	0.744099102990963\\
5155.875	0.754584792905131\\
5157	1.47660069482671\\
5158.125	0.751169688130408\\
5159.25	0.425402053818081\\
5160.375	0.828588156100209\\
5161.5	1.31842755426611\\
5162.625	0.331090199424732\\
5163.75	1.0152508249324\\
5164.875	0.698486318328774\\
5166	0.796396054818452\\
5167.125	0.483269491929921\\
5168.25	0.631550754914973\\
5169.375	0.907691060003524\\
5170.5	0.922679516846689\\
5171.625	0.806911916439153\\
5172.75	0.75694254463787\\
5173.875	0.464628195380938\\
5175	0.629336730181385\\
5176.125	1.12346515920509\\
5177.25	0.927014551331377\\
5178.375	0.876214485145519\\
5179.5	0.811225956576243\\
5180.625	0.203458509172958\\
5181.75	0.593810895682297\\
5182.875	1.3761788291473\\
5184	0.283040612761518\\
5185.125	1.27910682251781\\
5186.25	-0.219497199628465\\
5187.375	0.12187757964021\\
5188.5	0.731465306618197\\
5189.625	0.0961901516236423\\
5190.75	1.05223598194124\\
5191.875	0.583934384675744\\
5193	1.12497415614986\\
5194.125	0.482204419596363\\
5195.25	0.931233963488695\\
5196.375	0.456548021604949\\
5197.5	0.911558750313715\\
5198.625	0.315446658338891\\
5199.75	0.48655598772697\\
5200.875	0.878354196138991\\
5202	0.402440478703626\\
5203.125	0.471532620565224\\
5204.25	0.492084558871284\\
5205.375	0.556281677099473\\
5206.5	0.452074771308002\\
5207.625	0.189003601422989\\
5208.75	0.257474580429272\\
5209.875	0.289980660727767\\
5211	0.41741090560868\\
5212.125	-0.0185132885681789\\
5213.25	0.127266104837222\\
5214.375	0.140582338679034\\
5215.5	-0.0200183584423049\\
5216.625	-0.0126825009433606\\
5217.75	-0.132504561636013\\
5218.875	0.0886097600128126\\
5220	0.0298612847981776\\
5221.125	-0.0921508804117194\\
5222.25	0.148437656713303\\
5223.375	0.00222737643672412\\
5224.5	-0.0262636991295119\\
5225.625	-0.0344053188257217\\
5226.75	-0.0939774164630875\\
5227.875	-0.022920226399812\\
5229	0.0394184028097644\\
5230.125	0.087138007935245\\
5231.25	0.102591747078134\\
5232.375	-0.101825779750578\\
5233.5	-0.106007985187574\\
5234.625	0.1043715248167\\
5235.75	0.0273454781659755\\
5236.875	-0.100396039361203\\
5238	0.0355222607172528\\
5239.125	0.156499327149793\\
5240.25	0.170192718087622\\
5241.375	0.0422721671404449\\
5242.5	0.0630651490135132\\
5243.625	0.345713456172966\\
5244.75	0.16800633158001\\
5245.875	0.144974337988191\\
5247	-0.00847308684655468\\
5248.125	0.276436049012699\\
5249.25	0.453022222860611\\
5250.375	0.488613877151747\\
5251.5	0.178859107994106\\
5252.625	0.385630461179575\\
5253.75	0.641868100706104\\
5254.875	0.397564592489405\\
5256	0.536119721868644\\
5257.125	0.255707266903098\\
5258.25	0.189971090119987\\
5259.375	0.524124497414465\\
5260.5	1.12178597642611\\
5261.625	1.41993914127392\\
5262.75	0.501928532688069\\
5263.875	0.726359225770998\\
5265	0.56055481587114\\
5266.125	0.878080442380704\\
5267.25	0.738343325660329\\
5268.375	0.565029630492151\\
5269.5	1.40805934643352\\
5270.625	0.76937289700529\\
5271.75	0.702520516418284\\
5272.875	0.774840671987561\\
5274	0.708724372487494\\
5275.125	1.04776133867576\\
5276.25	0.0783584368981932\\
5277.375	0.916839212630633\\
5278.5	0.582066994822588\\
5279.625	0.242993092533389\\
5280.75	0.914764210582915\\
5281.875	0.919062572658616\\
5283	0.820832854048573\\
5284.125	0.994723382371802\\
5285.25	0.612605958447157\\
5286.375	1.57316827651616\\
5287.5	0.728333046912556\\
5288.625	0.367691500443664\\
5289.75	0.811964052990266\\
5290.875	0.443218110700419\\
5292	1.09450933222518\\
5293.125	0.757018659572785\\
5294.25	0.946746085118858\\
5295.375	1.28198711688069\\
5296.5	1.30458585159308\\
5297.625	0.917873027838248\\
5298.75	0.758668988727953\\
5299.875	0.930725420565082\\
5301	0.785377323661037\\
5302.125	1.4847065493886\\
5303.25	1.53751931877115\\
5304.375	1.2000660200011\\
5305.5	1.78585036128923\\
5306.625	0.832666707852489\\
5307.75	1.11292888357606\\
5308.875	0.765182285802816\\
5310	0.907887293770189\\
5311.125	1.48888626328994\\
5312.25	0.938459350133822\\
5313.375	1.11415991798929\\
5314.5	1.03269919037396\\
5315.625	1.29904143247372\\
5316.75	1.23045853131011\\
5317.875	1.05454073748025\\
5319	1.5631014317835\\
5320.125	1.58466887926712\\
5321.25	0.672438973874316\\
5322.375	1.12841180832626\\
5323.5	0.507112738547411\\
5324.625	2.25171702083038\\
5325.75	1.27096989524707\\
5326.875	1.12120198615439\\
5328	1.44943459817987\\
5329.125	1.43509970470337\\
5330.25	1.0412357320717\\
5331.375	1.3476321538909\\
5332.5	0.962950003016203\\
5333.625	2.06195458546302\\
5334.75	0.689976770425661\\
5335.875	1.51215198762018\\
5337	1.60047556152842\\
5338.125	0.548693019465593\\
5339.25	1.66321238264695\\
5340.375	0.884064773051455\\
5341.5	1.62178279076776\\
5342.625	2.10533236177769\\
5343.75	2.15121574737282\\
5344.875	1.6370461646036\\
5346	1.32027316954164\\
5347.125	1.59315574207482\\
5348.25	1.4789360029635\\
5349.375	1.84163726597635\\
5350.5	1.20489135775115\\
5351.625	2.17947026943783\\
5352.75	1.44777684432946\\
5353.875	0.939139299041934\\
5355	1.89331556989241\\
};
\addlegendentry{$\vec{y}$ sample};

\addplot [color=mycolor2,solid]
  table[row sep=crcr]{%
5099.625	0.93981649406476\\
5100.75	0.942502798154639\\
5101.875	0.939681544273415\\
5103	0.935265304582024\\
5104.125	0.932077244952112\\
5105.25	0.930339934942234\\
5106.375	0.918258192987215\\
5107.5	0.918117985105692\\
5108.625	0.93064116270239\\
5109.75	0.928251812672438\\
5110.875	0.923309180846668\\
5112	0.929634579099122\\
5113.125	0.92647184497659\\
5114.25	0.908443893355475\\
5115.375	0.909545343165531\\
5116.5	0.913437468503391\\
5117.625	0.901688884968611\\
5118.75	0.899486057310037\\
5119.875	0.904559498506987\\
5121	0.895714257482298\\
5122.125	0.886089712513006\\
5123.25	0.879195403616134\\
5124.375	0.880941675097463\\
5125.5	0.882334665957526\\
5126.625	0.885770423592753\\
5127.75	0.880727488235288\\
5128.875	0.878007813079823\\
5130	0.883683679261498\\
5131.125	0.875885725877187\\
5132.25	0.873513703295892\\
5133.375	0.87346412973136\\
5134.5	0.871059496365632\\
5135.625	0.87349429652162\\
5136.75	0.871377116719132\\
5137.875	0.864298471581843\\
5139	0.857781565211435\\
5140.125	0.85274920968107\\
5141.25	0.846412873817347\\
5142.375	0.847055539539617\\
5143.5	0.846883256065768\\
5144.625	0.850526639513252\\
5145.75	0.849932998642254\\
5146.875	0.84410065926438\\
5148	0.830855852102981\\
5149.125	0.827389903576036\\
5150.25	0.827693244470506\\
5151.375	0.827912701575064\\
5152.5	0.827424591957949\\
5153.625	0.830260545790489\\
5154.75	0.830377694941089\\
5155.875	0.828901603028778\\
5157	0.82835868757116\\
5158.125	0.819562373672586\\
5159.25	0.819486463818482\\
5160.375	0.816595019704821\\
5161.5	0.813178916306125\\
5162.625	0.819043608056638\\
5163.75	0.814227272552607\\
5164.875	0.799034524257768\\
5166	0.794824667373918\\
5167.125	0.788567734292952\\
5168.25	0.777439735458026\\
5169.375	0.782398179081202\\
5170.5	0.777281620520228\\
5171.625	0.770380223694884\\
5172.75	0.765093523375909\\
5173.875	0.757443755647515\\
5175	0.752800155862851\\
5176.125	0.744670237919365\\
5177.25	0.7373850094161\\
5178.375	0.736374486506979\\
5179.5	0.727069415285412\\
5180.625	0.711203991437354\\
5181.75	0.702613476254488\\
5182.875	0.703384401453659\\
5184	0.693300938390366\\
5185.125	0.681938554844115\\
5186.25	0.673667616184659\\
5187.375	0.65835731162232\\
5188.5	0.651486608364785\\
5189.625	0.638137000903754\\
5190.75	0.621432212016531\\
5191.875	0.606555605109459\\
5193	0.581413507041422\\
5194.125	0.5627056004191\\
5195.25	0.551296914031791\\
5196.375	0.532396534661106\\
5197.5	0.511999186855757\\
5198.625	0.485995270662517\\
5199.75	0.457627183095391\\
5200.875	0.431024548630553\\
5202	0.406204259883582\\
5203.125	0.383481031758487\\
5204.25	0.354179370323579\\
5205.375	0.320748704406623\\
5206.5	0.286485408067535\\
5207.625	0.248502410500888\\
5208.75	0.214668666965591\\
5209.875	0.178679759862099\\
5211	0.142992908225218\\
5212.125	0.107711692223796\\
5213.25	0.0772492759655981\\
5214.375	0.0507299407589032\\
5215.5	0.029679994660383\\
5216.625	0.0150160639345507\\
5217.75	0.00621574010049879\\
5218.875	0.00201814153867498\\
5220	0.000471854403197846\\
5221.125	7.28353616281231e-05\\
5222.25	6.72050691394045e-06\\
5223.375	3.24932918503828e-07\\
5224.5	6.12814010096858e-09\\
5225.625	1.18338295596829e-11\\
5226.75	6.08035246985456e-17\\
5227.875	6.27343956823174e-18\\
5229	4.10189482154438e-12\\
5230.125	3.37496115628254e-09\\
5231.25	2.12737663443417e-07\\
5232.375	4.80435605974871e-06\\
5233.5	5.55435722696483e-05\\
5234.625	0.000382617452952525\\
5235.75	0.00173366221848337\\
5236.875	0.00557132038415485\\
5238	0.0138929005615271\\
5239.125	0.0282774309118419\\
5240.25	0.0487435898902951\\
5241.375	0.0750649287137112\\
5242.5	0.106993519222473\\
5243.625	0.14216829717016\\
5244.75	0.180139451148563\\
5245.875	0.217959164367709\\
5247	0.256952260802669\\
5248.125	0.295725357546505\\
5249.25	0.333983768968038\\
5250.375	0.368555826932377\\
5251.5	0.400935561698611\\
5252.625	0.431890176807176\\
5253.75	0.462718826595856\\
5254.875	0.491605430713675\\
5256	0.518996739687555\\
5257.125	0.546657049691707\\
5258.25	0.575326198028092\\
5259.375	0.602765384530213\\
5260.5	0.62570721339263\\
5261.625	0.64507473066848\\
5262.75	0.664516038444124\\
5263.875	0.679606206877324\\
5265	0.694373546304618\\
5266.125	0.71877593096334\\
5267.25	0.740852883014733\\
5268.375	0.759456152603978\\
5269.5	0.769030917930189\\
5270.625	0.779035855284184\\
5271.75	0.792910111818378\\
5272.875	0.802405661884183\\
5274	0.813609991722298\\
5275.125	0.822746080998394\\
5276.25	0.832227521497617\\
5277.375	0.841512891522007\\
5278.5	0.851240716412928\\
5279.625	0.855746148699176\\
5280.75	0.873784095451292\\
5281.875	0.888719908661365\\
5283	0.900032355904407\\
5284.125	0.911646273392157\\
5285.25	0.92059093784054\\
5286.375	0.927170294735588\\
5287.5	0.930788635432304\\
5288.625	0.935516875906522\\
5289.75	0.949340438668413\\
5290.875	0.953292871501698\\
5292	0.961275047730952\\
5293.125	0.970955053138078\\
5294.25	0.97055364040468\\
5295.375	0.968401171411782\\
5296.5	0.971544746715425\\
5297.625	0.975193072182422\\
5298.75	0.970998358474403\\
5299.875	0.97953821492989\\
5301	0.992404188185605\\
5302.125	0.99694181022596\\
5303.25	1.00261330905967\\
5304.375	1.01783733575258\\
5305.5	1.03284093582689\\
5306.625	1.03994489982499\\
5307.75	1.05488419543125\\
5308.875	1.06432518349205\\
5310	1.06497797247379\\
5311.125	1.0735709690647\\
5312.25	1.08893363414267\\
5313.375	1.09368793326485\\
5314.5	1.10517839827853\\
5315.625	1.11503882780589\\
5316.75	1.12251406858547\\
5317.875	1.12446987440528\\
5319	1.13629271626263\\
5320.125	1.15144058157223\\
5321.25	1.15686970950627\\
5322.375	1.16475304473148\\
5323.5	1.18401325507754\\
5324.625	1.18868556723786\\
5325.75	1.19106821513507\\
5326.875	1.20557314480686\\
5328	1.22178663341466\\
5329.125	1.23827429491959\\
5330.25	1.24702994118339\\
5331.375	1.24883776957325\\
5332.5	1.26300381500778\\
5333.625	1.27578830025202\\
5334.75	1.29247173425368\\
5335.875	1.31290402356308\\
5337	1.32796468823301\\
5338.125	1.3425403782512\\
5339.25	1.35366344962835\\
5340.375	1.3702514344173\\
5341.5	1.38682170253807\\
5342.625	1.39889378845329\\
5343.75	1.41269938724992\\
5344.875	1.42937643946039\\
5346	1.43617099737363\\
5347.125	1.44642476343176\\
5348.25	1.46687370990364\\
5349.375	1.47318465327612\\
5350.5	1.48445933787898\\
5351.625	1.49380460698921\\
5352.75	1.49923710143306\\
5353.875	1.50213608359759\\
5355	1.51345251938187\\
};
\addlegendentry{$\vec{f}$ sample};

\addplot [color=mycolor2,dashed]
  table[row sep=crcr]{%
5099.625	0.968924351406191\\
5100.75	0.972222383315031\\
5101.875	0.969853652704584\\
5103	0.965849562987891\\
5104.125	0.963124835065803\\
5105.25	0.96191226529492\\
5106.375	0.950012041571572\\
5107.5	0.950475544158815\\
5108.625	0.964074974741538\\
5109.75	0.962251750249667\\
5110.875	0.95779591724636\\
5112	0.965050253099656\\
5113.125	0.962478369310879\\
5114.25	0.944468734835959\\
5115.375	0.946356075481489\\
5116.5	0.951174535574757\\
5117.625	0.939723615941681\\
5118.75	0.938234080009706\\
5119.875	0.944363161026114\\
5121	0.935984820255386\\
5122.125	0.926802612830752\\
5123.25	0.920488940891096\\
5124.375	0.92324697874689\\
5125.5	0.925670103318274\\
5126.625	0.93027524195513\\
5127.75	0.926008903930135\\
5128.875	0.924212796588934\\
5130	0.931296248500937\\
5131.125	0.924217415953323\\
5132.25	0.922892736816793\\
5133.375	0.924062688399134\\
5134.5	0.922784009012374\\
5135.625	0.926680997591624\\
5136.75	0.925800572378511\\
5137.875	0.919687934081332\\
5139	0.914206887350025\\
5140.125	0.91034719489041\\
5141.25	0.905136915165798\\
5142.375	0.907444413724477\\
5143.5	0.908948491991602\\
5144.625	0.914627823825135\\
5145.75	0.915834419428989\\
5146.875	0.911463508319759\\
5148	0.899130256748459\\
5149.125	0.897429626598373\\
5150.25	0.899904988214607\\
5151.375	0.902392086299067\\
5152.5	0.90421534863927\\
5153.625	0.909793503042659\\
5154.75	0.912524668785462\\
5155.875	0.913632372508359\\
5157	0.915902730345357\\
5158.125	0.909164225898904\\
5159.25	0.912227084414097\\
5160.375	0.912315482192839\\
5161.5	0.911975385057577\\
5162.625	0.922252815878299\\
5163.75	0.920721041966132\\
5164.875	0.907585870316188\\
5166	0.90707041547498\\
5167.125	0.904423849630622\\
5168.25	0.896371080938244\\
5169.375	0.907134276940617\\
5170.5	0.906546513686723\\
5171.625	0.904152997051091\\
5172.75	0.903954736579587\\
5173.875	0.901286034196169\\
5175	0.902553252599178\\
5176.125	0.900029294304204\\
5177.25	0.898928107307036\\
5178.375	0.905999717667806\\
5179.5	0.903419281839551\\
5180.625	0.893110350690855\\
5181.75	0.892420391075372\\
5182.875	0.904414973881337\\
5184	0.903313881414401\\
5185.125	0.901299756941204\\
5186.25	0.904260687579847\\
5187.375	0.89868906978968\\
5188.5	0.905725498930127\\
5189.625	0.905043975885762\\
5190.75	0.900794037424388\\
5191.875	0.900521577028317\\
5193	0.886212609957502\\
5194.125	0.88296512314097\\
5195.25	0.893316887090745\\
5196.375	0.894046050850146\\
5197.5	0.894710232516459\\
5198.625	0.887975632999268\\
5199.75	0.879115838367283\\
5200.875	0.876246442357118\\
5202	0.880625772935248\\
5203.125	0.894689884491294\\
5204.25	0.899024654052881\\
5205.375	0.897528559572966\\
5206.5	0.897995829186896\\
5207.625	0.889874032593057\\
5208.75	0.899868112375041\\
5209.875	0.903985061920895\\
5211	0.907560918422296\\
5212.125	0.901190995066805\\
5213.25	0.908458085361453\\
5214.375	0.911966152581433\\
5215.5	0.911049158342426\\
5216.625	0.911268956630154\\
5217.75	0.905273418454944\\
5218.875	0.909485605278538\\
5220	0.912913213971138\\
5221.125	0.914052597948806\\
5222.25	0.910585545258716\\
5223.375	0.908977800307646\\
5224.5	0.91389056776506\\
5225.625	0.911534472446989\\
5226.75	0.917542774938404\\
5227.875	0.915529524150708\\
5229	0.914999596565398\\
5230.125	0.912743184380868\\
5231.25	0.911718213924595\\
5232.375	0.907430570329747\\
5233.5	0.905911883757184\\
5234.625	0.908777023431197\\
5235.75	0.916233105728407\\
5236.875	0.9179866232783\\
5238	0.927844893081177\\
5239.125	0.935420003173781\\
5240.25	0.929500679002179\\
5241.375	0.925275114914435\\
5242.5	0.92978511386931\\
5243.625	0.930679996203628\\
5244.75	0.934922953757048\\
5245.875	0.933293224410943\\
5247	0.936734456593254\\
5248.125	0.941131270376287\\
5249.25	0.946772237638493\\
5250.375	0.946038644359426\\
5251.5	0.944533082446092\\
5252.625	0.944275933837013\\
5253.75	0.947700876925863\\
5254.875	0.950608699645905\\
5256	0.953810819042244\\
5257.125	0.960242866025133\\
5258.25	0.970632267182233\\
5259.375	0.98079699257929\\
5260.5	0.985521406616682\\
5261.625	0.986596924077146\\
5262.75	0.989622000511686\\
5263.875	0.987891344712234\\
5265	0.987330790509947\\
5266.125	1.0016223620686\\
5267.25	1.0134798407797\\
5268.375	1.02143548710247\\
5269.5	1.0182649088062\\
5270.625	1.01673279039694\\
5271.75	1.02112037776165\\
5272.875	1.02064734087467\\
5274	1.02308400853604\\
5275.125	1.02358251615486\\
5276.25	1.02512794302579\\
5277.375	1.02698582323723\\
5278.5	1.02988159840876\\
5279.625	1.02695680938396\\
5280.75	1.04064985137054\\
5281.875	1.05090036245946\\
5283	1.05715218103393\\
5284.125	1.06404611748526\\
5285.25	1.06810557901852\\
5286.375	1.06971254595613\\
5287.5	1.0682048669253\\
5288.625	1.06825988024679\\
5289.75	1.07891190689986\\
5290.875	1.07854370088472\\
5292	1.08294809156556\\
5293.125	1.08943662004404\\
5294.25	1.08480895784982\\
5295.375	1.07845524358318\\
5296.5	1.0782007722281\\
5297.625	1.07867215322179\\
5298.75	1.07064845543054\\
5299.875	1.07681890396112\\
5301	1.08783309552747\\
5302.125	1.08981260350543\\
5303.25	1.09314171630889\\
5304.375	1.10696008077126\\
5305.5	1.12058384944121\\
5306.625	1.12570007890156\\
5307.75	1.13935815701425\\
5308.875	1.14712912881139\\
5310	1.14550853308292\\
5311.125	1.15250668254845\\
5312.25	1.16681631385081\\
5313.375	1.16980789608559\\
5314.5	1.18005866361615\\
5315.625	1.18861113553539\\
5316.75	1.19466812174667\\
5317.875	1.19490872140249\\
5319	1.20568273117598\\
5320.125	1.22001049717553\\
5321.25	1.22407463523674\\
5322.375	1.23077839935112\\
5323.5	1.24952601533052\\
5324.625	1.2529036989249\\
5325.75	1.25391374073133\\
5326.875	1.26771739896008\\
5328	1.28333154479171\\
5329.125	1.29924480569469\\
5330.25	1.30706433527068\\
5331.375	1.30763555656333\\
5332.5	1.32117396650486\\
5333.625	1.33328211524774\\
5334.75	1.34947694603722\\
5335.875	1.36959046565022\\
5337	1.38410630866997\\
5338.125	1.39812764264211\\
5339.25	1.40856737866253\\
5340.375	1.42470558593314\\
5341.5	1.44083250711411\\
5342.625	1.45229651550901\\
5343.75	1.46557247614279\\
5344.875	1.48183586854901\\
5346	1.48786726245789\\
5347.125	1.49749964053614\\
5348.25	1.51769473787565\\
5349.375	1.52327180619823\\
5350.5	1.53399678038278\\
5351.625	1.54274093513902\\
5352.75	1.54746018443447\\
5353.875	1.54958369499049\\
5355	1.56040583875394\\
};
\addlegendentry{$\vec{f}$ sample (null model)};

\end{axis}
\end{tikzpicture}%

%% file: figures/lni_kde.tex
%
\tikzsetnextfilename{lni_kde}
\definecolor{mycolor1}{rgb}{0.12157,0.47059,0.70588}%
\begin{tikzpicture}

\begin{axis}[%
width=\figurewidth,
height=\figureheight,
at={(0\figurewidth,0\figureheight)},
scale only axis,
xmin=20,
xmax=23,
xlabel={$\lni$},
ymin=0,
ymax=1.2,
ylabel={$p(\lni)$},
axis background/.style={fill=white},
axis x line*=bottom,
axis y line*=left,
legend columns=1
]
\addplot [color=mycolor1,solid,forget plot]
  table[row sep=crcr]{%
20	1.1307893318219\\
20.003003003003	1.13104834750541\\
20.006006006006	1.13128228401642\\
20.009009009009	1.13149112529617\\
20.012012012012	1.13167485700675\\
20.015015015015	1.13183346653253\\
20.018018018018	1.13196694298179\\
20.021021021021	1.13207527718799\\
20.024024024024	1.1321584617104\\
20.027027027027	1.13221649083568\\
20.03003003003	1.13224936057778\\
20.033033033033	1.13225706867887\\
20.036036036036	1.13223961460954\\
20.039039039039	1.13219699956866\\
20.042042042042	1.1321292264835\\
20.045045045045	1.1320363000092\\
20.048048048048	1.13191822652824\\
20.0510510510511	1.13177501414984\\
20.0540540540541	1.13160667270909\\
20.0570570570571	1.13141321376545\\
20.0600600600601	1.13119465060167\\
20.0630630630631	1.13095099822273\\
20.0660660660661	1.13068227335321\\
20.0690690690691	1.13038849443585\\
20.0720720720721	1.13006968162974\\
20.0750750750751	1.12972585680739\\
20.0780780780781	1.12935704355265\\
20.0810810810811	1.12896326715798\\
20.0840840840841	1.12854455462131\\
20.0870870870871	1.12810093464341\\
20.0900900900901	1.12763243762409\\
20.0930930930931	1.12713909565899\\
20.0960960960961	1.12662094253593\\
20.0990990990991	1.12607801373108\\
20.1021021021021	1.1255103464048\\
20.1051051051051	1.12491797939732\\
20.1081081081081	1.12430095322467\\
20.1111111111111	1.12365931007381\\
20.1141141141141	1.12299309379755\\
20.1171171171171	1.12230234991026\\
20.1201201201201	1.12158712558187\\
20.1231231231231	1.12084746963325\\
20.1261261261261	1.12008343253012\\
20.1291291291291	1.11929506637761\\
20.1321321321321	1.1184824249142\\
20.1351351351351	1.11764556350562\\
20.1381381381381	1.11678453913856\\
20.1411411411411	1.11589941041429\\
20.1441441441441	1.11499023754199\\
20.1471471471471	1.11405708233174\\
20.1501501501502	1.11310000818798\\
20.1531531531532	1.11211908010174\\
20.1561561561562	1.11111436464359\\
20.1591591591592	1.11008592995615\\
20.1621621621622	1.10903384574636\\
20.1651651651652	1.10795818327747\\
20.1681681681682	1.1068590153612\\
20.1711711711712	1.10573641634938\\
20.1741741741742	1.10459046212552\\
20.1771771771772	1.10342123009651\\
20.1801801801802	1.10222879918358\\
20.1831831831832	1.10101324981376\\
20.1861861861862	1.09977466391052\\
20.1891891891892	1.09851312488478\\
20.1921921921922	1.09722871762557\\
20.1951951951952	1.09592152849035\\
20.1981981981982	1.09459164529536\\
20.2012012012012	1.09323915730599\\
20.2042042042042	1.09186415522656\\
20.2072072072072	1.09046673119022\\
20.2102102102102	1.0890469787488\\
20.2132132132132	1.08760499286227\\
20.2162162162162	1.08614086988832\\
20.2192192192192	1.08465470757124\\
20.2222222222222	1.08314660503143\\
20.2252252252252	1.08161666275437\\
20.2282282282282	1.08006498257906\\
20.2312312312312	1.07849166768732\\
20.2342342342342	1.07689682259192\\
20.2372372372372	1.07528055312506\\
20.2402402402402	1.07364296642706\\
20.2432432432432	1.07198417093391\\
20.2462462462462	1.07030427636584\\
20.2492492492492	1.0686033937149\\
20.2522522522523	1.06688163523297\\
20.2552552552553	1.06513911441923\\
20.2582582582583	1.06337594600783\\
20.2612612612613	1.06159224595541\\
20.2642642642643	1.059788131428\\
20.2672672672673	1.05796372078859\\
20.2702702702703	1.05611913358413\\
20.2732732732733	1.05425449053238\\
20.2762762762763	1.05236991350868\\
20.2792792792793	1.05046552553304\\
20.2822822822823	1.04854145075643\\
20.2852852852853	1.04659781444743\\
20.2882882882883	1.04463474297857\\
20.2912912912913	1.04265236381285\\
20.2942942942943	1.04065080548971\\
20.2972972972973	1.03863019761138\\
20.3003003003003	1.03659067082886\\
20.3033033033033	1.03453235682776\\
20.3063063063063	1.03245538831418\\
20.3093093093093	1.03035989900061\\
20.3123123123123	1.02824602359142\\
20.3153153153153	1.02611389776864\\
20.3183183183183	1.02396365817735\\
20.3213213213213	1.02179544241094\\
20.3243243243243	1.01960938899715\\
20.3273273273273	1.01740563738243\\
20.3303303303303	1.01518432791774\\
20.3333333333333	1.01294560184362\\
20.3363363363363	1.01068960127513\\
20.3393393393393	1.00841646918684\\
20.3423423423423	1.00612634939796\\
20.3453453453453	1.00381938655695\\
20.3483483483483	1.00149572612668\\
20.3513513513514	0.999155514368911\\
20.3543543543544	0.996798898329101\\
20.3573573573574	0.994426025821036\\
20.3603603603604	0.99203704541135\\
20.3633633633634	0.989632106404449\\
20.3663663663664	0.987211358826373\\
20.3693693693694	0.98477495341013\\
20.3723723723724	0.982323041579141\\
20.3753753753754	0.979855775432492\\
20.3783783783784	0.977373307728637\\
20.3813813813814	0.97487579187008\\
20.3843843843844	0.972363381887609\\
20.3873873873874	0.969836232424378\\
20.3903903903904	0.967294498720348\\
20.3933933933934	0.964738336596264\\
20.3963963963964	0.962167902438041\\
20.3993993993994	0.959583353180756\\
20.4024024024024	0.956984846292778\\
20.4054054054054	0.954372539759987\\
20.4084084084084	0.951746592069876\\
20.4114114114114	0.949107162195474\\
20.4144144144144	0.94645440957964\\
20.4174174174174	0.943788494118964\\
20.4204204204204	0.941109576147893\\
20.4234234234234	0.938417816422676\\
20.4264264264264	0.935713376105737\\
20.4294294294294	0.932996416749371\\
20.4324324324324	0.930267100280108\\
20.4354354354354	0.927525588982428\\
20.4384384384384	0.924772045483133\\
20.4414414414414	0.922006632735385\\
20.4444444444444	0.91922951400269\\
20.4474474474474	0.916440852843013\\
20.4504504504505	0.913640813092948\\
20.4534534534535	0.910829558852063\\
20.4564564564565	0.908007254466655\\
20.4594594594595	0.905174064514137\\
20.4624624624625	0.902330153787296\\
20.4654654654655	0.899475687278576\\
20.4684684684685	0.896610830163992\\
20.4714714714715	0.893735747787921\\
20.4744744744745	0.890850605647029\\
20.4774774774775	0.887955569374878\\
20.4804804804805	0.885050804726179\\
20.4834834834835	0.882136477561222\\
20.4864864864865	0.879212753830503\\
20.4894894894895	0.876279799559096\\
20.4924924924925	0.873337780831164\\
20.4954954954955	0.870386863774866\\
20.4984984984985	0.867427214546718\\
20.5015015015015	0.864458999316358\\
20.5045045045045	0.861482384251444\\
20.5075075075075	0.85849753550233\\
20.5105105105105	0.85550461918694\\
20.5135135135135	0.852503801375789\\
20.5165165165165	0.849495248076925\\
20.5195195195195	0.846479125221077\\
20.5225225225225	0.843455598646488\\
20.5255255255255	0.840424834084421\\
20.5285285285285	0.837386997144179\\
20.5315315315315	0.834342253298599\\
20.5345345345345	0.831290767869261\\
20.5375375375375	0.828232706012041\\
20.5405405405405	0.825168232702863\\
20.5435435435435	0.822097512722729\\
20.5465465465465	0.819020710644079\\
20.5495495495495	0.81593799081602\\
20.5525525525526	0.812849517350484\\
20.5555555555556	0.809755454108072\\
20.5585585585586	0.806655964684112\\
20.5615615615616	0.803551212394666\\
20.5645645645646	0.800441360262764\\
20.5675675675676	0.797326571004637\\
20.5705705705706	0.794207007016167\\
20.5735735735736	0.791082830359349\\
20.5765765765766	0.787954202748749\\
20.5795795795796	0.784821285538115\\
20.5825825825826	0.781684239707412\\
20.5855855855856	0.778543225849236\\
20.5885885885886	0.775398404156096\\
20.5915915915916	0.77224993440724\\
20.5945945945946	0.76909797595592\\
20.5975975975976	0.765942687716679\\
20.6006006006006	0.76278422815245\\
20.6036036036036	0.759622755262216\\
20.6066066066066	0.756458426568696\\
20.6096096096096	0.753291399105642\\
20.6126126126126	0.750121829405948\\
20.6156156156156	0.746949873489373\\
20.6186186186186	0.743775686850502\\
20.6216216216216	0.740599424447146\\
20.6246246246246	0.737421240688162\\
20.6276276276276	0.73424128942204\\
20.6306306306306	0.731059723925484\\
20.6336336336336	0.727876696891663\\
20.6366366366366	0.724692360419073\\
20.6396396396396	0.721506866000424\\
20.6426426426426	0.718320364511423\\
20.6456456456456	0.715133006199905\\
20.6486486486486	0.711944940675167\\
20.6516516516517	0.708756316896959\\
20.6546546546547	0.705567283165065\\
20.6576576576577	0.702377987108931\\
20.6606606606607	0.699188575677226\\
20.6636636636637	0.695999195127607\\
20.6666666666667	0.692809991016875\\
20.6696696696697	0.689621108190773\\
20.6726726726727	0.686432690774394\\
20.6756756756757	0.683244882162241\\
20.6786786786787	0.680057825008894\\
20.6816816816817	0.676871661219556\\
20.6846846846847	0.673686531940682\\
20.6876876876877	0.670502577550793\\
20.6906906906907	0.667319937651668\\
20.6936936936937	0.664138751059228\\
20.6966966966967	0.660959155794863\\
20.6996996996997	0.657781289076761\\
20.7027027027027	0.654605287311541\\
20.7057057057057	0.651431286085746\\
20.7087087087087	0.64825942015771\\
20.7117117117117	0.645089823449642\\
20.7147147147147	0.64192262903924\\
20.7177177177177	0.638757969152399\\
20.7207207207207	0.635595975155134\\
20.7237237237237	0.632436777546289\\
20.7267267267267	0.629280505950109\\
20.7297297297297	0.626127289108891\\
20.7327327327327	0.622977254875965\\
20.7357357357357	0.619830530208721\\
20.7387387387387	0.61668724116171\\
20.7417417417417	0.613547512879975\\
20.7447447447447	0.610411469592465\\
20.7477477477477	0.607279234605718\\
20.7507507507508	0.604150930297449\\
20.7537537537538	0.601026678110517\\
20.7567567567568	0.59790659854689\\
20.7597597597598	0.594790811161856\\
20.7627627627628	0.591679434558111\\
20.7657657657658	0.588572586380413\\
20.7687687687688	0.58547038330998\\
20.7717717717718	0.582372941059325\\
20.7747747747748	0.579280374366975\\
20.7777777777778	0.576192796992515\\
20.7807807807808	0.57311032171187\\
20.7837837837838	0.570033060312196\\
20.7867867867868	0.566961123587597\\
20.7897897897898	0.5638946213346\\
20.7927927927928	0.560833662347752\\
20.7957957957958	0.55777835441553\\
20.7987987987988	0.554728804316334\\
20.8018018018018	0.551685117814403\\
20.8048048048048	0.548647399656237\\
20.8078078078078	0.54561575356694\\
20.8108108108108	0.542590282246674\\
20.8138138138138	0.539571087367313\\
20.8168168168168	0.536558269569261\\
20.8198198198198	0.533551928458489\\
20.8228228228228	0.530552162603368\\
20.8258258258258	0.527559069531982\\
20.8288288288288	0.524572745729517\\
20.8318318318318	0.521593286635617\\
20.8348348348348	0.518620786642097\\
20.8378378378378	0.515655339090542\\
20.8408408408408	0.512697036270397\\
20.8438438438438	0.509745969416697\\
20.8468468468468	0.50680222870842\\
20.8498498498498	0.503865903266649\\
20.8528528528529	0.500937081152968\\
20.8558558558559	0.498015849368049\\
20.8588588588589	0.49510229385031\\
20.8618618618619	0.492196499474602\\
20.8648648648649	0.489298550051333\\
20.8678678678679	0.486408528325284\\
20.8708708708709	0.48352651597489\\
20.8738738738739	0.480652593611588\\
20.8768768768769	0.477786840779145\\
20.8798798798799	0.474929335953313\\
20.8828828828829	0.472080156541408\\
20.8858858858859	0.46923937888216\\
20.8888888888889	0.466407078245762\\
20.8918918918919	0.463583328833599\\
20.8948948948949	0.460768203778783\\
20.8978978978979	0.457961775146152\\
20.9009009009009	0.455164113932854\\
20.9039039039039	0.452375290068844\\
20.9069069069069	0.449595372417442\\
20.9099099099099	0.446824428776288\\
20.9129129129129	0.444062525877981\\
20.9159159159159	0.441309729391275\\
20.9189189189189	0.438566103922088\\
20.9219219219219	0.435831713014734\\
20.9249249249249	0.433106619153304\\
20.9279279279279	0.430390883763096\\
20.9309309309309	0.427684567212223\\
20.9339339339339	0.424987728813145\\
20.9369369369369	0.422300426824613\\
20.9399399399399	0.41962271845344\\
20.9429429429429	0.416954659856569\\
20.9459459459459	0.414296306143151\\
20.9489489489489	0.411647711376793\\
20.951951951952	0.409008928577743\\
20.954954954955	0.406380009725452\\
20.957957957958	0.403761005761008\\
20.960960960961	0.401151966589726\\
20.963963963964	0.39855294108402\\
20.966966966967	0.395963977085999\\
20.96996996997	0.393385121410517\\
20.972972972973	0.390816419848234\\
20.975975975976	0.388257917168542\\
20.978978978979	0.385709657122925\\
20.981981981982	0.383171682448174\\
20.984984984985	0.380644034869769\\
20.987987987988	0.378126755105417\\
20.990990990991	0.375619882868537\\
20.993993993994	0.373123456871954\\
20.996996996997	0.370637514831721\\
21	0.368162093470789\\
21.003003003003	0.365697228523004\\
21.006006006006	0.363242954737135\\
21.009009009009	0.360799305880887\\
21.012012012012	0.358366314745136\\
21.015015015015	0.355944013148079\\
21.018018018018	0.353532431939704\\
21.021021021021	0.351131601006009\\
21.024024024024	0.348741549273628\\
21.027027027027	0.346362304714309\\
21.03003003003	0.343993894349527\\
21.033033033033	0.341636344255312\\
21.036036036036	0.339289679566828\\
21.039039039039	0.336953924483414\\
21.042042042042	0.334629102273372\\
21.045045045045	0.332315235278993\\
21.048048048048	0.33001234492159\\
21.0510510510511	0.327720451706647\\
21.0540540540541	0.325439575229024\\
21.0570570570571	0.323169734178066\\
21.0600600600601	0.32091094634315\\
21.0630630630631	0.318663228618835\\
21.0660660660661	0.316426597010385\\
21.0690690690691	0.314201066639284\\
21.0720720720721	0.311986651748723\\
21.0750750750751	0.309783365709253\\
21.0780780780781	0.307591221024447\\
21.0810810810811	0.30541022933658\\
21.0840840840841	0.303240401432397\\
21.0870870870871	0.301081747248981\\
21.0900900900901	0.298934275879575\\
21.0930930930931	0.2967979955795\\
21.0960960960961	0.294672913772158\\
21.0990990990991	0.292559037055024\\
21.1021021021021	0.29045637120574\\
21.1051051051051	0.28836492118813\\
21.1081081081081	0.286284691158435\\
21.1111111111111	0.284215684471491\\
21.1141141141141	0.282157903686927\\
21.1171171171171	0.28011135057544\\
21.1201201201201	0.278076026125142\\
21.1231231231231	0.276051930547844\\
21.1261261261261	0.274039063285541\\
21.1291291291291	0.272037423016698\\
21.1321321321321	0.270047007662811\\
21.1351351351351	0.268067814394822\\
21.1381381381381	0.26609983963969\\
21.1411411411411	0.264143079086882\\
21.1441441441441	0.262197527695006\\
21.1471471471471	0.260263179698347\\
21.1501501501502	0.258340028613602\\
21.1531531531532	0.256428067246418\\
21.1561561561562	0.254527287698131\\
21.1591591591592	0.252637681372478\\
21.1621621621622	0.250759238982329\\
21.1651651651652	0.248891950556407\\
21.1681681681682	0.247035805446047\\
21.1711711711712	0.245190792332071\\
21.1741741741742	0.243356899231469\\
21.1771771771772	0.241534113504314\\
21.1801801801802	0.239722421860591\\
21.1831831831832	0.237921810366979\\
21.1861861861862	0.236132264453817\\
21.1891891891892	0.234353768921952\\
21.1921921921922	0.232586307949574\\
21.1951951951952	0.2308298650992\\
21.1981981981982	0.229084423324536\\
21.2012012012012	0.227349964977404\\
21.2042042042042	0.225626471814692\\
21.2072072072072	0.223913925005272\\
21.2102102102102	0.222212305136958\\
21.2132132132132	0.220521592223427\\
21.2162162162162	0.218841765711176\\
21.2192192192192	0.217172804486466\\
21.2222222222222	0.215514686882291\\
21.2252252252252	0.213867390685337\\
21.2282282282282	0.212230893142899\\
21.2312312312312	0.210605170969876\\
21.2342342342342	0.208990200355677\\
21.2372372372372	0.207385956971218\\
21.2402402402402	0.205792415975787\\
21.2432432432432	0.204209552024087\\
21.2462462462462	0.202637339273048\\
21.2492492492492	0.20107575138887\\
21.2522522522523	0.199524761553815\\
21.2552552552553	0.197984342473264\\
21.2582582582583	0.196454466382449\\
21.2612612612613	0.194935105053487\\
21.2642642642643	0.193426229802149\\
21.2672672672673	0.19192781149479\\
21.2702702702703	0.190439820555186\\
21.2732732732733	0.188962226971361\\
21.2762762762763	0.187495000302392\\
21.2792792792793	0.186038109685312\\
21.2822822822823	0.184591523841764\\
21.2852852852853	0.183155211084885\\
21.2882882882883	0.181729139326006\\
21.2912912912913	0.180313276081425\\
21.2942942942943	0.178907588479131\\
21.2972972972973	0.17751204326544\\
21.3003003003003	0.176126606811795\\
21.3033033033033	0.174751245121312\\
21.3063063063063	0.173385923835466\\
21.3093093093093	0.172030608240715\\
21.3123123123123	0.170685263275066\\
21.3153153153153	0.169349853534661\\
21.3183183183183	0.168024343280322\\
21.3213213213213	0.166708696444055\\
21.3243243243243	0.165402876635545\\
21.3273273273273	0.164106847148617\\
21.3303303303303	0.162820570967686\\
21.3333333333333	0.161544010774171\\
21.3363363363363	0.160277128952842\\
21.3393393393393	0.159019887598241\\
21.3423423423423	0.157772248520901\\
21.3453453453453	0.156534173253742\\
21.3483483483483	0.155305623058275\\
21.3513513513514	0.154086558930837\\
21.3543543543544	0.152876941608815\\
21.3573573573574	0.151676731576773\\
21.3603603603604	0.150485889072636\\
21.3633633633634	0.149304374093756\\
21.3663663663664	0.148132146402956\\
21.3693693693694	0.14696916553463\\
21.3723723723724	0.145815390800667\\
21.3753753753754	0.144670781296455\\
21.3783783783784	0.143535295906806\\
21.3813813813814	0.142408893311815\\
21.3843843843844	0.141291531992766\\
21.3873873873874	0.14018317023788\\
21.3903903903904	0.13908376614815\\
21.3933933933934	0.137993277643051\\
21.3963963963964	0.136911662466245\\
21.3993993993994	0.135838878191253\\
21.4024024024024	0.134774882227087\\
21.4054054054054	0.133719631823788\\
21.4084084084084	0.13267308407805\\
21.4114114114114	0.131635195938654\\
21.4144144144144	0.130605924211952\\
21.4174174174174	0.129585225567313\\
21.4204204204204	0.128573056542473\\
21.4234234234234	0.127569373548905\\
21.4264264264264	0.126574132877085\\
21.4294294294294	0.125587290701791\\
21.4324324324324	0.12460880308729\\
21.4354354354354	0.123638625992496\\
21.4384384384384	0.122676715276132\\
21.4414414414414	0.121723026701806\\
21.4444444444444	0.120777515943042\\
21.4474474474474	0.119840138588293\\
21.4504504504505	0.1189108501459\\
21.4534534534535	0.117989606048978\\
21.4564564564565	0.117076361660311\\
21.4594594594595	0.116171072277172\\
21.4624624624625	0.115273693136088\\
21.4654654654655	0.114384179417575\\
21.4684684684685	0.113502486250853\\
21.4714714714715	0.112628568718462\\
21.4744744744745	0.111762381860869\\
21.4774774774775	0.110903880681022\\
21.4804804804805	0.110053020148867\\
21.4834834834835	0.109209755205796\\
21.4864864864865	0.108374040769077\\
21.4894894894895	0.107545831736223\\
21.4924924924925	0.106725082989314\\
21.4954954954955	0.105911749399274\\
21.4984984984985	0.105105785830105\\
21.5015015015015	0.104307147143096\\
21.5045045045045	0.103515788200927\\
21.5075075075075	0.102731663871793\\
21.5105105105105	0.101954729033442\\
21.5135135135135	0.101184938577187\\
21.5165165165165	0.100422247411852\\
21.5195195195195	0.0996666104676713\\
21.5225225225225	0.0989179827001873\\
21.5255255255255	0.0981763190940393\\
21.5285285285285	0.0974415746667398\\
21.5315315315315	0.0967137044724099\\
21.5345345345345	0.0959926636054398\\
21.5375375375375	0.0952784072041456\\
21.5405405405405	0.0945708904543373\\
21.5435435435435	0.0938700685928481\\
21.5465465465465	0.0931758969110626\\
21.5495495495495	0.0924883307583257\\
21.5525525525526	0.0918073255453682\\
21.5555555555556	0.0911328367476548\\
21.5585585585586	0.0904648199086965\\
21.5615615615616	0.0898032306433072\\
21.5645645645646	0.0891480246408216\\
21.5675675675676	0.0884991576682763\\
21.5705705705706	0.0878565855735264\\
21.5735735735736	0.0872202642883264\\
21.5765765765766	0.0865901498313684\\
21.5795795795796	0.0859661983112844\\
21.5825825825826	0.0853483659295667\\
21.5855855855856	0.0847366089834797\\
21.5885885885886	0.084130883868921\\
21.5915915915916	0.0835311470832193\\
21.5945945945946	0.0829373552279119\\
21.5975975975976	0.0823494650114512\\
21.6006006006006	0.0817674332519013\\
21.6036036036036	0.0811912168795559\\
21.6066066066066	0.0806207729395192\\
21.6096096096096	0.080056058594278\\
21.6126126126126	0.0794970311261722\\
21.6156156156156	0.0789436479398719\\
21.6186186186186	0.0783958665647859\\
21.6216216216216	0.0778536446574354\\
21.6246246246246	0.0773169400037772\\
21.6276276276276	0.0767857105214917\\
21.6306306306306	0.0762599142622233\\
21.6336336336336	0.0757395094137863\\
21.6366366366366	0.0752244543023233\\
21.6396396396396	0.0747147073944168\\
21.6426426426426	0.0742102272991623\\
21.6456456456456	0.0737109727702157\\
21.6486486486486	0.0732169027077718\\
21.6516516516517	0.0727279761605203\\
21.6546546546547	0.072244152327554\\
21.6576576576577	0.071765390560246\\
21.6606606606607	0.0712916503640702\\
21.6636636636637	0.0708228914004005\\
21.6666666666667	0.0703590734882494\\
21.6696696696697	0.0699001566059958\\
21.6726726726727	0.0694461008930354\\
21.6756756756757	0.068996866651429\\
21.6786786786787	0.0685524143474885\\
21.6816816816817	0.0681127046133312\\
21.6846846846847	0.0676776982484012\\
21.6876876876877	0.0672473562209454\\
21.6906906906907	0.0668216396694574\\
21.6936936936937	0.0664005099040767\\
21.6966966966967	0.0659839284079574\\
21.6996996996997	0.0655718568385962\\
21.7027027027027	0.0651642570291309\\
21.7057057057057	0.0647610909895882\\
21.7087087087087	0.0643623209081174\\
21.7117117117117	0.0639679091521661\\
21.7147147147147	0.0635778182696249\\
21.7177177177177	0.0631920109899597\\
21.7207207207207	0.0628104502252705\\
21.7237237237237	0.0624330990713572\\
21.7267267267267	0.0620599208087084\\
21.7297297297297	0.0616908789035041\\
21.7327327327327	0.0613259370085366\\
21.7357357357357	0.0609650589641315\\
21.7387387387387	0.0606082087990207\\
21.7417417417417	0.0602553507311906\\
21.7447447447447	0.0599064491686932\\
21.7477477477477	0.0595614687104203\\
21.7507507507508	0.0592203741468635\\
21.7537537537538	0.0588831304608168\\
21.7567567567568	0.0585497028280683\\
21.7597597597598	0.0582200566180575\\
21.7627627627628	0.0578941573944933\\
21.7657657657658	0.0575719709159531\\
21.7687687687688	0.0572534631364412\\
21.7717717717718	0.0569386002059257\\
21.7747747747748	0.0566273484708485\\
21.7777777777778	0.0563196744745901\\
21.7807807807808	0.0560155449579228\\
21.7837837837838	0.0557149268594314\\
21.7867867867868	0.0554177873159004\\
21.7897897897898	0.0551240936626757\\
21.7927927927928	0.0548338134340091\\
21.7957957957958	0.0545469143633519\\
21.7987987987988	0.0542633643836564\\
21.8018018018018	0.0539831316276103\\
21.8048048048048	0.0537061844278797\\
21.8078078078078	0.0534324913173097\\
21.8108108108108	0.0531620210291002\\
21.8138138138138	0.0528947424969587\\
21.8168168168168	0.0526306248552292\\
21.8198198198198	0.052369637439\\
21.8228228228228	0.0521117497841744\\
21.8258258258258	0.0518569316275298\\
21.8288288288288	0.0516051529067518\\
21.8318318318318	0.0513563837604384\\
21.8348348348348	0.0511105945280884\\
21.8378378378378	0.0508677557500644\\
21.8408408408408	0.0506278381675308\\
21.8438438438438	0.0503908127223726\\
21.8468468468468	0.0501566505570951\\
21.8498498498498	0.049925323014695\\
21.8528528528529	0.0496968016385189\\
21.8558558558559	0.0494710581720919\\
21.8588588588589	0.0492480645589378\\
21.8618618618619	0.0490277929423639\\
21.8648648648649	0.0488102156652376\\
21.8678678678679	0.0485953052697408\\
21.8708708708709	0.0483830344971041\\
21.8738738738739	0.0481733762873199\\
21.8768768768769	0.0479663037788387\\
21.8798798798799	0.0477617903082523\\
21.8828828828829	0.0475598094099487\\
21.8858858858859	0.0473603348157567\\
21.8888888888889	0.0471633404545699\\
21.8918918918919	0.0469688004519568\\
21.8948948948949	0.0467766891297493\\
21.8978978978979	0.0465869810056196\\
21.9009009009009	0.046399650792635\\
21.9039039039039	0.0462146733988062\\
21.9069069069069	0.0460320239266059\\
21.9099099099099	0.0458516776724863\\
21.9129129129129	0.0456736101263714\\
21.9159159159159	0.0454977969711409\\
21.9189189189189	0.0453242140820946\\
21.9219219219219	0.0451528375264048\\
21.9249249249249	0.0449836435625557\\
21.9279279279279	0.0448166086397701\\
21.9309309309309	0.0446517093974153\\
21.9339339339339	0.0444889226644042\\
21.9369369369369	0.0443282254585806\\
21.9399399399399	0.0441695949860915\\
21.9429429429429	0.0440130086407464\\
21.9459459459459	0.0438584440033643\\
21.9489489489489	0.0437058788411139\\
21.951951951952	0.0435552911068328\\
21.954954954955	0.0434066589383445\\
21.957957957958	0.0432599606577588\\
21.960960960961	0.0431151747707642\\
21.963963963964	0.0429722799659076\\
21.966966966967	0.0428312551138657\\
21.96996996997	0.0426920792667056\\
21.972972972973	0.0425547316571347\\
21.975975975976	0.0424191916977413\\
21.978978978979	0.0422854389802284\\
21.981981981982	0.0421534532746347\\
21.984984984985	0.0420232145285482\\
21.987987987988	0.0418947028663126\\
21.990990990991	0.0417678985882255\\
21.993993993994	0.0416427821697238\\
21.996996996997	0.0415193342605662\\
22	0.0413975356840073\\
22.003003003003	0.0412773674359633\\
22.006006006006	0.0411588106841675\\
22.009009009009	0.0410418467673275\\
22.012012012012	0.0409264571942651\\
22.015015015015	0.0408126236430559\\
22.018018018018	0.0407003279601625\\
22.021021021021	0.0405895521595586\\
22.024024024024	0.0404802784218503\\
22.027027027027	0.0403724890933891\\
22.03003003003	0.0402661666853801\\
22.033033033033	0.040161293872989\\
22.036036036036	0.0400578534944346\\
22.039039039039	0.0399558285500865\\
22.042042042042	0.0398552022015525\\
22.045045045045	0.0397559577707624\\
22.048048048048	0.0396580787390479\\
22.0510510510511	0.0395615487462187\\
22.0540540540541	0.0394663515896343\\
22.0570570570571	0.0393724712232724\\
22.0600600600601	0.0392798917567918\\
22.0630630630631	0.0391885974545951\\
22.0660660660661	0.0390985727348868\\
22.0690690690691	0.0390098021687261\\
22.0720720720721	0.0389222704790822\\
22.0750750750751	0.0388359625398796\\
22.0780780780781	0.0387508633750501\\
22.0810810810811	0.0386669581575727\\
22.0840840840841	0.0385842322085183\\
22.0870870870871	0.0385026709960888\\
22.0900900900901	0.0384222601346582\\
22.0930930930931	0.0383429853838062\\
22.0960960960961	0.0382648326473552\\
22.0990990990991	0.0381877879724055\\
22.1021021021021	0.0381118375483651\\
22.1051051051051	0.038036967705984\\
22.1081081081081	0.0379631649163835\\
22.1111111111111	0.0378904157900865\\
22.1141141141141	0.0378187070760476\\
22.1171171171171	0.0377480256606804\\
22.1201201201201	0.0376783585668892\\
22.1231231231231	0.0376096929530931\\
22.1261261261261	0.0375420161122577\\
22.1291291291291	0.0374753154709215\\
22.1321321321321	0.0374095785882256\\
22.1351351351351	0.0373447931549418\\
22.1381381381381	0.0372809469925023\\
22.1411411411411	0.0372180280520297\\
22.1441441441441	0.0371560244133673\\
22.1471471471471	0.0370949242841104\\
22.1501501501502	0.0370347159986392\\
22.1531531531532	0.0369753880171517\\
22.1561561561562	0.0369169289246984\\
22.1591591591592	0.0368593274302184\\
22.1621621621622	0.0368025723655763\\
22.1651651651652	0.0367466526846015\\
22.1681681681682	0.0366915574621282\\
22.1711711711712	0.0366372758930373\\
22.1741741741742	0.0365837972913006\\
22.1771771771772	0.0365311110890267\\
22.1801801801802	0.0364792068355085\\
22.1831831831832	0.0364280741962738\\
22.1861861861862	0.0363777029521365\\
22.1891891891892	0.0363280829982525\\
22.1921921921922	0.0362792043431752\\
22.1951951951952	0.0362310571079151\\
22.1981981981982	0.0361836315250029\\
22.2012012012012	0.0361369179375523\\
22.2042042042042	0.0360909067983288\\
22.2072072072072	0.0360455886688188\\
22.2102102102102	0.0360009542183044\\
22.2132132132132	0.0359569942229369\\
22.2162162162162	0.0359136995648176\\
22.2192192192192	0.0358710612310801\\
22.2222222222222	0.0358290703129749\\
22.2252252252252	0.0357877180049587\\
22.2282282282282	0.0357469956037861\\
22.2312312312312	0.0357068945076061\\
22.2342342342342	0.0356674062150596\\
22.2372372372372	0.035628522324383\\
22.2402402402402	0.0355902345325139\\
22.2432432432432	0.035552534634201\\
22.2462462462462	0.0355154145211178\\
22.2492492492492	0.0354788661809802\\
22.2522522522523	0.035442881696667\\
22.2552552552553	0.0354074532453454\\
22.2582582582583	0.0353725730976001\\
22.2612612612613	0.0353382336165666\\
22.2642642642643	0.0353044272570676\\
22.2672672672673	0.0352711465647552\\
22.2702702702703	0.0352383841752559\\
22.2732732732733	0.0352061328133198\\
22.2762762762763	0.0351743852919753\\
22.2792792792793	0.0351431345116865\\
22.2822822822823	0.0351123734595163\\
22.2852852852853	0.0350820952082928\\
22.2882882882883	0.0350522929157812\\
22.2912912912913	0.0350229598238585\\
22.2942942942943	0.0349940892576955\\
22.2972972972973	0.0349656746249393\\
22.3003003003003	0.0349377094149045\\
22.3033033033033	0.0349101871977665\\
22.3063063063063	0.0348831016237594\\
22.3093093093093	0.0348564464223804\\
22.3123123123123	0.0348302154015966\\
22.3153153153153	0.0348044024470581\\
22.3183183183183	0.0347790015213154\\
22.3213213213213	0.0347540066630413\\
22.3243243243243	0.0347294119862576\\
22.3273273273273	0.034705211679567\\
22.3303303303303	0.0346814000053897\\
22.3333333333333	0.0346579712992044\\
22.3363363363363	0.0346349199687946\\
22.3393393393393	0.0346122404935002\\
22.3423423423423	0.0345899274234729\\
22.3453453453453	0.0345679753789373\\
22.3483483483483	0.0345463790494567\\
22.3513513513514	0.0345251331932043\\
22.3543543543544	0.0345042326362383\\
22.3573573573574	0.0344836722717831\\
22.3603603603604	0.0344634470595152\\
22.3633633633634	0.0344435520248535\\
22.3663663663664	0.0344239822582547\\
22.3693693693694	0.0344047329145153\\
22.3723723723724	0.0343857992120757\\
22.3753753753754	0.0343671764323324\\
22.3783783783784	0.0343488599189531\\
22.3813813813814	0.0343308450771974\\
22.3843843843844	0.0343131273732429\\
22.3873873873874	0.0342957023335165\\
22.3903903903904	0.0342785655440295\\
22.3933933933934	0.0342617126497191\\
22.3963963963964	0.0342451393537946\\
22.3993993993994	0.0342288414170882\\
22.4024024024024	0.0342128146574109\\
22.4054054054054	0.0341970549489142\\
22.4084084084084	0.0341815582214557\\
22.4114114114114	0.0341663204599708\\
22.4144144144144	0.0341513377038485\\
22.4174174174174	0.0341366060463131\\
22.4204204204204	0.0341221216338101\\
22.4234234234234	0.0341078806653978\\
22.4264264264264	0.0340938793921433\\
22.4294294294294	0.0340801141165238\\
22.4324324324324	0.0340665811918331\\
22.4354354354354	0.0340532770215926\\
22.4384384384384	0.0340401980589675\\
22.4414414414414	0.0340273408061881\\
22.4444444444444	0.0340147018139758\\
22.4474474474474	0.0340022776809741\\
22.4504504504505	0.0339900650531846\\
22.4534534534535	0.0339780606234079\\
22.4564564564565	0.0339662611306891\\
22.4594594594595	0.0339546633597687\\
22.4624624624625	0.033943264140538\\
22.4654654654655	0.0339320603474993\\
22.4684684684685	0.0339210488992313\\
22.4714714714715	0.0339102267578591\\
22.4744744744745	0.0338995909285285\\
22.4774774774775	0.0338891384588863\\
22.4804804804805	0.0338788664385641\\
22.4834834834835	0.0338687719986677\\
22.4864864864865	0.0338588523112706\\
22.4894894894895	0.0338491045889132\\
22.4924924924925	0.0338395260841052\\
22.4954954954955	0.033830114088834\\
22.4984984984985	0.0338208659340772\\
22.5015015015015	0.0338117789893199\\
22.5045045045045	0.0338028506620762\\
22.5075075075075	0.033794078397416\\
22.5105105105105	0.0337854596774956\\
22.5135135135135	0.0337769920210936\\
22.5165165165165	0.0337686729831507\\
22.5195195195195	0.0337605001543144\\
22.5225225225225	0.0337524711604881\\
22.5255255255255	0.0337445836623844\\
22.5285285285285	0.0337368353550834\\
22.5315315315315	0.0337292239675947\\
22.5345345345345	0.0337217472624245\\
22.5375375375375	0.0337144030351466\\
22.5405405405405	0.0337071891139777\\
22.5435435435435	0.0337001033593577\\
22.5465465465465	0.0336931436635331\\
22.5495495495495	0.0336863079501461\\
22.5525525525526	0.0336795941738267\\
22.5555555555556	0.0336730003197896\\
22.5585585585586	0.0336665244034355\\
22.5615615615616	0.0336601644699561\\
22.5645645645646	0.0336539185939436\\
22.5675675675676	0.0336477848790038\\
22.5705705705706	0.033641761457374\\
22.5735735735736	0.0336358464895446\\
22.5765765765766	0.0336300381638848\\
22.5795795795796	0.033624334696272\\
22.5825825825826	0.0336187343297257\\
22.5855855855856	0.0336132353340451\\
22.5885885885886	0.0336078360054506\\
22.5915915915916	0.0336025346662292\\
22.5945945945946	0.0335973296643842\\
22.5975975975976	0.033592219373288\\
22.6006006006006	0.0335872021913392\\
22.6036036036036	0.0335822765416236\\
22.6066066066066	0.0335774408715786\\
22.6096096096096	0.0335726936526616\\
22.6126126126126	0.0335680333800222\\
22.6156156156156	0.0335634585721778\\
22.6186186186186	0.0335589677706928\\
22.6216216216216	0.033554559539862\\
22.6246246246246	0.033550232466397\\
22.6276276276276	0.0335459851591167\\
22.6306306306306	0.0335418162486405\\
22.6336336336336	0.0335377243870861\\
22.6366366366366	0.0335337082477703\\
22.6396396396396	0.0335297665249129\\
22.6426426426426	0.0335258979333445\\
22.6456456456456	0.0335221012082177\\
22.6486486486486	0.0335183751047216\\
22.6516516516517	0.0335147183977994\\
22.6546546546547	0.0335111298818699\\
22.6576576576577	0.0335076083705516\\
22.6606606606607	0.0335041526963908\\
22.6636636636637	0.0335007617105923\\
22.6666666666667	0.0334974342827536\\
22.6696696696697	0.0334941693006025\\
22.6726726726727	0.0334909656697374\\
22.6756756756757	0.0334878223133712\\
22.6786786786787	0.0334847381720778\\
22.6816816816817	0.0334817122035424\\
22.6846846846847	0.0334787433823138\\
22.6876876876877	0.0334758306995609\\
22.6906906906907	0.0334729731628314\\
22.6936936936937	0.0334701697958137\\
22.6966966966967	0.0334674196381017\\
22.6996996996997	0.0334647217449627\\
22.7027027027027	0.0334620751871079\\
22.7057057057057	0.0334594790504661\\
22.7087087087087	0.0334569324359599\\
22.7117117117117	0.0334544344592849\\
22.7147147147147	0.0334519842506915\\
22.7177177177177	0.0334495809547699\\
22.7207207207207	0.0334472237302373\\
22.7237237237237	0.0334449117497281\\
22.7267267267267	0.0334426441995865\\
22.7297297297297	0.0334404202796623\\
22.7327327327327	0.0334382392031089\\
22.7357357357357	0.0334361001961834\\
22.7387387387387	0.033434002498051\\
22.7417417417417	0.0334319453605894\\
22.7447447447447	0.0334299280481985\\
22.7477477477477	0.0334279498376099\\
22.7507507507508	0.0334260100177013\\
22.7537537537538	0.0334241078893115\\
22.7567567567568	0.0334222427650588\\
22.7597597597598	0.0334204139691617\\
22.7627627627628	0.0334186208372614\\
22.7657657657658	0.0334168627162474\\
22.7687687687688	0.0334151389640852\\
22.7717717717718	0.0334134489496456\\
22.7747747747748	0.0334117920525375\\
22.7777777777778	0.0334101676629418\\
22.7807807807808	0.0334085751814484\\
22.7837837837838	0.033407014018895\\
22.7867867867868	0.0334054835962075\\
22.7897897897898	0.0334039833442442\\
22.7927927927928	0.0334025127036402\\
22.7957957957958	0.0334010711246552\\
22.7987987987988	0.0333996580670229\\
22.8018018018018	0.0333982729998025\\
22.8048048048048	0.0333969154012325\\
22.8078078078078	0.033395584758586\\
22.8108108108108	0.0333942805680284\\
22.8138138138138	0.033393002334477\\
22.8168168168168	0.0333917495714627\\
22.8198198198198	0.033390521800993\\
22.8228228228228	0.0333893185534177\\
22.8258258258258	0.0333881393672959\\
22.8288288288288	0.0333869837892653\\
22.8318318318318	0.0333858513739131\\
22.8348348348348	0.0333847416836486\\
22.8378378378378	0.0333836542885778\\
22.8408408408408	0.0333825887663799\\
22.8438438438438	0.0333815447021851\\
22.8468468468468	0.0333805216884546\\
22.8498498498498	0.0333795193248623\\
22.8528528528529	0.0333785372181776\\
22.8558558558559	0.0333775749821505\\
22.8588588588589	0.0333766322373985\\
22.8618618618619	0.0333757086112942\\
22.8648648648649	0.0333748037378555\\
22.8678678678679	0.0333739172576372\\
22.8708708708709	0.0333730488176232\\
22.8738738738739	0.0333721980711219\\
22.8768768768769	0.033371364677662\\
22.8798798798799	0.0333705483028897\\
22.8828828828829	0.0333697486184686\\
22.8858858858859	0.0333689653019794\\
22.8888888888889	0.0333681980368228\\
22.8918918918919	0.0333674465121222\\
22.8948948948949	0.0333667104226294\\
22.8978978978979	0.0333659894686302\\
22.9009009009009	0.0333652833558529\\
22.9039039039039	0.0333645917953768\\
22.9069069069069	0.0333639145035427\\
22.9099099099099	0.0333632512018653\\
22.9129129129129	0.0333626016169455\\
22.9159159159159	0.0333619654803856\\
22.9189189189189	0.0333613425287045\\
22.9219219219219	0.0333607325032549\\
22.9249249249249	0.0333601351501418\\
22.9279279279279	0.0333595502201413\\
22.9309309309309	0.0333589774686222\\
22.9339339339339	0.0333584166554674\\
22.9369369369369	0.0333578675449971\\
22.9399399399399	0.0333573299058931\\
22.9429429429429	0.0333568035111246\\
22.9459459459459	0.0333562881378743\\
22.9489489489489	0.0333557835674665\\
22.951951951952	0.0333552895852958\\
22.954954954955	0.0333548059807574\\
22.957957957958	0.0333543325471774\\
22.960960960961	0.0333538690817459\\
22.963963963964	0.0333534153854495\\
22.966966966967	0.0333529712630057\\
22.96996996997	0.0333525365227982\\
22.972972972973	0.0333521109768133\\
22.975975975976	0.033351694440577\\
22.978978978979	0.0333512867330932\\
22.981981981982	0.0333508876767831\\
22.984984984985	0.0333504970974256\\
22.987987987988	0.0333501148240978\\
22.990990990991	0.0333497406891178\\
22.993993993994	0.0333493745279874\\
22.996996996997	0.033349016179336\\
23	0.0333486654848652\\
};
\end{axis}
\end{tikzpicture}%

%% file: figures/dla_prior.tex
%
\tikzsetnextfilename{dla_prior}
\definecolor{mycolor1}{rgb}{0.12157,0.47059,0.70588}%
\begin{tikzpicture}

\begin{axis}[%
width=\figurewidth,
height=\figureheight,
at={(0\figurewidth,0\figureheight)},
scale only axis,
xmin=2,
xmax=6.5,
xlabel={$\zq$},
ymin=0,
ymax=11,
ylabel={$\Pr(\Md \given \zq)$ (\%)},
axis background/.style={fill=white},
axis x line*=bottom,
axis y line*=left
]
\addplot [color=mycolor1,solid,forget plot]
  table[row sep=crcr]{%
2.15	1.42070165265294\\
2.15094947814941	1.39920045688178\\
2.15116119384766	1.41307450756494\\
2.15193200111389	1.48667601683029\\
2.15394353866577	1.5199233191839\\
2.15427565574646	1.53193817535221\\
2.15494012832642	1.55531512036786\\
2.15569186210632	1.56566338237279\\
2.1563868522644	1.57922912205567\\
2.1567907333374	1.59069643095843\\
2.15685033798218	1.60363490578645\\
2.15693092346191	1.62741465996295\\
2.15894985198975	1.63997933884297\\
2.15906310081482	1.65161290322581\\
2.15994310379028	1.67406467977172\\
2.16078472137451	1.69298799747315\\
2.16093230247498	1.70768458061276\\
2.16194605827332	1.69385975837589\\
2.16393661499023	1.70426679744973\\
2.16563820838928	1.71916326039914\\
2.16593647003174	1.70400381315539\\
2.16836380958557	1.69412314522725\\
2.16993355751038	1.67642668503846\\
2.17093443870544	1.66419696796991\\
2.17294263839722	1.71779141104294\\
2.17313957214355	1.72817482439514\\
2.174640417099	1.74047146948667\\
2.17466735839844	1.7511013215859\\
2.17493271827698	1.7979731938542\\
2.17498683929443	1.80847586883103\\
2.177983045578	1.8193424832429\\
2.17994546890259	1.85978476648208\\
2.18093132972717	1.87409401532408\\
2.18196582794189	1.86399016796395\\
2.18294763565063	1.9059205190592\\
2.18456053733826	1.89587721938008\\
2.18532943725586	1.88585607940447\\
2.18684434890747	1.89624680683828\\
2.1869330406189	1.92345245069322\\
2.18811297416687	1.93592101442261\\
2.18993330001831	1.96711964268745\\
2.19040560722351	1.98216995447648\\
2.19493412971497	2.00625805264127\\
2.19696831703186	2.02175883952856\\
2.19795727729797	2.0340203402034\\
2.19895219802856	2.04462618899458\\
2.19993209838867	2.07436055199086\\
2.20076775550842	2.08662107662634\\
2.20095300674438	2.10351750021821\\
2.20193886756897	2.12286630274673\\
2.20293378829956	2.13608957795004\\
2.20493245124817	2.1551724137931\\
2.20573687553406	2.1678616617461\\
2.20593953132629	2.18267293485561\\
2.20660758018494	2.19596010393093\\
2.20993566513062	2.21438530304273\\
2.21148228645325	2.22439501344414\\
2.21193242073059	2.24116930572473\\
2.21272277832031	2.25320149132761\\
2.21383285522461	2.2427652733119\\
2.21769428253174	2.23336729096466\\
2.21793079376221	2.25052746737517\\
2.21927332878113	2.24048375843089\\
2.21993160247803	2.27516470047495\\
2.22024726867676	2.28820693349659\\
2.22154474258423	2.29876337151961\\
2.22232961654663	2.30995697138975\\
2.22294878959656	2.33641349260011\\
2.22384405136108	2.34650273633706\\
2.22394347190857	2.36226358675338\\
2.22516107559204	2.37707345971564\\
2.22556447982788	2.38905325443787\\
2.22672319412231	2.40253519050777\\
2.22693705558777	2.41255408080956\\
2.22787427902222	2.42636848644303\\
2.22993159294128	2.44200244200244\\
2.23085737228394	2.45561282932417\\
2.2319347858429	2.46904795787676\\
2.23290777206421	2.48156551332955\\
2.23293924331665	2.49787715822247\\
2.2349328994751	2.52235885969815\\
2.23593664169312	2.53754171301446\\
2.23693251609802	2.55063247390613\\
2.23793315887451	2.579936873885\\
2.23893690109253	2.5938566552901\\
2.24046444892883	2.58387445887446\\
2.24105882644653	2.57377711898666\\
2.24193286895752	2.56221074518747\\
2.24309277534485	2.55214233357766\\
2.24493384361267	2.58047528141663\\
2.24691224098206	2.57064364207221\\
2.24893569946289	2.58670988654781\\
2.24993276596069	2.61463163317878\\
2.2522029876709	2.62873146203297\\
2.25321745872498	2.6187614649883\\
2.25393867492676	2.60836693548387\\
2.25485157966614	2.62000502638854\\
2.25493764877319	2.63865441130495\\
2.25593829154968	2.65729043499907\\
2.25852608680725	2.66872110939908\\
2.25981760025024	2.65866209262436\\
2.25993275642395	2.6700804681785\\
2.2644579410553	2.68098122713369\\
2.26693344116211	2.69292366095261\\
2.26787614822388	2.70462212227023\\
2.26793336868286	2.71530606221593\\
2.26893424987793	2.70429159318048\\
2.26994371414185	2.72584308928883\\
2.27232575416565	2.73941068139963\\
2.27290105819702	2.75050244042492\\
2.27431273460388	2.76243093922652\\
2.27508234977722	2.77541772868876\\
2.27695107460022	2.76044302018328\\
2.2799334526062	2.77499721944166\\
2.28411769866943	2.78537813286637\\
2.28494000434875	2.80785764814714\\
2.29074025154114	2.81847133757962\\
2.29100561141968	2.83188651280965\\
2.29194736480713	2.86572196175526\\
2.29295706748962	2.88022664078485\\
2.29393315315247	2.89104976997072\\
2.29693102836609	2.90124412781994\\
2.29920029640198	2.9145110894842\\
2.29961395263672	2.92657968789972\\
2.29995608329773	2.9515216393944\\
2.30154204368591	2.96176082767015\\
2.30248689651489	2.95173110942633\\
2.30293250083923	2.97559007464192\\
2.30493187904358	2.99098196392785\\
2.30587673187256	3.00334815851282\\
2.30638766288757	3.01580180449629\\
2.30693960189819	3.02789240789539\\
2.30993628501892	3.04072309279363\\
2.31093192100525	3.05224026609274\\
2.31171488761902	3.06595713518528\\
2.3125011920929	3.07714744057168\\
2.3129403591156	3.08953341740227\\
2.31493306159973	3.10032736375891\\
2.31514954566956	3.1134209133343\\
2.31794095039368	3.12410569493465\\
2.31993317604065	3.14009661835749\\
2.32293272018433	3.16426026626664\\
2.32693243026733	3.18030973451327\\
2.32793474197388	3.19559228650138\\
2.32993221282959	3.20612077602696\\
2.3312656879425	3.21819254686578\\
2.33293390274048	3.23687418175252\\
2.3349883556366	3.24710320668283\\
2.33693194389343	3.26227823211333\\
2.33909869194031	3.27868852459017\\
2.33993291854858	3.29986287433096\\
2.34193229675293	3.31472282059173\\
2.34493088722229	3.33637429949173\\
2.34693312644958	3.36748368132106\\
2.34782981872559	3.37995337995338\\
2.34938025474548	3.37006111732805\\
2.35021638870239	3.3832810957022\\
2.3519332408905	3.37294801641587\\
2.35296702384949	3.38658146964856\\
2.35412049293518	3.39933712925979\\
2.3549325466156	3.4303138101893\\
2.35730481147766	3.420565466173\\
2.3598301410675	3.43053173241853\\
2.35993909835815	3.44108446298227\\
2.36037611961365	3.451438099208\\
2.36114239692688	3.46246570787264\\
2.36319804191589	3.47276706445885\\
2.36521148681641	3.4831183767354\\
2.36593270301819	3.497563173199\\
2.36994409561157	3.52702976310994\\
2.3743155002594	3.51799109470898\\
2.37593293190002	3.52880124545926\\
2.37695240974426	3.53929452981925\\
2.37886500358582	3.55243311899659\\
2.37993359565735	3.57368753953194\\
2.38194155693054	3.58465295832349\\
2.3839476108551	3.60307378656002\\
2.38468241691589	3.61898793670688\\
2.38493537902832	3.63139398672394\\
2.3859338760376	3.64429216632924\\
2.38692831993103	3.65763594667081\\
2.38844847679138	3.66798357734913\\
2.39093160629272	3.6793287919024\\
2.39294958114624	3.69036213834068\\
2.40012764930725	3.69783412572636\\
2.40145707130432	3.71137125004705\\
2.40606641769409	3.72054876230242\\
2.4069550037384	3.73748278301008\\
2.40796542167664	3.74786420028229\\
2.40842819213867	3.76034745165002\\
2.40904188156128	3.77099632911862\\
2.40993428230286	3.78544231266866\\
2.4124915599823	3.79593187920697\\
2.41493105888367	3.81227384041815\\
2.41594099998474	3.82300368438332\\
2.41693425178528	3.83663366336633\\
2.41796517372131	3.84992554389278\\
2.41907715797424	3.8605140102222\\
2.4199423789978	3.88008967961235\\
2.42296481132507	3.88984756974403\\
2.42414927482605	3.90215909906032\\
2.42490768432617	3.91257613758509\\
2.425	3.9257085599771\\
2.42745161056519	3.91590949581329\\
2.42793679237366	3.9382404212174\\
2.42853116989136	3.94825686769253\\
2.4293692111969	3.95941247427801\\
2.43293571472168	3.97437251381701\\
2.43394088745117	3.98551998031842\\
2.43494343757629	3.99677453283315\\
2.4384970664978	4.00641025641026\\
2.43993926048279	4.01736111111111\\
2.44193196296692	4.03348554033485\\
2.44310164451599	4.04486626402071\\
2.44442629814148	4.05591516320066\\
2.44710493087769	4.06701843349551\\
2.44882750511169	4.07765662918276\\
2.44993591308594	4.11000920151314\\
2.45310592651367	4.11979943081718\\
2.45393228530884	4.1304421365989\\
2.45493626594543	4.16048049669321\\
2.45794606208801	4.1708610160392\\
2.4590904712677	4.18117300482057\\
2.45993494987488	4.19491666944992\\
2.46117997169495	4.20624604206246\\
2.46212244033813	4.21728855307393\\
2.46493172645569	4.23439827843073\\
2.46689295768738	4.24534530569127\\
2.46774125099182	4.25644066679845\\
2.46837997436523	4.26889429717818\\
2.46958351135254	4.28074322106231\\
2.47015690803528	4.29353347972862\\
2.47171330451965	4.30388854367662\\
2.47245764732361	4.31800365821793\\
2.47381591796875	4.32983645012055\\
2.47784280776978	4.33923771898579\\
2.47928881645203	4.34979171375981\\
2.47994804382324	4.3618427415455\\
2.48293328285217	4.37351486738167\\
2.48493576049805	4.38663851027068\\
2.48824119567871	4.39644461435534\\
2.49099016189575	4.4072851884757\\
2.49364972114563	4.41720117539258\\
2.49426507949829	4.42726412117387\\
2.49593329429626	4.44192777148609\\
2.4962694644928	4.45199019053008\\
2.49686050415039	4.46392108817893\\
2.49695563316345	4.47779940965899\\
2.49829244613647	4.49033309309685\\
2.49993491172791	4.504842236801\\
2.50077748298645	4.51651164242462\\
2.50329375267029	4.52778555213484\\
2.50802230834961	4.53857323817468\\
2.50994205474854	4.55273533760473\\
2.5118727684021	4.56542443700268\\
2.51266574859619	4.57558460688639\\
2.51376008987427	4.58611604681086\\
2.5141863822937	4.59749005203551\\
2.51568531990051	4.60767538499145\\
2.51880598068237	4.61772305821553\\
2.52185440063477	4.6278640394685\\
2.52236437797546	4.63822453360748\\
2.52335023880005	4.64905660377359\\
2.5239360332489	4.65950901743168\\
2.52536344528198	4.66991635072516\\
2.527015209198	4.67999519692603\\
2.52904725074768	4.6914763031047\\
2.53155851364136	4.70153038393843\\
2.53282380104065	4.71411554496725\\
2.53320074081421	4.72590917207309\\
2.5349326133728	4.73659202945893\\
2.53689312934875	4.74666666666666\\
2.53711891174316	4.75682720217996\\
2.53805828094482	4.76894420956154\\
2.53978824615479	4.78045325779037\\
2.5417902469635	4.79119456134671\\
2.54378795623779	4.80216038511213\\
2.54457688331604	4.81369646155199\\
2.54785418510437	4.82420278004906\\
2.54910349845886	4.83602973327503\\
2.54993224143982	4.84659719392211\\
2.55195212364197	4.85871112014637\\
2.55355262756348	4.87055344562665\\
2.55594944953918	4.88249067730465\\
2.55666589736938	4.89354941213854\\
2.5581386089325	4.90413723511605\\
2.55879020690918	4.91486848549451\\
2.55994081497192	4.9324129997124\\
2.56228184700012	4.94464521310159\\
2.56721901893616	4.95394005076576\\
2.56824111938477	4.96524612579763\\
2.56994700431824	4.97768428234358\\
2.57125735282898	4.98765151730206\\
2.57295322418213	4.99914980445503\\
2.57523822784424	5.00932888562221\\
2.57993507385254	5.02052060493619\\
2.58337712287903	5.03038960311459\\
2.58495354652405	5.04290945685293\\
2.58561182022095	5.05335493602994\\
2.58689093589783	5.06555090655509\\
2.5909481048584	5.07471807121827\\
2.5921151638031	5.08780203623048\\
2.59293913841248	5.09935426655211\\
2.59383487701416	5.10995402426189\\
2.59493708610535	5.12551144531681\\
2.59581518173218	5.13564285319631\\
2.60102367401123	5.14441177278848\\
2.60205340385437	5.15495045428344\\
2.60256743431091	5.16612252736701\\
2.60390615463257	5.17619334685153\\
2.60605883598328	5.18688650097066\\
2.6077938079834	5.1979251979252\\
2.60918045043945	5.20978163081705\\
2.6173939704895	5.2167325071703\\
2.6204822063446	5.22778797365864\\
2.62149620056152	5.23973895690632\\
2.62462878227234	5.25112312700078\\
2.62661218643188	5.26103770544634\\
2.6274995803833	5.27233700026832\\
2.62833905220032	5.28404064234203\\
2.62945699691772	5.29416490373028\\
2.63349175453186	5.30387919534711\\
2.63403415679932	5.31634094968939\\
2.63493657112122	5.32736668264877\\
2.63661289215088	5.33780367739018\\
2.63838171958923	5.34831818906424\\
2.64206337928772	5.35889221319072\\
2.64769697189331	5.3674104552408\\
2.65037107467651	5.37724928969799\\
2.65186619758606	5.38778942941995\\
2.65447044372559	5.39902408311034\\
2.65512704849243	5.41135755859683\\
2.65825867652893	5.42272679711207\\
2.65957641601562	5.43316346530773\\
2.66225862503052	5.44402148406945\\
2.66524839401245	5.45468733742587\\
2.66940665245056	5.46530336845383\\
2.67084670066833	5.47664714115391\\
2.67512130737305	5.4872059963815\\
2.67693829536438	5.50077439339184\\
2.6783709526062	5.51114091190427\\
2.68004035949707	5.52193061529348\\
2.68293690681458	5.53156487764754\\
2.6830689907074	5.54327713815789\\
2.6872296333313	5.55427399718058\\
2.69135904312134	5.56379063586178\\
2.69648337364197	5.57267656572345\\
2.69993257522583	5.58365313746918\\
2.70106863975525	5.59534694706898\\
2.70296335220337	5.60572240259741\\
2.70688223838806	5.61513557264265\\
2.70736455917358	5.62703809499734\\
2.70785927772522	5.63768225810528\\
2.70893883705139	5.64870461087824\\
2.71092844009399	5.65852182464634\\
2.71387600898743	5.66831184528606\\
2.71493339538574	5.67941620533467\\
2.71555018424988	5.69056603773585\\
2.71693706512451	5.70135746606334\\
2.7199432849884	5.71256614088322\\
2.72037291526794	5.72295197031987\\
2.72235870361328	5.73316976691785\\
2.72354578971863	5.74344606764059\\
2.72517490386963	5.75376005596362\\
2.72693490982056	5.76534305039059\\
2.72914242744446	5.77531842767766\\
2.73269295692444	5.78524724660021\\
2.73293590545654	5.79580475196342\\
2.73394250869751	5.80754135823936\\
2.73484992980957	5.81782079920575\\
2.73495817184448	5.8287841191067\\
2.73593759536743	5.83947035631928\\
2.73747944831848	5.85029969782533\\
2.74011206626892	5.85995450499456\\
2.74293804168701	5.87291399229781\\
2.74493885040283	5.88597918824283\\
2.74534296989441	5.89627292447249\\
2.74783754348755	5.90683015708869\\
2.74889874458313	5.91796057973868\\
2.7499463558197	5.92820260634374\\
2.75187611579895	5.93928080172922\\
2.75360417366028	5.94960622193872\\
2.75529909133911	5.95950186311041\\
2.75648307800293	5.97139218183599\\
2.75936007499695	5.98146382021373\\
2.75991868972778	5.99222550913136\\
2.76205587387085	6.00253869062149\\
2.76296830177307	6.01336846213896\\
2.76452302932739	6.02530041192386\\
2.76555371284485	6.03691798168712\\
2.76693558692932	6.04696435089427\\
2.76798057556152	6.05817403570213\\
2.76968884468079	6.06973636253189\\
2.77165508270264	6.08347488473671\\
2.77315616607666	6.09395452562176\\
2.77552843093872	6.10367285669322\\
2.77731585502625	6.11381170040388\\
2.77799367904663	6.12491540172097\\
2.77880048751831	6.13541389888645\\
2.77994084358215	6.14653216854095\\
2.78392124176025	6.15695909813557\\
2.78734970092773	6.16649043545132\\
2.78848338127136	6.17648470881944\\
2.78994536399841	6.18749550327362\\
2.79096984863281	6.19818870094398\\
2.79360055923462	6.20843102540791\\
2.79636454582214	6.22015281757402\\
2.79752326011658	6.23151064032827\\
2.79967451095581	6.24300045273667\\
2.80092668533325	6.2529764739499\\
2.80306887626648	6.26352459990963\\
2.803959608078	6.27628983578507\\
2.8062469959259	6.2899052953882\\
2.80985760688782	6.29975360121304\\
2.81092977523804	6.30993252042145\\
2.81258797645569	6.319922420114\\
2.8137378692627	6.33022101406453\\
2.81496691703796	6.34122148221625\\
2.81681776046753	6.35190297538991\\
2.82090163230896	6.36303692539564\\
2.82270860671997	6.37422086322474\\
2.82338309288025	6.38547985141299\\
2.82518315315247	6.39622286949168\\
2.82646512985229	6.40700568155139\\
2.83014512062073	6.41758859928746\\
2.83290123939514	6.42760170404007\\
2.8345513343811	6.43765784304555\\
2.83593320846558	6.44784487793482\\
2.83743143081665	6.45861537743336\\
2.83996748924255	6.46853146853147\\
2.84094548225403	6.47874199184624\\
2.84285163879395	6.48966865227103\\
2.84296703338623	6.500093057882\\
2.84684634208679	6.51083979388144\\
2.84968328475952	6.52204144068975\\
2.85023808479309	6.53261675315048\\
2.85058331489563	6.54283530583412\\
2.85306859016418	6.55256158205157\\
2.85478711128235	6.5627961811415\\
2.85542893409729	6.57286757231309\\
2.85681056976318	6.582775340568\\
2.8578245639801	6.59297549268473\\
2.85858774185181	6.60377358490566\\
2.86139035224915	6.6138358404939\\
2.86293649673462	6.62538272059671\\
2.8638596534729	6.63613980993581\\
2.86495447158813	6.64781089036056\\
2.86674094200134	6.658548353744\\
2.86852860450745	6.66896156052782\\
2.86988830566406	6.67920115561874\\
2.87096309661865	6.69034253637301\\
2.873215675354	6.70084218235079\\
2.87465381622314	6.71163960667733\\
2.87636923789978	6.7224011512369\\
2.87723112106323	6.73621256391527\\
2.87794828414917	6.74621281255704\\
2.87971258163452	6.75709563433261\\
2.88051581382751	6.76857884904543\\
2.88159227371216	6.78051001821494\\
2.88313841819763	6.79057281950953\\
2.88506889343262	6.8019635471115\\
2.88686466217041	6.81364090454999\\
2.88719415664673	6.82509419401697\\
2.88995981216431	6.83599945620157\\
2.8924024105072	6.84714457095001\\
2.89445972442627	6.85842208805373\\
2.89493298530579	6.87340370228059\\
2.89588904380798	6.88642371387903\\
2.89635825157166	6.8979536522564\\
2.89902973175049	6.90891866287725\\
2.90096569061279	6.92009554283654\\
2.90295910835266	6.93303320202588\\
2.90594577789307	6.94637961256685\\
2.90779995918274	6.95710756793173\\
2.90932846069336	6.96766375692839\\
2.90960168838501	6.97778774960736\\
2.91192626953125	6.98774119809059\\
2.91231346130371	6.9991486310884\\
2.91494131088257	7.01114443002282\\
2.91595864295959	7.02225204070223\\
2.91874384880066	7.03212259303937\\
2.91938233375549	7.04253656358155\\
2.92019176483154	7.05278428746792\\
2.92261576652527	7.06386014983945\\
2.92397356033325	7.07633349672475\\
2.92552542686462	7.08752559878907\\
2.92717862129211	7.0978935427186\\
2.92789101600647	7.10791942727555\\
2.92873764038086	7.11856872985888\\
2.92994451522827	7.13143517079021\\
2.93093085289001	7.14143006193531\\
2.93288040161133	7.15157161553649\\
2.93379855155945	7.16154494257461\\
2.93546533584595	7.17244434599624\\
2.93712186813354	7.18285853420989\\
2.93829154968262	7.1929552846428\\
2.94040727615356	7.20394228006983\\
2.94194078445435	7.21430937396489\\
2.94269967079163	7.2251655629139\\
2.94466400146484	7.23502812396603\\
2.94638156890869	7.24548259144998\\
2.94797348976135	7.25847298993591\\
2.94859218597412	7.26864488861494\\
2.94950103759766	7.27912861700958\\
2.95029997825623	7.2908419107944\\
2.95263695716858	7.30212429428176\\
2.95345902442932	7.31364584476891\\
2.95479941368103	7.32403019132877\\
2.95512723922729	7.33681347604844\\
2.95792126655579	7.34810986026545\\
2.96	7.35998949510867\\
2.96149253845215	7.37079241486405\\
2.96266937255859	7.38255033557046\\
2.96315860748291	7.39340428786853\\
2.96444845199585	7.40433339157785\\
2.96538662910461	7.41622093657898\\
2.96595168113708	7.42656365937759\\
2.9668083190918	7.43706095379379\\
2.96734094619751	7.44771334481931\\
2.96922492980957	7.45832879398627\\
2.97082948684692	7.47022578327418\\
2.97199535369873	7.48155724327031\\
2.97393941879272	7.49429285791933\\
2.97534251213074	7.50564824469934\\
2.97765040397644	7.51568027432342\\
2.97875428199768	7.52662516539054\\
2.97993898391724	7.53939866900783\\
2.98194861412048	7.5493360483504\\
2.9843373298645	7.56064310908187\\
2.98565769195557	7.57454537591627\\
2.98717617988586	7.58525305787268\\
2.98793244361877	7.59676861931915\\
2.99044632911682	7.60714748154862\\
2.9920871257782	7.61849312114546\\
2.99412536621094	7.62982115923293\\
2.99493861198425	7.64044459762192\\
2.99725341796875	7.65075539103862\\
2.99878835678101	7.66070929307268\\
2.99996066093445	7.67461852568236\\
3.00136089324951	7.68536815877653\\
3.00270843505859	7.69561017494903\\
3.00417947769165	7.70616621983913\\
3.00594139099121	7.71785477062844\\
3.00693416595459	7.73102045186851\\
3.00840258598328	7.74350962052951\\
3.01036930084229	7.75399354190279\\
3.01086640357971	7.76483164083378\\
3.012	7.775807175367\\
3.01410984992981	7.78674523870527\\
3.01493716239929	7.79705876075217\\
3.017	7.80863407559082\\
3.02188444137573	7.81792245949455\\
3.02294230461121	7.83062027875306\\
3.02506899833679	7.84080278935283\\
3.02647805213928	7.85084457664932\\
3.02734327316284	7.86108751062022\\
3.02877855300903	7.87194293478261\\
3.03217720985413	7.88413418720046\\
3.03424263000488	7.89395609053148\\
3.03547024726868	7.90460254590897\\
3.03724694252014	7.91473508181452\\
3.0387110710144	7.92485561361568\\
3.03993487358093	7.93513193504736\\
3.04129600524902	7.94623610465363\\
3.0432710647583	7.9567531041473\\
3.04451274871826	7.9678331715142\\
3.04549884796143	7.97822968525863\\
3.04712200164795	7.98920719239445\\
3.04808354377747	7.99957863689034\\
3.04956245422363	8.01069496199919\\
3.05093312263489	8.02095165972485\\
3.05194807052612	8.03229740527355\\
3.05343699455261	8.04303063411355\\
3.05583906173706	8.05456453305352\\
3.0569441318512	8.06485443409683\\
3.05885982513428	8.07695531897058\\
3.05926561355591	8.08715692436623\\
3.06193256378174	8.09753034742571\\
3.06394481658936	8.10850378550215\\
3.06575679779053	8.12019391507857\\
3.06662559509277	8.13018320068518\\
3.06828737258911	8.14004467734191\\
3.06997776031494	8.15045060080107\\
3.07423734664917	8.16075669819575\\
3.07567000389099	8.17214599512794\\
3.07875561714172	8.18223405140148\\
3.07955813407898	8.19372271876948\\
3.08208203315735	8.20458509843011\\
3.08332204818726	8.21553853498557\\
3.08493137359619	8.22753298481453\\
3.08594608306885	8.24071001990711\\
3.08642983436584	8.251269824816\\
3.08798718452454	8.26203153162486\\
3.09068727493286	8.27279312842801\\
3.0919725894928	8.28419506341424\\
3.09295177459717	8.29782833505688\\
3.09486413002014	8.30818831893523\\
3.09575748443604	8.31870106798322\\
3.09829592704773	8.33126960541522\\
3.1	8.34141209958541\\
3.10265779495239	8.35153963477472\\
3.10468125343323	8.36182394859332\\
3.10603642463684	8.37175792507205\\
3.10747051239014	8.38185371875321\\
3.10893201828003	8.39211170289334\\
3.10963654518127	8.40272209543782\\
3.11093926429749	8.41278974190367\\
3.11273121833801	8.42368983078692\\
3.11394476890564	8.4351199885041\\
3.11493277549744	8.44567559248999\\
3.11616921424866	8.45587481285507\\
3.11805868148804	8.46637837394709\\
3.11929225921631	8.47707624861719\\
3.12018394470215	8.48811254684333\\
3.1219334602356	8.50236404200012\\
3.12392568588257	8.51233582913957\\
3.12438535690308	8.52269240995645\\
3.12493705749512	8.53626121981639\\
3.12896656990051	8.54674672221541\\
3.12996339797974	8.55719768106475\\
3.13304996490479	8.56822353661024\\
3.13418030738831	8.57935083999347\\
3.1363844871521	8.59058214642297\\
3.13774490356445	8.60063949817722\\
3.14124798774719	8.61118457188194\\
3.14425325393677	8.62072469364115\\
3.14558887481689	8.63144209478538\\
3.14697670936584	8.64144890459077\\
3.14871621131897	8.65189513838163\\
3.15074563026428	8.6635570878118\\
3.15240788459778	8.67441059011941\\
3.15431642532349	8.68534613748252\\
3.15611672401428	8.69644448044059\\
3.16037249565125	8.70655823926851\\
3.16247797012329	8.71709861318886\\
3.16405916213989	8.72842649852472\\
3.16464400291443	8.73878606643498\\
3.16790747642517	8.74954576654418\\
3.16895771026611	8.75978690774074\\
3.17147326469421	8.77051494011855\\
3.17327976226807	8.78097924642353\\
3.17498564720154	8.79196117521498\\
3.17701625823975	8.80198059659434\\
3.17984676361084	8.8118752137096\\
3.18279242515564	8.8219332274728\\
3.18437933921814	8.83357441337191\\
3.18709254264832	8.84325062233999\\
3.18866086006165	8.85321561151801\\
3.19135355949402	8.86286619478254\\
3.19423842430115	8.87343978522629\\
3.195	8.88515002203261\\
3.19629883766174	8.89618448973055\\
3.19894361495972	8.90835633425338\\
3.20296573638916	8.91886490807354\\
3.20544457435608	8.93024741897477\\
3.20776200294495	8.94065512904965\\
3.20887875556946	8.95141175296818\\
3.21062088012695	8.96233942392085\\
3.214	8.97458893871451\\
3.21524715423584	8.98585375572824\\
3.21616721153259	8.99729105250578\\
3.21811199188232	9.00836320191159\\
3.22093963623047	9.01981852913085\\
3.22508454322815	9.03001391373483\\
3.2271580696106	9.04070402669898\\
3.22884035110474	9.05192097686886\\
3.23187661170959	9.06349206349206\\
3.23362708091736	9.07342414688795\\
3.23438954353333	9.08496213773145\\
3.23614621162415	9.09523243371775\\
3.23954820632935	9.10513079466921\\
3.24080920219421	9.1157432151553\\
3.24639892578125	9.12524221932218\\
3.25056219100952	9.1356317297244\\
3.25174474716187	9.14657028302817\\
3.25479674339294	9.15730669772399\\
3.25578880310059	9.16877170824123\\
3.25820207595825	9.17896315000394\\
3.26176881790161	9.189466214618\\
3.26380276679993	9.1998187227838\\
3.26494002342224	9.21195170471331\\
3.26796841621399	9.22407513142093\\
3.26898145675659	9.23443374869584\\
3.27251315116882	9.24505291317518\\
3.27593564987183	9.2556914245272\\
3.28080725669861	9.2659541024835\\
3.28252935409546	9.27733799524642\\
3.28446292877197	9.28798649020147\\
3.28871369361877	9.29785313395346\\
3.29151177406311	9.30793675703245\\
3.29512047767639	9.31851154766106\\
3.29965090751648	9.3304857260419\\
3.30117797851562	9.34145577058685\\
3.30431461334229	9.35168686730239\\
3.30893325805664	9.36081102238923\\
3.31278848648071	9.37310513447431\\
3.31898212432861	9.38214655694767\\
3.3218367099762	9.39252427753571\\
3.32495737075806	9.40344571629487\\
3.32762289047241	9.41454460240588\\
3.32893681526184	9.42515718358261\\
3.33284306526184	9.43532488825561\\
3.33641219139099	9.44566935279312\\
3.34114933013916	9.45582211819779\\
3.34493660926819	9.46633649852789\\
3.348	9.4770286338031\\
3.35252046585083	9.48623531474661\\
3.35876321792603	9.49506357953772\\
3.36236619949341	9.50443786982249\\
3.36549949645996	9.51528477749022\\
3.36897373199463	9.52594004707529\\
3.373206615448	9.53501487719025\\
3.37631773948669	9.54621456624023\\
3.37985181808472	9.55666556529514\\
3.38154768943787	9.56665371162067\\
3.38431024551392	9.57626624701287\\
3.39058375358582	9.58539664045053\\
3.3953800201416	9.59701840204985\\
3.39999866485596	9.60706316095857\\
3.405	9.61672744200481\\
3.40975546836853	9.62713047524867\\
3.41819882392883	9.63252965346151\\
3.42043042182922	9.64385366609827\\
3.42316126823425	9.65360920609091\\
3.428	9.66361330054031\\
3.43420505523682	9.67148692335986\\
3.43631839752197	9.68241373304174\\
3.44426894187927	9.68877541151667\\
3.44768714904785	9.69912599582335\\
3.45320463180542	9.71065195121479\\
3.46048045158386	9.71755341756501\\
3.46623134613037	9.72718321008631\\
3.46828961372375	9.73745173745173\\
3.473	9.74663759334659\\
3.47816562652588	9.75600347188736\\
3.4827606678009	9.76611457108151\\
3.48891687393188	9.77550823778785\\
3.49205207824707	9.78503736805609\\
3.49764847755432	9.79384419934168\\
3.50497794151306	9.80150792429604\\
3.50974774360657	9.81063154859175\\
3.51161241531372	9.82163450449757\\
3.51792812347412	9.83231853715691\\
3.52207064628601	9.84295807095684\\
3.52604484558105	9.85355929600983\\
3.53427505493164	9.85999232834676\\
3.53684663772583	9.87384485601441\\
3.54055285453796	9.88385464581417\\
3.54365563392639	9.89423666462293\\
3.54878807067871	9.90290518413543\\
3.55573225021362	9.91101330016266\\
3.56022191047668	9.92176590982996\\
3.56258416175842	9.93211588105937\\
3.56756114959717	9.94190046632519\\
3.57131505012512	9.95243462148275\\
3.57761740684509	9.96047883613035\\
3.58325958251953	9.96927539550773\\
3.58916044235229	9.97997520739963\\
3.59206581115723	9.98970526556602\\
3.59391498565674	9.99980941853595\\
3.60065197944641	10.0072364411944\\
3.60534381866455	10.01655786689\\
3.61140537261963	10.0245382435183\\
3.61592698097229	10.0349903016012\\
3.61993098258972	10.0463825419154\\
3.62863492965698	10.0514729624494\\
3.63151359558105	10.0613287257676\\
3.63499808311462	10.0715600857962\\
3.64065837860107	10.080048560264\\
3.64436793327332	10.090821182761\\
3.64951992034912	10.1008186779867\\
3.652	10.113296199462\\
3.65562915802002	10.1238495625497\\
3.66137480735779	10.132884076626\\
3.66689133644104	10.1419110690634\\
3.67429971694946	10.1487909553249\\
3.67879986763	10.157800245677\\
3.68362522125244	10.16718617172\\
3.68824815750122	10.1763395385719\\
3.69242167472839	10.1866354664\\
3.69565677642822	10.1967664648065\\
3.69964528083801	10.2074297567415\\
3.70526599884033	10.2163506841054\\
3.711	10.2252608581007\\
3.71927976608276	10.2322779806498\\
3.72437572479248	10.2417458376446\\
3.72921133041382	10.2518290045326\\
3.73576068878174	10.2593984962406\\
3.74437236785889	10.2645025548542\\
3.75372791290283	10.2681085952461\\
3.76027154922485	10.2764276465179\\
3.76529335975647	10.2851247420747\\
3.77246522903442	10.2921323033076\\
3.779	10.2999062792877\\
3.78826928138733	10.303677476161\\
3.79537963867188	10.3123478521404\\
3.80189538002014	10.3206244033915\\
3.80717515945435	10.3292797006548\\
3.81596541404724	10.3341873468853\\
3.821	10.3439895307534\\
3.82964468002319	10.3511427556951\\
3.836	10.359451481523\\
3.84371089935303	10.365819499169\\
3.85359716415405	10.3680615177872\\
3.86242294311523	10.3760913364674\\
3.87170028686523	10.3800678825855\\
3.879	10.390142971648\\
3.887	10.397376902584\\
3.894	10.4050656485706\\
3.90274786949158	10.4114690001862\\
3.91198658943176	10.4154258487195\\
3.921	10.4203494538621\\
3.93037486076355	10.4303835137448\\
3.93969798088074	10.4341359114995\\
3.94770240783691	10.4403263879854\\
3.95737862586975	10.4452372988442\\
3.965	10.4518094671918\\
3.97320628166199	10.4577955241898\\
3.9818811416626	10.4630867498514\\
3.98828458786011	10.4718942600431\\
3.99835777282715	10.4728851935711\\
4.00806331634521	10.4771976696724\\
4.01620578765869	10.4830631098579\\
4.02223014831543	10.4916634210575\\
4.02959775924683	10.4992862042754\\
4.03988122940063	10.5012419827235\\
4.049	10.5061250208492\\
4.05853986740112	10.5095423383361\\
4.06859683990479	10.5100320506882\\
4.079	10.5121793090673\\
4.08812379837036	10.5169571579396\\
4.09824800491333	10.5203703703704\\
4.10774850845337	10.5237821925163\\
4.11758756637573	10.5261209226554\\
4.12783718109131	10.5334271118679\\
4.13827657699585	10.5386850237791\\
4.14816951751709	10.5424807578449\\
4.158	10.5466654333549\\
4.167	10.5539628225285\\
4.176	10.5597987905239\\
4.185	10.5643699840971\\
4.195	10.565046963982\\
4.205	10.5657237936772\\
4.215	10.5645519077196\\
4.225	10.5676629882253\\
4.235	10.5677533821246\\
4.246	10.5686249145307\\
4.25609922409058	10.5698869096016\\
4.2667350769043	10.5707581160732\\
4.275	10.5763901718086\\
4.28451299667358	10.5822158188334\\
4.29459762573242	10.5832810018101\\
4.30716037750244	10.5843460274074\\
4.31751918792725	10.5864756066034\\
4.328	10.5889955686854\\
4.334	10.5970535021969\\
4.34393119812012	10.6008306414398\\
4.354	10.6004393008103\\
4.363	10.6051937025895\\
4.375	10.6073188285445\\
4.38634490966797	10.6087726744293\\
4.39672613143921	10.609834855614\\
4.40716791152954	10.6103095826722\\
4.41607093811035	10.6152568951204\\
4.42634153366089	10.614473611393\\
4.4363579750061	10.617575444551\\
4.44893264770508	10.6204810388077\\
4.4591121673584	10.6229943565342\\
4.47	10.6277431490429\\
4.48	10.6308432757381\\
4.49	10.6306472432233\\
4.5	10.635198672444\\
4.51	10.6348064974095\\
4.52	10.6389646781211\\
4.53	10.6420631555662\\
4.54	10.6437103672025\\
4.549973487854	10.6480628156449\\
4.56	10.6497097041747\\
4.57	10.6528069593424\\
4.579	10.6573539981203\\
4.59177780151367	10.6596646397642\\
4.602	10.6605251036389\\
4.612	10.6617782526991\\
4.623	10.6617782526991\\
4.63596868515015	10.6640876853643\\
4.649	10.6655368688636\\
4.66	10.6663965225079\\
4.67	10.6672560178279\\
4.679	10.6738365775952\\
4.69	10.6787713612257\\
4.7	10.6783780797702\\
4.71	10.6796295273343\\
4.72	10.6812741668201\\
4.73	10.6825253622521\\
4.74	10.685814231796\\
4.75	10.6868683149542\\
4.765	10.6868683149542\\
4.77567863464355	10.686474843983\\
4.78602647781372	10.6877254988587\\
4.798	10.6873320325443\\
4.808	10.6904231625835\\
4.818	10.6937107728553\\
4.829	10.693317136915\\
4.84	10.694960798027\\
4.85059452056885	10.694960798027\\
4.861	10.6980510517695\\
4.876	10.6980510517695\\
4.888	10.6993006993007\\
4.899	10.7009440385712\\
4.91	10.7007471200913\\
4.925	10.7005502088585\\
4.938	10.7001564081332\\
4.94864988327026	10.6997626363921\\
4.96	10.7014057555016\\
4.98	10.7030488141456\\
4.994	10.7046918123275\\
5.00510787963867	10.7046918123275\\
5.019	10.7046918123275\\
5.03	10.7046918123275\\
5.04	10.7044948575003\\
5.06	10.7075836185009\\
5.07	10.7075836185009\\
5.085	10.7073866249655\\
5.11	10.7073866249655\\
5.12011957168579	10.7073866249655\\
5.135	10.7071896386783\\
5.14777517318726	10.7071896386783\\
5.16	10.7071896386783\\
5.17057371139526	10.7088323490995\\
5.185	10.7088323490995\\
5.215	10.7104749990802\\
5.23	10.71027796685\\
5.25	10.71027796685\\
5.263	10.71027796685\\
5.302	10.71027796685\\
5.33	10.7119205298013\\
5.3770956993103	10.7119205298013\\
5.42286586761475	10.7119205298013\\
5.47	10.7119205298013\\
5.855	10.7119205298013\\
6.00990009307861	10.7119205298013\\
6.44	10.7119205298013\\
};
\end{axis}
\end{tikzpicture}%

%% file: figures/hat_z_dlas_kde.tex
%
\tikzsetnextfilename{hat_z_dlas_kde}
\definecolor{mycolor1}{rgb}{0.12157,0.47059,0.70588}%
\begin{tikzpicture}

\begin{axis}[%
width=\figurewidth,
height=\figureheight,
at={(0\figurewidth,0\figureheight)},
scale only axis,
xmin=-0.01,
xmax=0.01,
xlabel={difference between estimated $\zd$ (new method/concordance)},
xtick={-0.01, -0.005, 0, 0.005, 0.01},
xticklabels={{$-0.01$}, {$-0.005$}, {$0$}, {$0.005$}, {$0.01$}},
ymin=0,
ymax=300,
ylabel={$p(\text{difference})$},
axis background/.style={fill=white},
axis x line*=bottom,
axis y line*=left,
legend columns=1
]
\addplot [color=mycolor1,solid,forget plot]
  table[row sep=crcr]{%
-0.01	4.24072164578862\\
-0.00997997997997998	4.23676994985538\\
-0.00995995995995996	4.23303572622503\\
-0.00993993993993994	4.22969829635789\\
-0.00991991991991992	4.22676973649377\\
-0.0098998998998999	4.22432776766903\\
-0.00987987987987988	4.2223166933522\\
-0.00985985985985986	4.22082584176996\\
-0.00983983983983984	4.21971016054868\\
-0.00981981981981982	4.21924898739168\\
-0.0097997997997998	4.21915609384208\\
-0.00977977977977978	4.21948474608767\\
-0.00975975975975976	4.22023647537997\\
-0.00973973973973974	4.22140271289372\\
-0.00971971971971972	4.22298125539432\\
-0.0096996996996997	4.2249782413948\\
-0.00967967967967968	4.227391660469\\
-0.00965965965965966	4.23031353762537\\
-0.00963963963963964	4.23373715167778\\
-0.00961961961961962	4.23786952884905\\
-0.0095995995995996	4.24257755336303\\
-0.00957957957957958	4.24795555082547\\
-0.00955955955955956	4.25426846485421\\
-0.00953953953953954	4.26143076097521\\
-0.00951951951951952	4.26960698430856\\
-0.0094994994994995	4.27886970722372\\
-0.00947947947947948	4.28942600891982\\
-0.00945945945945946	4.30135136022357\\
-0.00943943943943944	4.31474094424485\\
-0.00941941941941942	4.32960462522971\\
-0.0093993993993994	4.34621976191817\\
-0.00937937937937938	4.36464227334036\\
-0.00935935935935936	4.38486816396004\\
-0.00933933933933934	4.40693647519251\\
-0.00931931931931932	4.430785196057\\
-0.0092992992992993	4.45662115227963\\
-0.00927927927927928	4.48428012728994\\
-0.00925925925925926	4.51399991024721\\
-0.00923923923923924	4.5454628973307\\
-0.00921921921921922	4.57857060216995\\
-0.0091991991991992	4.61322644292947\\
-0.00917917917917918	4.64928662802991\\
-0.00915915915915916	4.68670192161771\\
-0.00913913913913914	4.72508154744521\\
-0.00911911911911912	4.76436897103812\\
-0.0090990990990991	4.8042512540646\\
-0.00907907907907908	4.84456370908449\\
-0.00905905905905906	4.8850191669252\\
-0.00903903903903904	4.92544373438975\\
-0.00901901901901902	4.96566706253741\\
-0.008998998998999	5.00541314801759\\
-0.00897897897897898	5.0443602830811\\
-0.00895895895895896	5.08253824960097\\
-0.00893893893893894	5.11952499438064\\
-0.00891891891891892	5.1550628864132\\
-0.0088988988988989	5.18919209531645\\
-0.00887887887887888	5.22171504549421\\
-0.00885885885885886	5.25252948749629\\
-0.00883883883883884	5.28155891053574\\
-0.00881881881881882	5.30881304765241\\
-0.0087987987987988	5.33413144294243\\
-0.00877877877877878	5.3575944879679\\
-0.00875875875875876	5.37933748136254\\
-0.00873873873873874	5.39929918843127\\
-0.00871871871871872	5.41761800280626\\
-0.0086986986986987	5.43433295616188\\
-0.00867867867867868	5.4495103183947\\
-0.00865865865865866	5.46328501253857\\
-0.00863863863863864	5.47573652507755\\
-0.00861861861861862	5.48714128828705\\
-0.0085985985985986	5.49755671764679\\
-0.00857857857857858	5.5070961317198\\
-0.00855855855855856	5.51596969380545\\
-0.00853853853853854	5.52419615387753\\
-0.00851851851851852	5.53199745234025\\
-0.0084984984984985	5.53958674089985\\
-0.00847847847847848	5.54685073275565\\
-0.00845845845845846	5.55414616513617\\
-0.00843843843843844	5.56142762580345\\
-0.00841841841841842	5.56867489249417\\
-0.0083983983983984	5.57627587083733\\
-0.00837837837837838	5.58394713979074\\
-0.00835835835835836	5.59196923459939\\
-0.00833833833833834	5.60021154438814\\
-0.00831831831831832	5.60874670377346\\
-0.0082982982982983	5.61748795041995\\
-0.00827827827827828	5.6266400542821\\
-0.00825825825825826	5.63607855031117\\
-0.00823823823823824	5.64573769748545\\
-0.00821821821821822	5.65574114613117\\
-0.0081981981981982	5.66590389653866\\
-0.00817817817817818	5.67631428234205\\
-0.00815815815815816	5.68688920465873\\
-0.00813813813813814	5.69762240514162\\
-0.00811811811811812	5.70836530196646\\
-0.0080980980980981	5.71891095482914\\
-0.00807807807807808	5.72939214905549\\
-0.00805805805805806	5.73970716231379\\
-0.00803803803803804	5.74966308113064\\
-0.00801801801801802	5.759114462834\\
-0.007997997997998	5.76807159644306\\
-0.00797797797797798	5.77627336438151\\
-0.00795795795795796	5.78372170238168\\
-0.00793793793793794	5.79025817908524\\
-0.00791791791791792	5.79583347446906\\
-0.0078978978978979	5.8003016936426\\
-0.00787787787787788	5.80352674806454\\
-0.00785785785785786	5.80568162986419\\
-0.00783783783783784	5.80641324461328\\
-0.00781781781781782	5.80574290070459\\
-0.0077977977977978	5.80378991329897\\
-0.00777777777777778	5.80039055838634\\
-0.00775775775775776	5.79587069649047\\
-0.00773773773773774	5.79000190222341\\
-0.00771771771771772	5.78306413669198\\
-0.0076976976976977	5.77514327683434\\
-0.00767767767767768	5.76621307894664\\
-0.00765765765765766	5.75670128862021\\
-0.00763763763763764	5.74686649312847\\
-0.00761761761761762	5.73689326947131\\
-0.0075975975975976	5.72692297909614\\
-0.00757757757757758	5.71729283349516\\
-0.00755755755755756	5.70822157333538\\
-0.00753753753753754	5.70019514931001\\
-0.00751751751751752	5.69324320525988\\
-0.0074974974974975	5.68785339427767\\
-0.00747747747747748	5.68411192043118\\
-0.00745745745745746	5.68263413522817\\
-0.00743743743743744	5.68347522496452\\
-0.00741741741741742	5.68693593039595\\
-0.0073973973973974	5.69299451683664\\
-0.00737737737737738	5.70216169074955\\
-0.00735735735735736	5.71438947983933\\
-0.00733733733733734	5.72987596487525\\
-0.00731731731731732	5.74869199181323\\
-0.0072972972972973	5.7710656157736\\
-0.00727727727727728	5.79696350920177\\
-0.00725725725725726	5.82618198664689\\
-0.00723723723723724	5.85880664129177\\
-0.00721721721721722	5.89480693433086\\
-0.0071971971971972	5.93412499713356\\
-0.00717717717717718	5.97662836796691\\
-0.00715715715715716	6.02243935724907\\
-0.00713713713713714	6.07121212406884\\
-0.00711711711711712	6.12293003790804\\
-0.0070970970970971	6.17749665946731\\
-0.00707707707707708	6.2348099169583\\
-0.00705705705705706	6.29462161521221\\
-0.00703703703703704	6.35675881842354\\
-0.00701701701701702	6.42131903436763\\
-0.006996996996997	6.48796586097844\\
-0.00697697697697698	6.55673824182802\\
-0.00695695695695696	6.62744508938774\\
-0.00693693693693694	6.7001402034587\\
-0.00691691691691692	6.77452672244116\\
-0.0068968968968969	6.8505862900365\\
-0.00687687687687688	6.92824592552881\\
-0.00685685685685686	7.00741268091231\\
-0.00683683683683684	7.08774781937687\\
-0.00681681681681682	7.16941219059681\\
-0.0067967967967968	7.25213521540564\\
-0.00677677677677678	7.33568274048245\\
-0.00675675675675676	7.41996369847673\\
-0.00673673673673674	7.50489744393621\\
-0.00671671671671672	7.59026744035838\\
-0.0066966966966967	7.6758685232396\\
-0.00667667667667668	7.76145628681649\\
-0.00665665665665666	7.8469155037339\\
-0.00663663663663664	7.93200639403319\\
-0.00661661661661662	8.01651536331677\\
-0.0065965965965966	8.10016878670188\\
-0.00657657657657658	8.18292953660901\\
-0.00655655655655656	8.26435231785046\\
-0.00653653653653654	8.34423410580231\\
-0.00651651651651652	8.42247357097041\\
-0.0064964964964965	8.49883325027368\\
-0.00647647647647648	8.57298210979058\\
-0.00645645645645646	8.64520797103988\\
-0.00643643643643644	8.71470681677698\\
-0.00641641641641642	8.78193558067465\\
-0.0063963963963964	8.84630940803163\\
-0.00637637637637638	8.90794961442429\\
-0.00635635635635636	8.96664702551777\\
-0.00633633633633634	9.02239332883875\\
-0.00631631631631632	9.07526146245269\\
-0.0062962962962963	9.12551849132065\\
-0.00627627627627628	9.17299044644786\\
-0.00625625625625626	9.21773988302631\\
-0.00623623623623624	9.25990730842489\\
-0.00621621621621622	9.29944505243414\\
-0.0061961961961962	9.3366592795803\\
-0.00617617617617618	9.371618066962\\
-0.00615615615615616	9.40482960448421\\
-0.00613613613613614	9.43644385556753\\
-0.00611611611611612	9.4668578751066\\
-0.0060960960960961	9.49603566390138\\
-0.00607607607607608	9.52431581903608\\
-0.00605605605605606	9.55212243425523\\
-0.00603603603603604	9.57989495043373\\
-0.00601601601601602	9.60790123198261\\
-0.005995995995996	9.6364014552619\\
-0.00597597597597598	9.6656960511367\\
-0.00595595595595596	9.69617868099097\\
-0.00593593593593594	9.72789236180505\\
-0.00591591591591592	9.76113619224351\\
-0.0058958958958959	9.79636013303872\\
-0.00587587587587588	9.83360393934676\\
-0.00585585585585586	9.87313980032564\\
-0.00583583583583584	9.91496160181104\\
-0.00581581581581582	9.95932737334653\\
-0.0057957957957958	10.0061667170921\\
-0.00577577577577578	10.0556421001644\\
-0.00575575575575576	10.1077329438222\\
-0.00573573573573574	10.1622692701115\\
-0.00571571571571572	10.2191879891744\\
-0.0056956956956957	10.2781716815756\\
-0.00567567567567568	10.3391996821348\\
-0.00565565565565566	10.4021857996501\\
-0.00563563563563564	10.4663731992748\\
-0.00561561561561562	10.5317930659683\\
-0.0055955955955956	10.5984076212596\\
-0.00557557557557558	10.6653721503375\\
-0.00555555555555556	10.7323689247707\\
-0.00553553553553554	10.7991432536836\\
-0.00551551551551552	10.8652364210923\\
-0.0054954954954955	10.9306741675924\\
-0.00547547547547548	10.994727185558\\
-0.00545545545545546	11.0571549599471\\
-0.00543543543543543	11.1175246013973\\
-0.00541541541541542	11.175924575365\\
-0.0053953953953954	11.2316670080186\\
-0.00537537537537538	11.2849637250107\\
-0.00535535535535536	11.3350371794407\\
-0.00533533533533534	11.3820858579154\\
-0.00531531531531532	11.426041242783\\
-0.0052952952952953	11.4669830577511\\
-0.00527527527527528	11.5050064513362\\
-0.00525525525525526	11.5396135901273\\
-0.00523523523523524	11.57146133695\\
-0.00521521521521522	11.600084090243\\
-0.0051951951951952	11.6264001948183\\
-0.00517517517517518	11.6504072611866\\
-0.00515515515515516	11.6722896075802\\
-0.00513513513513513	11.6927661845404\\
-0.00511511511511512	11.7117998657151\\
-0.00509509509509509	11.7301743256976\\
-0.00507507507507508	11.7484479058583\\
-0.00505505505505506	11.7667813668621\\
-0.00503503503503503	11.786154547787\\
-0.00501501501501501	11.8069684105919\\
-0.004994994994995	11.8296892810787\\
-0.00497497497497497	11.8550953048798\\
-0.00495495495495496	11.8839559354356\\
-0.00493493493493493	11.9166225995134\\
-0.00491491491491491	11.9540176048581\\
-0.00489489489489489	11.9962351670817\\
-0.00487487487487488	12.0443544796751\\
-0.00485485485485486	12.0989834768208\\
-0.00483483483483483	12.1605415543023\\
-0.00481481481481482	12.2288776463999\\
-0.00479479479479479	12.3051511770897\\
-0.00477477477477478	12.3891810367191\\
-0.00475475475475476	12.482025894891\\
-0.00473473473473473	12.5829889962161\\
-0.00471471471471471	12.6928058690044\\
-0.0046946946946947	12.8114901408835\\
-0.00467467467467467	12.9390000610442\\
-0.00465465465465466	13.0753768060515\\
-0.00463463463463463	13.2198967837731\\
-0.00461461461461461	13.3734408430129\\
-0.00459459459459459	13.5352934074989\\
-0.00457457457457458	13.7048828110162\\
-0.00455455455455455	13.8820477167119\\
-0.00453453453453453	14.0660073197444\\
-0.00451451451451451	14.2566593339814\\
-0.00449449449449449	14.4534551984226\\
-0.00447447447447447	14.6559501835874\\
-0.00445445445445446	14.8634892475908\\
-0.00443443443443443	15.0749779975261\\
-0.00441441441441441	15.2905832305111\\
-0.00439439439439439	15.5093015384408\\
-0.00437437437437437	15.7306918671749\\
-0.00435435435435436	15.9545072106186\\
-0.00433433433433433	16.1796704279709\\
-0.00431431431431431	16.4062505666516\\
-0.00429429429429429	16.6334305359137\\
-0.00427427427427427	16.8610853068208\\
-0.00425425425425425	17.0890468043403\\
-0.00423423423423423	17.3171665567332\\
-0.00421421421421421	17.5446894339502\\
-0.00419419419419419	17.7720606472115\\
-0.00417417417417417	17.9990917243164\\
-0.00415415415415415	18.225581743052\\
-0.00413413413413413	18.4524871070588\\
-0.00411411411411411	18.6791789776867\\
-0.00409409409409409	18.9057617934026\\
-0.00407407407407407	19.1326288862062\\
-0.00405405405405405	19.3604120941151\\
-0.00403403403403403	19.5892824880343\\
-0.00401401401401401	19.8188228753165\\
-0.00399399399399399	20.0500402351347\\
-0.00397397397397397	20.2833234084683\\
-0.00395395395395395	20.5191086172171\\
-0.00393393393393393	20.7570646581674\\
-0.00391391391391391	20.9981387098686\\
-0.00389389389389389	21.2424603731402\\
-0.00387387387387387	21.4900258786808\\
-0.00385385385385385	21.7407810603332\\
-0.00383383383383383	21.9953791557544\\
-0.00381381381381381	22.2531379797204\\
-0.00379379379379379	22.5145809616025\\
-0.00377377377377377	22.7791419891457\\
-0.00375375375375375	23.0473722048151\\
-0.00373373373373373	23.3187843111978\\
-0.00371371371371371	23.5925922280289\\
-0.00369369369369369	23.868845463982\\
-0.00367367367367367	24.1468222123261\\
-0.00365365365365365	24.4257231155301\\
-0.00363363363363363	24.706054489715\\
-0.00361361361361361	24.9865553891118\\
-0.00359359359359359	25.2669685869915\\
-0.00357357357357357	25.5471527496324\\
-0.00355355355355355	25.8258154934237\\
-0.00353353353353353	26.1028945975679\\
-0.00351351351351351	26.3777785625765\\
-0.00349349349349349	26.6501767645232\\
-0.00347347347347347	26.919724835305\\
-0.00345345345345345	27.1867952409179\\
-0.00343343343343343	27.4512684579795\\
-0.00341341341341341	27.7127445585948\\
-0.00339339339339339	27.971161543382\\
-0.00337337337337337	28.2273237806215\\
-0.00335335335335335	28.4816056788768\\
-0.00333333333333333	28.7340022727337\\
-0.00331331331331331	28.9855929566295\\
-0.00329329329329329	29.2372268114748\\
-0.00327327327327327	29.489523376677\\
-0.00325325325325325	29.7430508357773\\
-0.00323323323323323	29.9991697206763\\
-0.00321321321321321	30.2587381767649\\
-0.00319319319319319	30.5228001215263\\
-0.00317317317317317	30.7923195534411\\
-0.00315315315315315	31.0687185897815\\
-0.00313313313313313	31.3530699987924\\
-0.00311311311311311	31.6464980008245\\
-0.00309309309309309	31.9497990339042\\
-0.00307307307307307	32.2638938123463\\
-0.00305305305305305	32.5892490131247\\
-0.00303303303303303	32.9275286330778\\
-0.00301301301301301	33.2790857415613\\
-0.00299299299299299	33.6438388794135\\
-0.00297297297297297	34.022732878181\\
-0.00295295295295295	34.4157917360985\\
-0.00293293293293293	34.8232690179047\\
-0.00291291291291291	35.2454791122564\\
-0.00289289289289289	35.682554556775\\
-0.00287287287287287	36.1337532589091\\
-0.00285285285285285	36.5979828170708\\
-0.00283283283283283	37.0757459513817\\
-0.00281281281281281	37.5664969642457\\
-0.00279279279279279	38.0690943627159\\
-0.00277277277277277	38.5829881045424\\
-0.00275275275275275	39.1078820788216\\
-0.00273273273273273	39.6417785599823\\
-0.00271271271271271	40.1846092067989\\
-0.00269269269269269	40.7353581557079\\
-0.00267267267267267	41.2928490606593\\
-0.00265265265265265	41.856525323049\\
-0.00263263263263263	42.4251421909997\\
-0.00261261261261261	42.9989566298496\\
-0.00259259259259259	43.5761444323781\\
-0.00257257257257257	44.1566257468017\\
-0.00255255255255255	44.7391475071377\\
-0.00253253253253253	45.3243413409642\\
-0.00251251251251251	45.9112477847292\\
-0.00249249249249249	46.4990509573297\\
-0.00247247247247247	47.0883585533649\\
-0.00245245245245245	47.6796024255539\\
-0.00243243243243243	48.2719498017081\\
-0.00241241241241241	48.8670629932081\\
-0.00239239239239239	49.4646890411097\\
-0.00237237237237237	50.0659534121886\\
-0.00235235235235235	50.6714658977734\\
-0.00233233233233233	51.2827789509745\\
-0.00231231231231231	51.9003024599977\\
-0.00229229229229229	52.5259559263103\\
-0.00227227227227227	53.1608221154976\\
-0.00225225225225225	53.807079600101\\
-0.00223223223223223	54.4673638018425\\
-0.00221221221221221	55.1417219759995\\
-0.00219219219219219	55.8328176022734\\
-0.00217217217217217	56.5431810798298\\
-0.00215215215215215	57.2740651961291\\
-0.00213213213213213	58.0283357072138\\
-0.00211211211211211	58.8078393456855\\
-0.00209209209209209	59.6144808604924\\
-0.00207207207207207	60.4511735521292\\
-0.00205205205205205	61.3190229164236\\
-0.00203203203203203	62.2200659236314\\
-0.00201201201201201	63.1576890847888\\
-0.00199199199199199	64.1328317523892\\
-0.00197197197197197	65.1467973600081\\
-0.00195195195195195	66.2011703452491\\
-0.00193193193193193	67.2987601197105\\
-0.00191191191191191	68.4395376074103\\
-0.00189189189189189	69.6249951595567\\
-0.00187187187187187	70.8571185038448\\
-0.00185185185185185	72.1356362953945\\
-0.00183183183183183	73.4617720995555\\
-0.00181181181181181	74.836431518945\\
-0.00179179179179179	76.2594989957803\\
-0.00177177177177177	77.7319626222832\\
-0.00175175175175175	79.253032118895\\
-0.00173173173173173	80.8236630877371\\
-0.00171171171171171	82.443481094434\\
-0.00169169169169169	84.1108834016539\\
-0.00167167167167167	85.8267651926371\\
-0.00165165165165165	87.5910543339079\\
-0.00163163163163163	89.4029920970505\\
-0.00161161161161161	91.2612730248074\\
-0.00159159159159159	93.1660109927676\\
-0.00157157157157157	95.11679324605\\
-0.00155155155155155	97.1121242938324\\
-0.00153153153153153	99.1520553363032\\
-0.00151151151151151	101.235366830531\\
-0.00149149149149149	103.361510558088\\
-0.00147147147147147	105.530878665884\\
-0.00145145145145145	107.742431289918\\
-0.00143143143143143	109.99407071878\\
-0.00141141141141141	112.285869396222\\
-0.00139139139139139	114.617538612389\\
-0.00137137137137137	116.989550453244\\
-0.00135135135135135	119.4003009063\\
-0.00133133133133133	121.846754023905\\
-0.00131131131131131	124.331126719396\\
-0.00129129129129129	126.852592981704\\
-0.00127127127127127	129.408619974882\\
-0.00125125125125125	131.999474554252\\
-0.00123123123123123	134.62382525306\\
-0.00121121121121121	137.28217776003\\
-0.00119119119119119	139.972287624395\\
-0.00117117117117117	142.693444164257\\
-0.00115115115115115	145.444448151378\\
-0.00113113113113113	148.223973494558\\
-0.00111111111111111	151.032016301558\\
-0.00109109109109109	153.866173431001\\
-0.00107107107107107	156.725249494916\\
-0.00105105105105105	159.60857647431\\
-0.00103103103103103	162.515127343804\\
-0.00101101101101101	165.443201327899\\
-0.000990990990990991	168.391006088281\\
-0.000970970970970972	171.357575663991\\
-0.000950950950950951	174.341193450617\\
-0.000930930930930931	177.339687583532\\
-0.000910910910910912	180.351736721384\\
-0.000890890890890891	183.375187709434\\
-0.000870870870870869	186.409056733182\\
-0.00085085085085085	189.450469415653\\
-0.000830830830830831	192.497075492929\\
-0.000810810810810811	195.547072332797\\
-0.00079079079079079	198.59675208606\\
-0.000770770770770771	201.643966223988\\
-0.000750750750750751	204.684693301398\\
-0.000730730730730732	207.716262658283\\
-0.000710710710710712	210.73448351984\\
-0.000690690690690689	213.73558864629\\
-0.00067067067067067	216.714431504521\\
-0.000650650650650651	219.665985075458\\
-0.000630630630630631	222.585579002775\\
-0.00061061061061061	225.467422018328\\
-0.000590590590590591	228.305994694187\\
-0.000570570570570571	231.093651970951\\
-0.000550550550550552	233.824869436356\\
-0.000530530530530531	236.492767456271\\
-0.000510510510510511	239.090384661873\\
-0.00049049049049049	241.610484163701\\
-0.00047047047047047	244.045703647576\\
-0.000450450450450449	246.388371864992\\
-0.00043043043043043	248.632537482388\\
-0.00041041041041041	250.770058785961\\
-0.000390390390390391	252.793018531313\\
-0.00037037037037037	254.695849277539\\
-0.00035035035035035	256.471755774203\\
-0.000330330330330331	258.114877334958\\
-0.000310310310310312	259.618371175941\\
-0.000290290290290289	260.977510627721\\
-0.000270270270270269	262.187934628392\\
-0.00025025025025025	263.244412651248\\
-0.00023023023023023	264.143726075871\\
-0.000210210210210211	264.883701762973\\
-0.00019019019019019	265.460793589243\\
-0.00017017017017017	265.873533716974\\
-0.000150150150150151	266.121338612711\\
-0.000130130130130131	266.202870605669\\
-0.00011011011011011	266.119488011243\\
-9.00900900900892e-05	265.872136265201\\
-7.00700700700697e-05	265.461569153847\\
-5.00500500500503e-05	264.890399309253\\
-3.00300300300291e-05	264.161258092325\\
-1.00100100100097e-05	263.277154037249\\
1.00100100100097e-05	262.242353939054\\
3.00300300300291e-05	261.060757409088\\
5.00500500500503e-05	259.737656063614\\
7.00700700700697e-05	258.276951826998\\
9.00900900900892e-05	256.68514438643\\
0.00011011011011011	254.965807383371\\
0.000130130130130131	253.126338106883\\
0.000150150150150151	251.173047920127\\
0.00017017017017017	249.110588615863\\
0.00019019019019019	246.945197182821\\
0.000210210210210211	244.683614694438\\
0.00023023023023023	242.331058920721\\
0.00025025025025025	239.894344830926\\
0.000270270270270269	237.378286212896\\
0.000290290290290289	234.789516688548\\
0.000310310310310312	232.133428678353\\
0.000330330330330331	229.415491878243\\
0.00035035035035035	226.641369146738\\
0.00037037037037037	223.815992881387\\
0.000390390390390391	220.943755980329\\
0.00041041041041041	218.029618713517\\
0.00043043043043043	215.078173496773\\
0.000450450450450449	212.093839041899\\
0.00047047047047047	209.080665871658\\
0.00049049049049049	206.042431763626\\
0.000510510510510511	202.982489827331\\
0.000530530530530531	199.904614306085\\
0.000550550550550552	196.811544056342\\
0.000570570570570571	193.706539667669\\
0.000590590590590591	190.592638167843\\
0.00061061061061061	187.472595828168\\
0.000630630630630631	184.347862020545\\
0.000650650650650651	181.222228857274\\
0.00067067067067067	178.097019827647\\
0.000690690690690689	174.975989939104\\
0.000710710710710712	171.859416607301\\
0.000730730730730732	168.750157190235\\
0.000750750750750751	165.650059482333\\
0.000770770770770771	162.560536157733\\
0.00079079079079079	159.483282081617\\
0.000810810810810811	156.420954663393\\
0.000830830830830831	153.373966703755\\
0.00085085085085085	150.345186076694\\
0.000870870870870869	147.335021256382\\
0.000890890890890891	144.346156689588\\
0.000910910910910912	141.379213560981\\
0.000930930930930931	138.436536739892\\
0.000950950950950951	135.520145748634\\
0.000970970970970972	132.631447295371\\
0.000990990990990991	129.770730928394\\
0.00101101101101101	126.941349577686\\
0.00103103103103103	124.145046084697\\
0.00105105105105105	121.382570267482\\
0.00107107107107107	118.65682431995\\
0.00109109109109109	115.969802073788\\
0.00111111111111111	113.32258270851\\
0.00113113113113113	110.718166734487\\
0.00115115115115115	108.158228225075\\
0.00117117117117117	105.643796599771\\
0.00119119119119119	103.178292172319\\
0.00121121121121121	100.764620482935\\
0.00123123123123123	98.4036146336712\\
0.00125125125125125	96.0962009260475\\
0.00127127127127127	93.8458605582077\\
0.00129129129129129	91.6521497440405\\
0.00131131131131131	89.5203341654723\\
0.00133133133133133	87.4485714010705\\
0.00135135135135135	85.4380645368708\\
0.00137137137137137	83.4911192214146\\
0.00139139139139139	81.6081683689995\\
0.00141141141141141	79.789151146657\\
0.00143143143143143	78.0330447196741\\
0.00145145145145145	76.3406853478309\\
0.00147147147147147	74.7108493465463\\
0.00149149149149149	73.143133723538\\
0.00151151151151151	71.6360701725199\\
0.00153153153153153	70.1872489926593\\
0.00155155155155155	68.795623362322\\
0.00157157157157157	67.4590332418781\\
0.00159159159159159	66.1747686715497\\
0.00161161161161161	64.9395989820864\\
0.00163163163163163	63.7506176932672\\
0.00165165165165165	62.6052349796072\\
0.00167167167167167	61.5002612516794\\
0.00169169169169169	60.4326713625266\\
0.00171171171171171	59.3989709850199\\
0.00173173173173173	58.396325567155\\
0.00175175175175175	57.4207015555416\\
0.00177177177177177	56.4688189197208\\
0.00179179179179179	55.5382241506588\\
0.00181181181181181	54.6256445374963\\
0.00183183183183183	53.7279919752365\\
0.00185185185185185	52.8439929903731\\
0.00187187187187187	51.9704068408619\\
0.00189189189189189	51.1061122169339\\
0.00191191191191191	50.24881552583\\
0.00193193193193193	49.3964984052097\\
0.00195195195195195	48.548939218191\\
0.00197197197197197	47.7050719855693\\
0.00199199199199199	46.864397178263\\
0.00201201201201201	46.0261948063852\\
0.00203203203203203	45.1907291162141\\
0.00205205205205205	44.357966546535\\
0.00207207207207207	43.5287878182093\\
0.00209209209209209	42.7034029427466\\
0.00211211211211211	41.8824855977612\\
0.00213213213213213	41.0666366622069\\
0.00215215215215215	40.2572281068835\\
0.00217217217217217	39.4552825099111\\
0.00219219219219219	38.6625446808014\\
0.00221221221221221	37.8790328276484\\
0.00223223223223223	37.1073043830002\\
0.00225225225225225	36.3482737693204\\
0.00227227227227227	35.6025603497231\\
0.00229229229229229	34.8724490159901\\
0.00231231231231231	34.158151262147\\
0.00233233233233233	33.46117330164\\
0.00235235235235235	32.782620131651\\
0.00237237237237237	32.1228585463198\\
0.00239239239239239	31.4828225314537\\
0.00241241241241241	30.8638184160137\\
0.00243243243243243	30.2656076451905\\
0.00245245245245245	29.6887895875056\\
0.00247247247247247	29.1334509816819\\
0.00249249249249249	28.599658126879\\
0.00251251251251251	28.087228246754\\
0.00253253253253253	27.5953888022539\\
0.00255255255255255	27.1248324319027\\
0.00257257257257257	26.6742763508744\\
0.00259259259259259	26.2435256372427\\
0.00261261261261261	25.8312629398969\\
0.00263263263263263	25.4368782953724\\
0.00265265265265265	25.0594068050249\\
0.00267267267267267	24.6973410309106\\
0.00269269269269269	24.3501722504912\\
0.00271271271271271	24.016644415671\\
0.00273273273273273	23.695431518108\\
0.00275275275275275	23.3857247971868\\
0.00277277277277277	23.0860641001724\\
0.00279279279279279	22.7944710894448\\
0.00281281281281281	22.5110083386255\\
0.00283283283283283	22.2340201266473\\
0.00285285285285285	21.9623605120346\\
0.00287287287287287	21.6951583024466\\
0.00289289289289289	21.4312477782097\\
0.00291291291291291	21.1700230600611\\
0.00293293293293293	20.9109760482157\\
0.00295295295295295	20.6523864997348\\
0.00297297297297297	20.3946491713842\\
0.00299299299299299	20.1364628600609\\
0.00301301301301301	19.8776584561164\\
0.00303303303303303	19.6181066726468\\
0.00305305305305305	19.356916619445\\
0.00307307307307307	19.0943965807141\\
0.00309309309309309	18.8305820899803\\
0.00311311311311311	18.5645566884978\\
0.00313313313313313	18.2970499678696\\
0.00315315315315315	18.0281115271349\\
0.00317317317317317	17.7572179764762\\
0.00319319319319319	17.4849488339632\\
0.00321321321321321	17.2117753641415\\
0.00323323323323323	16.9373161241536\\
0.00325325325325325	16.6623590334307\\
0.00327327327327327	16.3867629644051\\
0.00329329329329329	16.1109744684926\\
0.00331331331331331	15.835092175822\\
0.00333333333333333	15.5596610873641\\
0.00335335335335335	15.2850008570069\\
0.00337337337337337	15.0117934539055\\
0.00339339339339339	14.7399183695526\\
0.00341341341341341	14.4696781210565\\
0.00343343343343343	14.2022928490636\\
0.00345345345345345	13.9372651390579\\
0.00347347347347347	13.6752036628638\\
0.00349349349349349	13.4168282023516\\
0.00351351351351351	13.1625076691102\\
0.00353353353353353	12.9124804456588\\
0.00355355355355355	12.6672462243543\\
0.00357357357357357	12.427244316507\\
0.00359359359359359	12.1930553609087\\
0.00361361361361361	11.965029614571\\
0.00363363363363363	11.7439562242776\\
0.00365365365365365	11.529911635371\\
0.00367367367367367	11.3235276038269\\
0.00369369369369369	11.1254121575676\\
0.00371371371371372	10.9356373449713\\
0.00373373373373373	10.7552123272685\\
0.00375375375375375	10.5843140542102\\
0.00377377377377377	10.4233781066446\\
0.00379379379379379	10.2726092695901\\
0.00381381381381381	10.1325439972132\\
0.00383383383383383	10.0035827832307\\
0.00385385385385385	9.88585760089032\\
0.00387387387387387	9.77984651410442\\
0.00389389389389389	9.68546316704744\\
0.00391391391391391	9.60304826610943\\
0.00393393393393393	9.53289958180094\\
0.00395395395395395	9.47488099349239\\
0.00397397397397397	9.42876342094386\\
0.00399399399399399	9.39440662569855\\
0.00401401401401401	9.3716375005333\\
0.00403403403403403	9.36013981622349\\
0.00405405405405405	9.35979619638323\\
0.00407407407407407	9.36996113991867\\
0.00409409409409409	9.39015687724442\\
0.00411411411411411	9.41992814120833\\
0.00413413413413414	9.45810985188236\\
0.00415415415415415	9.50434247828707\\
0.00417417417417417	9.55721508596771\\
0.00419419419419419	9.61601100789658\\
0.00421421421421421	9.67968405896858\\
0.00423423423423423	9.74737657631661\\
0.00425425425425425	9.81787862945604\\
0.00427427427427427	9.88997194753162\\
0.00429429429429429	9.96247861296794\\
0.00431431431431432	10.0342463414155\\
0.00433433433433433	10.1040960088369\\
0.00435435435435435	10.1707296320621\\
0.00437437437437437	10.2333302330589\\
0.00439439439439439	10.2908827014141\\
0.00441441441441441	10.3421996849952\\
0.00443443443443443	10.3863637696678\\
0.00445445445445445	10.4225875400222\\
0.00447447447447447	10.4503141748005\\
0.00449449449449449	10.4686008531039\\
0.00451451451451452	10.4769783074596\\
0.00453453453453453	10.4751666365961\\
0.00455455455455456	10.4630653122787\\
0.00457457457457458	10.4405676511184\\
0.00459459459459459	10.4073938204198\\
0.00461461461461461	10.3639202843197\\
0.00463463463463464	10.3105436208506\\
0.00465465465465465	10.2474157889643\\
0.00467467467467467	10.175297212542\\
0.00469469469469469	10.0948537271267\\
0.00471471471471472	10.0069365392\\
0.00473473473473474	9.91229802098279\\
0.00475475475475476	9.81173420369003\\
0.00477477477477477	9.70630538362224\\
0.0047947947947948	9.59712057584077\\
0.00481481481481482	9.48503975722631\\
0.00483483483483483	9.37139453746953\\
0.00485485485485485	9.25701000062706\\
0.00487487487487487	9.14285168417887\\
0.00489489489489489	9.03014078780589\\
0.00491491491491492	8.91959410174709\\
0.00493493493493494	8.81213855921981\\
0.00495495495495495	8.70847593188867\\
0.00497497497497498	8.60916440126749\\
0.004994994994995	8.51501799959966\\
0.00501501501501501	8.4264873262071\\
0.00503503503503503	8.34392428867314\\
0.00505505505505506	8.26760984403189\\
0.00507507507507507	8.19749506074351\\
0.00509509509509509	8.13374455643575\\
0.00511511511511512	8.07638402593253\\
0.00513513513513514	8.02498534437788\\
0.00515515515515516	7.97972882857028\\
0.00517517517517518	7.93999102370336\\
0.00519519519519519	7.90522246490009\\
0.00521521521521522	7.87502949895259\\
0.00523523523523524	7.84886978760082\\
0.00525525525525525	7.82617813376512\\
0.00527527527527527	7.80628884650389\\
0.0052952952952953	7.78856258217083\\
0.00531531531531532	7.77217770200363\\
0.00533533533533534	7.75671511136707\\
0.00535535535535536	7.74143605979043\\
0.00537537537537537	7.72576276129557\\
0.0053953953953954	7.70895610066268\\
0.00541541541541542	7.69054303493046\\
0.00543543543543543	7.67005103537756\\
0.00545545545545545	7.64687651829385\\
0.00547547547547548	7.62078839743173\\
0.00549549549549549	7.59131342110301\\
0.00551551551551552	7.55804119140539\\
0.00553553553553554	7.52083999319078\\
0.00555555555555556	7.47946040523636\\
0.00557557557557558	7.4336888564166\\
0.0055955955955956	7.38354710307306\\
0.00561561561561561	7.32903968860596\\
0.00563563563563564	7.27003941062297\\
0.00565565565565566	7.20656594319482\\
0.00567567567567567	7.1387736300389\\
0.00569569569569569	7.06678014485577\\
0.00571571571571572	6.99084417373498\\
0.00573573573573574	6.91099492758032\\
0.00575575575575576	6.82765354629832\\
0.00577577577577578	6.74100426393258\\
0.00579579579579579	6.65104972615279\\
0.00581581581581582	6.55830469704139\\
0.00583583583583584	6.46320464156016\\
0.00585585585585586	6.36576702445663\\
0.00587587587587588	6.2667116593825\\
0.0058958958958959	6.16593953340337\\
0.00591591591591592	6.0640397802248\\
0.00593593593593593	5.96141193147107\\
0.00595595595595595	5.85815358112638\\
0.00597597597597598	5.75464042352926\\
0.005995995995996	5.65135056939467\\
0.00601601601601602	5.54849053293206\\
0.00603603603603604	5.44656149854288\\
0.00605605605605605	5.34554573347739\\
0.00607607607607607	5.2460144879146\\
0.0060960960960961	5.14824381441571\\
0.00611611611611612	5.05241263067288\\
0.00613613613613614	4.95877691700473\\
0.00615615615615616	4.86757754440479\\
0.00617617617617617	4.77913854767859\\
0.00619619619619619	4.6935097443242\\
0.00621621621621622	4.61071169691735\\
0.00623623623623624	4.53124618773796\\
0.00625625625625626	4.45496690351837\\
0.00627627627627628	4.38212996307694\\
0.0062962962962963	4.31280600849224\\
0.00631631631631632	4.24706640503465\\
0.00633633633633634	4.18479005178122\\
0.00635635635635636	4.12608178915606\\
0.00637637637637638	4.07098455095439\\
0.0063963963963964	4.01939773478983\\
0.00641641641641641	3.97137127875246\\
0.00643643643643644	3.92672589016298\\
0.00645645645645646	3.88549520072647\\
0.00647647647647648	3.84742099268158\\
0.0064964964964965	3.81263409006028\\
0.00651651651651652	3.7808843776239\\
0.00653653653653653	3.75211445398956\\
0.00655655655655655	3.72602200722681\\
0.00657657657657657	3.70260178263037\\
0.0065965965965966	3.6816500542437\\
0.00661661661661662	3.66310478839358\\
0.00663663663663664	3.64662744289332\\
0.00665665665665666	3.63209004710591\\
0.00667667667667668	3.61920261538807\\
0.0066966966966967	3.60813724299017\\
0.00671671671671672	3.59844662486174\\
0.00673673673673674	3.59010689267069\\
0.00675675675675676	3.58271681167354\\
0.00677677677677678	3.57646504343375\\
0.0067967967967968	3.57089011534194\\
0.00681681681681682	3.56637293949623\\
0.00683683683683684	3.5624248058245\\
0.00685685685685686	3.55910377130006\\
0.00687687687687688	3.55626725398391\\
0.00689689689689689	3.55381153610136\\
0.00691691691691691	3.55177213716319\\
0.00693693693693694	3.55019251581507\\
0.00695695695695696	3.54899942186205\\
0.00697697697697698	3.5481447046164\\
0.006996996996997	3.54766763855439\\
0.00701701701701702	3.54775246965319\\
0.00703703703703704	3.54841558105564\\
0.00705705705705706	3.54950748804727\\
0.00707707707707708	3.55130555055284\\
0.0070970970970971	3.55389648433841\\
0.00711711711711712	3.55715109021743\\
0.00713713713713714	3.561247034545\\
0.00715715715715716	3.56611836590609\\
0.00717717717717718	3.57173289868383\\
0.0071971971971972	3.57818590323816\\
0.00721721721721722	3.58532293416653\\
0.00723723723723724	3.59339113395404\\
0.00725725725725726	3.60221033752042\\
0.00727727727727728	3.61156649947661\\
0.0072972972972973	3.62141730433087\\
0.00731731731731732	3.63167364742595\\
0.00733733733733734	3.64209765710272\\
0.00735735735735736	3.6524311926138\\
0.00737737737737738	3.6626038261961\\
0.00739739739739739	3.67226490106568\\
0.00741741741741742	3.68123995373615\\
0.00743743743743744	3.68920623414145\\
0.00745745745745746	3.6958169123465\\
0.00747747747747748	3.70097646709366\\
0.0074974974974975	3.70423669136826\\
0.00751751751751752	3.70536653506557\\
0.00753753753753754	3.7039326494224\\
0.00755755755755756	3.69997323016571\\
0.00757757757757758	3.69308561696799\\
0.0075975975975976	3.68293721488404\\
0.00761761761761762	3.66960222501713\\
0.00763763763763764	3.65278178347383\\
0.00765765765765766	3.63235587430197\\
0.00767767767767768	3.60819671677677\\
0.0076976976976977	3.58032290119596\\
0.00771771771771772	3.54876718815277\\
0.00773773773773774	3.5136362417003\\
0.00775775775775776	3.47495370405609\\
0.00777777777777778	3.43286614174912\\
0.0077977977977978	3.38756269984855\\
0.00781781781781782	3.33921488163628\\
0.00783783783783784	3.28821781353713\\
0.00785785785785786	3.2347983143389\\
0.00787787787787788	3.17928592406779\\
0.0078978978978979	3.12203188563296\\
0.00791791791791792	3.06342331653481\\
0.00793793793793794	3.00387864806029\\
0.00795795795795796	2.94375773096496\\
0.00797797797797798	2.8833136985564\\
0.007997997997998	2.82321265033195\\
0.00801801801801802	2.76368029153413\\
0.00803803803803804	2.70507866708655\\
0.00805805805805806	2.6477473131873\\
0.00807807807807808	2.59199846512325\\
0.0080980980980981	2.53812268109013\\
0.00811811811811812	2.48628716332766\\
0.00813813813813814	2.4367753543491\\
0.00815815815815816	2.38977228894923\\
0.00817817817817818	2.34544533176652\\
0.0081981981981982	2.30380782476585\\
0.00821821821821822	2.26490441187255\\
0.00823823823823824	2.22868411806375\\
0.00825825825825826	2.19521002681663\\
0.00827827827827828	2.16441372912365\\
0.0082982982982983	2.13614453238527\\
0.00831831831831832	2.1101512560043\\
0.00833833833833834	2.08655111020245\\
0.00835835835835836	2.0649485936406\\
0.00837837837837838	2.04528459658812\\
0.0083983983983984	2.02722541060329\\
0.00841841841841842	2.01068414093251\\
0.00843843843843844	1.99539489367649\\
0.00845845845845846	1.98112382533503\\
0.00847847847847848	1.96768084665817\\
0.0084984984984985	1.95487462414224\\
0.00851851851851852	1.94246362551985\\
0.00853853853853854	1.93042787334518\\
0.00855855855855856	1.91847791900483\\
0.00857857857857858	1.90667381625268\\
0.0085985985985986	1.89473838033772\\
0.00861861861861862	1.88273197360412\\
0.00863863863863864	1.87052076958483\\
0.00865865865865866	1.85807253677512\\
0.00867867867867868	1.84548940798111\\
0.0086986986986987	1.83258229003261\\
0.00871871871871872	1.81946670776229\\
0.00873873873873874	1.80621968202002\\
0.00875875875875876	1.79283778239163\\
0.00877877877877878	1.77935871186766\\
0.0087987987987988	1.76597330552104\\
0.00881881881881882	1.75269244527086\\
0.00883883883883884	1.73973130981207\\
0.00885885885885886	1.72696646187645\\
0.00887887887887888	1.71471097793761\\
0.0088988988988989	1.70291444660815\\
0.00891891891891892	1.69169072967729\\
0.00893893893893894	1.68105655133166\\
0.00895895895895896	1.67115757015192\\
0.00897897897897898	1.66212585672298\\
0.008998998998999	1.6540270090685\\
0.00901901901901902	1.64684856757993\\
0.00903903903903904	1.6406818105838\\
0.00905905905905906	1.63546349252884\\
0.00907907907907908	1.63131734134655\\
0.0090990990990991	1.62834405437901\\
0.00911911911911912	1.62648503234831\\
0.00913913913913914	1.6257011011071\\
0.00915915915915916	1.62615357988685\\
0.00917917917917918	1.62778040621602\\
0.0091991991991992	1.63054050941681\\
0.00921921921921922	1.63457849839377\\
0.00923923923923924	1.63984932678894\\
0.00925925925925926	1.64629024574656\\
0.00927927927927928	1.65405718586916\\
0.0092992992992993	1.66304000371999\\
0.00931931931931932	1.67332495794491\\
0.00933933933933934	1.68498396227224\\
0.00935935935935936	1.69796745720625\\
0.00937937937937938	1.71224806853823\\
0.0093993993993994	1.72795062717155\\
0.00941941941941942	1.74502483555024\\
0.00943943943943944	1.7634795720148\\
0.00945945945945946	1.78331716941178\\
0.00947947947947948	1.80453181281488\\
0.0094994994994995	1.82710788895574\\
0.00951951951951952	1.851018338136\\
0.00953953953953954	1.87624249225264\\
0.00955955955955956	1.90269260564432\\
0.00957957957957958	1.93031346823825\\
0.0095995995995996	1.95901802897336\\
0.00961961961961962	1.98870148882752\\
0.00963963963963964	2.01924074442219\\
0.00965965965965966	2.05051050635877\\
0.00967967967967968	2.08232292432143\\
0.0096996996996997	2.11451287771376\\
0.00971971971971972	2.14688575404928\\
0.00973973973973974	2.17916559402881\\
0.00975975975975976	2.21121024443006\\
0.00977977977977978	2.2427264777404\\
0.0097997997997998	2.27364603045843\\
0.00981981981981982	2.30356979696467\\
0.00983983983983984	2.33224295710818\\
0.00985985985985986	2.35934516188015\\
0.00987987987987988	2.38476516096033\\
0.0098998998998999	2.40817348099898\\
0.00991991991991992	2.42933149639968\\
0.00993993993993994	2.44801300443058\\
0.00995995995995996	2.46400758604633\\
0.00997997997997998	2.47712380047641\\
0.01	2.48722229844518\\
};
\end{axis}
\end{tikzpicture}%

%% file: figures/hat_log_nhis_kde.tex
%
\tikzsetnextfilename{hat_log_nhis_kde}
\definecolor{mycolor1}{rgb}{0.12157,0.47059,0.70588}%
\begin{tikzpicture}

\begin{axis}[%
width=\figurewidth,
height=\figureheight,
at={(0\figurewidth,0\figureheight)},
scale only axis,
xmin=-1,
xmax=1,
xlabel={difference between estimated $\lni$ (new method/concordance)},
ymin=0,
ymax=2.5,
ylabel={$p(\text{difference})$},
axis background/.style={fill=white},
axis x line*=bottom,
axis y line*=left,
legend columns=1
]
\addplot [color=mycolor1,solid,forget plot]
  table[row sep=crcr]{%
-1	0.0174723987108811\\
-0.997997997997998	0.0176536106555805\\
-0.995995995995996	0.0178401151511117\\
-0.993993993993994	0.0180322521791321\\
-0.991991991991992	0.018230147630604\\
-0.98998998998999	0.0184348507076526\\
-0.987987987987988	0.0186473142529273\\
-0.985985985985986	0.0188681005437621\\
-0.983983983983984	0.0190981711259475\\
-0.981981981981982	0.0193383750251889\\
-0.97997997997998	0.0195895594079856\\
-0.977977977977978	0.0198520187639808\\
-0.975975975975976	0.0201277560848997\\
-0.973973973973974	0.0204171738244069\\
-0.971971971971972	0.0207205222377167\\
-0.96996996996997	0.0210387277807608\\
-0.967967967967968	0.0213724306111898\\
-0.965965965965966	0.0217222393231998\\
-0.963963963963964	0.0220885848972988\\
-0.961961961961962	0.0224719060717685\\
-0.95995995995996	0.0228719267355848\\
-0.957957957957958	0.0232895777944715\\
-0.955955955955956	0.0237255319775103\\
-0.953953953953954	0.0241790708888317\\
-0.951951951951952	0.0246500848823494\\
-0.94994994994995	0.0251383422044643\\
-0.947947947947948	0.0256434853140361\\
-0.945945945945946	0.026165027754958\\
-0.943943943943944	0.0267023515550846\\
-0.941941941941942	0.0272547051265641\\
-0.93993993993994	0.0278223705668511\\
-0.937937937937938	0.0284022619388035\\
-0.935935935935936	0.0289946488546052\\
-0.933933933933934	0.02959740664059\\
-0.931931931931932	0.0302106474262002\\
-0.92992992992993	0.0308308009647474\\
-0.927927927927928	0.0314575172689752\\
-0.925925925925926	0.0320883014012307\\
-0.923923923923924	0.0327220026872171\\
-0.921921921921922	0.0333557974185486\\
-0.91991991991992	0.0339897121088759\\
-0.917917917917918	0.0346183294081231\\
-0.915915915915916	0.0352420989501154\\
-0.913913913913914	0.0358564236439196\\
-0.911911911911912	0.0364609913853899\\
-0.90990990990991	0.0370509024100045\\
-0.907907907907908	0.0376245433766412\\
-0.905905905905906	0.0381802981598683\\
-0.903903903903904	0.0387159773814969\\
-0.901901901901902	0.039227484560128\\
-0.8998998998999	0.0397124522491664\\
-0.897897897897898	0.0401692074998051\\
-0.895895895895896	0.0405946924662431\\
-0.893893893893894	0.0409873944172066\\
-0.891891891891892	0.0413445721347527\\
-0.88988988988989	0.041664143335244\\
-0.887887887887888	0.0419474868008122\\
-0.885885885885886	0.0421894591670198\\
-0.883883883883884	0.042389715998345\\
-0.881881881881882	0.0425482158350117\\
-0.87987987987988	0.0426638601833354\\
-0.877877877877878	0.0427361875457436\\
-0.875875875875876	0.0427656025333963\\
-0.873873873873874	0.0427531116026861\\
-0.871871871871872	0.0426986442630164\\
-0.86986986986987	0.0426046038735518\\
-0.867867867867868	0.0424717103898182\\
-0.865865865865866	0.0423028351248309\\
-0.863863863863864	0.0420996537890461\\
-0.861861861861862	0.0418665790495639\\
-0.85985985985986	0.0416053990862209\\
-0.857857857857858	0.0413205775847771\\
-0.855855855855856	0.0410142803927511\\
-0.853853853853854	0.0406927084123063\\
-0.851851851851852	0.040359693965125\\
-0.84984984984985	0.0400204148470828\\
-0.847847847847848	0.0396773867763612\\
-0.845845845845846	0.0393396836024635\\
-0.843843843843844	0.0390099978202842\\
-0.841841841841842	0.0386936623345641\\
-0.83983983983984	0.0383957521780803\\
-0.837837837837838	0.0381224669571501\\
-0.835835835835836	0.037877045220503\\
-0.833833833833834	0.037665209817182\\
-0.831831831831832	0.0374898178279508\\
-0.82982982982983	0.0373570528575948\\
-0.827827827827828	0.0372696934817876\\
-0.825825825825826	0.0372303939067775\\
-0.823823823823824	0.0372425454735577\\
-0.821821821821822	0.0373083125640806\\
-0.81981981981982	0.0374303060841889\\
-0.817817817817818	0.0376097247493438\\
-0.815815815815816	0.0378464002585221\\
-0.813813813813814	0.0381406476690215\\
-0.811811811811812	0.0384934467147224\\
-0.80980980980981	0.0389027245975482\\
-0.807807807807808	0.0393681581650533\\
-0.805805805805806	0.0398875293841558\\
-0.803803803803804	0.0404578576342736\\
-0.801801801801802	0.0410765016401217\\
-0.7997997997998	0.041740134706936\\
-0.797797797797798	0.0424455106019026\\
-0.795795795795796	0.0431877396321693\\
-0.793793793793794	0.0439641585121417\\
-0.791791791791792	0.0447681184807596\\
-0.78978978978979	0.045596162429765\\
-0.787787787787788	0.0464419579154972\\
-0.785785785785786	0.0473034828480145\\
-0.783783783783784	0.0481734271494555\\
-0.781781781781782	0.0490469727228322\\
-0.77977977977978	0.0499204359593618\\
-0.777777777777778	0.0507881519324993\\
-0.775775775775776	0.0516472916354852\\
-0.773773773773774	0.0524921357967351\\
-0.771771771771772	0.0533189388918991\\
-0.76976976976977	0.0541252188775412\\
-0.767767767767768	0.054907154368891\\
-0.765765765765766	0.0556611575453741\\
-0.763763763763764	0.0563852060148757\\
-0.761761761761762	0.0570787038952404\\
-0.75975975975976	0.0577387598659445\\
-0.757757757757758	0.0583632844539533\\
-0.755755755755756	0.0589541260694409\\
-0.753753753753754	0.0595090362703282\\
-0.751751751751752	0.0600271780207041\\
-0.74974974974975	0.0605090258801627\\
-0.747747747747748	0.0609570576562883\\
-0.745745745745746	0.0613722558713255\\
-0.743743743743744	0.0617528432748889\\
-0.741741741741742	0.0621048157950047\\
-0.73973973973974	0.0624241838287061\\
-0.737737737737738	0.0627158601231069\\
-0.735735735735736	0.0629795704729474\\
-0.733733733733734	0.0632202216087898\\
-0.731731731731732	0.0634399452720814\\
-0.72972972972973	0.0636389161400339\\
-0.727727727727728	0.063819677366446\\
-0.725725725725726	0.0639839500351321\\
-0.723723723723724	0.0641359747123658\\
-0.721721721721722	0.0642735888006752\\
-0.71971971971972	0.0644050978890185\\
-0.717717717717718	0.0645266406336571\\
-0.715715715715716	0.0646434149973109\\
-0.713713713713714	0.0647553796258959\\
-0.711711711711712	0.0648656898474422\\
-0.70970970970971	0.0649751061840019\\
-0.707707707707708	0.065083205403342\\
-0.705705705705706	0.0651956384522285\\
-0.703703703703704	0.0653119449613904\\
-0.701701701701702	0.0654341993378319\\
-0.6996996996997	0.0655630943445474\\
-0.697697697697698	0.0656996262129058\\
-0.695695695695696	0.0658463081180819\\
-0.693693693693694	0.0660058490330064\\
-0.691691691691692	0.0661777086609233\\
-0.68968968968969	0.0663665387836986\\
-0.687687687687688	0.0665671324964576\\
-0.685685685685686	0.0667859546616486\\
-0.683683683683684	0.0670226171919921\\
-0.681681681681682	0.0672800856313255\\
-0.67967967967968	0.0675601994758017\\
-0.677677677677678	0.0678633875284502\\
-0.675675675675676	0.0681920300862914\\
-0.673673673673674	0.0685470214350675\\
-0.671671671671672	0.0689310422322715\\
-0.66966966966967	0.0693465925099466\\
-0.667667667667668	0.0697943228295431\\
-0.665665665665666	0.0702752452594454\\
-0.663663663663664	0.0707942831694428\\
-0.661661661661662	0.0713485931019518\\
-0.65965965965966	0.0719411601181053\\
-0.657657657657658	0.0725738756061579\\
-0.655655655655656	0.0732465929871337\\
-0.653653653653654	0.0739615069512133\\
-0.651651651651652	0.0747182713542594\\
-0.64964964964965	0.075517910099194\\
-0.647647647647648	0.0763608589040795\\
-0.645645645645646	0.0772461859705098\\
-0.643643643643644	0.0781766413828143\\
-0.641641641641642	0.079149885698817\\
-0.63963963963964	0.0801653767523278\\
-0.637637637637638	0.0812236417755464\\
-0.635635635635636	0.0823216765663823\\
-0.633633633633634	0.0834603633047984\\
-0.631631631631632	0.0846390553984585\\
-0.62962962962963	0.0858556341814815\\
-0.627627627627628	0.0871081010523688\\
-0.625625625625626	0.0883943217000053\\
-0.623623623623624	0.0897128370809626\\
-0.621621621621622	0.0910610303019785\\
-0.61961961961962	0.0924385143254587\\
-0.617617617617618	0.0938394786101927\\
-0.615615615615616	0.0952653164601579\\
-0.613613613613614	0.096714062976969\\
-0.611611611611612	0.0981834085900559\\
-0.60960960960961	0.0996680802290178\\
-0.607607607607608	0.101167837928425\\
-0.605605605605606	0.102683523630366\\
-0.603603603603604	0.104205835216802\\
-0.601601601601602	0.105738263173766\\
-0.5995995995996	0.107276389214683\\
-0.597597597597598	0.108820742047695\\
-0.595595595595596	0.110367169752094\\
-0.593593593593594	0.111913747112028\\
-0.591591591591592	0.113459268196198\\
-0.58958958958959	0.115004911903656\\
-0.587587587587588	0.116543593139031\\
-0.585585585585586	0.118076873813881\\
-0.583583583583584	0.119603225640568\\
-0.581581581581582	0.121121681609916\\
-0.57957957957958	0.122630635925631\\
-0.577577577577578	0.124127559227558\\
-0.575575575575576	0.12561310141591\\
-0.573573573573574	0.127086079550173\\
-0.571571571571572	0.128545083849694\\
-0.56956956956957	0.129989317789872\\
-0.567567567567568	0.131417726419309\\
-0.565565565565566	0.132830684976521\\
-0.563563563563564	0.134224164006524\\
-0.561561561561562	0.135598582923838\\
-0.559559559559559	0.136956305896358\\
-0.557557557557558	0.13829313562498\\
-0.555555555555556	0.139608288804253\\
-0.553553553553554	0.140902480690471\\
-0.551551551551552	0.142177082124541\\
-0.549549549549549	0.143431001572242\\
-0.547547547547548	0.144663038333643\\
-0.545545545545546	0.145868294266541\\
-0.543543543543544	0.1470530863009\\
-0.541541541541542	0.148214452792027\\
-0.539539539539539	0.149354407777268\\
-0.537537537537538	0.150472431830358\\
-0.535535535535536	0.151566062294132\\
-0.533533533533534	0.152640163031748\\
-0.531531531531532	0.15369439659085\\
-0.529529529529529	0.154730569821363\\
-0.527527527527528	0.155751318033611\\
-0.525525525525526	0.156757474061607\\
-0.523523523523524	0.157749175000695\\
-0.521521521521522	0.158730716264889\\
-0.519519519519519	0.159701930221492\\
-0.517517517517518	0.160668415200039\\
-0.515515515515516	0.161630504483349\\
-0.513513513513513	0.162597382524305\\
-0.511511511511512	0.163566819722922\\
-0.509509509509509	0.164543783935783\\
-0.507507507507508	0.165531733855306\\
-0.505505505505506	0.166537128083748\\
-0.503503503503503	0.167558084244223\\
-0.501501501501502	0.16860756999575\\
-0.499499499499499	0.169684248206412\\
-0.497497497497497	0.170793124219057\\
-0.495495495495496	0.171937453651524\\
-0.493493493493493	0.173120560676012\\
-0.491491491491492	0.174349395829437\\
-0.489489489489489	0.175630089265482\\
-0.487487487487487	0.176965199374094\\
-0.485485485485486	0.178355110293846\\
-0.483483483483483	0.179804566523671\\
-0.481481481481482	0.181317806625516\\
-0.479479479479479	0.18289604912418\\
-0.477477477477477	0.184545721363135\\
-0.475475475475476	0.186264962374959\\
-0.473473473473473	0.188057678283439\\
-0.471471471471472	0.189926083860848\\
-0.469469469469469	0.191870540978616\\
-0.467467467467467	0.193891355444191\\
-0.465465465465465	0.195992756500612\\
-0.463463463463463	0.198173760473179\\
-0.461461461461461	0.200432242254449\\
-0.459459459459459	0.202768121758628\\
-0.457457457457457	0.205185971000088\\
-0.455455455455455	0.207677522583689\\
-0.453453453453453	0.210247622534062\\
-0.451451451451451	0.212890120963257\\
-0.449449449449449	0.215608434886882\\
-0.447447447447447	0.218397157337963\\
-0.445445445445445	0.221250255790485\\
-0.443443443443443	0.224171892925407\\
-0.441441441441441	0.227156603710174\\
-0.439439439439439	0.230202360584842\\
-0.437437437437437	0.233305746443703\\
-0.435435435435435	0.236462537558669\\
-0.433433433433433	0.239666751180476\\
-0.431431431431431	0.242918301478096\\
-0.429429429429429	0.246210018844486\\
-0.427427427427427	0.249541577389852\\
-0.425425425425425	0.252904498057185\\
-0.423423423423423	0.256291929648062\\
-0.421421421421421	0.259704562576145\\
-0.419419419419419	0.263135556772442\\
-0.417417417417417	0.266580544765288\\
-0.415415415415415	0.270033333135397\\
-0.413413413413413	0.273486755915212\\
-0.411411411411411	0.276940718534317\\
-0.409409409409409	0.280389848489582\\
-0.407407407407407	0.283827729286009\\
-0.405405405405405	0.287250149257121\\
-0.403403403403403	0.290649936410977\\
-0.401401401401401	0.29402440004569\\
-0.399399399399399	0.297370768790576\\
-0.397397397397397	0.300683520509887\\
-0.395395395395395	0.303962610352574\\
-0.393393393393393	0.307204243794307\\
-0.391391391391391	0.310404151350154\\
-0.389389389389389	0.31356265086519\\
-0.387387387387387	0.316679757975652\\
-0.385385385385385	0.319750222604651\\
-0.383383383383383	0.322780669912349\\
-0.381381381381381	0.325762397631345\\
-0.379379379379379	0.328700949729837\\
-0.377377377377377	0.331597187530774\\
-0.375375375375375	0.334454616786845\\
-0.373373373373373	0.33727207766484\\
-0.371371371371371	0.340053594137336\\
-0.369369369369369	0.342801421829095\\
-0.367367367367367	0.345517148761744\\
-0.365365365365365	0.348206626004713\\
-0.363363363363363	0.350868084804481\\
-0.361361361361361	0.3535112734045\\
-0.359359359359359	0.356138544692567\\
-0.357357357357357	0.358758478486989\\
-0.355355355355355	0.361372054748079\\
-0.353353353353353	0.363980825427543\\
-0.351351351351351	0.366589970846388\\
-0.349349349349349	0.369205225050416\\
-0.347347347347347	0.371830392668072\\
-0.345345345345345	0.374468121965392\\
-0.343343343343343	0.377123766113329\\
-0.341341341341341	0.379797334801851\\
-0.339339339339339	0.382492273570077\\
-0.337337337337337	0.385215117538161\\
-0.335335335335335	0.387970081761145\\
-0.333333333333333	0.390759598037626\\
-0.331331331331331	0.393583088016739\\
-0.329329329329329	0.396440240203833\\
-0.327327327327327	0.399335906719431\\
-0.325325325325325	0.402273128525842\\
-0.323323323323323	0.405256638541734\\
-0.321321321321321	0.408288327443448\\
-0.319319319319319	0.411367535538594\\
-0.317317317317317	0.414496798478401\\
-0.315315315315315	0.417678625281071\\
-0.313313313313313	0.420914097009364\\
-0.311311311311311	0.424214277352935\\
-0.309309309309309	0.427568462730762\\
-0.307307307307307	0.430993008124919\\
-0.305305305305305	0.434483021599221\\
-0.303303303303303	0.438038303073923\\
-0.301301301301301	0.441667102253842\\
-0.299299299299299	0.445374423302993\\
-0.297297297297297	0.449160896948592\\
-0.295295295295295	0.453028833965923\\
-0.293293293293293	0.456985542604256\\
-0.291291291291291	0.461036472625091\\
-0.289289289289289	0.465178919467723\\
-0.287287287287287	0.469421690755274\\
-0.285285285285285	0.473767195009811\\
-0.283283283283283	0.478224472258911\\
-0.281281281281281	0.482789114704199\\
-0.279279279279279	0.487462346904402\\
-0.277277277277277	0.492254003605019\\
-0.275275275275275	0.49716479665024\\
-0.273273273273273	0.502197136747694\\
-0.271271271271271	0.507353723478315\\
-0.269269269269269	0.512635134761798\\
-0.267267267267267	0.518043458011369\\
-0.265265265265265	0.523583928014778\\
-0.263263263263263	0.52925227741657\\
-0.261261261261261	0.535049668283181\\
-0.259259259259259	0.540981478936895\\
-0.257257257257257	0.547040407658602\\
-0.255255255255255	0.553230747780954\\
-0.253253253253253	0.559548929284278\\
-0.251251251251251	0.565991815098186\\
-0.249249249249249	0.572557498502347\\
-0.247247247247247	0.579251610568975\\
-0.245245245245245	0.586067889033081\\
-0.243243243243243	0.593000497267884\\
-0.241241241241241	0.600053545711322\\
-0.239239239239239	0.607226568521822\\
-0.237237237237237	0.614512205942969\\
-0.235235235235235	0.621908982176672\\
-0.233233233233233	0.629415713018266\\
-0.231231231231231	0.637040166491896\\
-0.229229229229229	0.644771819244865\\
-0.227227227227227	0.652607640998361\\
-0.225225225225225	0.660556206687067\\
-0.223223223223223	0.668620101978024\\
-0.221221221221221	0.676797150536506\\
-0.219219219219219	0.685089184127229\\
-0.217217217217217	0.693492960439106\\
-0.215215215215215	0.702022724211377\\
-0.213213213213213	0.710669030255122\\
-0.211211211211211	0.71944069786149\\
-0.209209209209209	0.728345975669145\\
-0.207207207207207	0.737395394844398\\
-0.205205205205205	0.746573627204903\\
-0.203203203203203	0.755899469180388\\
-0.201201201201201	0.765379979069674\\
-0.199199199199199	0.775007300256317\\
-0.197197197197197	0.784796310375844\\
-0.195195195195195	0.794752718725231\\
-0.193193193193193	0.804876930713248\\
-0.191191191191191	0.815174988175705\\
-0.189189189189189	0.825653568907694\\
-0.187187187187187	0.836305731402042\\
-0.185185185185185	0.84713632849228\\
-0.183183183183183	0.85815343664758\\
-0.181181181181181	0.869361158733811\\
-0.179179179179179	0.880745778535539\\
-0.177177177177177	0.892315877268783\\
-0.175175175175175	0.904066455239016\\
-0.173173173173173	0.915994498784741\\
-0.171171171171171	0.92809326788346\\
-0.169169169169169	0.940355810512222\\
-0.167167167167167	0.95278208959266\\
-0.165165165165165	0.965360289579694\\
-0.163163163163163	0.978084946327066\\
-0.161161161161161	0.990945393976611\\
-0.159159159159159	1.00393425494476\\
-0.157157157157157	1.01704059633076\\
-0.155155155155155	1.0302553388929\\
-0.153153153153153	1.04357026524985\\
-0.151151151151151	1.05696962471193\\
-0.149149149149149	1.07044349441683\\
-0.147147147147147	1.08397389591986\\
-0.145145145145145	1.09756545698515\\
-0.143143143143143	1.11119846586031\\
-0.141141141141141	1.12486500149244\\
-0.139139139139139	1.13856026416197\\
-0.137137137137137	1.1522674253189\\
-0.135135135135135	1.1659723465163\\
-0.133133133133133	1.17968271705115\\
-0.131131131131131	1.19337989866073\\
-0.129129129129129	1.2070678144861\\
-0.127127127127127	1.22072602947411\\
-0.125125125125125	1.2343562938205\\
-0.123123123123123	1.24795878331572\\
-0.121121121121121	1.2615296223251\\
-0.119119119119119	1.2750650227839\\
-0.117117117117117	1.2885624229223\\
-0.115115115115115	1.30202152260156\\
-0.113113113113113	1.31543832798404\\
-0.111111111111111	1.32882252438891\\
-0.109109109109109	1.34216813056719\\
-0.107107107107107	1.3554735734087\\
-0.105105105105105	1.36875260198103\\
-0.103103103103103	1.38200253392005\\
-0.101101101101101	1.39521825965117\\
-0.0990990990990991	1.40841277849058\\
-0.0970970970970971	1.42158353845667\\
-0.0950950950950951	1.4347321221957\\
-0.0930930930930931	1.44786353781229\\
-0.091091091091091	1.4609757719403\\
-0.0890890890890891	1.47407742041016\\
-0.0870870870870871	1.487168375487\\
-0.0850850850850851	1.50025729413592\\
-0.0830830830830831	1.51334887093268\\
-0.081081081081081	1.52644170970053\\
-0.0790790790790791	1.53954276289525\\
-0.0770770770770771	1.55264404970338\\
-0.075075075075075	1.56575295848275\\
-0.0730730730730731	1.57887338534488\\
-0.071071071071071	1.59201088365294\\
-0.0690690690690691	1.60516352295633\\
-0.0670670670670671	1.6183275098066\\
-0.065065065065065	1.63151821169194\\
-0.0630630630630631	1.64472967265065\\
-0.061061061061061	1.65796193815519\\
-0.0590590590590591	1.67121621892109\\
-0.0570570570570571	1.68448552134988\\
-0.055055055055055	1.69777155381275\\
-0.0530530530530531	1.71107419519686\\
-0.051051051051051	1.72439327495789\\
-0.0490490490490491	1.73773155693499\\
-0.0470470470470471	1.75107446437229\\
-0.045045045045045	1.76441615294229\\
-0.0430430430430431	1.7777577904316\\
-0.041041041041041	1.79108408740615\\
-0.039039039039039	1.80439701538706\\
-0.0370370370370371	1.8176765040683\\
-0.035035035035035	1.83091451457493\\
-0.0330330330330331	1.84410082186439\\
-0.031031031031031	1.85721562694102\\
-0.029029029029029	1.87024749882819\\
-0.027027027027027	1.88317921299184\\
-0.025025025025025	1.89599511241416\\
-0.0230230230230231	1.90867315417794\\
-0.021021021021021	1.92118243028379\\
-0.019019019019019	1.93350509955965\\
-0.017017017017017	1.94561899761237\\
-0.015015015015015	1.9575004206074\\
-0.0130130130130131	1.96912311002792\\
-0.011011011011011	1.98045486247356\\
-0.00900900900900903	1.99146882544689\\
-0.00700700700700696	2.00213907425274\\
-0.005005005005005	2.01243292932064\\
-0.00300300300300305	2.0223188722365\\
-0.00100100100100098	2.03177750641751\\
0.00100100100100109	2.04078285730936\\
0.00300300300300305	2.04929707273452\\
0.005005005005005	2.05729482923857\\
0.00700700700700696	2.06476223150676\\
0.00900900900900892	2.07165745233419\\
0.0110110110110111	2.07797486387043\\
0.0130130130130131	2.08368348952181\\
0.015015015015015	2.08876927398473\\
0.017017017017017	2.09321122830688\\
0.0190190190190189	2.09700651389066\\
0.0210210210210211	2.10013941150462\\
0.0230230230230231	2.10260023660035\\
0.025025025025025	2.10438321385026\\
0.027027027027027	2.10548364425066\\
0.0290290290290289	2.10589974093587\\
0.0310310310310311	2.10563420680804\\
0.0330330330330331	2.10469297881318\\
0.035035035035035	2.1030807586125\\
0.037037037037037	2.10081423631097\\
0.0390390390390389	2.09790617444512\\
0.0410410410410411	2.09435977168705\\
0.0430430430430431	2.09020649027561\\
0.045045045045045	2.08546222580834\\
0.047047047047047	2.08014590148042\\
0.0490490490490489	2.07428220062517\\
0.0510510510510511	2.06789521365738\\
0.0530530530530531	2.0610055407932\\
0.055055055055055	2.05364548162377\\
0.057057057057057	2.04584536943829\\
0.0590590590590589	2.03763436339224\\
0.0610610610610611	2.02903053630512\\
0.0630630630630631	2.02006787814544\\
0.065065065065065	2.010779137992\\
0.067067067067067	2.00119135064971\\
0.0690690690690692	1.99133068804114\\
0.0710710710710711	1.98121996672852\\
0.0730730730730731	1.97089274809688\\
0.075075075075075	1.9603686404071\\
0.077077077077077	1.94967258270857\\
0.0790790790790792	1.93881716846991\\
0.0810810810810811	1.92783943129182\\
0.0830830830830831	1.91674416004572\\
0.0850850850850851	1.90555228159196\\
0.087087087087087	1.89427738352265\\
0.0890890890890892	1.88293224349971\\
0.0910910910910911	1.87152516297904\\
0.0930930930930931	1.8600587685006\\
0.0950950950950951	1.84854855036686\\
0.097097097097097	1.8369959709134\\
0.0990990990990992	1.82539773239751\\
0.101101101101101	1.8137567467885\\
0.103103103103103	1.80206929663478\\
0.105105105105105	1.7903326007822\\
0.107107107107107	1.77853312118485\\
0.109109109109109	1.76666838465101\\
0.111111111111111	1.75473135332398\\
0.113113113113113	1.74271553073286\\
0.115115115115115	1.73061289384732\\
0.117117117117117	1.71841137834276\\
0.119119119119119	1.70609254930006\\
0.121121121121121	1.69365052129765\\
0.123123123123123	1.68107923602015\\
0.125125125125125	1.66836478279949\\
0.127127127127127	1.65549829123447\\
0.129129129129129	1.64247754774229\\
0.131131131131131	1.62929473520328\\
0.133133133133133	1.61594358069968\\
0.135135135135135	1.60241618530068\\
0.137137137137137	1.58870928008948\\
0.139139139139139	1.57482472317334\\
0.141141141141141	1.56076252947172\\
0.143143143143143	1.54652743135689\\
0.145145145145145	1.53211951746783\\
0.147147147147147	1.51755025666337\\
0.149149149149149	1.50282466596931\\
0.151151151151151	1.48794997035154\\
0.153153153153153	1.47294836329964\\
0.155155155155155	1.45781773968016\\
0.157157157157157	1.4425754466401\\
0.159159159159159	1.42723723978284\\
0.161161161161161	1.41181532403573\\
0.163163163163163	1.39632628907156\\
0.165165165165165	1.38079328088996\\
0.167167167167167	1.36521955038854\\
0.169169169169169	1.34962681206988\\
0.171171171171171	1.33403542513322\\
0.173173173173173	1.31844801387765\\
0.175175175175175	1.30289290920269\\
0.177177177177177	1.28737612287848\\
0.179179179179179	1.27190599325292\\
0.181181181181181	1.25649530681935\\
0.183183183183183	1.24115627173039\\
0.185185185185185	1.2258945591561\\
0.187187187187187	1.21072070278027\\
0.189189189189189	1.19563461830558\\
0.191191191191191	1.18064515627976\\
0.193193193193193	1.16574989209154\\
0.195195195195195	1.15095573144798\\
0.197197197197197	1.13624900441802\\
0.199199199199199	1.12164402429342\\
0.201201201201201	1.10713476534985\\
0.203203203203203	1.09271146602447\\
0.205205205205205	1.07837624463671\\
0.207207207207207	1.06412579296261\\
0.209209209209209	1.04994687221651\\
0.211211211211211	1.03583857764205\\
0.213213213213213	1.02180118767672\\
0.215215215215215	1.00781611080532\\
0.217217217217217	0.993888012626969\\
0.219219219219219	0.980011468782373\\
0.221221221221221	0.966175614849287\\
0.223223223223223	0.952385287245659\\
0.225225225225225	0.938647645151563\\
0.227227227227227	0.924952768676147\\
0.229229229229229	0.911305152799994\\
0.231231231231231	0.89769625694978\\
0.233233233233233	0.884147746275472\\
0.235235235235235	0.870651408693562\\
0.237237237237237	0.857216563263416\\
0.239239239239239	0.843853881481244\\
0.241241241241241	0.830573107311293\\
0.243243243243243	0.817384614343238\\
0.245245245245245	0.804304618574078\\
0.247247247247247	0.791341868998439\\
0.249249249249249	0.778511302304056\\
0.251251251251251	0.765820309350118\\
0.253253253253253	0.753290424805655\\
0.255255255255255	0.740936242028176\\
0.257257257257257	0.728771523251489\\
0.259259259259259	0.716810013150404\\
0.261261261261261	0.705068371329135\\
0.263263263263263	0.693559433556353\\
0.265265265265265	0.68229531147721\\
0.267267267267267	0.671281472539325\\
0.269269269269269	0.660534823279732\\
0.271271271271271	0.650061768060466\\
0.273273273273273	0.639872458038258\\
0.275275275275275	0.629966331487791\\
0.277277277277277	0.620344801133723\\
0.279279279279279	0.61101476267477\\
0.281281281281281	0.601971073656769\\
0.283283283283283	0.59321855709447\\
0.285285285285285	0.584749356107148\\
0.287287287287287	0.576548771765254\\
0.289289289289289	0.568608738947063\\
0.291291291291291	0.560931356884525\\
0.293293293293293	0.553502684607868\\
0.295295295295295	0.546303565147986\\
0.297297297297297	0.539323306434325\\
0.299299299299299	0.532551049033768\\
0.301301301301301	0.525962599841733\\
0.303303303303303	0.519541048150689\\
0.305305305305305	0.513277441701916\\
0.307307307307307	0.507151441453202\\
0.309309309309309	0.501147915983471\\
0.311311311311311	0.495252236539665\\
0.313313313313313	0.489446862979583\\
0.315315315315315	0.483725212849716\\
0.317317317317317	0.478066674956793\\
0.319319319319319	0.47245996560195\\
0.321321321321321	0.466891581337796\\
0.323323323323323	0.461357538931448\\
0.325325325325325	0.455855785346223\\
0.327327327327327	0.450365942899255\\
0.329329329329329	0.444900303779351\\
0.331331331331331	0.439444316995837\\
0.333333333333333	0.43399692898193\\
0.335335335335335	0.428569765010745\\
0.337337337337337	0.423153666616237\\
0.339339339339339	0.417749577078713\\
0.341341341341341	0.412372177976452\\
0.343343343343343	0.40702221600716\\
0.345345345345345	0.401704172567323\\
0.347347347347347	0.396432663179916\\
0.349349349349349	0.391206184751471\\
0.351351351351351	0.386043975930319\\
0.353353353353353	0.380942365842427\\
0.355355355355355	0.375929212420554\\
0.357357357357357	0.371002737131324\\
0.359359359359359	0.366173013188988\\
0.361361361361361	0.36144970508112\\
0.363363363363363	0.356840970562368\\
0.365365365365365	0.352360210601907\\
0.367367367367367	0.348010911452435\\
0.369369369369369	0.343801647010567\\
0.371371371371371	0.339739647388499\\
0.373373373373373	0.335831765803852\\
0.375375375375375	0.332080864086892\\
0.377377377377377	0.328488901593394\\
0.379379379379379	0.325062863451797\\
0.381381381381381	0.321806371557579\\
0.383383383383383	0.318709878518922\\
0.385385385385385	0.315780983221652\\
0.387387387387387	0.313012563603413\\
0.389389389389389	0.310408756387336\\
0.391391391391391	0.30795723379593\\
0.393393393393393	0.305659851576819\\
0.395395395395395	0.303507256120659\\
0.397397397397397	0.301489310872233\\
0.399399399399399	0.299604396507066\\
0.401401401401401	0.297840114478635\\
0.403403403403403	0.29618675241221\\
0.405405405405405	0.294633803831876\\
0.407407407407407	0.293177735613158\\
0.409409409409409	0.291800389622189\\
0.411411411411411	0.290491267173312\\
0.413413413413413	0.289238568860982\\
0.415415415415415	0.288024844086811\\
0.417417417417417	0.286845806823061\\
0.419419419419419	0.285688298116394\\
0.421421421421421	0.284539190276618\\
0.423423423423423	0.283388670236935\\
0.425425425425425	0.282224633919363\\
0.427427427427427	0.281031484687309\\
0.429429429429429	0.279809591547434\\
0.431431431431431	0.2785401874521\\
0.433433433433433	0.277220583410486\\
0.435435435435436	0.275839452349189\\
0.437437437437437	0.274391912661055\\
0.439439439439439	0.272875002400398\\
0.441441441441441	0.271280371357981\\
0.443443443443444	0.269602824664687\\
0.445445445445446	0.267843462252501\\
0.447447447447447	0.26599828261864\\
0.449449449449449	0.264069224172658\\
0.451451451451451	0.26205529734101\\
0.453453453453454	0.259958457087601\\
0.455455455455456	0.25777986412932\\
0.457457457457457	0.255524832548887\\
0.459459459459459	0.253196979424597\\
0.461461461461461	0.25080161859064\\
0.463463463463464	0.248345419927662\\
0.465465465465466	0.245833913786641\\
0.467467467467467	0.243276686312771\\
0.469469469469469	0.240678516042926\\
0.471471471471471	0.23804759324727\\
0.473473473473474	0.23538933718556\\
0.475475475475476	0.232714115063004\\
0.477477477477477	0.230028056022307\\
0.479479479479479	0.227338103699138\\
0.481481481481481	0.224655272718707\\
0.483483483483484	0.221981367696254\\
0.485485485485486	0.219325278806483\\
0.487487487487487	0.216695111828491\\
0.489489489489489	0.214090321547553\\
0.491491491491491	0.21152069209992\\
0.493493493493494	0.208988018604803\\
0.495495495495496	0.206493888099413\\
0.497497497497497	0.204042211407026\\
0.499499499499499	0.201636526980864\\
0.501501501501501	0.199277731797974\\
0.503503503503504	0.196969350252804\\
0.505505505505506	0.194705929921074\\
0.507507507507508	0.192489367617896\\
0.509509509509509	0.190318759002177\\
0.511511511511511	0.18819004125612\\
0.513513513513514	0.186103952477866\\
0.515515515515516	0.184055481920752\\
0.517517517517518	0.182043716450731\\
0.519519519519519	0.180065351504094\\
0.521521521521521	0.178116683876636\\
0.523523523523524	0.176196364047153\\
0.525525525525526	0.174300621374283\\
0.527527527527528	0.172424958947364\\
0.529529529529529	0.170564523960259\\
0.531531531531531	0.168717094925772\\
0.533533533533534	0.166884817772334\\
0.535535535535536	0.165060843606351\\
0.537537537537538	0.163245189725662\\
0.539539539539539	0.161433371387334\\
0.541541541541541	0.159625299709154\\
0.543543543543544	0.157820025457296\\
0.545545545545546	0.156018419459406\\
0.547547547547548	0.154217346192932\\
0.549549549549549	0.152417346761034\\
0.551551551551551	0.150617258900688\\
0.553553553553554	0.14882529868246\\
0.555555555555556	0.147035041758097\\
0.557557557557558	0.14525152635009\\
0.559559559559559	0.143476192276452\\
0.561561561561561	0.141710168289196\\
0.563563563563564	0.139955420422103\\
0.565565565565566	0.13821489729709\\
0.567567567567568	0.136493318450472\\
0.56956956956957	0.134792815466778\\
0.571571571571572	0.133116701889664\\
0.573573573573574	0.131467176968336\\
0.575575575575576	0.129847093498634\\
0.577577577577578	0.128260981514853\\
0.57957957957958	0.126712362739071\\
0.581581581581582	0.125201391641243\\
0.583583583583584	0.123732695205141\\
0.585585585585586	0.122305677466582\\
0.587587587587588	0.120927303516712\\
0.58958958958959	0.119601262618352\\
0.591591591591592	0.118324754415521\\
0.593593593593594	0.117101619657257\\
0.595595595595596	0.115930901200485\\
0.597597597597598	0.114819011393811\\
0.5995995995996	0.113763621626374\\
0.601601601601602	0.112764006641538\\
0.603603603603604	0.111822531295884\\
0.605605605605606	0.110937030736011\\
0.607607607607608	0.110112012401852\\
0.60960960960961	0.109341708300046\\
0.611611611611612	0.108624539846276\\
0.613613613613614	0.107963130299682\\
0.615615615615616	0.107353165746926\\
0.617617617617618	0.106795872600474\\
0.61961961961962	0.106287008408011\\
0.621621621621622	0.105827341519599\\
0.623623623623624	0.105410404996596\\
0.625625625625626	0.105036724886686\\
0.627627627627628	0.104704107691902\\
0.62962962962963	0.104407761333835\\
0.631631631631632	0.104146749207066\\
0.633633633633634	0.103921416706111\\
0.635635635635636	0.103726483253262\\
0.637637637637638	0.103556909725835\\
0.63963963963964	0.103414167684833\\
0.641641641641642	0.103293713522117\\
0.643643643643644	0.103193332410392\\
0.645645645645646	0.103107488440928\\
0.647647647647648	0.103036927364253\\
0.64964964964965	0.102978113081091\\
0.651651651651652	0.102932265006334\\
0.653653653653654	0.102893793525225\\
0.655655655655656	0.102860511263802\\
0.657657657657658	0.102829617823515\\
0.65965965965966	0.102803943405663\\
0.661661661661662	0.102774686025068\\
0.663663663663664	0.102746691118269\\
0.665665665665666	0.102715728871495\\
0.667667667667668	0.102678385577672\\
0.66966966966967	0.102633901317659\\
0.671671671671672	0.102582324726891\\
0.673673673673674	0.102520797532508\\
0.675675675675676	0.102449016260861\\
0.677677677677678	0.10236520146129\\
0.67967967967968	0.102269714600707\\
0.681681681681682	0.102159524455648\\
0.683683683683684	0.102034354045109\\
0.685685685685686	0.101893304431643\\
0.687687687687688	0.10173627284372\\
0.68968968968969	0.101563935911097\\
0.691691691691692	0.101373650868561\\
0.693693693693694	0.101166958866793\\
0.695695695695696	0.100941474589223\\
0.697697697697698	0.100699477517954\\
0.6996996996997	0.100439381176631\\
0.701701701701702	0.100162408871169\\
0.703703703703704	0.099866044717114\\
0.705705705705706	0.0995532502656386\\
0.707707707707708	0.0992248862785218\\
0.70970970970971	0.0988810091593494\\
0.711711711711712	0.0985200937205646\\
0.713713713713714	0.0981461066925813\\
0.715715715715716	0.0977590297713617\\
0.717717717717718	0.097359094778235\\
0.71971971971972	0.096947293312983\\
0.721721721721722	0.0965249462849927\\
0.723723723723724	0.0960930328361839\\
0.725725725725726	0.0956528063569869\\
0.727727727727728	0.0952077494250028\\
0.72972972972973	0.0947569052336415\\
0.731731731731732	0.0943015353403438\\
0.733733733733734	0.093844485042633\\
0.735735735735736	0.0933852285533621\\
0.737737737737738	0.092927868452808\\
0.73973973973974	0.092469160520802\\
0.741741741741742	0.092009905796299\\
0.743743743743744	0.0915552776750648\\
0.745745745745746	0.0911024766464747\\
0.747747747747748	0.0906532071091678\\
0.74974974974975	0.090206219669197\\
0.751751751751752	0.089763693822909\\
0.753753753753754	0.0893250906283442\\
0.755755755755756	0.0888879685935151\\
0.757757757757758	0.0884543356568723\\
0.75975975975976	0.0880228229035329\\
0.761761761761762	0.0875944936216334\\
0.763763763763764	0.0871646747582478\\
0.765765765765766	0.0867352216394899\\
0.767767767767768	0.0863033759738999\\
0.76976976976977	0.085868989863522\\
0.771771771771772	0.0854306722068474\\
0.773773773773774	0.0849884563330736\\
0.775775775775776	0.084539729036393\\
0.777777777777778	0.0840853730586939\\
0.77977977977978	0.0836200548017557\\
0.781781781781782	0.0831455112007377\\
0.783783783783784	0.082660877506217\\
0.785785785785786	0.0821657631851198\\
0.787787787787788	0.0816592886075108\\
0.78978978978979	0.0811440845948985\\
0.791791791791792	0.0806180317654016\\
0.793793793793794	0.0800817149841197\\
0.795795795795796	0.0795353926495392\\
0.797797797797798	0.0789799608147719\\
0.7997997997998	0.0784168615181997\\
0.801801801801802	0.0778488449136554\\
0.803803803803804	0.0772758175716975\\
0.805805805805806	0.0766994149636846\\
0.807807807807808	0.0761250203842799\\
0.80980980980981	0.0755538783903849\\
0.811811811811812	0.0749878662533603\\
0.813813813813814	0.0744322470463006\\
0.815815815815816	0.0738893997845353\\
0.817817817817818	0.0733613327051436\\
0.81981981981982	0.072852665780026\\
0.821821821821822	0.0723647382021479\\
0.823823823823824	0.071903565518595\\
0.825825825825826	0.0714717904357022\\
0.827827827827828	0.0710685154520268\\
0.82982982982983	0.0707013427766082\\
0.831831831831832	0.0703691707084649\\
0.833833833833834	0.070077706221503\\
0.835835835835836	0.0698262922982058\\
0.837837837837838	0.0696161162041799\\
0.83983983983984	0.0694482627270777\\
0.841841841841842	0.069324985575691\\
0.843843843843844	0.0692449449597408\\
0.845845845845846	0.0692079924759238\\
0.847847847847848	0.0692144365518661\\
0.84984984984985	0.0692613386709485\\
0.851851851851852	0.0693475325467696\\
0.853853853853854	0.0694706383849871\\
0.855855855855856	0.0696269409857757\\
0.857857857857858	0.0698169133924296\\
0.85985985985986	0.0700338892932146\\
0.861861861861862	0.0702747189380033\\
0.863863863863864	0.0705363284777844\\
0.865865865865866	0.0708127730671253\\
0.867867867867868	0.0710998770350421\\
0.86986986986987	0.0713922056062783\\
0.871871871871872	0.0716845372781304\\
0.873873873873874	0.0719709766429341\\
0.875875875875876	0.0722497212446898\\
0.877877877877878	0.0725133452821637\\
0.87987987987988	0.07275804624349\\
0.881881881881882	0.0729781553882804\\
0.883883883883884	0.0731703695215782\\
0.885885885885886	0.0733292809065421\\
0.887887887887888	0.0734508987412269\\
0.88988988988989	0.0735309994063226\\
0.891891891891892	0.0735667744004129\\
0.893893893893894	0.0735553347090249\\
0.895895895895896	0.0734945178922712\\
0.897897897897898	0.0733833155383523\\
0.8998998998999	0.0732194291895472\\
0.901901901901902	0.0730005243112017\\
0.903903903903904	0.0727267420517282\\
0.905905905905906	0.0723982810639902\\
0.907907907907908	0.0720158817657061\\
0.90990990990991	0.0715797466970451\\
0.911911911911912	0.071091736912381\\
0.913913913913914	0.0705529954979208\\
0.915915915915916	0.0699651150286491\\
0.917917917917918	0.0693297149792865\\
0.91991991991992	0.0686504775365394\\
0.921921921921922	0.067931216239518\\
0.923923923923924	0.0671731219341752\\
0.925925925925926	0.0663809733824672\\
0.927927927927928	0.0655583610811095\\
0.92992992992993	0.0647075834669043\\
0.931931931931932	0.0638331331443901\\
0.933933933933934	0.0629394078581631\\
0.935935935935936	0.0620288610837103\\
0.937937937937938	0.0611062235743172\\
0.93993993993994	0.0601755899583267\\
0.941941941941942	0.0592382255819648\\
0.943943943943944	0.0583013566696062\\
0.945945945945946	0.0573649931894691\\
0.947947947947948	0.0564346119549802\\
0.94994994994995	0.0555126963734129\\
0.951951951951952	0.0546011849551667\\
0.953953953953954	0.0537020109048909\\
0.955955955955956	0.0528197339005619\\
0.957957957957958	0.0519554986787737\\
0.95995995995996	0.0511115344656697\\
0.961961961961962	0.0502891470634044\\
0.963963963963964	0.0494922047934337\\
0.965965965965966	0.0487194922910797\\
0.967967967967968	0.0479745665111312\\
0.96996996996997	0.0472573819171111\\
0.971971971971972	0.0465698019376811\\
0.973973973973974	0.0459125130375706\\
0.975975975975976	0.0452848758177389\\
0.977977977977978	0.0446897600991632\\
0.97997997997998	0.0441264837836858\\
0.981981981981982	0.0435956693364676\\
0.983983983983984	0.0430970637200885\\
0.985985985985986	0.0426317995479973\\
0.987987987987988	0.0421987225441581\\
0.98998998998999	0.041798932619519\\
0.991991991991992	0.0414335203941847\\
0.993993993993994	0.0411000660168559\\
0.995995995995996	0.0408005532603435\\
0.997997997997998	0.0405346595336287\\
1	0.0403000729628079\\
};
\end{axis}
\end{tikzpicture}%